\documentstyle[12pt,epsf,epsfig]{article}
\def\bra{\langle}
\def\ket{\rangle}
\def\ketc{\rangle^\ast}
\def\bq{\begin{equation}}
\def\eq{\end{equation}}
\def\ba{\begin{eqnarray}}
\def\ea{\end{eqnarray}}
\def\O{{\cal{O}}}
\def\smin{s_{\mbox{\scriptsize min}}}
\def\lsimfig{\mathrel{\raise.2ex\hbox{$<$}\hskip-.8em\lower.9ex\hbox{$\sim$}}}
\def\gsimfig{\mathrel{\raise.2ex\hbox{$>$}\hskip-.8em\lower.9ex\hbox{$\sim$}}}

\def\mboxsc#1{\mbox{\scriptsize #1}}
\def\Lms#1{\Lambda_{\overline{MS}}^{(#1)}}
\def\docuname{{\large \sc  mepjet}}                           
  
\newcommand{\oas}{\mbox{$\mbox{$\cal{O}$}(\alpha_{s})$}}
\newcommand{\oasz}{\mbox{$\mbox{$\cal{O}$}(\alpha_{s}^{2})$}}
\newcommand{\as}{\mbox{$\alpha_{s}$}}

\newcommand{\version}{\mbox{2.1}}

\newcounter{bean}
\begin{document}
\thispagestyle{empty}
\vspace*{-2mm}
\noindent
\pagenumbering{roman}
\begin{center}
\begin{Large}
\begin{bf}
Theory of Jets in Deep Inelastic Scattering\footnote{
Habilitationsschrift, 
Universit\"at Karlsruhe, Germany, 1997\\
TTP97-39, October 1997.}\\[2cm]
\end{bf}
\end{Large}
\vspace{0.8cm}
\begin{large}
Erwin Mirkes\\[5mm]
\end{large}
{\it 
Institut f\"ur Theoretische Teilchenphysik, Universit\"at Karlsruhe,\\
D-76128 Karlsruhe, Germany
}\\[2cm]
{\bf Abstract}
\end{center}
\begin{quotation}
\noindent
The large center of mass energy and increasing statistical
precision for a wide range of hadronic
final state observables  at the HERA lepton-proton collider
has provided a detailed testing ground for QCD  dynamics.
Fully flexible next-to-leading order calculations are mandatory on the
theoretical side for such tests and will be discussed in detail.
Next-to-leading order QCD predictions for one- and two-jet cross sections 
in deep inelastic  scattering  with complete neutral 
current ($\gamma^\ast$ and/or $Z$) and charged current ($W^\pm$) exchange
together with leading order results for three- and four-jet final states 
are presented. The theoretical framework, based on the phase space 
slicing method and the use of universal crossing functions, is described 
in detail. All analytical formulae necessary for the next-to-leading order 
calculations are provided. The  numerical results are based on the
fully differential $ep \rightarrow n$ jets event generator \docuname\ 
which allows to analyze any infrared and collinear safe observable and 
general cuts in terms of parton 4-momenta. The importance of higher order 
corrections is studied for various jet algorithms. Implications 
and comparisons with (ongoing) experimental analyses for 
jet cross sections at high $Q^2$, the determination of $\alpha_s(\mu_R)$, 
the gluon density,  power corrections in event shapes and the associated 
forward jet production in the low $x$ regime at HERA are discussed.
A study of jet cross sections in polarized electron and polarized 
proton collisions shows that dijet events provide a good measurement of the
polarized gluon distribution $\Delta g(x_g)$, in a region, 
where $x_g \Delta g(x_g)$ is expected to show a maximum.
\end{quotation}
\newpage
\thispagestyle{empty}
\vspace*{4.8cm}
\begin{center}{\large \hspace{-2cm} To Caroline}\end{center}
\newpage
\tableofcontents
\newpage
\pagenumbering{arabic}
\setcounter{page}{1}
\section{Introduction}
\label{intro}
Deep inelastic lepton-nucleon scattering has played
an important role in our present understanding
of the structure of matter. Early fixed target
experiments \cite{taylor} 
have established the partonic structure of the nucleon
and contributed essentially to the development of Quantum Chromodynamics (QCD),
the theory that describes the strong interaction of quarks and
gluons, collectively known as partons. The start-up of the HERA 
lepton-proton collider in 1992 with a center of mass energy  
of $\sqrt{s}\approx 300$ GeV (27.5 GeV positrons on 820 GeV protons) 
marked the beginning of a new 
era of experiments exploring deep inelastic scattering (DIS).
While the traditional method of obtaining information
on the parton structure of the nucleon is through measurements of
inclusive (w.r.t. the hadronic final state) structure
functions, increasingly precise data from HERA 
for a wide range of hadronic final state variables
became availabe and provide a  considerably more detailed testing ground 
for the strong interaction.

The physics of the hadronic final state in DIS,
and in particular the study of multi-jet events and
event shapes,  has in fact become
one of the main interests at HERA.
Topics to be studied include
the measurement of the strong coupling constant from jet rates
and event shapes,
the measurement of the gluon density from dijet events,
the study of power-suppressed corrections to event shapes and
the search for new physics at small $x$ in 
associated forward jet production
and in 1-jet\footnote{In the following 
the jet due to the beam remnant is not 
included in the number of jets.}   inclusive events at very high $Q^2$.

On the theoretical side, versatile
next-to-leading order (NLO) QCD calculations are
mandatory  for these studies.
Based on  significant recent theoretical  progress,
such a fully flexible NLO calculation became available
for arbitrary infrared-safe observables (1- and 2-jet-like
quantities) in DIS \cite{plb1,krakau}.
In general, such calculations are highly non-trivial.
The main theoretical problem is the occurance
of severe infrared and/or collinear divergencies.
However, 
any physical hadronic final state observable must be
infrared safe and either collinear safe or collinear factorizable,
{\it i.e.} the singularities ultimately cancel or are factorized 
into process-independent physical parton distribution or fragmentation
functions.
In this review, we  present  a detailed description of the
theoretical framework for the calculation
together with a comparison of the theoretical expectations with
a variety of recent experimental results at HERA.


Why is it essential to work at least to NLO
to make quantitative predictions in perturbative QCD,
and in particular in jet physics?
Leading order (LO)  calculations rely on tree level matrix elements
and therefore provide only  a basic description
of cross sections and distributions, but are sensitive to
potentially large, but uncalculated ultraviolet and infrared logarithms.
In a NLO calculation, the virtual corrections 
(initial state collinear factorization in the
bremsstrahlung contributions) introduce an explicit logarithmic dependence 
on the renormalization scale $\mu_R$ (factorization scale $\mu_F$),
which substantially reduces the $\mu_R$ ($\mu_F$)
scale dependence  present in the LO calculation.
In addition, infrared logarithms from the NLO bremsstrahlung contributions
introduce  an explicit logarithmic
dependence on the jet-defining parameters through the 
presence of real radiation inside a jet or soft real radiation outside a jet.

NLO corrections in jet physics imply furthermore that a jet 
(in a given jet definition scheme) may consist of two partons.
Thus first sensitivity to the partition of particles
in a jet and therefore the internal jet structure is obtained, 
such as the dependence on the cone size or on recombination prescriptions.
Such studies  give interesting information
about the process by which hard partons are confined into
jets of hadrons. 

Full NLO corrections for 1-jet and 2-jet 
cross sections
and distributions in DIS  $e^\pm p$ scattering
with complete neutral current ($\gamma^\ast$ and/or $Z$) and
charged current ($W^\pm$) exchange are discussed.
All analytical formulae necessary for the NLO
order QCD calculations are provided.
The numerical results are
based on the $ep \rightarrow n$ jets event generator \docuname\ 
\cite{plb1,krakau}, which also allows for the calculation of LO 3-jet 
and 4-jet differential cross sections including 
$W$ and $Z$ exchange. In addition,
charm and bottom quark mass effects can be taken into account for
LO 1-, 2- and 3-jet results with $\gamma^\ast$ exchange.
The matrix elements are derived by using spinor helicity methods
\cite{HZ,heraii,giele1}.
Thus the full spin structure is kept in the calculation and
so \docuname\ allows for the calculation of all 
possible jet-jet and jet-lepton correlations
as well as for the calculation of jet cross sections
in  polarized electron and polarized proton collisions.

Jet studies on the  experimental (hadronic) and
theoretical (partonic) level require an exact definition
of resolvable jets, which is usually given in terms of
one or more resolution parameters and a recombination
scheme description of how to combine
clusters of particles or jets which do not fulfill the resolution criteria.
A good jet algorithm will have small higher order corrections,
small hadronization corrections and a small recombination scheme
dependence.
An important goal of a versatile NLO calculation is to allow
for an easy implementation of arbitrary jet algorithms 
together with the chosen recombination scheme 
and to impose any kinematical resolution
and acceptance cuts on the final state particles.
This is best achieved by performing all hard phase space
integrals numerically, with a Monte Carlo integration technique,
which in fact allows to analyze
any infrared and collinear safe observable and 
general cuts in terms of parton 4-momenta.

The results for the NLO jet cross sections in this paper
are based on the ``phase space slicing'' 
method ($\smin$-technique)
\cite{giele1,kramerps,jim}
and on the technique of universal crossing functions \cite{giele2}.
An invariant theoretical resolution parameter $\smin$ is 
introduced\footnote{The theoretical cutoff parameter $\smin$ is a completely
unphysical parameter and the numerical results for any infrared safe
observable are insensitive to a reasonable variation for
sufficiently small $\smin$ values.}
to isolate the infrared (soft) as well as collinear 
divergencies associated with the unresolved regions
where at least one pair of partons, 
including initial ones, has $s_{ij}=2p_i\cdot p_j<\smin$.
Integrating analytically over the soft and/or collinear final state 
parton allows to cancel the soft and collinear divergencies
against the corresponding divergencies in the virtual contributions
and to factorize the remaining collinear {\it initial state} divergencies into 
the bare parton densities (see section~\ref{sec_struc} 
for a detailed discussion). 
The integration over the finite resolved phase space with $s_{ij}>\smin$ 
is done numerically, by Monte-Carlo techniques, and, thus,  
the parton 4-momenta are available at each phase space point.
As a result the program is flexible enough to implement 
arbitrary jet algorithms and
kinematical resolution and acceptance cuts. 
Upon adding the resolved and unresolved contributions the dependence
on the resolution parameter $\smin$ disappears 
(in the limit $\smin\rightarrow 0$)
and one obtains the NLO fully differential cross section.

The 1-jet final state is the most basic
high transverse momentum 
event at HERA, with a lepton and a jet back-to-back in the
transverse plane. 
In lowest order ($\alpha_s^0$), the underlying partonic
process, $e^\pm$-quark scattering, is 
completely fixed by the  Standard Model electroweak interactions.
Although the NLO matrix elements for this parton model
process have been known for a long time \cite{aem},
including electroweak exchange \cite{herai},
\docuname\ is currently the only program that allows to calculate
the NLO 1-jet or NLO total inclusive 
cross sections including all electroweak effects, in the
presense of arbitrary acceptance cuts on the final state lepton
(or jet).

One-jet and total cross sections at \oas\ are of particular interest
as a normalizer for jet rates. 
In addition, both HERA experiments have recently reported an excess of 
$e^+p$  neutral current
 1-jet (inclusive) events above Standard Model predictions.
The excess amounts to about a factor two increase in rate
at large values of $Q^2$ ($Q^2>1.5\times 10^4$ GeV$^2$)~\cite{highq2},
and has led to speculations on evidence for new physics
(see {\it e.g.} Ref.~\cite{dieterhighq2} and references therein).
The full 1-loop corrections in this high $Q^2$-region are investigated 
in section~\ref{onejetz}.

The properties of jet events can be studied in much greater depth in 
final states with 2 or more jets.  
In Born approximation, the subprocesses  
$eq \rightarrow eqg$, $e\bar q \rightarrow e\bar qg$
and $eg \rightarrow eq\bar{q}$  contribute to the 2-jet cross section
\cite{georgie,herai}. A large variety of issues can be adressed with 
these processes, and in all cases the availability of NLO corrections 
is necessary to move such studies from the qualitative level 
to precision studies of QCD effects. The list of topics which can
be studied quantitatively, once NLO corrections for the complete 
neutral current and charged current exchange 
processes are available, include the following:\\[1mm]
{\it i) The
determination of  $\alpha_s(\mu_R)$ over a range of scales $\mu_R$
from dijet production:}\\[1mm]
The dijet cross section is proportional to $\alpha_s(\mu_R)$ 
at LO, thus suggesting 
a direct measurement of the strong coupling constant. 
However, the LO calculation leaves the 
renormalization scale $\mu_R$ undetermined.
The NLO corrections substantially reduce the dependence of the cross 
section on the renormalization scale
and thus reliable cross section predictions in terms of $\alpha_s(m_Z)$
are made possible.
An important issue which must be addressed in such 
an $\alpha_s$ determination is the appropriate choice of the
renormalization scale $\mu_R$ and the factorization scale $\mu_F$
in DIS jet production.
The chosen scale should be characteristic for the QCD
portion of the process at hand, 
and this typically is not the momentum transfer to the scattered 
lepton\footnote{Obviously, in the limit of large jet transverse momenta
(with respect to boson-proton direction) and vanishing $Q^2$,
{\it i.e.} in the photoproduction limit, $Q^2$ cannot be chosen
as the hard scattering scale.}. 
The ``natural'' scale choice for $n$-jet production in DIS 
appears to be the average $k_T^B$ of the jets 
in the Breit frame \cite{rheinsberg,rom}, as will be discussed in  
sections~\ref{scalechoice}-\ref{sec_gluon}.\\[1mm]
{\it ii) The measurement of the 
gluon density in the proton (via $e g\to eq\bar q$):}\\[1mm]
The boson gluon fusion subprocess $\mbox{boson}+g\rightarrow
q+\bar{q}$, which enters already
at LO,  dominates the 2-jet cross section
at low and medium Bjorken-$x$ and allows for a direct measurement 
of the gluon density $g(\xi,\mu_F)$. The momentum fraction
$\xi$ is related to $x$ by
$\xi=x(1+m_{jj}^2/Q^2)$, where $m_{jj}$ is the invariant mass of the
produced dijet system.
NLO corrections reduce the factorization scale 
$\mu_F$ dependence in the LO calculation 
due to the initial state collinear factorization,
which introduces a mixture of the quark and gluon densities 
according to the Altarelli-Parisi evolution.
Thus reliable cross section predictions in terms of the
scale dependent  parton distributions are made possible \cite{krakau}.
The ``natural'' choice for $\mu_F$ is again given by the average
$k_T^B$ of the jets in the Breit frame
(see section~\ref{sec_gluon}).\\[1mm]
{\it iii) Power corrections in event shapes:}\\[1mm]
Recent theoretical developments in the understanding
of infrared renormalon contributions, which lead to
deviations from the perturbative calculation,
allow the first steps  towards a direct
comparison of theory and data without invoking hadronization models.
These power corrections, with a characteristic $1/Q$ dependence,
can be calculated for event shape variables \cite{dasgupta} and
differential jet rates.
Comparing data directly with NLO theory, which is augmented by the calculated
power corrections, allows to fix the nonperturbative parameters
which characterize the size of these power corrections \cite{h1event}.
Having constrained the hadronization corrections in this way
allows also for a precise extraction of $\alpha_s$,
which is expected to be less sensitive to hadronization uncertainties
(see section~\ref{eventshape}).\\[1mm]
{\it iv) Probing the full hadronic structure via lepton-hadron
correlations:}\\[1mm]
Lepton-hadron correlations provide a  unique opportunity to
analyze the hadronic production mechanism in more detail than
is possible by analyzing jet production cross sections alone.
In the absence of jet cuts in the laboratory frame,
the full QCD matrix elements predict a typical
azimuthal distribution of the jets of the form 
\begin{equation}
\frac{d\sigma}{d\phi}=A+B\cos\phi+C\cos 2\phi
+D\sin\phi+E\sin 2\phi\; .
\label{dphi1}
\end{equation}
Here $\phi$ denotes the azimuthal angle of the jets around the
virtual boson direction 
(in the Breit or hadronic center of mass frame),
where the lepton plane defines $\phi=0^\circ$.
This angular distribution is determined
by the gauge boson polarization: the coefficients $A,B,C,D,E$
are linearly related  to the nine polarization density 
elements of the exchanged gauge boson and a measurement
of the coefficients reveals first sensitivity to
the non-diagonal   polarization density elements
(see sections~\ref{twojetintro} and \ref{nloeffects}).\\[1mm]
{\it v) Associated forward jet production in the low $x$ regime as a signal 
of BFKL dynamics:}\\[1mm]
Recently, much interest has been focused on the small Bjorken-$x$
region, where one would like to distinguish the
Balitsky-Fadin-Kuraev-Lipatov (BFKL) from the traditional
Altarelli-Parisi  (DGLAP) evolution.
BFKL evolution can be enhanced and DGLAP evolution
suppressed by studying DIS events which contain
an identified jet with large longitudinal momentum
fraction compared to Bjorken-$x$.
BFKL evolution leads to a larger cross section for such
events than the DGLAP evolution. 
A conventional fixed order QCD calculation up to ${\cal O}(\alpha_s^2)$ 
does not yet contain any BFKL resummation and must be 
considered a background for its detection.
Clearly, NLO QCD corrections
for fixed order QCD, with DGLAP evolution, 
are mandatory on the theoretical side in order to establish a 
signal for BFKL evolution in the data \cite{prl,hera_forward}.
A detailed discussion of these fixed order effects is presented in 
section~\ref{sec_forward}.\\[1mm]
{\it vi)  The determination of the polarized gluon structure function
(via $e g\to eq\bar q$) in polarized electron on 
polarized proton scattering:}\\[1mm]
The measurement of the polarized parton densities
and in particular the polarized gluon density would allow 
to discriminate between the different pictures of the proton
spin underlying these parametrizations.
The measurement of the 2-jet final state at the 
HERA collider, in a scenario where both the electron
and  proton beams are polarized,
would allow for a unique determination of the polarized
gluon distribution $\Delta g(\xi,\mu_F)$.
As in the unpolarized case, the polarized gluon distribution
enters the 2-jet cross section at LO  suggesting such a direct measurement.
The size of the expected 2-jet spin asymmetry, which is dominated by the 
polarized gluon initiated subprocess, can reach a few percent
\cite{feltesse} and is thus
much larger than asymmetries based on more inclusive observables.
The prospects for such a measurement are discussed in section~\ref{poljets}.
\\[5mm]
Beyond these studies, which concern the 2-jet final state, 
higher jet multiplicities, {\it i.e.} 3-jet and 4-jet final states in DIS,
provide further interesting laboratories of perturbative QCD.
$\O(\alpha_s^3)$ 4-jet events, for example,  are first sensitive to
BFKL resummation effects (see section~\ref{sec_forward}).
Results for $n$-jet rates ($n=1,2,3,4$)
as a function of $y_{\mboxsc{cut}}$  in the 
JADE  and the $k_T$ jet  algorithms are presented
in section~\ref{sec_34jet}.

The numerical results in this paper can be easily reproduced
by running the \docuname-program, which can be obtained
from the author (see Appendix~\ref{sec_docu}).
A documentation of \docuname\ version
2.1 is given in Appendix~\ref{sec_docu}.

A second fully flexible NLO calculation for dijet production
in the one-photon exchange approximation 
has become available with the {\large \sc disent} program \cite{disent},
which is based on the ``exact subtraction method''
as described in Ref.~\cite{dipol}.
Very recently   a third calculation, {\large \sc disaster}++, has been
provided in Ref.~\cite{disaster}, but is again restricted to the
one-photon approximation.
Comparisons between these calculations  in the 
one-photon approximation yielded  so far
satisfactory agreement \cite{compare,disaster}.

The overall organization of this paper is as follows:

The general structure of NLO $n$-jet cross sections in DIS based on the
crossing function and phase space slicing techniques is described 
in detail in section~\ref{sec_struc},
which also establishes our basic notation.
Some commonly used jet definitions in DIS are introduced in 
section~\ref{sec_jetdef}.
Analytical formulae necessary for the NLO calculations
for 1-jet cross section are presented in section~\ref{sec_onejet}
together with a discussion of numerical results.
Special emphasis is put on the calculation of 
the total inclusive (w.r.t. the hadronic activity)
cross section and its relation to the
1-jet inclusive cross section  when typical acceptance cuts on the
scattered lepton are imposed.
We also show that
choosing $Q$ for the hard scattering scale in
the $k_T$ jet algorithm (see section~\ref{sec_jetdef} for the definition) is 
{\it not} an infrared safe choice in the 1-jet case.

Two-jet cross sections are presented in detail in section~\ref{sec_twojet}.
After a discussion of the exchanged gauge boson  
polarization effects in section~\ref{twojetintro}, 
we present all analytical formulae
necessary for the NLO 2-jet calculations
in sections~\ref{sec_ana_twojet} and \ref{twojetz}.
A careful investigation   of numerical effects in
section~\ref{num2jet} includes 
a discussion of charm and bottom quark mass effects,
single jet mass effects and  recombination scheme dependences,
dijet azimuthal angular decorrelations,
the determination of $\alpha_s(\mu_R)$, 
the gluon density, and  power corrections in event shapes.
Three-jet and 4-jet cross section are presented
in section~\ref{sec_34jet}.
A full ${\cal O}(\alpha_s^2)$ calculation of the inclusive forward
jet cross section is presented and compared to the expected BFKL
cross section in section~\ref{sec_forward}.
Prospects for measuring the polarized gluon density
$\Delta g$ from jets at HERA with polarized electrons and protons
are presented in section~\ref{poljets}.
Conclusions and outlook are given in section~\ref{conclusions}.
Finally, there are three Appendices.
In the first part we provide  tree level and one-loop   
helicity amplitudes for the  2-parton
final state processes in the Weyl-van der Waerden formalism.
The second one explains the crossing function
technique for the simplest example of the NLO 1-jet cross section.
The last part of the appendix contains a documentation
of the \docuname\ program (version 2.1).

\newpage
\section{NLO Jet Cross Sections in DIS}
\label{sec_struc}
\subsection{Introductory Remarks}
\begin{figure}[hbt]
\vspace*{5cm}            
\begin{picture}(0,0)(0,0)
\includegraphics{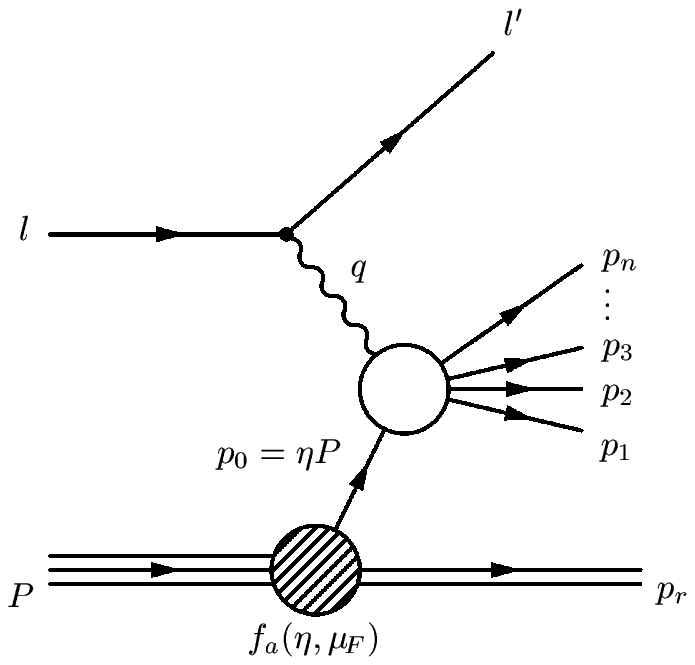}
\end{picture}
\vspace{2cm}
\caption{
DIS $n$ parton production in the parton model.
}
\label{d_proton}
\end{figure}

%
Deep inelastic neutral current (NC) 
lepton proton scattering with several partons
in the final state (see Fig.~\ref{d_proton}),
\begin{equation}
e^\pm(l) + \mbox{proton}(P) \rightarrow  e^\pm(l^\prime)+   
\mbox{proton remnant}(p_r ) +
\mbox{parton} \,\,1 (p_1)
\ldots
+\mbox{parton}\,\, n (p_n)
\label{eq2}
\end{equation}
proceeds via the exchange of an
intermediate vector boson $\gamma^\ast, Z$.
The charged current (CC) reaction with $e^-$ ($e^+$)
on the r.h.s. of Eq.~(\ref{eq2}) replaced by 
$\nu_e$ ($\bar{\nu}_e$) 
is mediated by the exchange of a $W^-$ ($W^+$) boson.
We denote the exchanged boson-momentum by $q$,
its absolute square by $Q^2$, the center of mass energy by $\sqrt{s}$, 
the square of the final hadronic mass by 
$W^2$ and use the standard scaling variables 
$x$ and $y$ (the proton mass is neglected throughout):
\begin{eqnarray} 
q   & = & l-l' \nonumber \\
Q^2 & \equiv & -q^2=xys>0 \nonumber \\
s & = & (P+l)^2 \nonumber \\ 
W^2 & \equiv & P_f^2=(P+q)^2 
\label{variables} \\
x & = & \frac{Q^2}{2Pq} \hspace{1cm} (0<x \le 1) \nonumber \\
y & = & \frac{Pq}{Pl} \hspace{1cm} (0<y \le 1) \nonumber 
\end{eqnarray}
At fixed s, only two variables are independent, since
\begin{displaymath} 
xW^2=(1-x)Q^2,\hspace{1cm}  Q^2=xys. 
\end{displaymath}
%
%
The NLO $n$-jet cross section can be 
described in an intuitively appealing way
in the framework of the  QCD  improved parton model \cite{partonmodel}
\begin{equation}
d\sigma_{\mboxsc{had}}[n\mbox{-jet}] =
\sum_a \int d\eta \,\,f_a(\eta,\mu_F)\,\,\,
\as^n(\mu_R) \,\, \hat{\sigma}_a(p_0=\eta P, \mu_R, \mu_F)
\label{sighaddef}
\end{equation}
where one sums over $a=q,\bar{q},g$. $f_a(\eta,\mu_F)$ 
is the probability density to find a parton $a$ with fraction $\eta$ 
in the proton if the proton is probed at a scale $\mu_F$.
$\hat{\sigma}_a$ denotes the NLO differential  partonic cross section 
with $\as$ set to one from which collinear initial state 
singularities have been factorized out 
at a scale $\mu_F$ and have been
implicitly included in the scale dependent parton densities 
$f_a(\eta,\mu_F)$.
The following tree level and one loop subprocesses contribute to 
$n$-jet production (up to NLO for $n=1,2$ and LO for $n=1,2,3,4$):
\begin{equation}
\begin{array}{lllll}
\mbox{1-jet:} 
& \mbox{LO} 
& \O(\alpha_s^0): 
& e+q\rightarrow e + q 
&  \mbox{}   \\[1mm]
\mbox{1-jet:}
& \mbox{NLO} 
& \O(\alpha_s):
& e+q\rightarrow e + q 
& \hspace{-4cm} \mbox{1-loop corrections}  \\
 \mbox{}
& 
&
& \mbox{+ unresolved contributions from the $\O(\as)$}
& \hspace{-2mm}\mbox{2-parton final states }  \\
\mbox{2-jets:} 
& \mbox{LO} 
& \O(\alpha_s): 
& e+q\rightarrow e + q + g 
&   \\
  \mbox{}
&
& 
& e+g\rightarrow e + q + \bar{q} 
&   \\
  \mbox{2-jets:}
& \mbox{NLO} 
& \O(\alpha_s^2):
& e+q\rightarrow e + q + g
&\hspace{-4cm} \mbox{1-loop corrections}  \\
\mbox{}
&
& 
& e+g\rightarrow e + q + \bar{q} 
&\hspace{-4cm} \mbox{1-loop corrections}  \\
  \mbox{}
&
& 
& \mbox{+ unresolved contributions from the $\O(\alpha_s^2)$}
& \hspace{-2mm}\mbox{3-parton final states }  \\
  \mbox{3-jets:} 
& \mbox{LO} 
& \O(\alpha_s^2): 
& e+q\rightarrow e + q + g + g 
&    \\
  \mbox{}
&
& 
& e+q\rightarrow e + q + q + \bar{q}
&   \\
  \mbox{}
&
& 
& e+g\rightarrow e + q + \bar{q} + g
    \\
  \mbox{4-jets:} 
& \mbox{LO} 
& \O(\alpha_s^3): 
& e+q\rightarrow e + q + g + g + g
&    \\
  \mbox{}
&
& 
& e+q\rightarrow e + q + q + \bar{q} + g
&   \\
  \mbox{}
& 
&
& e+g\rightarrow e + q + \bar{q} + g   + g
& \mbox{}  \\
  \mbox{}
& 
&
& e+g\rightarrow e + q + \bar{q} + q   + \bar{q}
&   \\
\end{array}
\label{processes}
\end{equation}
and the crossing related anti-quark processes with $q\leftrightarrow \bar{q}$.
The results for the NLO $n$-jet cross sections in DIS presented
in this paper are based on the ``phase space slicing'' 
method ($\smin$-technique) \cite{kramerps,giele1,jim}
and on the technique of universal crossing functions \cite{giele2}.
This $\smin$-technique considerably simplifies the structure 
of NLO QCD corrections to hadronic processes and has already been applied 
to the calculation of NLO jet cross sections at LEP and the TEVATRON
\cite{giele2,giele1}. 
The general structure
of a NLO $n$-jet cross section in this framework is 
briefly described in the remaining part of this section.
A  detailed discussion of the structure of NLO
jet cross sections in DIS  is given in section~\ref{genstruc}.

\begin{figure}[t]
  \centering
  \mbox{\epsfig{file=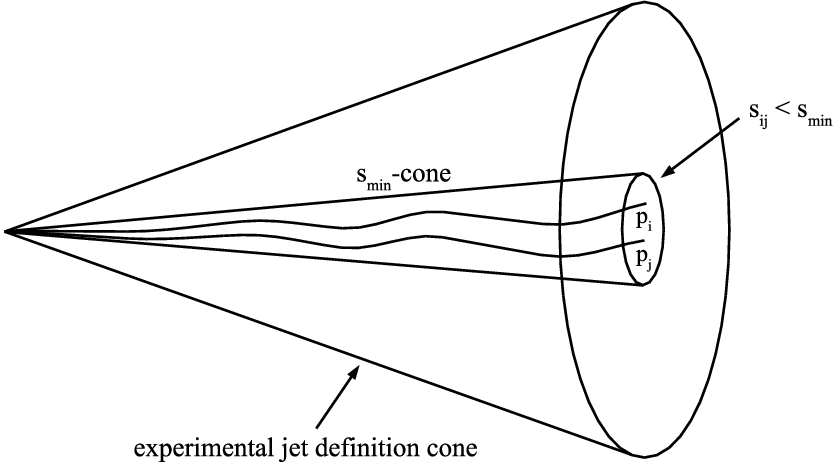,
         bbllx=0,bblly=520,bburx=570,bbury=840,
         width=0.7\linewidth}} 
\vspace*{-3mm}
\caption{
Two unresolved partons according to a physical
jet definition (outer cone).
Infrared and collinear divergencies in the jet definition cone
are isolated by a theoretical ``cone''
defined by the resolution parameter $\protect\smin$.
}\label{fig_cones}
\end{figure}

As listed  in Eq.~(\ref{processes})
a NLO $n$-jet cross section receives
contributions from
1-loop corrections
to $n$-parton final states
and from 
$(n+1)$-parton final states at  tree level.
Both contributions are divergent.
The real $(n+1)$-parton tree level matrix elements 
need to be integrated over the entire phase space
where only $n$ jets are reconstructed according to a given jet 
definition scheme, including the unresolved regions.
The physical situation of two unresolved partons according
to a physical jet definition scheme is shown 
by the ``experimental jet definition cone'' in Fig.~\ref{fig_cones}. 
This outer ``cone'' represents the boundaries
given by any arbitrary jet algorithm 
(or any infrared and collinear safe observable)
including arbitrary experimental cuts.
Infrared as well as collinear divergencies associated with
two partons within this jet definition ``cone'' are further
isolated by introducing a purely  theoretical parton resolution
parameter $\smin$ (shown by the  inner cone in Fig.~\ref{fig_cones}).

Soft and collinear approximations to the $(n+1)$-final state parton 
matrix elements
are used in the phase space region 
inside the $\smin$ cone,
where at least one 
pair of partons, including initial ones, has $s_{ij}=2p_i.p_j<\smin$.
The soft and/or collinear final state parton is then integrated over 
analytically. Factorizing the collinear initial state divergencies into the 
bare parton distribution functions and adding this soft+collinear part to the
virtual contributions for the $n$-parton final state 
gives a finite result for, effectively, $n$-parton final states. 
In general 
this $n$-parton contribution is negative and
grows logarithmically in magnitude as $\smin$ is decreased. This logarithmic 
growth is exactly cancelled by the increase in the $n+1$ parton cross 
section with $s_{ij}>\smin$
({\it i.e.} the region between the two cones in Fig.~\ref{fig_cones}), 
once $\smin$ is small enough for the approximations
made within the $\smin$-cone to be valid.
The integration over the $n+1$-parton phase space with $s_{ij}>\smin$
is done by Monte-Carlo techniques in \docuname\ without using
any approximations. Since, at each phase space point, the parton 4-momenta 
are available, the program is flexible enough to implement 
any jet definition algorithms or to impose arbitrary
kinematical resolution and acceptance cuts on the final state particles
\cite{jim}.

As mentioned before the 
collinear initial state divergencies are factorized into 
the bare parton densities introducing a dependence on the factorization 
scale $\mu_F$. In order to handle these singularities we follow 
Ref.~\cite{giele2} and use the technique of universal ``crossing
functions'' (see section~\ref{crossing}) for the definition and details.
The idea is to start with the result of the NLO calculation
with all partons in the final state,
{\it i.e.} $e^+e^-\rightarrow n+1$ ``jets'', where no such singularities 
occur after adding real and virtual contributions.
This NLO $e^+e^-\rightarrow n+1$ ``jets'' cross section needs only 
been known in the soft and collinear limit, where the
$e^+e^-\rightarrow n+2$ parton matrix elements
have been integrated analytically
over the unresolved region, where one pair of partons has
$s_{ij}<\smin$.
Let us now specify the general structure of the NLO jet cross sections
in DIS within the
framework of the phase space slicing and 
the crossing function technique in full detail.

\subsection{The General Structure of the NLO Jet Cross Section:
\protect\newline
Crossing Functions and Phase Space Slicing Technique
\protect\vspace*{1mm}
}
\label{genstruc}
According to the general discussion in the previous section
the structure of the  $n$-jet cross section up to NLO 
can be summarized as:
\begin{eqnarray}
\sigma_{\mboxsc{had}}^{\mboxsc{NLO}}[n\mbox{-jet}]
&=& 
\sigma_{\mboxsc{had}}^{\mboxsc{LO}}[n\mbox{-jet}] \nonumber \\[2mm]
&+& 
\sigma_{\mboxsc{had}}^{\mboxsc{NLO, virtual+soft+collinear}}[n\mbox{-jet}]
                                               \label{sig_sum} \\[2mm]
&+&
\sigma_{\mboxsc{had}}^{\mboxsc{NLO, hard}}[n\mbox{-jet}]
\nonumber
\end{eqnarray}
where the virtual + soft + collinear piece is given by
\begin{eqnarray}
\sigma_{\mboxsc{had}}^{\mboxsc{NLO, virtual+soft+collinear}}
&=& 
\sigma_{\mboxsc{had}}^{\mboxsc{NLO, final}} \nonumber \\[2mm]
&+&
\sigma_{\mboxsc{had}}^{\mboxsc{NLO, crossing}}
\label{sig_virt}
\end{eqnarray}
We will now specify the individual pieces in 
Eqs.~(\ref{sig_sum},\ref{sig_virt}) for $n$-jet production in DIS
in  detail.
%
%
\subsubsection{The LO Part
$\sigma_{\protect\mboxsc{had}}^{\protect\mboxsc{LO}}$[$n$-jet]
\protect\vspace{1mm}}
At LO each jet is modelled by a single outgoing parton:
\begin{eqnarray}
\sigma_{\mboxsc{had}}^{\mboxsc{LO}}[n\mbox{-jet}] 
&=& \sum_a\int_0^1d\eta
\int d{\mbox{PS}}^{(l^\prime+n)} 
f_a(\eta,\mu_F)\,\,\label{hadlo} \\
&& \alpha_s^{(n-1)}(\mu_R)\,\,
\hat{\sigma}^{\mboxsc{LO}}_{a\rightarrow n\,\, \mboxsc{partons}}
(l+p_0\rightarrow l^\prime+p_1\ldots p_n)\,\,
J_{n\leftarrow n}(\{p_i\})\nonumber
\end{eqnarray}
where $a=q,\bar{q},g$ and
$d\mbox{PS}^{(l^\prime+n)}$
stands for the  $n+1$ particle Lorentz-invariant  phase-space measure
\begin{equation}
d\mbox{PS}^{(l^\prime+n)}=\delta^4(p_0+l-l^\prime-\sum_{i=1}^{n}p_i)
\,2\pi\frac{d^3l^\prime}{2E^\prime}
\prod_{i=1}^{n}\frac{d^3 p_i}{(2\pi)^3\,2E_i}\>.
\label{phasespace}
\end{equation}
$\hat{\sigma}^{\mboxsc{LO}}$ denotes the LO $n$-parton final
state differential cross section with $\as$ set to one.

The jet algorithm $J_{n \leftarrow n}$, which yields one if the
original final state $n$-parton configuration yields $n$ jets satisfying the
experimental cuts can be expressed as a product of a clustering
part and an acceptance part 
(jet momenta are denoted by $k_i$ in the following):
\bq
J_{n\leftarrow n} = 
J_{n\leftarrow n}^{\mboxsc{cluster}}
(p_1,\ldots,p_n\rightarrow k_1,\ldots, k_n)\cdot
J_{n\leftarrow n}^{\mboxsc{accept}}
(k_1,\ldots,k_n\rightarrow k_1,\ldots, k_n)\>.
\label{jnndef}
\eq
$J_{n\leftarrow n}^{\mboxsc{cluster}}$ evaluates to one if 
all final state partons are
well separated according to a given jet algorithm,
and vanishes otherwise:
\bq
J_{n\leftarrow n}^{\mboxsc{cluster}}
(p_1,\ldots,p_n\rightarrow k_1,\ldots, k_n)
= \prod_{i<j;\,1}^n (1-\Theta_{\mboxsc{cluster}}(p_i,p_j))\>.
\eq
The clustering is defined by the $\Theta_{\mboxsc{cluster}}$ function,
\bq
\Theta_{\mboxsc{cluster}}(p_i,p_j)=
\left\{
\begin{array}{ll}
1,\hspace{1cm}& p_i,p_j \hspace{1cm}\mbox{clustered into one jet,} \\
0,            & \mbox{otherwise}\>. \\
\end{array}
\right.
\eq
The jet momenta $k_i$ are functions of the parton momenta $p_i$
defined by the recombination algorithm
\bq
k_{\mboxsc{jet}}=  {p_i \otimes p_j \otimes \ldots}
\eq
or simpler in the $E$ scheme
\bq
k_{\mboxsc{jet}}=\sum_{i\in \mboxsc{jet}} p_i
\eq
Similarly, $J_{n\leftarrow n}^{\mboxsc{accept}}$
evaluates to one, if all final state jets with momenta $k_i$
(=$p_i$ in the LO case) pass all acceptance
and detector resolution criteria,  for example
cuts on the transverse momenta, pseudo-rapidities,$\ldots$
\begin{eqnarray}
J_{n\leftarrow n}^{\mboxsc{accept}}(k_1,\ldots,k_n\rightarrow k_1,\ldots, k_n)
&=& \\
&&\hspace{-2cm} \prod_{i=1}^n\,\Theta(p_T(k_i)>p_{T, \mboxsc{min}})\,\,
  \prod_{i=1}^n\,\Theta(|\eta(k_i)|<\eta_{\mboxsc{max}})\,\,
  \prod_{i=1}^n\,\Theta(\ldots)\,\,\ldots\nonumber
\end{eqnarray}
%
%
\subsubsection{The NLO Resolved Part
$\sigma_{\protect\mboxsc{had}}^{\protect\mboxsc{NLO, hard}}$[$n$-jet]
\protect\vspace{1mm}}
This part calculates the finite  contributions from the
real emission outside the soft and collinear region.
\begin{eqnarray}
\sigma_{\mboxsc{had}}^{\mboxsc{NLO, hard}}[n\mbox{-jet}] 
&=& \sum_a\int_0^1d\eta\,
\int \,d{\mbox{PS}}^{(l^\prime+n+1)}\,\,
f_a(\eta,\mu_F)\,\,
\label{hadlo1}\\
&& \hspace{-2.2cm}\alpha_s^{n}(\mu_R)\,\,
\hat{\sigma}^{\mboxsc{LO}}_{a\rightarrow (n+1)\,\,\mboxsc{partons}}
(l+p_0\rightarrow l^\prime+p_1\ldots p_{n+1})\,\,
\prod_{i<j;\,0}^{n+1}\Theta(|s_{ij}| - \smin)\,\,
J_{n\leftarrow n+1}(\{p_i\})
 \nonumber
\end{eqnarray}
where
$d\mbox{PS}^{(l^\prime+n+1)}$
stands for the  $n+2$ particle  phase-space measure
\begin{equation}
d\mbox{PS}^{(l^\prime+n+1)}=\delta^4(p_0+l-l^\prime-\sum_{i=1}^{n+1}p_i)
\,2\pi\frac{d^3l^\prime}{2E^\prime}
\prod_{i=1}^{n+1}\frac{d^3 p_i}{(2\pi)^3\,2E_i}
\label{phasespace1}
\end{equation}
and $\hat{\sigma}^{\mboxsc{LO}}$ denotes the LO $(n+1)$-parton final
state differential cross section with $\as$ set to one.
The $(n+1)$-parton final state cross section needs to be integrated
over the phase space with $s_{ij}=2p_i.p_j
>\smin$ ($i,j\in\{0,\ldots,n+1\},\, i<j$),
where only $n$ jets are reconstructed according to a given jet 
definition scheme.
The jet algorithm $J_{n\leftarrow n+1}$ in Eq.~(\ref{hadlo1})
evaluates to one if the $(n+1)$-
parton configuration yields $n$ detected jets and vanishs otherwise.
More precisely, $J_{n\leftarrow n+1}(\{p_i\})$ 
evaluates to one either if one pair
of partons is clustered into one jet and the remaining $(n-1)$ partons
are well separated from this jet and pass all acceptance criteria 
(together with the jet) or all $(n+1)$ partons are resolved but one parton 
does not pass the acceptance cut:
\begin{eqnarray}
J_{n\leftarrow n+1} &=& 
J_{n\leftarrow n+1}^{\mboxsc{cluster}}
(p_1,\ldots,p_{n+1}\rightarrow k_1,\ldots, k_n)\cdot
J_{n\leftarrow n}^{\mboxsc{accept}}
(k_1,\ldots,k_n\rightarrow k_1,\ldots, k_n) 
\label{jnn1def} \\
&+&
J_{n+1\leftarrow n+1}^{\mboxsc{cluster}}
(p_1,\ldots,p_{n+1}\rightarrow k_1,\ldots, k_{n+1})\cdot
J_{n\leftarrow n+1}^{\mboxsc{accept}}
(k_1,\ldots,k_{n+1}\rightarrow k_1,\ldots, k_n)
\nonumber
\end{eqnarray}
with
\begin{eqnarray}
J_{n\leftarrow n+1}^{\mboxsc{cluster}}
(p_1,\ldots,p_{n+1}\rightarrow k_1,\ldots, k_n)
&=&  \\
&&\hspace{-6cm} 
\sum_{i<j}^{n+1}\Theta_{\mboxsc{cluster}}(p_i,p_j)
\prod_{l<k; l \neq i,j; k \neq i,j}
(1-\Theta_{\mboxsc{cluster}}(p_l,p_k))
\prod_{l; l \neq i,j}
(1-\Theta_{\mboxsc{cluster}}(p_l,p_i\otimes p_j))
\nonumber
\end{eqnarray}
and
\begin{eqnarray}
J_{n\leftarrow n+1}^{\mboxsc{accept}}(k_1,\ldots,k_{n+1}\rightarrow 
                                      k_1,\ldots, k_n)
&=& \\
&&\hspace{-6cm}
  \sum_i^{n+1} \Theta[(p_T(k_i)<p_{T, \mboxsc{min}})\,\,\,\,\mbox{or}\,\,\,\,
                      (|\eta(k_i)|>\eta_{\mboxsc{max}})\cdot
  \prod_{j\neq i}^n\,\Theta(p_T(k_i)>p_{T, \mboxsc{min}})
                     \Theta(|\eta(k_i)|<\eta_{\mboxsc{max}})
\nonumber
\end{eqnarray}
%
%
\subsubsection{The NLO Unresolved Parts 
$\sigma_{\protect\mboxsc{had}}^{\protect\mboxsc{NLO, final}}$[$n$-jet]
and $\sigma_{\protect\mboxsc{had}}^{\protect\mboxsc{NLO, crossing}}$[$n$-jet]
\protect\vspace{1mm}}
\label{unresolved}
%
%
The sum of these two pieces is the virtual+soft+collinear part of the
NLO $n$-jet cross section in DIS.
The first term 
summarizes the contributions of the integration over the soft and 
{\em final state} collinear regions combined with the corresponding virtual
singularities, where
all soft and collinear poles cancel according to the
Bloch-Nordsiek \cite{bloch} and Kinoshita-Lee-Nauenberg 
\cite{kinoshita} theorems.
Here, the soft and final state collinear regions 
(referred to as a final state cluster in the following)
are defined
by the phase space requirement $s_{ij}=2p_i.p_j<\smin$.
Absorbing the ultraviolet divergencies in the virtual corrections 
(for $n\geq 2$ ) into the bare coupling constant 
according to the modified minimal subtraction
($\overline{\mbox{MS}}$) renormalization scheme \cite{msbar}
yields a finite result. The corresponding finite
NLO partonic cross section can be written in the form
\begin{eqnarray}
\alpha_s^n(\mu_R)
\hat{\sigma}^{\mboxsc{NLO}}_{a\rightarrow n\,\,\mboxsc{partons}}
&\sim& 
\label{finalstruc}\\[1mm]
&&\hspace{-2cm} \alpha_s^{n}(\mu_R)
\left[
|M^{\mboxsc{LO}}_{a\rightarrow n\,\,\mboxsc{partons}}|^2\,
{\cal{K}}_{a\rightarrow n\,\,\mboxsc{partons}}(\smin,s_{ij},\mu_R)
\,+\,
{{\cal{F}}_{a\rightarrow n\,\,\mboxsc{partons}}}  \right]
\nonumber
\end{eqnarray}
where $|M^{\mboxsc{LO}}_{a\rightarrow n\,\,\mboxsc{partons}}|^2$
denotes the LO matrix element squared with $\alpha_s^{(n-1)}$ set to one.
The dynamical ${\cal {K}}-$factor, which multiplies the LO matrix element
squared depends on the resolution parameter $\smin$,
the invariant masses of the hard partons and (for $n\geq 2$) 
on the renormalization scale $\mu_R$.
The function ${\cal{F}}$ in Eq.~(\ref{finalstruc}) denotes the
finite part of the virtual corrections, which does not factorize
the Born term amplitude.
Both functions ${\cal{K}}_{a\rightarrow n\,\,\mboxsc{partons}}$
and ${\cal {F}}_{a\rightarrow n\,\,\mboxsc{partons}}$
can be obtained from the NLO $e^+e^-\rightarrow (n+1)$ partons
result as presented in  \cite{giele1}.
The hadronic cross section for the virtual + soft + {\em final state}
collinear divergencies is thus given by
\begin{eqnarray}
\sigma_{\mboxsc{had}}^{\mboxsc{NLO, final}}[n\mbox{-jet}] 
&=& \sum_a\int_0^1d\eta\,
\int \,d{\mbox{PS}}^{(l^\prime+n)}\,\,
f_a(\eta,\mu_F)\,\, \nonumber \\
&&\alpha_s^{n}(\mu_R)\,\,
\hat{\sigma}^{\mboxsc{NLO}}_{a\rightarrow n\,\,\mboxsc{partons}}
(l+p_0\rightarrow l^\prime+p_1\ldots p_{n}; \smin, \mu_R)\,\,
J_{n\leftarrow n}(\{p_i\})
\label{hadfinal}
\end{eqnarray}
where $\alpha_s^{n}$ is set to one  in the partonic NLO cross section.
The crossing of a final state cluster
to the initial state, which is effectively done by crossing the
function ${\cal{K}}_{a\rightarrow n\,\,\mboxsc{partons}}$ 
in Eq.~({\ref{finalstruc}) from
the corresponding $e^+e^-$ results,  becomes  possible through
the introduction of the crossing functions $C_{a}$ \cite{giele2},
which essentially  contain the convolution of the parton distribution
function with the Altarelli-Parisi kernels.
They also take into account the difference between the
initial state collinear cluster and the final state collinear
cluster together with the factorization
of the initial state mass singularities.
In the case of DIS the crossing function contribution
$\sigma_{\mboxsc{had}}^{\mboxsc{NLO, crossing}}$
has the form:
\begin{eqnarray}
\sigma_{\mboxsc{had}}^{\mboxsc{NLO, crossing}}[n\mbox{-jet}] 
&=& \sum_a\int_0^1d\eta\,
\int \,d{\mbox{PS}}^{(l^\prime+n)}\,\,
\alpha_s(\mu_R)\,\, C_a(\eta,\mu_F,\smin)\,\, \nonumber \\
&& \alpha_s^{n-1}(\mu_R)\,\,
\hat{\sigma}^{\mboxsc{LO}}_{a\rightarrow n\,\,\mboxsc{partons}}
(l+p_0\rightarrow l^\prime+p_1\ldots, p_{n})\,\,
J_{n\leftarrow n}(\{p_i\})
\label{hadcross}
\end{eqnarray}
The explicit dependence on the factorization scale $\mu_F$ is effectively
introduced by the crossing functions $C_a(\eta,\mu_F)$.
$J_{n\leftarrow n}(\{p_i\})$ in Eqs.~(\ref{hadfinal},
\ref{hadcross}) represents again
the jet algorithm and cuts as in Eq.~(\ref{hadlo}) for the LO case.

\subsubsection{Crossing Functions (Unpolarized Case)}
\label{crossing}
Consider the case where an initial
parton $p$  splits into an (unobserved) collinear
parton $u$ with momentum $p_u=(1-z)p_p$ and a parton
$a$ with $p_a=zp_p$ (which participates in the hard scattering):
$p\rightarrow ua$.
The region where parton $u$ is collinear with $p$
is defined by the invariant mass criterion $|s_{pu}|<\smin$.
This configuration is indistinguishable from the leading order
configuration where parton $a$ comes directly from the proton.
After removing the mass singularity at $|s_{pu}|\rightarrow 0$
by mass factorization into the ``bare'' parton densities
the remaining part of the initial state collinear radiation 
in the phase space region $|s_{pu}|<\smin$ is absorbed into
effective parton distribution functions, 
called crossing functions $C_a(x,\mu_F,\smin)$ \cite{giele2}. 
This part of the crossing function is essentially
a convolution of the parton densities with
the Altarelli-Parisi splitting functions $P_{p\rightarrow a}(z)$
and depends on $\smin$, 
the factorization scale $\mu_F$ and on the factorization scheme.

A second contribution to the crossing functions arises from
the crossing of a pair of collinear partons 
$u$ and $p$ (which originates from the splitting $a\rightarrow up$)
from the final state to the initial state,
which is done in the functions 
${\cal{K}}_{a\rightarrow n\,\,\mboxsc{partons}}$
in Eqs.~(\ref{finalstruc},\ref{hadfinal}).
The crossed pair of collinear final state partons $up$
has been integrated over the final state collinear phase space region
defined by the invariant mass criterion $s_{up}<\smin$.
This ``wrong'' contribution, which also depends on $\smin$,
can effectively be subtracted from the parton densities.
In fact, the crossing of the final state collinear pair of partons
$a\rightarrow up$ to the initial state 
corresponds to a two parton incoming state
with invariant mass smaller than $\smin$, which cannot be 
distinguished from a single incoming parton $a$.
The relevant  subtraction from the parton distribution function
$f_a(x,\mu_F)$ has been performed in the corresponding
crossing function $C_a(x,\mu_F,\smin)$ for the initial state parton $a$.

The  crossing functions
for an initial state parton $a$, which participates in the
hard scattering process, can then be written in the form
(for a detailed derivation of the 
unpolarized crossing functions we refer the reader to
\cite{giele2}):

\begin{equation}
C_{a}^{\overline{\mboxsc{MS}}}(x,\mu_F,\smin)=
\left(\frac{N}{2\pi}\right)
\left[ A_{a}(x,\mu_F)\ln\left(\smin/\mu_F\right)
+      B_{a}^{\overline{\mboxsc{MS}}}(x,\mu_F)\right]
\label{crossf}
\end{equation}
with
\begin{equation}
A_a(x,\mu_F) = \sum_p A_{p\rightarrow a}(x,\mu_F)
\end{equation}
\begin{equation}
B_a^{{\overline{\mboxsc{MS}}}}(x,\mu_F) = \sum_p
B_{p\rightarrow a}^{{\overline{\mboxsc{MS}}}}(x,\mu_F)
\end{equation}
$N$ denotes the number of colors.
The sum runs over $p=q,\bar{q},g$.
To be more specific, the crossing functions for valence quarks,
for example $u_v$ quarks reads:
\begin{equation}
C_{u_v}^{\overline{\mboxsc{MS}}}(x,\mu_F,\smin)=
\left(\frac{N}{2\pi}\right)
\bigg[ A_{u_v\rightarrow u_v}(x,\mu_F)\ln\left(\frac{\smin}{\mu_F}\right)
+      B_{u_v\rightarrow u_v}^{\overline{\mboxsc{MS}}}(x,\mu_F)\bigg]
\label{crossf_uv}
\end{equation}
For sea quarks, for example $s$ quarks:
\begin{eqnarray}
C_{s}^{\overline{\mboxsc{MS}}}(x,\mu_F,\smin)&=&
\left(\frac{N}{2\pi}\right)
  \bigg[ \bigg(
          A_{s\rightarrow s}(x,\mu_F)
       +  A_{g\rightarrow s}(x,\mu_F)
         \bigg)
         \ln\left(\frac{\smin}{\mu_F}\right)
\label{crossf_s}\\
&&    \hspace{2cm}
       +    
         \left(
          B_{s\rightarrow s}^{\overline{\mboxsc{MS}}}(x,\mu_F)
       +  B_{g\rightarrow s}^{\overline{\mboxsc{MS}}}(x,\mu_F)
         \right)
         \bigg] \nonumber
\end{eqnarray}
For gluons
\begin{eqnarray}
C_{g}^{\overline{\mboxsc{MS}}}(x,\mu_F,\smin)&=&
\left(\frac{N}{2\pi}\right)
  \bigg[ \bigg(
          A_{g\rightarrow g}(x,\mu_F)
       +  \sum_{i=q,\bar{q}} A_{i\rightarrow g}(x,\mu_F)
         \bigg)
         \ln\left(\frac{\smin}{\mu_F}\right)
\label{crossf_g} \\
&&    \hspace{2cm}
       +    
         \bigg(
         B_{g\rightarrow g}^{\overline{\mboxsc{MS}}}(x,\mu_F)
  +  \sum_{i=q,\bar{q}} B_{i\rightarrow g}^{\overline{\mboxsc{MS}}}(x,\mu_F)
         \bigg)
         \bigg] \nonumber
\end{eqnarray}
where the sum $\sum_{i=q,\bar{q}}$
in Eq.~(\ref{crossf_g}) runs over all $n_f$ quark
(valence and sea) and antiquark flavors.
The functions
$A_{p\rightarrow a}$ and $B_{p\rightarrow a}^{{\overline{\mboxsc{MS}}}}$
are defined via  a one dimensional integration over the parton densities
$f_p$, which also involves the integration over $()_+$ prescriptions.
We have performed this numerical integration in a
separate program, which is provided together with the
\docuname\ program, and the results for 
$A_{p\rightarrow a}$ and $B_{p\rightarrow a}^{{\overline{\mboxsc{MS}}}}$
for different values of $x$ and $\mu_F$ are stored 
in an array in complete analogy to the usual parton densities,
{\em e.g.} in Ref.~\cite{mrsdmp}.
The finite scheme independent functions 
$A_{p\rightarrow a}(x,\mu_F)$ are:
\begin{eqnarray}
A_{g\rightarrow g} &=& \int_x^1\frac{dz}{z}\,\, f_g(x/z,\mu_F)
\left\{\frac{(11N-2n_f)}{6N}\delta(1-z)\right. \nonumber\\
&&\left.
\hspace{2cm}\,+ \, 
2\left( \frac{z}{(1-z)_+}+\frac{(1-z)}{z}+z(1-z)\right)\right\}
\label{aggdef}
\\
A_{q\rightarrow q} &=& \int_x^1 \frac{dz}{z}\,\,f_q(x/z,\mu_F)
\frac{2C_F}{3}\,\left\{\frac{3}{4}\delta(1-z)\,+\,
\frac{1}{2}\left(\frac{1+z^2}{(1-z)_+}\right)\right\}
\label{aqq}
\\
A_{g\rightarrow q} &=& \int_x^1 \frac{dz}{z}\,\, f_g(x/z,\mu_F)
\,\,\,\frac{1}{4}\,\,\, 
\hat{P}^{(4)}_{g\rightarrow q}(z)
\label{agq}
\\
A_{q\rightarrow g} &=& \int_x^1 \frac{dz}{z}\,\, f_q(x/z,\mu_F)
\,\,\,\frac{1}{4}\,\,\,
\hat{P}^{(4)}_{q\rightarrow g}(z)
\end{eqnarray}
and the scheme dependent functions 
$B_{p\rightarrow h}^{{\overline{\mboxsc{MS}}}}(x,\mu_F)$ by
\begin{eqnarray}
B_{g\rightarrow g}^{{\overline{\mboxsc{MS}}}} &=& \int_x^1 \frac{dz}{z}\,\,
f_g(x/z,\mu_F)
\left\{
\left(\frac{\pi^2}{3} -\frac{67}{18}+\frac{5n_f}{9N}\right)\delta(1-z)
+2z\left(\frac{\ln(1-z)}{(1-z)}\right)_+ \right.\nonumber\\
& &\hspace{3.5cm}\left.
+2\left( \frac{(1-z)}{z}+z(1-z)\right)\ln(1-z)\right\}
\label{bggdef}
\\
B_{q\rightarrow q}^{{\overline{\mboxsc{MS}}}} &=& \int_x^1 \frac{dz}{z}\,\,
f_q(x/z,\mu_F)
\frac{2C_F}{3}\,\left\{\left(\frac{\pi^2}{6}-\frac{7}{4}\right)\delta(1-z)
+\frac{1}{2}(1-z)\right.
\nonumber\\
& &\hspace{3.5cm}\left.
+\frac{1}{2}(1+z^2)
\left(\frac{\ln(1-z)}{(1-z)}\right)_+\right\}
\label{bqq}
\\
B_{g\rightarrow q}^{{\overline{\mboxsc{MS}}}} &=& \int_x^1 \frac{dz}{z}\,\,
f_g(x/z,\mu_F)
\,\,\,\frac{1}{4}\,\,\,\left\{ 
\hat{P}^{(4)}_{g\rightarrow q}(z)
\ln(1-z)-
\hat{P}^{(\epsilon)}_{g\rightarrow q}(z)\right\}
\label{bgq}
\\
B_{q\rightarrow g}^{{\overline{\mboxsc{MS}}}}
 &=& \int_x^1 \frac{dz}{z}\,\,f_q(x/z,\mu_F)
\,\,\,\frac{1}{4}\,\,\,\left\{ 
\hat{P}^{(4)}_{q\rightarrow g}(z)
\ln(1-z)-
\hat{P}^{(\epsilon)}_{q\rightarrow g}(z)\right\}
\end{eqnarray}
where $n_f$ is Eqs.~(\ref{aggdef},\ref{bggdef}) denotes the
number of flavors and $N=3$ is the number of colors.
The Altarelli-Parisi kernels in the previous equations are
\begin{eqnarray}
\hat{P}^{(n\neq 4)}_{g\rightarrow g}(z) &=& 
{P}^{(n\neq 4)}_{g\rightarrow g}(z)\,\,\,\, =
\,4\,
\left( \frac{z}{1-z}+\frac{1-z}{z}+z(1-z)  \right) \\
\hat{P}^{(n\neq 4)}_{q\rightarrow g}(z) &=& 
\frac{8}{9}
{P}^{(n\neq 4)}_{q\rightarrow g}(z) =
\frac{16}{9}
\left( \frac{1+(1-z)^2}{z}-\epsilon z  \right) \\
\hat{P}^{(n\neq 4)}_{g\rightarrow q}(z) &=& 
\frac{1}{3}
P^{(n\neq 4)}_{g\rightarrow q}(z) = 
\frac{2}{3}\,\left( \frac{z^2+(1-z)^2-\epsilon}{1-\epsilon}
                      \right)  \label{pgqdef}\\
\hat{P}^{(n\neq 4)}_{q\rightarrow q}(z) &=& 
\frac{8}{9}
P^{(n\neq 4)}_{q\rightarrow q}(z) = 
\frac{16}{9}
\left( \frac{1+z^2}{1-z}-\epsilon(1-z)   \right)  \label{pqqdef}
\end{eqnarray}
where  $P_{ij}^{(\epsilon)}$ is the $\epsilon$ dimensional part
of these $n-$dimensional splitting functions
\begin{eqnarray}
\hat{P}^{(\epsilon)}_{q\rightarrow g}(z) &=& 
\frac{8}{9}
{P}^{(\epsilon)}_{q\rightarrow g}(z) =
-\frac{8}{9}\,2z \\[2mm]
\hat{P}^{(\epsilon)}_{g\rightarrow q}(z) &=& 
\frac{1}{3}
P^{(\epsilon)}_{g\rightarrow q}(z) = 
-\frac{4}{3}\,z(1-z)      \label{pgqepsi}   
\end{eqnarray}
The $(\,\,\,)_+$ prescriptions in 
Eqs.~(\ref{aggdef},\ref{aqq},\ref{bggdef},\ref{bqq})
are defined for an arbitrary
test function $g(z)$ (which is well behaved at $z=1$) as
\begin{eqnarray}
\int_x^1 dz \frac{g(z)}{(1-z)_+} &=&
\int_x^1 dz \frac{g(z)-g(1)}{1-z}\,+\,
    g(1)\log(1-x)\\
\int_x^1 dz g(z)\left( \frac{\log(1-z)}{1-z}\right)_+ &=&
\int_x^1 dz \frac{g(z)-g(1)}{1-z}\,+\,
\frac{1}{2}g(1)\log^2(1-x)
\label{plusdef}
\end{eqnarray}
The structure and use
of the crossing functions are completely analog to the
usual parton distribution function.
Explicit examples of their appearance and use
in NLO 1-jet and 2-jet cross section in DIS 
 are given in Eqs.~(\ref{onejet},\ref{twojet}).
Note that a set of crossing functions
$C_a(\eta,\mu_F,\smin)$
has to be calculated (using an extra program
provided together with \docuname) for each chosen set of 
parton distribution functions $f_a(\eta,\mu_F)$ in a NLO calculation.

%
%
%
%
%
\newpage
\section{Jet Definitions in DIS}
\label{sec_jetdef}
Jet studies on the experimental (hadronic)
and theoretical (partonic)
level require an exact definition of
resolvable jets. The definition of resolvable jets is
given by a jet algorithm which organizes the sprays of hadrons
(or partons) in an event into a small number of jets.
Such a jet algorithm 
is usually defined in terms of one or more resolution
parameters, and a description of how to combine cluster of particles or
jets which do not fulfill the resolution criteria.
By identifying high transverse momentum clusters on the 
experimental and theoretical level w.r.t.
the proton direction in both the lab \underline{and} Breit (or HCM) 
frame\footnote{
Jet production in DIS is a multi-scale problem. 
For a discussion of the
characteristic  hard scale in DIS multi-jet production see 
section~\protect\ref{scalechoice}.}
one can make 
a connection with the underlying primordial partonic scattering and
apply perturbative QCD for the theoretical prediction.
Therefore, jet production provides an intuitive test of the underlying
parton structure of hadronic events.
Clearly, all jet definition schemes have to be infrared and collinear safe, or 
in other words the resulting jet cross sections are not affected
when an infinitely soft parton is added or when a massless parton is replaced 
by a collinear pair of massless partons.
Preferred jet algorithms are those with small higher-order corrections,
small hadronization corrections, and small recombination scheme dependences.

Jet algorithms are represented by  ``$\theta$-functions'' in this paper,
denoted by $J_{n\leftarrow n}$ for  LO (see Eq.~(\ref{jnndef})) and
$J_{n\leftarrow n+1}$ for NLO (see Eq.~(\ref{jnn1def})) calculations,
which can be expressed as a product of a resolution/clustering part
and an acceptance part.
For the appearance of these jet-algorithms in NLO
calculations  see  Eqs.~(\ref{onejet},\ref{twojet},\ref{onejetpol}).

Jet definitions in DIS are applied in the laboratory
frame at HERA (defined by the 27.5 GeV lepton and the 820 GeV proton beam),
in the hadronic center of mass (HCM) frame (=the virtual boson 
and proton rest frame) and the Breit frame. 
The Breit frame is characterized by the vanishing energy component of the 
momentum of the exchanged virtual boson, {\it i.e.} the momentum transfer $q$
is purely spacelike.
Both the boson momentum  
\begin{equation}
q=(0,0,0,-2xE),\,\hspace{1cm}\,\,-q^2=Q^2=4x^2E^2
\label{breitdef1}
\end{equation}
and the proton momentum
\begin{equation}
P=E(1,0,0,1)=\frac{Q}{2x}(1,0,0,-1)
\label{breitdef2}
\end{equation}
are chosen along the $z$-direction.  
Here, $x$ is the standard Bjorken scaling variable.
Unless stated otherwise and in all frames, 
the proton direction defines the $+z$ direction
in this paper.

The following jet algorithms have been used in DIS:
\begin{itemize}
\item[1)]  $W$-scheme:\\
In the $W$-scheme, which was introduced for DIS in Refs.~\cite{herai,heraii}
in analogy to the JADE scheme for $e^+e^-$ \cite{jade},
the invariant mass squared, 
$s_{ij}=(p_i+p_j)^2$, is calculated for each pair of final state 
particles (including the proton remnant).
If the pair with the smallest invariant mass squared is below $y_{cut}W^2$,
the pair is clustered according to a recombination scheme
(see below).
This process is repeated until all invariant 
masses are above $y_{cut} W^2$.        

\item[2)] JADE-scheme:\\ 
The experimental analyses in \cite{exp_as} are  based 
on a variant of the $W$-scheme, the ``JADE'' algorithm \cite{jade}. It is
obtained from the $W$-scheme by replacing the invariant
definition $s_{ij}=(p_i+p_j)^2$ by 
$M_{ij}^2=2E_iE_j(1-\cos\theta_{ij})$, where all quantities are defined 
in the laboratory frame\footnote{The JADE/W schemes can of course also
be applied in the HCM or the Breit frame. All options are 
implemented in \docuname (see Appendix~\ref{sec_docu}.)}.
At LO the $W$ and the JADE scheme are equivalent.
However, neglecting the explicit mass terms $p_i^2$ 
and $p_j^2$ in the definition of $M_{ij}^2$ causes
substantial differences in NLO dijet cross sections 
between the $W$ and the JADE scheme.
The NLO cross sections in the two schemes can differ 
dramatically \cite{plb1,krakau}
(see section~\ref{kfactors}).
One problem with the JADE- or $W$-scheme is that the resulting
jets can still have very low transverse momenta (see for example
Fig.~\ref{f_fig5}b in section~\ref{kfactors}).
We recomment therefore to impose additional cuts on the
jet transverse momenta after the clustering in these schemes.

\item[3)] cone schemes:\\
In the cone algorithm (which can be defined in the laboratory frame,
the HCM or the Breit frame) the distance 
$\Delta R=\sqrt{(\Delta\eta)^2+(\Delta\phi)^2}$ between two partons 
decides whether they should be recombined according to a given 
recombination scheme (see below) to a single jet. 
Here the variables are the pseudo-rapidity $\eta$ 
and the azimuthal angle $\phi$. 
Applicability of fixed order perturbation theory requires
sufficiently high transverse momenta of the jets 
in both the lab \underline{and} Breit frame.
Since the initial state collinear singularity
in DIS is restricted to the proton remnant direction, 
a minimal transverse momentum requirement on the jets
can alternatively be replaced by an effective transverse momentum
cut for the forward direction, {\it i.e.} a $k_{T, {\mboxsc{min}}}^B(j)$
cut, where $k_T^B(j)$ is defined in Eq.~(\ref{ktdef}).

\item[4)] $k_T$ scheme:\\
For the $k_T$ algorithm (which is implemented in the Breit frame), 
we follow the description introduced
in Ref.~\cite{kt}.

After defining
a  hard scattering scale $E_T$ 
and a particle resolution parameter
$y_{cut}$ the following quanities are calculated:
\bq
d_{kP}=2\min(E_i^2,E_P^2)(1-\cos\theta_{iP})
\eq
(the ``transverse energy'' of particle $k$ with respect to the
incoming proton)
where the subscripts $k$ and $P$
denote the final particle and proton, 
respectively, and
\bq
d_{ij}=2\min(E_i^2,E_j^2)(1-\cos\theta_{ij})
\eq
(the ``transverse energy'' of particle $i$
with respect to particle $j$).
If $d_{kP} < d_{ij}$ and $d_{kP}<E_T^2$, then particle
$k$ is eliminated from further clustering considerations 
and is associated with the proton remnant.
If $d_{ij}<d_{kP}$ amd $d_{ij}<E_T^2$,
particle $i$ and $j$ are recombined into a single pre-cluster
(or ``macro jet'') according to a recombination prescription.
This iteration continues for all particles
and pre-clusters until all objects have been formed into single 
pre-clusters or included into the proton remnant.
For $y_{cut}=1$ these final objects are the final jets.
The single pre-clusters can be further resolved 
if  $y_{cut}<1$. If any pair $k$ and $l$
in a pre cluster has $d_{kl}<y_{cut} E_T^2$,
these particles are further combined into a single jet until all
objects in a pre-cluster have $d_{kl}>y_{cut} E_T^2$.
\end{itemize}
Various other definitions could be chosen as for example the
``mixed scheme'' in Ref.~\cite{herai} or the ARCLUS algorithm
proposed in Ref.~\cite{arclus}.

On top of these jet resolution criteria, several
prescriptions of how to combine  a cluster of particles 
which do not fulfill the resolution criteria have been used,
{\it i.e.} how the momenta of  partons
(or hadrons/cluster of particles) are recombined to give a 
composite momentum in this case.
The recombined momentum is denoted by
$p^\mu_{rec} = (E_{rec},\vec{p}_{rec})$ with
$ E_{rec}=\alpha \,\,(E_1+E_2)$ and
$ \vec{p}_{rec} = \beta \,\,(\vec{p}_1+\vec{p}_2).$
The energy and momentum rescaling factors $\alpha$ and $\beta$ define the
$E$-scheme, $E0$-scheme and $P$-scheme as follows
\cite{bethke}:
\bq
\begin{array}{lll}
E-\mbox{scheme}: \hspace{5mm}
                 & \alpha = 1   & \beta = 1  \\
E0-\mbox{scheme}:& \alpha = 1   & 
                  \beta  = \frac{E_1+E_2}{|\vec{p_1}+\vec{p}_2|}  \\
P-\mbox{scheme}: & \alpha = \frac{|\vec{p_1}+\vec{p}_2|}{E_1+E_2} 
                   \hspace{1cm}
                 & \beta  = 1 \\
\end{array}
\label{recombination}
\eq
Obviously, the $E$ scheme conserves energy-momentum, while
the $E0$ ($P$) scheme conserves only energy (momentum).
The  rescaling factors for the $E0$ and $P$ scheme are chosen
such that the recombined four-vector has zero mass.
The recombined vector is only in the $E$ scheme not massless.

Another commonly used recombination scheme for jets defined in a cone scheme
has been proposed in Ref.~\cite{snowmass}.
Here, the transverse energy, pseudo-rapidity and azimuthal angle of the
jets are calculated by performing energy weighted sums over the particles
within the cone radius $R$,
\begin{eqnarray}
E_T &=&  \sum_{i\in R} E_{Ti}\nonumber\\
\eta_{\mboxsc{jet}}&=&\frac{1}{E_T}  \sum_{i\in R} E_{Ti}\,\,\eta_i
\label{conecluster}\\
\phi_{\mboxsc{jet}}&=&\frac{1}{E_T}  \sum_{i\in R} E_{Ti}\,\,\phi_i\nonumber
\end{eqnarray}
Several variants of Eq.~(\ref{conecluster})   have been used
for the cone scheme in hadron hadron collisions:
the ``fixed-cone'' algorithm used by UA2 \cite{ua2},
the ``iterative-cone'' algorithm used by both CDF and D0 
collaborations at the FERMILAB collider \cite{cdf,d0},
the ``EKS'' algorithm  introduced in Ref.~\cite{eks}.
For a recent  discussion of various definitions,
including a detailed discussion of problems related 
to overlapping cones\footnote{In fact, the problem of overlapping 
cones occurs already at \oasz\ in DIS dijet production, when
the cone scheme is defined in the laboratory frame.
Here, the scattered lepton  balances the transverse momenta of
all three partons in the NLO tree level contribution.
An equivalent  situation (with three partons balanced in $p_T$
by another parton) occurs at hadron-hadron collisions only at
${\cal{O}}(\alpha_s^3$).}, we refer the reader to \cite{mikejets,kilgore}.

It is clear that the algorithms (including the 
recombination prescriptions)  in the
theoretical calculations must be matched to the
chosen experimental definition.
Unless stated otherwise, and for all
jet algorithms, we use the $E$-scheme to recombine partons, i.e.
the cluster momentum is taken as $p_i+p_j$, the sum of the 4-momenta
of partons $i$ and $j$, if these
are unresolved according to a given jet definition scheme.
Large recombination scheme dependencies have been found in particular
for the $W$  scheme \cite{plb1} (see also section~\ref{kfactors}).

\newpage
\section{One-Jet Cross Sections
\protect\vspace{1mm}}
\label{sec_onejet}
%
%
\subsection[{NLO One-Jet and 
Total $\oas$ Cross Sections (One-Photon Exchange)\protect\vspace{1mm}
}]{NLO One-Jet and 
Total $\O({\protect\boldmath{\alpha_s}})$ 
Cross Sections\protect\\ (One-Photon Exchange)}
\label{sec_ana_onejet}
The 1-jet final state is the most basic
high transverse momentum 
event at HERA, with a lepton and a jet back-to-back in the
transverse plane. 
The lowest order $\O(\as^0)$ partonic contribution to the 1-jet
cross section arises from the quark parton model (QPM)
subprocess (see Fig.~\ref{f_qtoq}a)
\bq
e(l)+q(p_0)\rightarrow e(l^\prime) + q(p_1)
\label{qtoq}
\eq
and the corresponding anti-quark process with $q\leftrightarrow \bar{q}$.
Imposing no (or sufficiently weak cuts on the scattered parton)
yields directly the total DIS cross section in the parton model.
\begin{figure}[hb]
\vspace*{3cm}            
\begin{picture}(0,0)(0,0)
\includegraphics{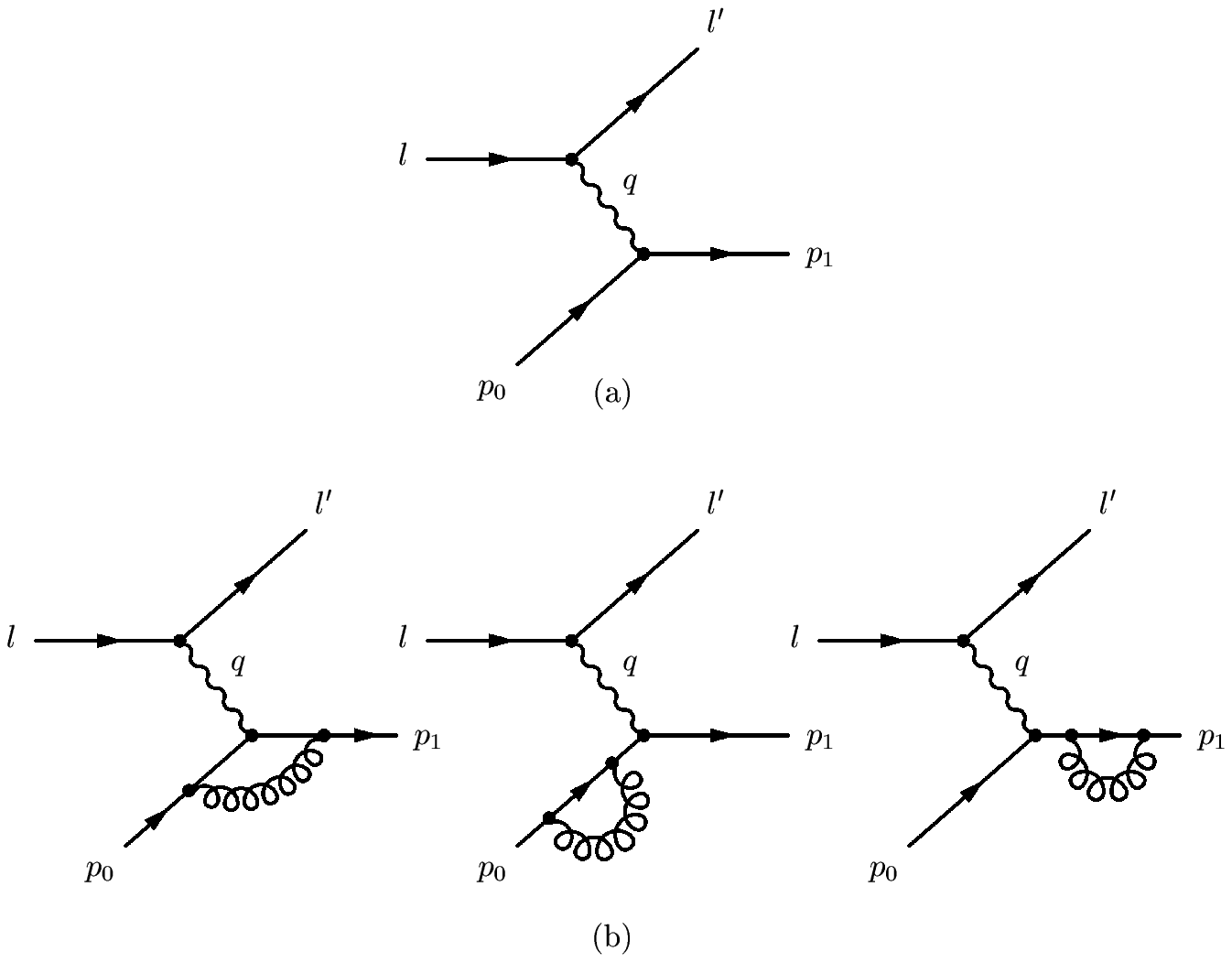}
\end{picture}
\vspace{7cm}
\caption{
Zeroth order Born diagram (a) for lepton parton scattering
and virtual gluon corrections (b) to the Born process.
}
\label{f_qtoq}
\end{figure}
The NLO $\O(\as)$ 1-jet cross section receives contributions
from the 1-loop corrections to the subprocess
in Eq.~(\ref{qtoq}) (see Fig.~\ref{f_qtoq}b)
and from the 2-parton tree level final
state matrix elements (see Fig.~\ref{f_atobc})
\begin{eqnarray}
e(l)+q(p_0)&\rightarrow& e(l^\prime) + q(p_1) + g(p_2)
\label{qtoqg}\\
e(l)+g(p_0)&\rightarrow& e(l^\prime) + q(p_1) + \bar{q}(p_2)
\label{gtoqqbar}
\end{eqnarray}
and the corresponding anti-quark processes with $q\leftrightarrow \bar{q}$.
\begin{figure}[hbt]
\vspace*{5cm}            
\begin{picture}(0,0)(0,0)
\includegraphics{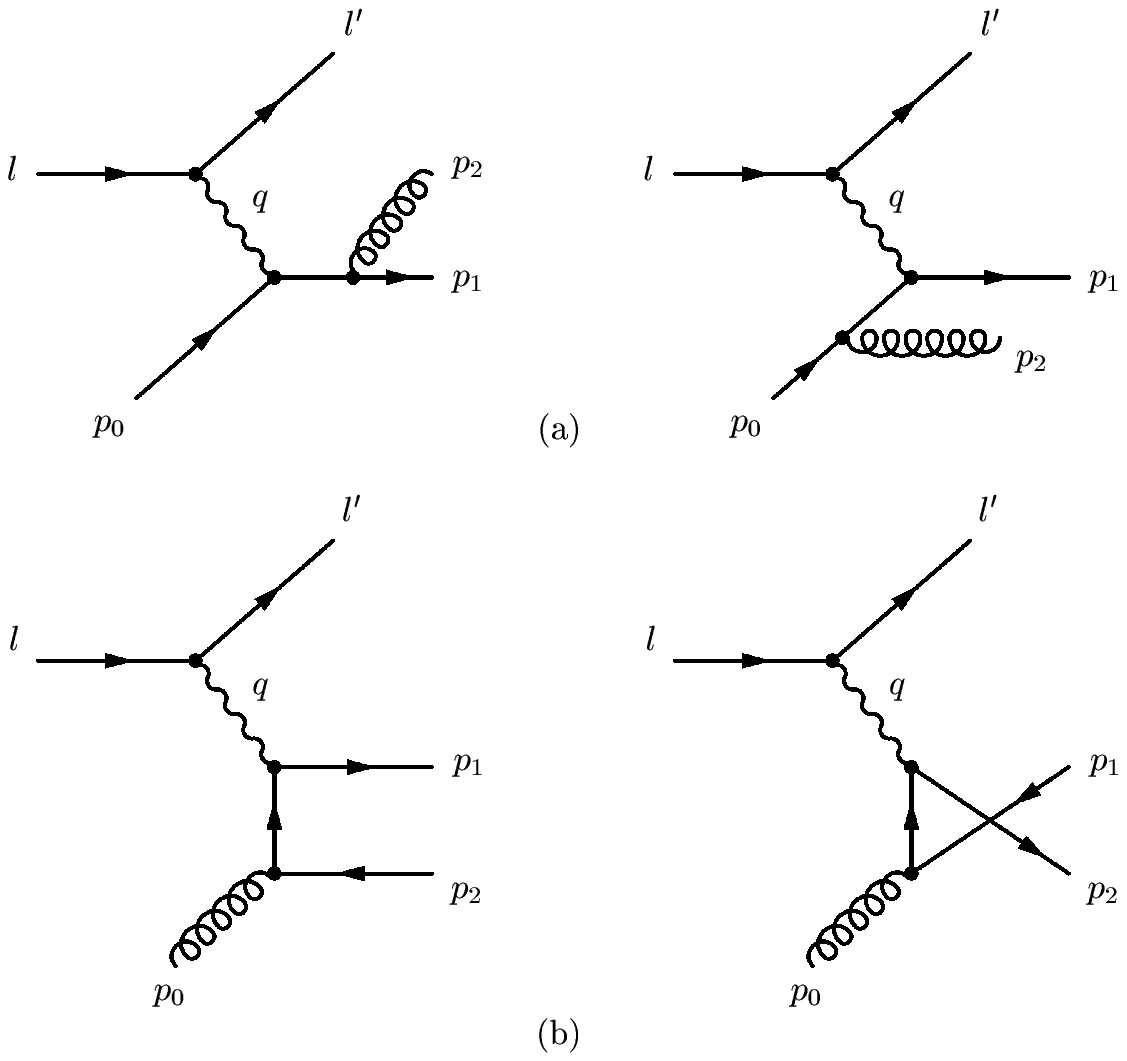}
\end{picture}
\vspace{7cm}
\caption{
First order \protect\oas\ tree graph two parton final state processes:
quark initiated ``Compton process'' (a) and 
boson-gluon fusion process (b).
}
\label{f_atobc}
\end{figure}

According  to Eqs.~(\ref{sig_sum},\ref{sig_virt}) 
the NLO  1-jet {\it exclusive} cross section is given by
\begin{eqnarray}
\sigma_{\mboxsc{had}}^{\mboxsc{NLO}}[1\mbox{-jet}]
&=& 
\sigma_{\mboxsc{had}}^{\mboxsc{LO}}[1\mbox{-jet}]\nonumber \\[1mm]
&+&
\sigma_{\mboxsc{had}}^{\mboxsc{NLO, final}}[1\mbox{-jet}]
\label{sig_sum1j} \\[1mm]
&+&
\sigma_{\mboxsc{had}}^{\mboxsc{NLO, crossing}}[1\mbox{-jet}]\nonumber\\[1mm]
&+&
\sigma_{\mboxsc{had}}^{\mboxsc{NLO, hard}}[1\mbox{-jet}]\nonumber
\end{eqnarray}
where the hadronic cross sections on the r.h.s. of Eq.~(\ref{sig_sum1j})
are defined in Eqs.~(\ref{hadlo},\ref{hadfinal},\ref{hadcross},\ref{hadlo1})
for $n=1$, respectively.
The required partonic cross sections
in the hadronic cross sections 
in Eq.~(\ref{sig_sum1j}) will be discussed in the following subsections.
The final formula for the hadronic NLO 1-jet {\it exclusive}
cross section in terms 
of these partonic results is  given in Eq.~(\ref{onejet}).

The total $\oas$ cross section can be obtained
from Eqs.~(\ref{sig_sum1j},\ref{onejet}) 
by adding the LO
two jet cross section, {\it i.e.} 
the total cross section is defined as the 
NLO 1-jet {\it inclusive} cross section,
provided the cuts on the jets are sufficiently weak
(see  section~\ref{numonejet}).

Let us now specify the relevant partonic cross sections
for the hadronic cross sections in Eq.~(\ref{sig_sum1j}).
 
\subsubsection[{Partonic cross sections for
$\sigma_{\protect\mboxsc{had}}^{\protect\mboxsc{LO}}$[1-jet] 
and
$\sigma_{\protect\mboxsc{had}}^{\protect\mboxsc{NLO, crossing}}$[1-jet]
\protect\vspace{1mm}}
]{
$\hat{\sigma}^{\protect\mboxsc{LO}}_{q\rightarrow q}$
in 
$\sigma_{\protect\mboxsc{had}}^{\protect\mboxsc{LO}}$[1-jet] 
and
$\sigma_{\protect\mboxsc{had}}^{\protect\mboxsc{NLO, crossing}}$[1-jet]
}
%
%
According to Eqs.~(\ref{hadlo},\ref{hadcross}) the
partonic cross section 
\bq
\hat{\sigma}^{\mboxsc{LO}}_{q\rightarrow q}
=\sigma_0\,\,e_q^2\,\,|M^{(\mboxsc{pc})}_{q\rightarrow q}|^2
\label{siglodef}
\eq
for the subprocess in  Eq.~(\protect\ref{qtoq}) 
enters the hadronic cross sections
$\sigma_{\protect\mboxsc{had}}^{\protect\mboxsc{LO}}$[1-jet]
and
$\sigma_{\protect\mboxsc{had}}^{\protect\mboxsc{NLO, crossing}}$[1-jet]
where
(a numerical factor $1/4$ for the initial state spin average
is included in $\sigma_0$)
\begin{equation}
\sigma_0 =\frac{1}{4p_0.l}\, \frac{1}{4}\,\frac{(4\pi\alpha)^2}{Q^4}
\label{sigma0def}
\end{equation}
and
\begin{equation}
|M^{(\mboxsc{pc})}_{q\rightarrow q}|^2=32\,
\left[(p_0.l)^2 \, +\,(p_0.l^\prime)^2 \,\right]
\,= \,8 \hat{s}^2\,\,(1+(1-y)^2)
\label{m_qtoq}
\end{equation}
The superscript (pc) ($\equiv$ parity conserving) refers to the vector
current coupling at the leptonic and hadronic vertices and
$\hat{s}=\eta s$ denotes the partonic center of mass energy squared.
%
%
\subsubsection[{Partonic cross sections for
$\sigma_{\protect\mboxsc{had}}^{\protect\mboxsc{NLO, final}}$
[1-jet]
\protect\vspace{1mm}}]{$\hat{\sigma}^{\protect\mboxsc{NLO}}_{q\rightarrow q}$
in $\sigma_{\protect\mboxsc{had}}^{\protect\mboxsc{NLO, final}}$[1-jet]
}
%
%
According to Eq.~(\ref{finalstruc}) the NLO (finite) partonic
cross section
\bq
\as \,\hat{\sigma}^{\mboxsc{NLO}}_{q\rightarrow q}\,\,
=
\alpha_s \,\,\sigma_0\,\,e_q^2\,\,
\left[ |M^{(\mboxsc{pc})}_{q\rightarrow q}|^2\,\,
{\cal{K}}_{q\rightarrow q}(\smin,Q^2)
+
{\cal{F}}_{q\rightarrow q} \right]
\label{nlo1j}
\eq
in the hadronic cross section
$\sigma_{\protect\mboxsc{had}}^{\protect\mboxsc{NLO, final}}$[1-jet]
combines the virtual 1-loop corrections in Fig.~\ref{f_qtoq}b
to the Born process in Eq.~(\ref{qtoq})
with the singular integrals over the two parton
{\it final state} unresolved phase space
region for the subprocesses in Eqs.~(\ref{qtoqg}) and (\ref{gtoqqbar})
(see Fig.~\ref{f_atobc}).
All finite parts of the virtual corrections factorize the Born matrix
element in the 1-jet case
and hence ${\cal{F}}_{q\rightarrow q}$ in Eq.~(\ref{nlo1j}) vanishes. 
The dynamical ${\cal{K}}_{q\rightarrow q}$ factor, which multiplies the
lowest order ${\cal{O}}(\alpha_s^0)$ matrix element squared,
depends on both $\smin$ and the invariant mass of the hard 
partons  $2p_0.p_1=Q^2$ ($N=3$ is the number of colors):
\begin{equation}
{\cal{K}}_{q\rightarrow q}(\smin,Q^2) = \frac{8}{9}\,
\left(\frac{N}{2\pi}\right)
\left[
-\ln^2\left( \frac{\smin}{Q^2}\right) 
- \frac{3}{2}\ln\left(\frac{\smin}{Q^2}\right)
-\frac{\pi^2 }{3}-\frac{1}{2}
\, + \, {\cal{O}}(\smin) \right]
\label{r_qtoq}
\end{equation}
Terms proportional to $\smin$ have been neglected 
in Eq.~(\ref{r_qtoq}).
${\cal{K}}_{q\rightarrow q}$ may be crossed in exactly the same
manner as the usual tree level crossing\footnote{Some care must be taken
in crossing logarithms with negative arguments}
from the ${\cal{K}}$ factor
in $e^+e^-\rightarrow$ 2 partons as given\footnote{The NLO
$e^+e^-\rightarrow 2$ parton ${\cal K}$-factor is also given in
Eq.~3.1.68 of \cite{kramer}.}
in Eq.~(4.31) with $n=0$
in Ref~\cite{giele1}.
Thus, Eq.~(\ref{r_qtoq}) includes also the crossing of a pair
of collinear partons with an invariant mass smaller than $\smin$
from the final state to the initial state. 
This ``wrong'' contribution is replaced by the
correct collinear initial state configuration by adding the appropriate
crossing function contribution to the hadronic cross section
as given in Eq.~(\ref{onejet}). The crossing function contribution
in Eq.~(\ref{onejet})
takes also into account the corresponding factorization of the
initial state singularities, which is encoded in the crossing functions
$C_q^{\overline{\mboxsc{MS}}}$ 
for valence and sea quark distributions as described in
Eqs.~({\ref{crossf_uv},\ref{crossf_s}).

\subsubsection[{Partonic cross sections for
$\sigma_{\protect\mboxsc{had}}^{\protect\mboxsc{NLO, hard}}$
[1-jet]\protect\vspace{1mm}}]{
$\hat{\sigma}^{\protect\mboxsc{LO}}_{q\rightarrow qg}$ and
$\hat{\sigma}^{\protect\mboxsc{LO}}_{g\rightarrow q\bar{q}}$ 
in
$\sigma_{\protect\mboxsc{had}}^{\protect\mboxsc{NLO, hard}}$[1-jet]
}
%
%
According to Eq.~(\ref{hadlo1})
$\sigma_{\protect\mboxsc{had}}^{\protect\mboxsc{NLO, hard}}$[1-jet]
calculates the finite contributions from the real emission outside 
the soft and collinear region 
$(s_{ij}>\smin$) for the 2-parton final state 
subprocesses in Eqs.~(\ref{qtoqg}) and (\ref{gtoqqbar}) (Fig.~\ref{f_atobc}).
The partonic cross sections in 1-photon exchange are:
\begin{eqnarray}
\as\, \hat{\sigma}^{\mboxsc{LO}}_{q\rightarrow qg}\,\,
&=&
\sigma_0\,\,e_q^2\,\,
(4\pi\alpha_s(\mu_R))\,\,|M^{(\mboxsc{pc})}_{q\rightarrow qg}|^2
\label{sig_qtoqg} \\[1mm]
\as \, \hat{\sigma}^{\mboxsc{LO}}_{g\rightarrow q\bar{q}}\,\,
&=&
\sigma_0\,\,e_q^2\,\,
(4\pi\alpha_s(\mu_R))\,\,
|M^{(\mboxsc{pc})}_{g\rightarrow q\bar{q}}|^2
\label{sig_gtoqqbar}
\end{eqnarray}
with [see appendix~\ref{app_born}]
\begin{eqnarray}
|M^{(\mboxsc{pc})}_{q\rightarrow qg}|^2
&=&
\frac{128}{3}\,\,(l.l^\prime)\,\,
\frac{(l.p_0)^2+(l^\prime.p_0)^2+(l.p_1)^2+(l^\prime.p_1)^2
     }{(p_1.p_2)(p_0.p_2)}
\label{m_qtoqg} \\[1mm]
|M^{(\mboxsc{pc})}_{g\rightarrow q\bar{q}}|^2
&=&
-\frac{3}{8}\,\,
|M^{(\mboxsc{pc})}_{q\rightarrow qg}|^2(p_0\leftrightarrow-p_2)\nonumber\\
&=&
16\,\,(l.l^\prime)\,\,
\frac{(l.p_2)^2+(l^\prime.p_2)^2+(l.p_1)^2+(l^\prime.p_1)^2
     }{(p_0.p_1)(p_0.p_2)}
\label{m_gtoqqbar}
\end{eqnarray}
Color factors (including the initial state color average) 
are included in these squared matrix elements.
The superscript (pc) refers again to the vector current
coupling of the virtual photon at the leptonic and hadronic vertex.
Note that the initial state spin average factors are included
in the definition of $\sigma_0$ in Eq.~(\ref{m_qtoq}).
These compact expressions for the squared matrix elements can be obtained
by analytically squaring the  helicity amplitudes
for the subprocesses in Eqs.~(\ref{qtoqg},\ref{gtoqqbar})
in the Weyl-van der Waerden formalism 
as shown in appendix \ref{app_born}
and therefore, the full spin structure is kept.
In fact, each addend in Eqs.~(\protect\ref{m_qtoqg}) 
and (\protect\ref{m_gtoqqbar}) corresponds to a specific helicity
configuration for the external particles as listed in 
table~\ref{helitab}.
%
%
\begin{table}[b]
\caption{
Helicity dependent contributions to the squared matrix elements
in Eqs.~(\protect\ref{m_qtoqg}) and (\protect\ref{m_gtoqqbar}).
The helicities $\lambda_i$ 
 are defined via
$e(\lambda_l)q(\lambda_0)\rightarrow 
e(\lambda_{l^\prime})q(\lambda_1)g(\lambda_2)$
and 
$e(\lambda_l)g(\lambda_0)\rightarrow 
e(\lambda_{l^\prime})q(\lambda_1)\bar{q}(\lambda_2).$
The identical contributions from the four nonvanishing 
remaining helicity combinations with all helicities reversed are
not listed in the table (see appendix~\protect\ref{app_born}).
}
\label{helitab}
\vspace{3mm}
\begin{center}
\begin{tabular}{|l|ccccc|c|l|ccccc|}
\hline\hline
$eq\rightarrow eqg$
     &  $\lambda_0$
     &  $\lambda_1$
     &  $\lambda_2$
     &  $\lambda_l$
     &  $\lambda_{l^\prime}$
     &  \hspace{0mm}
     &  $eg\rightarrow eq\bar{q}$      
     &  $\lambda_0$
     &  $\lambda_1$
     &  $\lambda_2$
     &  $\lambda_l$
     &  $\lambda_{l^\prime}$
 \\ \hline\\[-4mm]
$\sim(l.p_0)^2$
     &  $+$
     &  $+$ 
     &  $+$
     &  $+$
     &  $+$
     &
     &  $\sim(l.p_2)^2$
     &  $-$
     &  $+$ 
     &  $-$
     &  $+$
     &  $+$ \\[1mm]
$\sim(l^\prime.p_0)^2$
     &  $+$
     &  $+$ 
     &  $+$
     &  $-$
     &  $-$
     &
     &   $\sim(l^\prime.p_2)^2$
     &  $-$
     &  $+$ 
     &  $-$
     &  $-$
     &  $-$ \\[1mm]
$\sim(l.p_1)^2$
     &  $-$
     &  $-$ 
     &  $+$
     &  $+$
     &  $+$
     & 
     &  $\sim(l.p_1)^2$
     &  $-$
     &  $-$ 
     &  $+$
     &  $+$
     &  $+$   \\[1mm]
$\sim(l^\prime.p_1)^2$
     &  $-$
     &  $-$ 
     &  $+$
     &  $-$
     &  $-$
     &
     &  $\sim(l^\prime.p_1)^2$
     &  $-$
     &  $-$ 
     &  $+$
     &  $-$
     &  $-$ \\ \hline 
\end{tabular}
\end{center}
\end{table}
%
%
Thus, all lepton-hadron and jet-jet correlations are fully taken into account.
The squared matrix elements
$|M^{(\mboxsc{pc})}_{q\rightarrow qg}|^2$ and
$|M^{(\mboxsc{pc})}_{g\rightarrow q\bar{q}}|^2$
are expressed in terms of more DIS like (partonic)
variables in section~\ref{helicross}.
The resulting expressions naturally factorize the characteristic $y$ and 
$\phi$ dependencies for the helicity cross sections
$d\sigma^{F_i}$ (see Eq.~(\ref{sigphi})).

Note that the results in Eqs.~(\ref{m_qtoqg},\ref{m_gtoqqbar})
contain  the full polarization
dependence of the virtual boson, {\it i.e.}

These results can  also be expressed in terms
of helicity cross sections $d\sigma^{F_i}$ which correspond to certain 
polarization states of the exchanged virtual boson
(see section~\ref{twojetintro}).

\subsubsection{The Hadronic One-Jet Cross Section\protect\vspace{1mm}}
Based on Eq.~(\ref{sig_sum1j}) and the results in the previous
sections, the hadronic 1-jet {\it exclusive}
cross section in the 1-photon exchange up to $\O(\alpha_s)$ reads
\\[5mm]
\fbox{\rule[-5cm]{0cm}{8cm}\mbox{\hspace{16.2cm}}}
\\[-8.3cm]
\begin{eqnarray}
\displaystyle
\sigma_{\mboxsc{had}}[\mbox{1-jet}] &=&
\nonumber\\
&&\hspace{-15mm}
\int_0^1d\eta \int
\,d{\mbox{PS}}^{(l^\prime+1)}\,\,\sigma_0\,\,
\bigg[  \nonumber \\
&&  [\sum_{i=q,\bar{q}}e_i^2 f_i(\eta,\mu_F)\big]\,
\,\,|M^{(\mboxsc{pc})}_{q\rightarrow q}|^2\,
\left(1+\alpha_s(\mu_R)\,
{\cal{K}}_{q\rightarrow q}(\smin,Q^2)\right)\nonumber \\
&+&
\,
[\sum_{i=q,\bar{q}}e_i^2\, 
\,C_i^{\overline{\mboxsc{MS}}}(\eta,\mu_F,\smin)]\,\,
\alpha_s(\mu_R)\,\,\,|M^{(\mboxsc{pc})}_{q\rightarrow q}|^2\,
\,\bigg] J_{1\leftarrow 1}(\{p_i\}) 
\label{onejet}\\
&&\hspace{-20mm}
+
\int_0^1 d\eta \int
\,d\mbox{PS}^{(l^\prime+2)}\,\,\sigma_0\,\,
(4\pi\alpha_s(\mu_R))\,\,  
\bigg[    \nonumber \\
&&[\sum_{i=q,\bar{q}}e_i^2 f_i(\eta,\mu_F)] \,
\,\,|M^{(\mboxsc{pc})}_{q\rightarrow qg}|^2
\nonumber    \\ 
&+&
(\sum_{i=q}e_i^2 ) f_g(\eta,\mu_F)\,\,
\,|M^{(\mboxsc{pc})}_{g\rightarrow q\bar{q}}|^2
\bigg]
\,\,
\prod_{i<j;\,0}^{2}\Theta(|s_{ij}| - \smin)\,\,
J_{1\leftarrow 2}(\{p_i\})
\nonumber\\[5mm]\nonumber
\end{eqnarray}
where the Lorentz-invariant phase space measure
$d\mbox{PS}^{(l^\prime+n)}$
is defined in
Eq.~(\ref{phasespace}).
Analytical expressions for $|M^{(\mboxsc{pc})}_{q\rightarrow q}|^2, 
{\cal{K}}_{q\rightarrow q}, |M^{(\mboxsc{pc})}_{q\rightarrow qg}|^2$ and 
$|M^{(\mboxsc{pc})}_{g\rightarrow q\bar{q}}|^2$ are listed in 
Eqs.~(\ref{m_qtoq},\ref{r_qtoq},\ref{m_qtoqg},\ref{m_gtoqqbar}),
the crossing functions
$C_q^{\overline{\mboxsc{MS}}}$ 
for valence and sea quark distributions  in
Eqs.~({\ref{crossf_uv},\ref{crossf_s}),
and the general structure of the jet algorithms 
$J_{1\leftarrow 1}(\{p_i\})$ and
$J_{1\leftarrow 2}(\{p_i\})$ is described in Eqs.~(\ref{jnndef})
and (\ref{jnn1def}), respectively.
The 1-jet {\it inclusive} cross section is defined via
Eq.~(\ref{onejet}) by replacing
$J_{1\leftarrow 2}(\{p_i\})$ in the last line by 
($J_{1\leftarrow 2}(\{p_i\})+J_{2\leftarrow 2}(\{p_i\}))$,
{\it i.e.} the 1-jet inclusive cross section is defined
as the sum of the NLO 1-jet exclusive cross section (as defined
in Eq.~(\ref{onejet})) plus the LO two jet cross section.

Eq.~(\ref{onejet}) includes all relevant information  to construct
a Monte Carlo program for the numerical evaluation of the fully differential
NLO 1-jet cross section in DIS.
In particular  all ``plus prescriptions'' associated with the 
factorization of the initial state collinear
divergencies
are absorbed in the crossing functions $C_q^{\overline{\mboxsc{MS}}}$ 
which  is very useful for a Monte Carlo approach.

Note that the second integral over the bremsstrahlung matrix elements
is restricted to regions
where all partons are resolved, {\it i.e.} any pair of partons has
$s_{ij}>\smin$.
As mentioned before, $\smin$ is an arbitrary theoretical parameter 
and any measurable quantity should not depend on it.
The bremsstahlung contribution
grows with $\ln^2\smin$ and $\ln\smin$
with decreasing $\smin$.
This logarithmic growth is exactly cancelled by the
explicit   $-\ln^2\smin$ and $-\ln\smin$ terms
in ${\cal{K}}_{q\rightarrow q}(\smin,Q^2)$ (see Eq.~(\ref{r_qtoq})) and 
the $\smin$ dependence in the
the crossing functions $C_q^{\overline{\mboxsc{MS}}}(\eta,\mu_F,\smin)$
once $\smin$ is small enough for the soft and collinear approximations
in ${\cal{K}}_{q\rightarrow q}$  and $C_q^{\overline{\mboxsc{MS}}}$ 
to be valid.
The explicit $\smin$ dependence of 
$C_q^{\overline{\mboxsc{MS}}}$  is discussed in
sect.~\ref{crossing}.

A powerful test of the numerical program is the $\smin$ independence of 
the NLO 1-jet cross sections.
Fig.~\ref{fig_smin_one_jet} shows the inclusive 1-jet cross section
for HERA energies as a function of $\smin$ for 
jets defined in a cone scheme (in the laboratory frame) with
$\Delta R=1$ for different $Q^2$ bins.
\begin{figure}[tb]
  \centering                
  \mbox{\epsfig{file=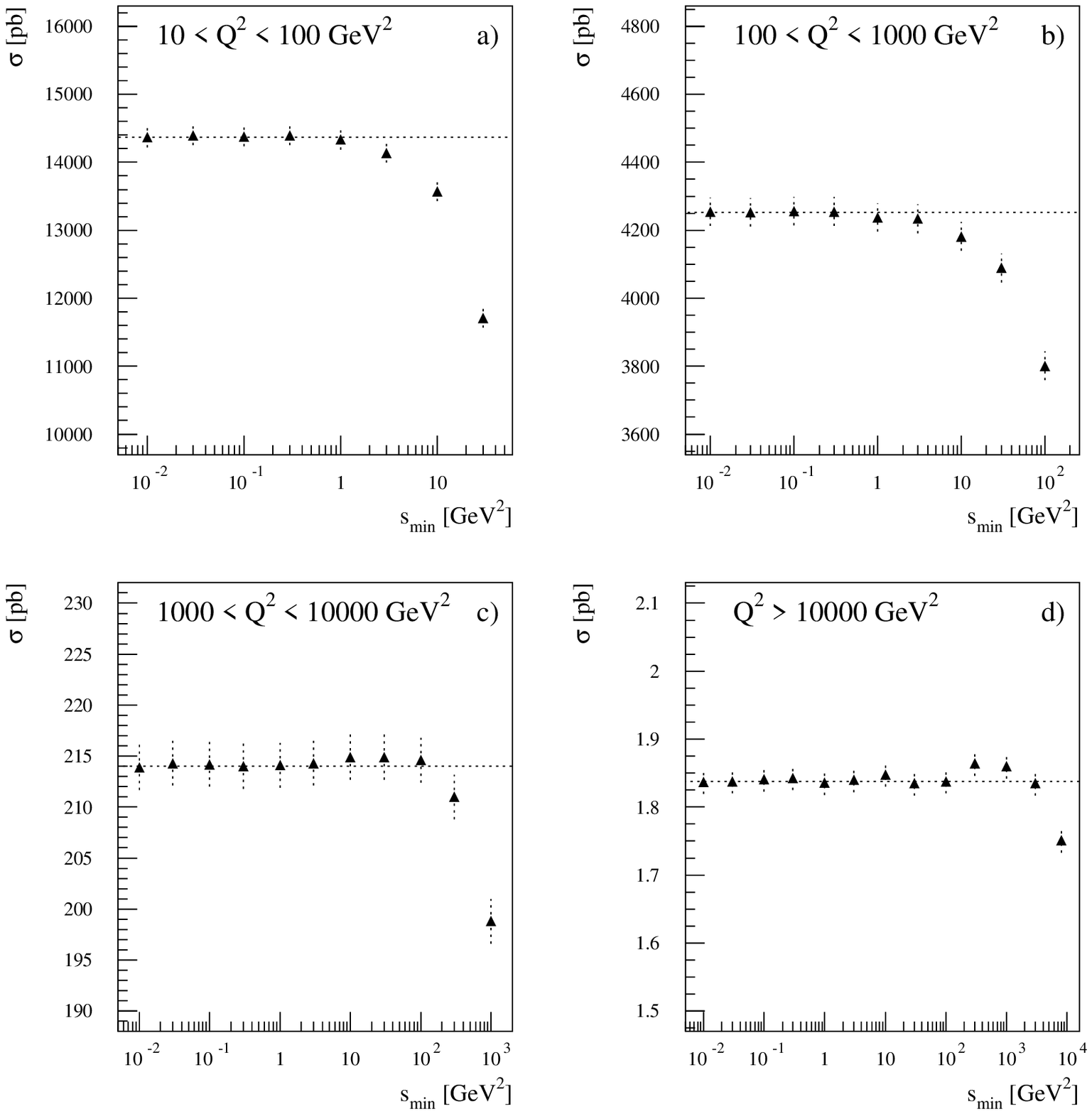,bbllx=0,bblly=0,
                bburx=550,bbury=550,width=0.80\linewidth}} 
\caption{
Dependence of the inclusive 1-jet cross section in a 
cone scheme on $\protect\smin$, the two-parton resolution parameter,
for four $Q^2$ bins.
Error bars represent statistical errors of the Monte Carlo program. 
$\protect\smin$ independence in the four $Q^2$ bins is achieved for 
$\protect\smin\lsimfig 1\, (a), 3 \,(b), 100 \,(c), 100\, (d) $~GeV$^2$. 
}\label{fig_smin_one_jet}
\end{figure}
One observes that for $\smin$ values $\lsimfig$ 1,3,100,100 GeV$^2$
the results are indeed independent 
of $\smin$  for
the four $Q^2$ bins in Fig.~\ref{fig_smin_one_jet}a,b,c,d, respectively
(within one percent, which is about the statistical error in
the Monte Carlo runs).
Since $Q^2$ sets the typical hard scale 
for 1-jet production,
$\smin$ independence of the results is indeed expected
to start at higher $\smin$ values with increasing $Q^2$.
The $\smin$ dependence of the NLO 
cross sections for larger $\smin$ values shows that the soft 
and collinear approximations used in the phase space region $s_{ij}<\smin$
are no longer valid, {\it i.e.} terms of ${\cal{O}}(\smin)$ and
${\cal{O}}(\smin\ln \smin)$ become important.
In general, one wants to choose $\smin$ as large as possible to avoid large
cancellations between the virtual+collinear+soft part 
(first integral in Eq.~(\ref{onejet}))
and the hard part of the phase space  
(second integral in Eq.~(\ref{onejet})).
Note that up to factor 5-10 cancellations 
occur between the effective 1-parton and 
2-parton final states at the lowest $\smin$ values in 
Fig.~\ref{fig_smin_one_jet}a,b,c,d, whereas typically cancellations of factors
2-3 occur for the largest possible  $\smin$ values
where the cross section is still independent on $\smin$.

\subsection{Numerical Results: One-Jet and Total $\oas$  Cross Sections
  \protect\vspace{1mm}}\label{numonejet}
Numerical results for LO and NLO (exclusive and inclusive) 1-jet 
cross sections with $\gamma^\ast$ exchange
are presented in this section. 
Electroweak effects through the additional exchange of a
$Z (W^\pm)$ boson in NC (CC) scattering will be discussed in 
section~\ref{onejetz}.
Special emphasis is put on the calculation of 
the total inclusive (w.r.t. the hadronic activity)
cross section  when typical acceptance cuts on the
scattered lepton are imposed.
The characteristics of the highest transverse momentum 
jet in a NLO inclusive calculation shows 
that the total \oas\ cross section can be obtained
by the sum of the NLO 1-jet exclusive cross section
plus the LO 2-jet cross section, {\it provided the 
acceptance requirements on the jets are sufficiently weak}
(see below).

For the following numerical studies, 
the lepton and hadron beam energies are 27.5 and 820 GeV, respectively.
Furthermore, we  require 40~GeV$^2<Q^2<2500$ GeV$^2$,
where the upper limit is imposed to suppress the additional 
$Z$ exchange contributions.
Unless stated otherwise,
the LO parton distributions of Gl\"uck, Reya and Vogt \cite{grv,pdflib}
together with the 1-loop formula for the strong coupling constant are used for
the parton model results.
For the NLO $\oas$  numerical studies,
we use the NLO GRV parton distribution functions and
the two loop formula for the strong coupling constant
\begin{equation}
\alpha_{s\,\overline{MS}}(\mu_R^2)= 
\frac{12\pi}{(33-2n_{f})\ln(\mu_R^2/\Lambda^{2})}
\left[1-\frac{6(153-19n_{f})}{(33-2n_{f})^{2}}
\frac{\ln\ln(\mu_R^2/\Lambda^{2})}{\ln(\mu_R^2/\Lambda^{2})}\right]
\label{asnlo}
\end{equation}
with 
${\Lambda_{\overline{MS}}^{(4)}}$ chosen according to the 
value from the parton distribution functions.
The value of $\alpha_s$ is matched at the thresholds $\mu_R=m_q$ and the
number of flavors is fixed to $n_f=5$ throughout. 
We work in the $\overline{MS}$  factorization scheme and
a running QED fine structure constant $\alpha(Q^2)$ is used.
In addition, we require
$0.04 < y < 1$, an energy cut of $E(e^\prime)>10$~GeV on the scattered 
electron, and a cut on the pseudo-rapidity 
%
%
$\eta=-\ln\tan(\theta/2)$
%
%
of the scattered lepton and jets of $|\eta|<3.5$. 
Within these general cuts the four different jet definition schemes
described in  section~\ref{sec_jetdef} are considered.
\begin{table}[htb]
\caption{One-jet  cross sections for 40 GeV$^2 < Q^2 <$ 2500 GeV$^2$ at HERA.
Results are given at LO and NLO for the cone, $k_T$ and JADE schemes
and acceptance cuts described in the text. Additional
parameters  are given in the first column.
}
\label{tab_j1}
\vspace{3mm}
\begin{tabular}{llll}
\hline\hline\\[-4mm]
       \mbox{jet scheme} 
     &  1-jet 
      &  1-jet exclusive
     &  1-jet inclusive \\
        \mbox{}
     &  LO
     &  NLO
     &  NLO\\
\hline\\[-3mm]
       \mbox{cone} ($p_T^{\mboxsc{lab}}(j)>5$ GeV)
     &  13210~pb                                 
     &  9670~pb                                 
     &  10750~pb  \\
       \mbox{cone} ($p_T^{\mboxsc{lab}}(j)>0.5$ GeV)
     &  13950~pb=$\sigma_{\mboxsc{tot}}$
     &  
     &  12645~pb=$\sigma_{\mboxsc{tot}}$  \\
       \mbox{$k_T$ scheme} ($E_T^2=Q^2$)
     &  13950~pb=$\sigma_{\mboxsc{tot}}$                                 
     &  not infrared safe
     &  not infrared safe  \\
       \mbox{$k_T$ scheme} ($E_T^2=Q^2/2$)
     &  13950~pb=$\sigma_{\mboxsc{tot}}$                                 
     &  11220~pb
     &  12200~pb  \\
       \mbox{$k_T$ scheme} ($E_T^2=40$ GeV$^2$)
     &  13950~pb=$\sigma_{\mboxsc{tot}}$                                 
     &  9050~pb
     &  10280~pb  \\
       \mbox{JADE} ($y_{\mboxsc{cut}}=0.04$)
     &  13950~pb=$\sigma_{\mboxsc{tot}}$                                 
     &  11304~pb
     &  12320~pb  \\[1mm]
\hline
\end{tabular}
\end{table}

Table~\ref{tab_j1} shows the effect of higher oder corrections
to the 1-jet cross section for the cone (defined in the lab frame), W, 
and $k_T$ (defined in the Breit frame) scheme. 
We find only small differences between the NLO cross sections in the
the $W$ and the JADE scheme and therefore
only results for the $W$ scheme are presented.

The LO 1-jet results listed
in the second column of table~\ref{tab_j1} are  identical for
all jet algorithm, besides the cone scheme with
a $p_T^{\mboxsc{lab}}(j)>5$ GeV cut in the first line.
This is due to the
$Q^2>40$ GeV$^2$ event selection cut, which 
implies that the  transverse momentum $p_T^{\mboxsc{lab}}$ 
of the scattered parton (= jet in LO) 
is always larger than 3 GeV with a peak around 6 GeV
(see Fig.~\ref{f_pt1j}a).
The parton's transverse momentum  is large enough to pass all jet
requirements $J_{1\leftarrow 1}$ (see Eq.~(\ref{jnndef}))
for the $k_T$, $W$ and also for the cone scheme with 
a $p_T^{\mboxsc{lab}}(j)>0.5$ GeV 
(or even $p_T^{\mboxsc{lab}}(j)>2$ GeV)  cut in table~\ref{tab_j1}.
For the $k_T$ scheme this is directly evident from the choices 
for the hard scattering scale (see below).
The small fraction of events with $p_T^{\mboxsc{lab}}(j)$ 
values below 5 GeV  in  Fig.~\ref{f_pt1j}a
corresponds to the difference  in the first two cross sections
for the cone scheme   in the second column  in table~\ref{tab_j1}.
Since the scattered parton falls into the central part
of the detector (dotted line in Fig.~\ref{f_eta1j}a)
the LO 1-jet cross sections of 13950 pb are identical to the total
LO parton model cross section (without imposing any cuts on
the hadronic activity).

\begin{figure}[hbt]
\vspace*{0.5in}            
\begin{picture}(0,0)(0,0)
\includegraphics{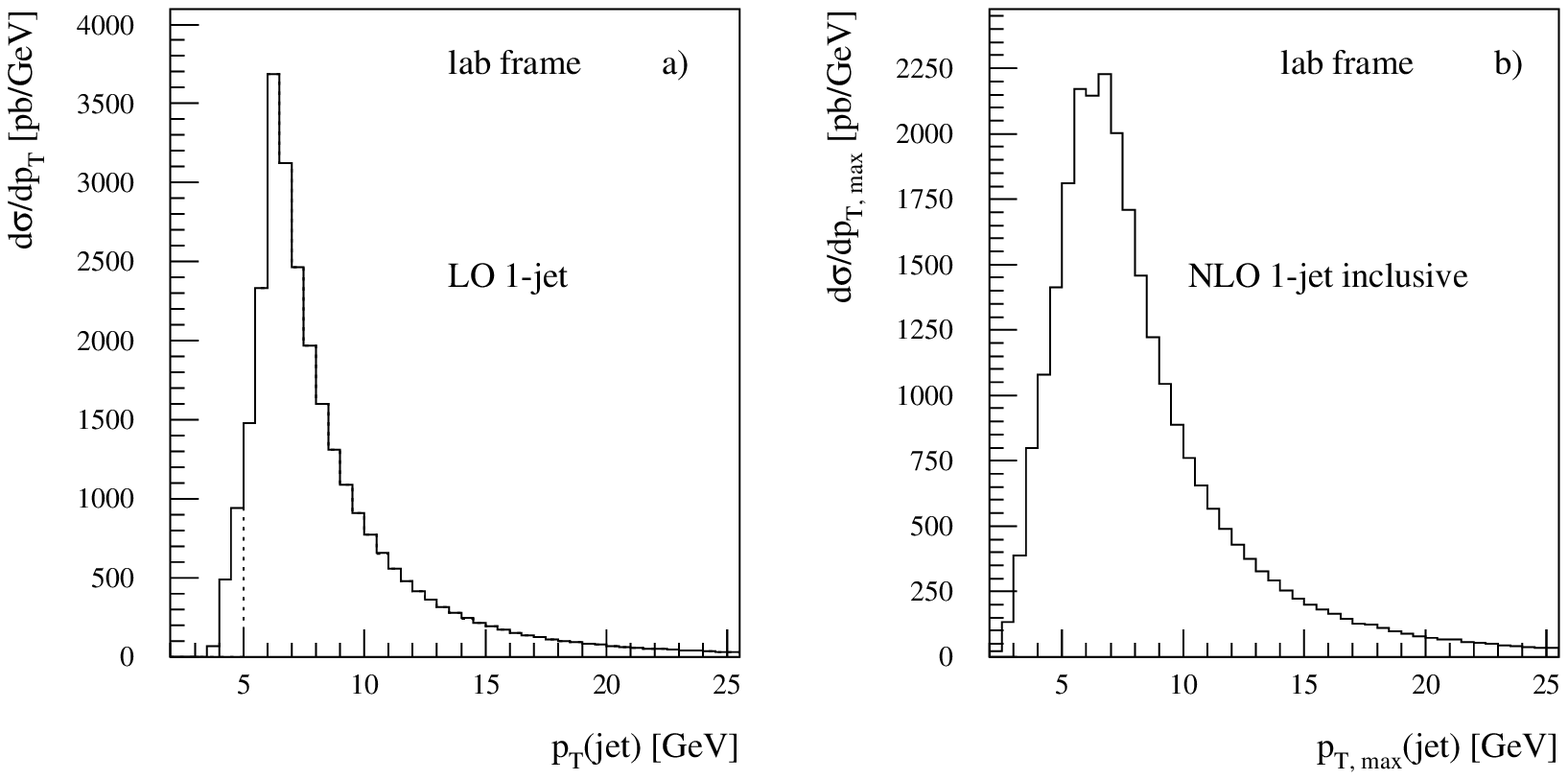}
\end{picture}
\vspace{6.cm}
\caption{
Transverse momentum distribution for jets in the lab frame.
Jets are defined in the cone scheme effectively without any
$p_T^{\protect\mboxsc{lab}}$ cut.
Results are shown for the  LO 1-jet cross section (a) and
for the jet with maximal transverse momentum in the
NLO 1-jet inclusive cross section (b).
}
\label{f_pt1j}
\end{figure}
\oas\ corrected 1-jet
inclusive cross sections, defined by the sum of the NLO
1-jet exclusive and the LO 2-jet cross sections,
are listed in the last column 
of table~\ref{tab_j1}.
Since the NLO 1-jet inclusive cross sections 
depend on the jet algorithms
the total inclusive \oas\ cross section 
is in general {\it not} given by the 1-jet inclusive cross section
\begin{equation}
\sigma_{\mboxsc{tot}}[\oas]\neq
\sigma_{\mboxsc{1-jet}}[\oas,\mbox{excl.}]+\sigma_{\mboxsc{2-jet}}[\oas]
\equiv \sigma_{\mboxsc{1-jet}}[\oas,\mbox{incl.}]
\label{sigtotdef}
\end{equation}
Similar to the discussion for the LO case
this is evident from the $p_T^{\mboxsc{lab}}$ distribution of the 
jet with maximum  transverse momentum
in the NLO inclusive calculation, which is shown 
by  in Fig.~\ref{f_pt1j}b.
About 85 \%  of the events in the inclusive cross section
contain at least one jet with
$p_T^{\mboxsc{lab}}>5$ GeV and all events contain at least one jet
with $p_T^{\mboxsc{lab}}>2$ GeV, which typically
falls into the central part of the detector
(see solid line in Fig.~\ref{f_eta1j}a).
Thus, the first equality in  Eq.~(\ref{sigtotdef}) is only correct
for sufficiently weak cuts on the jets, such as
nominal $p_T^{\mboxsc{lab}}>2$ GeV (or below) requirements 
for the jets in  cone scheme.

For completeness, LO and NLO results for 
the pseudo-rapidity distribution of the scattered lepton are shown 
in Fig.~\ref{f_eta1j}b.
%
\begin{figure}[hbt]
\vspace*{0.5in}            
\begin{picture}(0,0)(0,0)
\includegraphics{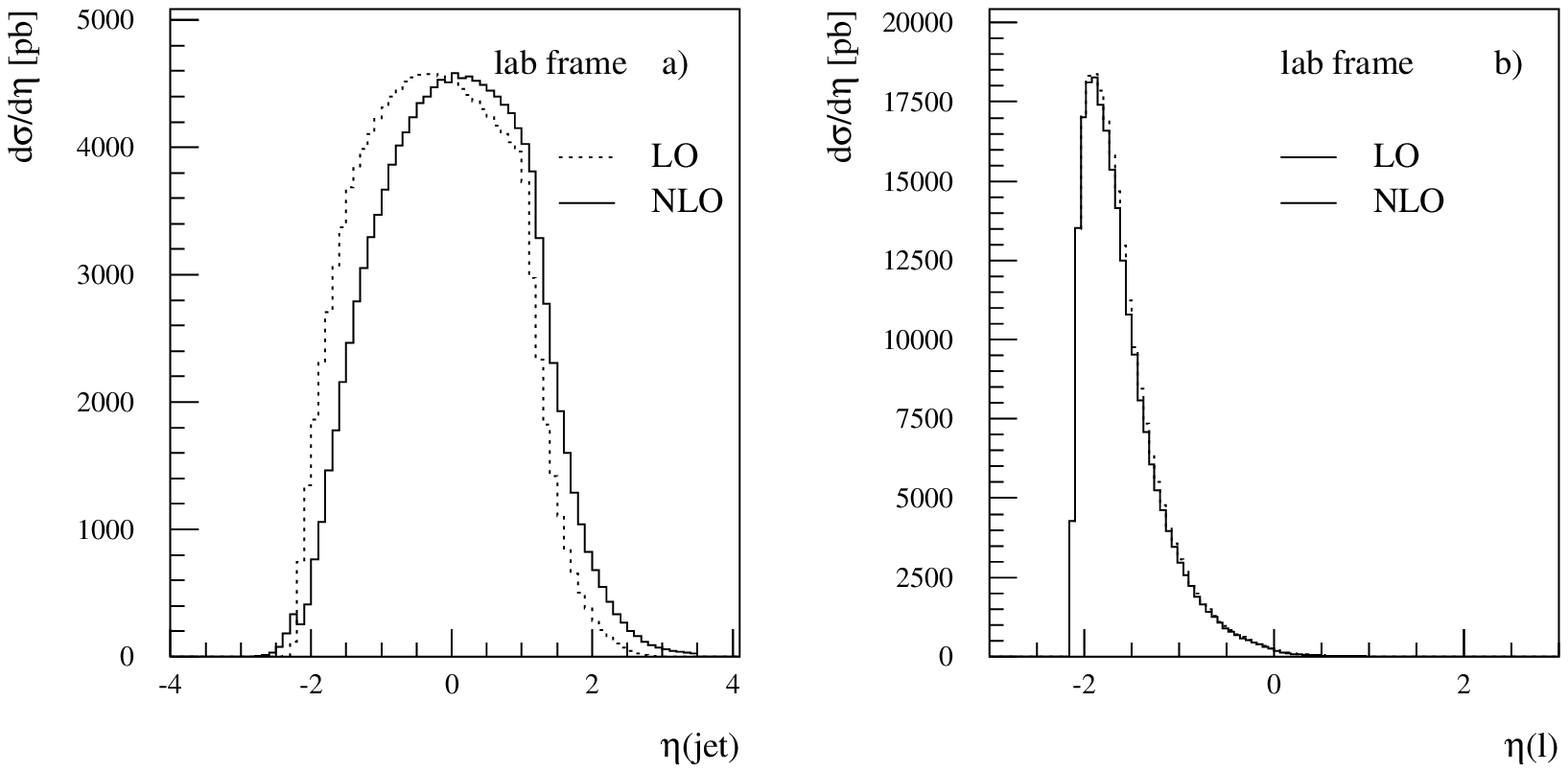}
\end{picture}
\vspace{6.5cm}
\caption{
a) Pseudo-rapidity distribution of the jet with the highest
$p_T^{\protect\mboxsc{lab}}$ in the total \protect\oas\ cross section (solid)
and pseudo-rapidity distribution of the parton
in the LO total cross section (dotted);
b) Pseudo-rapidity distribution of the scattered lepton
in the NLO (solid) and LO (dotted) 1-jet inclusive
cross section.
The $+z$ direction is defined in the direction of the proton.
}
\label{f_eta1j}
\end{figure}

Some comments are in order regarding 
1-jet cross sections 
in the $k_T$ algorithm,  when the hard scattering scale $E_T$
is chosen to be $Q$, as suggested in Ref.~\cite{kt}.
In LO, the parton model, $k_T^2(j)=Q^2$ and therefore
{\it every} event satisfies just the {\it minimum}
required jet criterion in the $k_T$ scheme.
In the NLO contribution from the two parton final 
state tree level matrix elements in Eqs.~(\ref{qtoqg},\ref{gtoqqbar}), 
however, almost none of the two partons has an
individual $k_T^2(j)\ge E_T^2=Q^2$ and hence there is almost
no  contribution from the two parton final state to the NLO 1-jet cross
section in the $k_T$ scheme with $E_T^2=Q^2$. 
The virtual corrections with the parton model kinematics
on the other hand give a negative contribution 
for {\it all} events, which finally
results in a unphysical {\it negative}  QCD corrected
1-jet cross section in the $k_T$ scheme, when $Q^2$ is chosen as the hard
scattering scale. Hence, the choice $k_T^2(j)\ge Q^2$ is {\it not}
an infrared safe jet definition in the 1-jet case.
However, the choice $k_T^2(j)\ge Q^2/2$ (fourth line in 
table~\ref{tab_j1}) or a fixed
scale like $k_T^2(j)\ge Q^2_{\mboxsc{min}}=40$ GeV$^2$ 
(fifth line in table~\ref{tab_j1}))
results in an infrared safe NLO cross section.

The difference between the 1-jet exclusive and 1-jet inclusive
NLO results in table~\ref{tab_j1} corresponds to the \oas\ 2-jet cross
section in the given jet algorithm.

Finally, Fig.~\ref{f_kfac1j} shows the LO and NLO $x, Q^2$ and $y$ dependence
of the total cross section  together with the resulting $K$-factors,
(defined as $K=\sigma_{NLO}/\sigma_{LO}$).
We find that $K$ is always smaller than one with a maximum deviation from
unity of less than 5 \%.
%
\begin{figure}[htb]
\vspace*{0.5in}            
\begin{picture}(0,0)(0,0)
\includegraphics{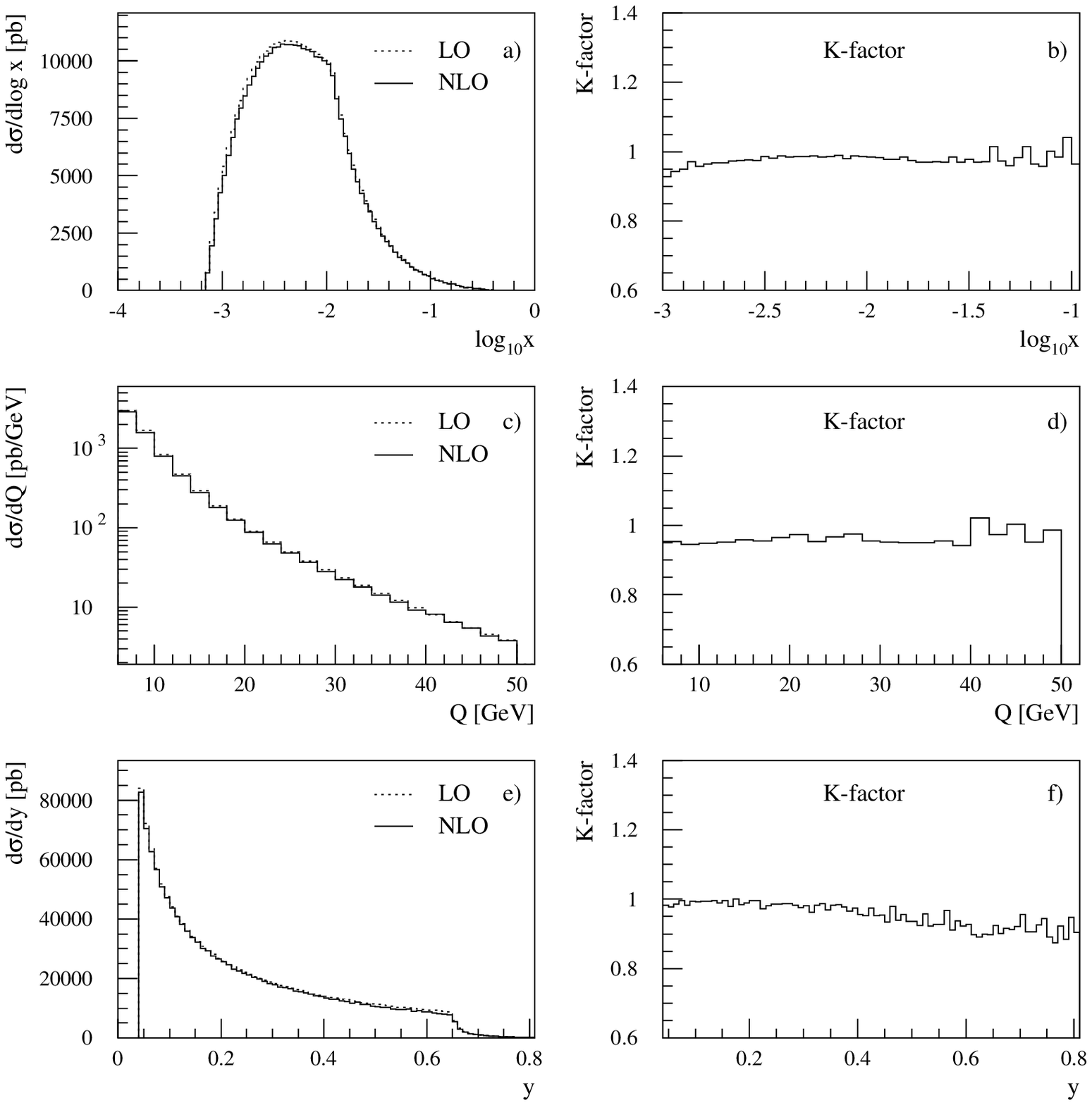}
\end{picture}
\vspace{13.5cm}
\caption{LO and NLO inclusive cross sections
and $K$-factors as a function of Bjorken $x$ (a,b),
$Q$ (c,d), and the leptonic scaling variable $y$ (e,f).
Results are shown for HERA energies with cuts on the
scattered lepton as described in the text.
LO (NLO) results are based on LO (NLO) GRV parton distributions
with the 1-loop (2-loop) formula for the strong coupling constant.
}
\label{f_kfac1j}
\end{figure}
\newpage
\subsection[{NLO One-Jet 
Cross Sections Including $Z$ and $W$ 
Exchange\protect\vspace{1mm}}]{NLO One-Jet 
Cross Sections Including {\boldmath{$Z$}} and {\boldmath{$W$}} Exchange}
\label{onejetz}
For very high  $Q^2$ ({\it e.g.} up to $10^5$ GeV$^2$ at HERA)
the $\gamma-Z$ interference term and the pure $Z$ and $W$ exchange
become also important.
Analytical and numerical results for LO and NLO 
1-jet and total cross sections including 
these electroweak effects are presented in this section.
The analytical results for the hadronic cross sections can
be obtained from the 1-photon exchange result in
Eq.~(\ref{onejet}) by suitable replacements as listed below.
 
\subsubsection{Matrix Elements and Coupling Factors\protect\vspace{1mm}}
\label{couplings}
Let us first specify the couplings of the quarks to the
weak current. The couplings of the photon to a quark with flavor $f$ via the
vector current is given by the minimal charge coupling $e_f\gamma^\mu$
of the quark (in units of $e$). The coupling factors of the $Z$
are specified in the following way:\\
{\it hadronic coupling to quark with flavor $f$:}
\begin{equation}
\Gamma^\mu=\frac{v_f}{2\sin 2 \theta_W}\gamma^\mu+
\frac{a_f}{2\sin 2\theta_W}\gamma^\mu\gamma^5
\label{hadcoupl}
\end{equation}
{\it leptonic coupling:}
\begin{equation}
\Gamma^\mu=\frac{v_e}{2\sin 2 \theta_W}\gamma^\mu+
\frac{a_e}{2\sin 2\theta_W}\gamma^\mu\gamma^5
\label{lepcoupl}
\end{equation}
where
\begin{eqnarray}
v_f=2t_f-4e_f\sin^2\theta_W, && \hspace{2cm} a_f=2t_f \nonumber \\
v_e=-1+4\sin^2\theta_W,    && \hspace{2cm} a_e=-1
\end{eqnarray}
$t_f$ denotes the third component of the weak isospin
of the $f$-type quark ($t_{u,c,t}=1/2,t_{d,s,b}=-1/2$) 
and $e_f$ represents its charge.
$\theta_W$ is the weak mixing angle (Weinberg-angle)
which is defined by the ratio of the charged and neutral weak 
gauge boson masses, $\cos\theta_W=M_W/M_Z$.

It is useful to define the following
$Q^2$ dependent electroweak factors 
for the NC $e^{\pm}p$ cross section as\footnote{We have corrected a sign error 
in Ref.~\cite{herai} in the coefficient $4v_ea_ev_fa_f$ on the r.h.s of 
Eq.~(\protect\ref{bfdef}).}
(see {\em e.g.} Ref.~\cite{herai})
\begin{eqnarray}
A_f(Q^2)&=& e_f^2-2e_fv_ev_f\,\chi_Z
+ (v_e^2+a_e^2)(v_f^2+a_f^2)\chi_Z^2
\label{afdef}\\
B_f(Q^2)&=& \pm(2e_fa_ea_f\,\chi_Z-4v_ea_ev_fa_f\chi_Z^2)
\label{bfdef}
\end{eqnarray}
Terms which are linear in $\chi_Z$ in Eqs.~(\ref{afdef},\ref{bfdef})
arise from $\gamma/Z$ interference while those which are quadratic in $\chi_Z$
are due to pure $Z$ exchange.
Here $\chi_Z(Q^2)$ is the ratio of the $Z$ propagator to the photon
propagator times the coupling strength factor $(2\sin 2\theta_W)^{-2}=
(4\sin\theta_W\cos\theta_W)^{-2}$:
\begin{equation}
\chi_Z(Q^2)=\frac{1}{(2\sin 2\theta_W)^2}
\frac{Q^2}{Q^2+M_Z^2}
\end{equation}
The NC NLO hadronic  1-jet cross section including
all $\gamma-\gamma, \gamma-Z$ and $Z-Z$ contributions
can be obtained from Eq.~(\ref{onejet})
by the following replacements:
\begin{eqnarray}
\mbox{third line:}\hspace{2cm}&& \nonumber  \\[2mm] \mbox{}
[\sum_{i=q,\bar{q}}e_i^2 f_i(\eta)] 
\,\, |M^{(\mboxsc{pc})}_{q\rightarrow q}|^2 
&\longrightarrow&
[\sum_{i=q,\bar{q}} A_i(Q^2) f_i(\eta)] \,
\,\,|M^{(\mboxsc{pc})}_{q\rightarrow q}|^2 
+
[\sum_{i=q} B_i(Q^2) 
\tilde\Delta f_i(\eta)]
\,\,|M^{(\mboxsc{pv})}_{q\rightarrow q}|^2 \nonumber \\[2mm]
\mbox{fourth line:}\hspace{2cm}&& \nonumber  \\[2mm] \mbox{}
[\sum_{i=q,\bar{q}}e_i^2 C_i^{\overline{\mboxsc{MS}}}(\eta)] 
\,\,|M^{(\mboxsc{pc})}_{q\rightarrow q}|^2
&\longrightarrow&\!
[\sum_{i=q,\bar{q}} A_i(Q^2) C_i^{\overline{\mboxsc{MS}}}(\eta)] 
\,\,|M^{(\mboxsc{pc})}_{q\rightarrow q}|^2 
+
[\sum_{i=q} B_i(Q^2) \tilde\Delta C_i^{\overline{\mboxsc{MS}}}(\eta)] 
\,\,|M^{(\mboxsc{pv})}_{q\rightarrow q}|^2 \nonumber \\[2mm]
\mbox{sixth line:}\hspace{2cm}&& \nonumber  \\[2mm] \mbox{}
[\sum_{i=q,\bar{q}}e_i^2 f_i(\eta]
\,\,|M^{(\mboxsc{pc})}_{q\rightarrow qg}|^2 
&\longrightarrow&
[\sum_{i=q,\bar{q}} A_i(Q^2) f_i(\eta)]
\,\,|M^{(\mboxsc{pc})}_{q\rightarrow qg}|^2 
+
[\sum_{i=q} B_i(Q^2) \tilde\Delta f_i(\eta)]
\,\,|M^{(\mboxsc{pv})}_{q\rightarrow qg}|^2 \nonumber \\[2mm]
\mbox{seventh line:}\hspace{2cm}&&\nonumber \\[2mm]
(\sum_{i=q}e_i^2) f_g(\eta)\,\,
\,\,|M^{(\mboxsc{pc})}_{g\rightarrow q\bar{q}}|^2
&\longrightarrow&
(\sum_{i=q}A_i(Q^2) ) f_g(\eta)\,\,
|M^{(\mboxsc{pc})}_{g\rightarrow q\bar{q}}|^2 
+
(\sum_{i=q}B_i(Q^2)) f_g(\eta)\,\,
\,|M^{(\mboxsc{pv})}_{g\rightarrow q\bar{q}}|^2  \nonumber 
\nonumber
\end{eqnarray}
where\footnote{We  use the notation 
 $\Delta f_q(\eta)
\equiv f_{q \uparrow}(\eta)-f_{q \downarrow}(\eta)$ (without tilde)
for the definition of the polarized parton densities,
where $f_{q \uparrow} (f_{q \downarrow})$ denotes 
the probability to find a quark $q$ 
in the longitudinally polarized  
proton whose spin is aligned (anti-aligned) to the proton's spin
(see section \protect\ref{poljets}).}
\begin{eqnarray}
\tilde\Delta f_q(\eta) &\equiv& f_q(\eta)-f_{\bar{q}}(\eta) 
\label{deltaqdef}\\[2mm]
\tilde\Delta C_q^{\overline{\mboxsc{MS}}}(\eta) 
&\equiv& C^{\overline{\mboxsc{MS}}}_q(\eta)
-\
C^{\overline{\mboxsc{MS}}}_{\bar{q}}(\eta) 
\label{deltacdef}
\end{eqnarray}
and 
\begin{eqnarray}
|M^{(\mboxsc{pv})}_{q\rightarrow q}|^2
&=&
32\,\left[
(p_0.l)^2 \, -\,
(p_0.l^\prime)^2 
\, \right]
= 8 \hat{s}^2\,\,(1-(1-y)^2)
\label{m_qtoqpv}\\[1mm]
|M^{(\mboxsc{pv})}_{q\rightarrow qg}|^2
&=&
\frac{128}{3}\,\,(l.l^\prime)\,
\frac{(l.p_0)^2-(l^\prime.p_0)^2-(l.p_1)^2+(l^\prime.p_1)^2
     }{(p_1.p_2)(p_0.p_2)}
\label{m_qtoqgpv}\\[1mm]
|M^{(\mboxsc{pv})}_{g\rightarrow q\bar{q}}|^2
&=&
-\frac{3}{8}\,\,
|M^{(\mboxsc{pv})}_{q\rightarrow qg}|^2(p_0\leftrightarrow-p_2)\nonumber\\
&=& 
16\,\,(l.l^\prime)\,
\frac{(l.p_2)^2-(l^\prime.p_2)^2-(l.p_1)^2+(l^\prime.p_1)^2
     }{(p_0.p_1)(p_0.p_2)}
\label{m_gtoqqbarpv}
\end{eqnarray}
The superscript pv ($\equiv$ parity violating) refers to the interference
term of the vector and axial vector current in these contributions 
(see Eq.~(\ref{sum_mqtoqgpv})) which factorize the coupling factor
$B_f(Q^2)$ defined in Eq.~(\ref{bfdef}).
Note that the parity violating squared matrix element
$|M^{(\mboxsc{pv})}_{g\rightarrow q\bar{q}}|^2$
is anti-symmetric under quark-antiquark ($p_1\leftrightarrow p_2$) 
exchange, which
implies that flavor tagging would be required to detect this
gluon initiated part of the cross section.
The matrix elements in Eqs.~(\ref{m_qtoqgpv},\ref{m_gtoqqbarpv}) 
are derived in sect.~\ref{app_born}.

For CC $e^\pm p$ scattering  one has to replace
$A_f(Q^2)$ and $B_f(Q^2)$ in Eqs.~(\ref{afdef},\ref{bfdef})
by
\begin{equation}
A_f(Q^2)\rightarrow \chi_W^2\,|V_{f,f^\prime}|^2,
\hspace{1cm}
B_f(Q^2)\rightarrow \pm \chi_W^2\,|V_{f,f^\prime}|^2,
\label{abccdef}
\end{equation}
where
\begin{equation}
\chi_W=\frac{1}{(4\sin\theta^2_W)}\frac{Q^2}{Q^2+M_W^2}
\label{chiwdef}
\end{equation}
and $V_{ff^\prime}$ denotes the Kobayashi-Maskawa matrix
element \cite{kmmat}
for the charged current $f\rightarrow f^\prime$ transitions.

With these coupling factors
the CC hadronic 1-jet cross section
for $e^-p$ scattering is obtained from Eq.~(\ref{onejet})
by the following replacements:
\begin{eqnarray}
\mbox{third line:}\hspace{2cm}&& \nonumber \\[2mm] \mbox{}
[\sum_{i=q,\bar{q}}e_i^2 f_i(\eta)] 
\,\, |M^{(\mboxsc{pc})}_{q\rightarrow q}|^2 
&\longrightarrow&
\chi_W^2 
\bigg[
(\sum_{i=u,c} f_i(\eta)) 
\,\,\,|M^{(\mboxsc{pc+pv})}_{q\rightarrow q}|^2 
+
(\sum_{i=\bar{d},\bar{s}} f_i(\eta)) 
\,\,\,|M^{(\mboxsc{pc--pv})}_{q\rightarrow q}|^2 
\bigg]\nonumber
\\[2mm]
\mbox{fourth line:}\hspace{2cm}&&  \nonumber\\[2mm] \mbox{}
[\sum_{i=q,\bar{q}}e_i^2 C_i^{\overline{\mboxsc{MS}}}(\eta)] 
\,\,|M^{(\mboxsc{pc})}_{q\rightarrow q}|^2
&\longrightarrow&
\chi_W^2
\bigg[
(\sum_{i=u,c} C_i^{\overline{\mboxsc{MS}}}(\eta)) 
\,\,\,|M^{(\mboxsc{pc+pv})}_{q\rightarrow q}|^2 
+
(\sum_{i=\bar{d},\bar{s}} C_i^{\overline{\mboxsc{MS}}}(\eta)) 
\,\,\,|M^{(\mboxsc{pc--pv})}_{q\rightarrow q}|^2
\bigg] \nonumber\\[2mm]
\mbox{sixth line:}\hspace{2cm}&&  \nonumber\\[2mm] \mbox{}
[\sum_{i=q,\bar{q}}e_i^2 f_i(\eta]
\,\,|M^{(\mboxsc{pc})}_{q\rightarrow qg}|^2 
&\longrightarrow&
\chi_W^2
\bigg[
(\sum_{i=u,c}  f_i(\eta))
\,\,|M^{(\mboxsc{pc+pv})}_{q\rightarrow qg}|^2 
+
(\sum_{i=\bar{d},\bar{s}} f_i(\eta))
\,\,\,|M^{(\mboxsc{pc--pv})}_{q\rightarrow qg}|^2 \bigg]
\nonumber\\[2mm]
\mbox{seventh line:}\hspace{2cm}&& \nonumber\\[2mm]
(\sum_{i=q}e_i^2) f_g(\eta)\,\,
\,\,|M^{(\mboxsc{pc})}_{g\rightarrow q\bar{q}}|^2
&\longrightarrow&
2\chi_W^2 f_g(\eta)\,\,
|M^{(\mboxsc{pc--pv})}_{g\rightarrow q\bar{q}}|^2 
\nonumber
\end{eqnarray}
For CC $e^+p$ scattering one has to replace
$\sum_{i=u,c}$ ($\sum_{i=\bar{d},\bar{s}}$) by
$\sum_{i=\bar{u},\bar{c}}$ ($\sum_{i=d,s}$), respectively, 
in the previous equations and 
$|M^{(\mboxsc{pc--pv})}_{g\rightarrow q\bar{q}}|^2$
by
$|M^{(\mboxsc{pc+pv})}_{g\rightarrow q\bar{q}}|^2$
in the last line.
The sum and differences of the parity conserved and parity violating
squared matrix elements 
\begin{equation}
|M^{(\mboxsc{pc$\pm$pv})}_{a\rightarrow \ldots}|^2 := 
|M^{(\mboxsc{pc})}_{a\rightarrow \ldots}|^2\pm 
|M^{(\mboxsc{pv})}_{a\rightarrow \ldots}|^2
\label{sumdiff}
\end{equation}
are
\begin{eqnarray}
|M^{(\mboxsc{pc+pv})}_{q\rightarrow q}|^2 &=&
64\,(p_0.l)^2\,=16\hat{s}^2
\hspace{3.7cm}\mbox{(from Eqs.~(\ref{m_qtoq},\ref{m_qtoqpv})})\\[2mm]
|M^{(\mboxsc{pc--pv})}_{q\rightarrow q}|^2 &=&
64\,(p_0.l^\prime)^2\,=16\hat{s}^2(1-y)^2
\hspace{2.2cm}\mbox{(from Eqs.~(\ref{m_qtoq},\ref{m_qtoqpv})})\\[1mm]
|M^{(\mboxsc{pv+pc})}_{q\rightarrow qg}|^2&=&
\frac{256}{3}\,\,(l.l^\prime)
\,\frac{(l.p_0)^2+(l^\prime.p_1)^2}{(p_1.p_2)(p_0.p_2)}
\label{qtoqgsum}
\hspace{2cm}\mbox{(from Eqs.~(\ref{m_qtoqg},\ref{m_qtoqgpv})})\\[1mm]
|M^{(\mboxsc{pv-pc})}_{q\rightarrow qg}|^2&=&
\frac{256}{3}\,\,(l.l^\prime)\,
\frac{(l^\prime.p_0)^2+(l.p_1)^2}{(p_1.p_2)(p_0.p_2)}
\label{qtoqgdiff}
\hspace{2cm}\mbox{(from Eqs.~(\ref{m_qtoqg},\ref{m_qtoqgpv})})\\[1mm]
|M^{(\mboxsc{pc--pv})}_{g\rightarrow q\bar{q}}|^2 &=&
32\,\,(l.l^\prime)\,
\frac{(l^\prime.p_2)^2+(l.p_1)^2}{(p_0.p_1)(p_0.p_2)}
\label{qtoqqbardiff}
\hspace{2.3cm}\mbox{(from Eqs.~(\ref{m_gtoqqbar},\ref{m_gtoqqbarpv}))}\\[1mm]
|M^{(\mboxsc{pc+pv})}_{g\rightarrow q\bar{q}}|^2 &=&
32\,\,(l.l^\prime)\,
\frac{(l.p_2)^2+(l^\prime.p_1)^2}{(p_0.p_1)(p_0.p_2)}
\label{qtoqqbarsum}
\hspace{2.3cm}\mbox{(from Eqs.~(\ref{m_gtoqqbar},\ref{m_gtoqqbarpv}))}
\end{eqnarray}
%
%
\newpage
\subsubsection{Numerical Results\protect\vspace{1mm}}
\label{numzw1}
The numerical studies in this section
show effects of the $Z$ and $W$ exchange in LO and NLO
1-jet cross sections.
In contrast to the 1-photon exchange case these
results differ for $e^+p$ and $e^-p$ scattering
(see Eqs.~(\ref{bfdef},\ref{abccdef}) and the replacements
listed before Eq.~(\ref{sumdiff})).
All results are given for HERA energies.

Jets are defined in a cone scheme 
(in the laboratory frame) with radius $R=1$
and $p_T^{\mboxsc{lab}}(j)>5$~GeV.
In addition, we require
$0.04 < y < 1$, an energy cut of $E(e^\prime)>10$~GeV on the scattered 
lepton (in NC scattering) and a cut on the pseudo-rapidity 
$
\eta=-\ln\tan(\theta/2)
$
of the scattered lepton and jets of $|\eta|<3.5$. 
The results in Figs.~\ref{f_zq1j} and \ref{f_wq1j}  are 
based on  MRS Set (R1) \cite{mrsr1}
parton distributions with the two-loop formula in Eq.~(\ref{asnlo})
for the strong coupling constant, 
whereas LO (NLO) GRV parton distributions \cite{grv}
are used for $\sigma^{\mboxsc{LO}}$ ($\sigma^{\mboxsc{NLO}}$)
in the results for the
$K$-factors (defined by $\sigma^{\mboxsc{NLO}}/\sigma^{\mboxsc{LO}}$)
in Fig.~\ref{f_zw1j_q_kfac}.

Fig.~\ref{f_zq1j}a shows the $Q^2$ distributions 
for NC $e^+p$ scattering.
Results are given for complete NC $\gamma^\ast$ and $Z$ exchange
(solid), for pure $\gamma^\ast$ (dot-dashed) and for
pure $Z$ (dotted) exchange. 
Sizable electroweak effects are observed for
$Q^2\gsimfig 2500$ GeV$^2$.
The electroweak effects in $e^+p$ scattering are dominated by the
{\it negative} $\gamma^\ast/Z$ interference
contribution.
The size of the effect is shown 
in Fig.~\ref{f_zq1j}b, where the ratio of the complete NC result and the
1-photon result is shown as a function of $Q^2$.
The $\gamma^\ast/Z$ interference
reduces the 1-photon result by about 27\% 
at $Q^2=10000$ GeV$^2$, which--together with the positive contribution 
of about 12\% from the pure $Z$ exchange (dotted line in
Fig.~\ref{f_zq1j}b)--results in a NC cross section 
which is about 15\% smaller compared to the 1-photon exchange result
at $Q^2=10000$ GeV$^2$.
For very high $Q^2$, the electroweak effects lower the 1-photon exchange
cross section result even by more than a factor two.


The situation is fairly different for $e^-p$ NC scattering.
Results are shown in ~Fig.~\ref{f_zq1j}c,d.
The $\gamma^\ast/Z$ interference is now positive and leads
together with the positive pure $Z$ exchange contribution
(dotted line in ~Fig.~\ref{f_zq1j}c) to a large
enhancement  of the NC cross section at high $Q^2$.
The ratio of the complete NC cross section
and the 1-photon exchange cross section, 
shown by the solid line in Fig.~\ref{f_zq1j}b,
is about 1.3 at $Q^2$=10000 GeV$^2$
and can reach more than a factor two at the upper kinematical limit.

Fig.~\ref{f_wq1j} compares the 
1-jet cross sections for CC $W^\pm$ exchange with the complete
NC ($\gamma^\ast$ and $Z$) results.
For $e^+p$ scattering, the CC cross section
is always considerably smaller than the
NC cross section, reaching a maximum of about  20\%  of the NC cross section
at $Q^2\approx 10000$ GeV$^2$ (solid curve in Fig.~\ref{f_wq1j}b).
For $e^-p$ scattering and  $Q^2\gsimfig 10000$ GeV$^2$, however, the CC
cross section becomes even larger than the NC cross section
(dashed curve in  Fig.~\ref{f_wq1j}b).
The relative importance of the $e^-p$ and $e^+p$ 1-jet
cross section is shown in Fig.~\ref{f_zwrel1j} as a function of $Q^2$ 
for the NC (solid) and CC (dashed) exchange.
The strong drop in the dashed curve with increasing $Q^2$
is caused by the vanishing $d(x)/u(x)$ ratio in the
contributing valence quark densities (see section~\ref{couplings})
for $x\rightarrow 1$.

We will show later that the relative importance of the electroweak
effects is largely {\it independent} on the jet multiplicity $n$.
Thus, the ratios
\begin{equation}
\frac{\sigma^{\gamma^\ast+Z}(Q^2)[\mbox{n-jet}]}{
      \sigma^{\gamma^\ast  }(Q^2)[\mbox{n-jet}]};
\hspace{1cm}
\hspace{1cm}
\frac{\sigma^{CC}(Q^2)[\mbox{n-jet}]}{
      \sigma^{NC}(Q^2)[\mbox{n-jet}]};
\hspace{1cm}
\hspace{1cm}
\frac{\sigma^{e^+p}(Q^2)[\mbox{n-jet}]}{
      \sigma^{e^-p}(Q^2)[\mbox{n-jet}]}
\label{ewratio}
\end{equation}
for $n=2,3,4$ 
are very similar to the results for $n=1$ in Fig.~\ref{f_zq1j}b,d,
Fig.~\ref{f_wq1j}b and Fig.~\ref{f_zwrel1j}.
\newpage
The QCD corrections to the electroweak cross sections are investigated in
Fig.~\ref{f_zw1j_q_kfac} where the $Q^2$ dependence of the
$K$-factor is shown for NC (a) and CC (b) scattering.
The $K$-factor in the NC case is very similar for $e^+p$ and $e^-p$ 
scattering. In the 1-jet-inclusive case $K$ is
always larger than 0.9 and approaches 1 in the high $Q^2$ limit.
In the high $Q^2$ regime, the 1-jet-inclusive cross section
is identical to the total $\oas$ cross section
whereas effects discussed in connection with
Eq.~(\ref{sigtotdef}) are responsible for the smaller $K$-factor
at lower $Q^2$ values.
The $K$-factor for the 1-jet-exclusive cross section,
defined as the difference of the 1-jet-inclusive cross section and the
$\oas$-2-jet cross section, is about 0.8 for the whole $Q^2$ range.

The $K$-factors for the CC 1-jet cross section differ for
$e^+p$ and $e^-p$ scattering at large $Q^2>10000$ GeV$^2$.
The CC cross section for $e^+p$ scattering is
considerably smaller than the corresponding  $e^-p$
cross sections (see Fig.~\ref{f_zwrel1j}) and resummation
effects due to the effectively strongly restricted phase space
are expected to become important.
Note that the $K$-factor for the 1-jet-exclusive
scross section in CC events decreases from about
0.88 at $Q^2=100$ GeV$^2$ to about 0.75 at $Q^2=10000$ GeV$^2$.


Both HERA experiments have recently reported an excess of 
$e^+p$  NC 1-jet events of about a factor two 
above the Standard Model predictions
at large values of $Q^2>1.5\times 10^4$ GeV$^2$ \cite{highq2},
which lead to many speculations  for new  physics (see {\it e.g.}
Ref.~\cite{dieterhighq2} and references therein).
Fig.~\ref{f_highq2} compares updated H1 and ZEUS
data \cite{eps} with the full 1-loop QCD corrected
total cross section (without any jet  requirement)
and the NLO 1-jet inclusive result, where
at least one jet with $E_T^{\mboxsc{lab}}>15$ GeV is required.
Additional cuts are given in the figure caption.
Taking  the  requirement of at least one jet into account in the 
calculation  increases the discrepancy between the
SM prediction and data. 
\begin{figure}[ht]
  \centering
 \mbox{\epsfig{file=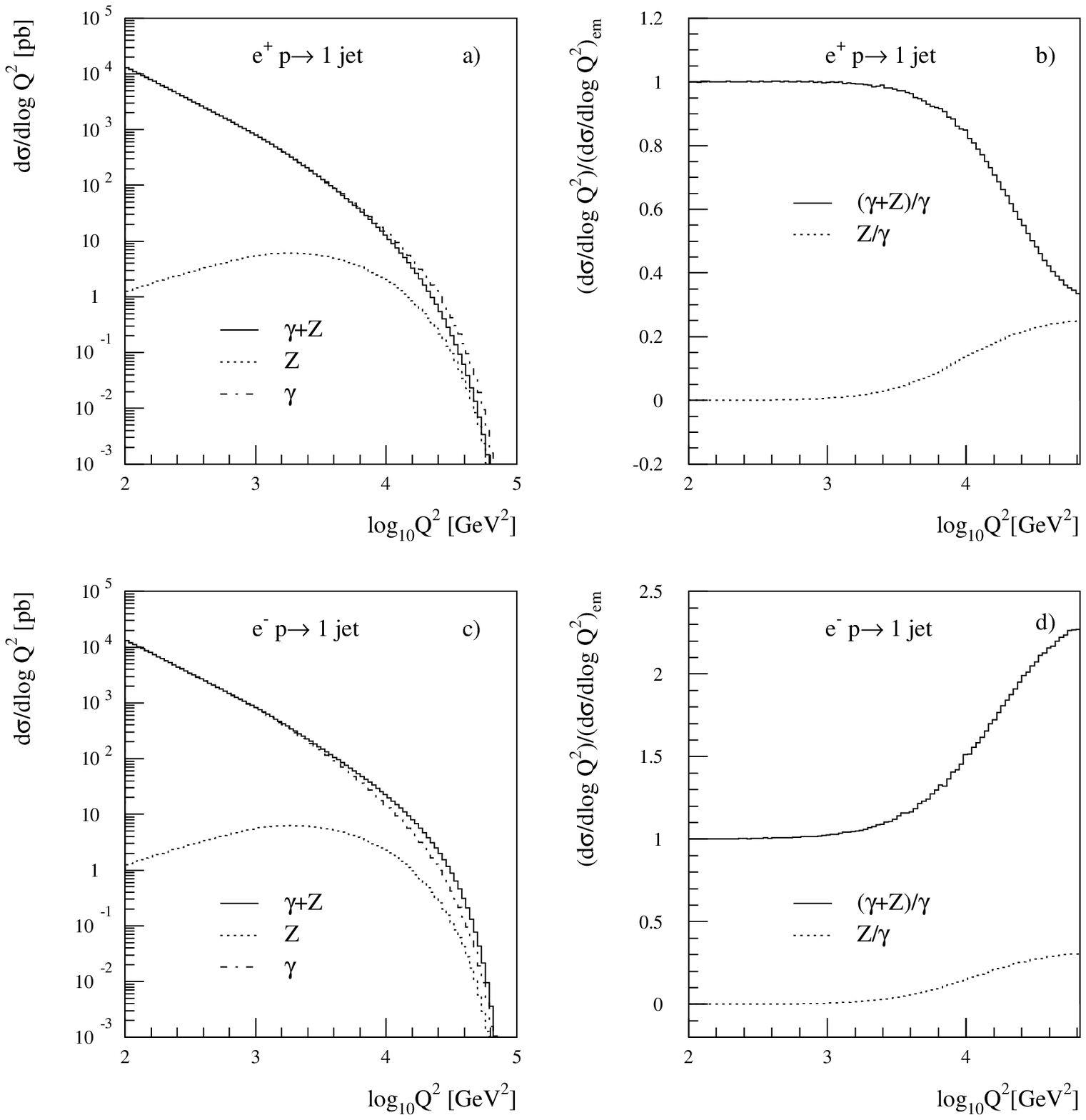,bbllx=0,bblly=30,
               bburx=540,bbury=540,width=0.95\linewidth}} 
\caption{
$Q^2$ dependence of the 1-jet cross section 
for  $e^+p$ (a) and $e^-p$ (c)  scattering in
NC $\gamma^\ast/Z$ (solid), $\gamma^\ast$ (dot-dashed) 
and pure $Z$ (dotted) exchange.
(b),(d) Ratio of the $\gamma^\ast/Z$ (solid)  and pure $Z$ exchange results
(dotted) and the 1-photon exchange results.
Jet are defined in a cone scheme in the lab frame
with radius $R=1$. 
Additional cuts are explained in the text.
}
\label{f_zq1j}
\end{figure}
%
\enlargethispage{1cm}
\begin{figure}[ht]
  \centering
 \mbox{\epsfig{file=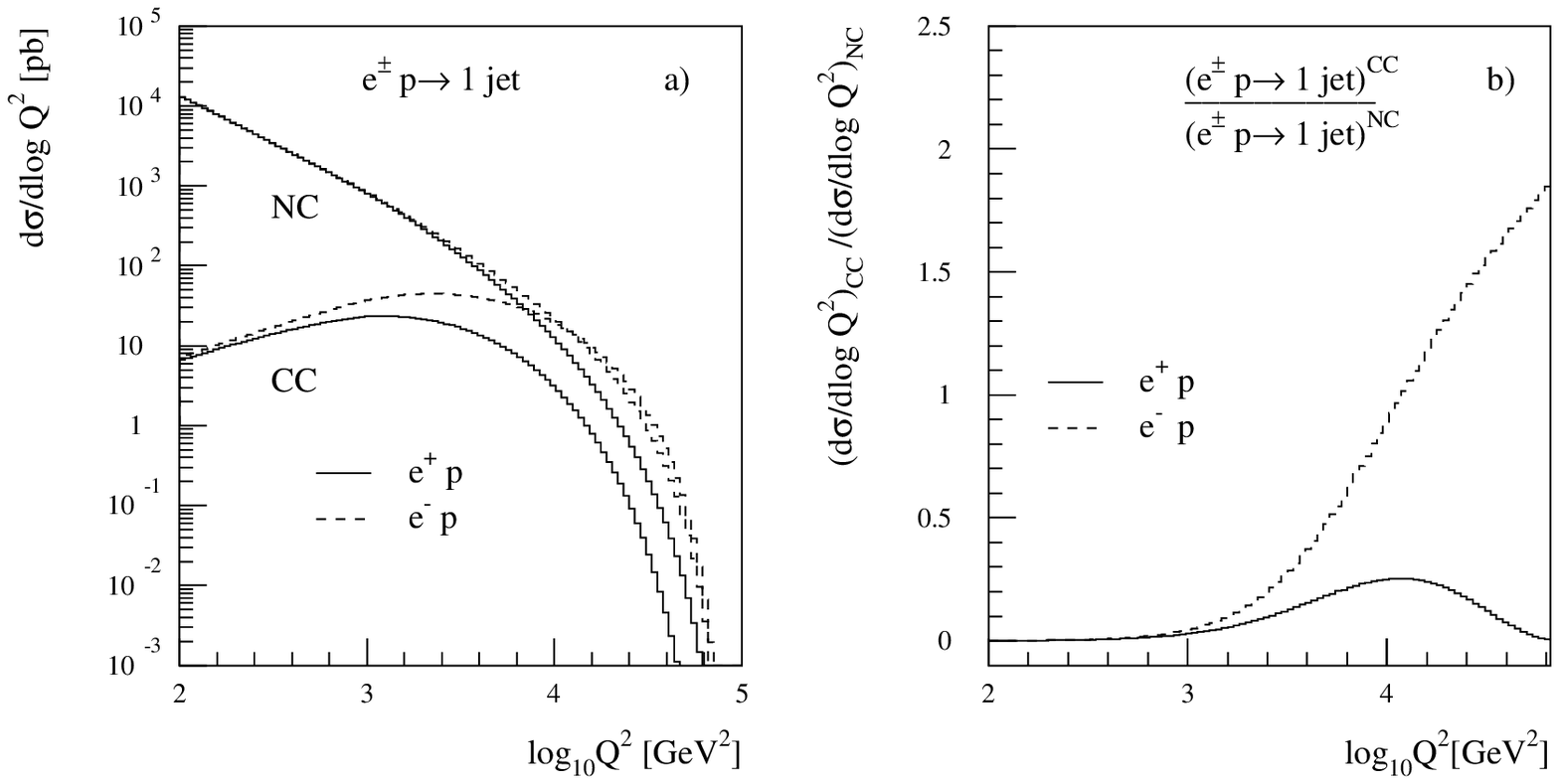,bbllx=10,bblly=290,
               bburx=540,bbury=540,width=0.95\linewidth}} 
\vspace*{5mm}
\caption{
$Q^2$ dependence of the  CC 
and NC 1-jet cross sections for
$e^+p$ (solid) and $e^-p$ (dashed) scattering (a) and 
the relative contributions of CC and NC 1-jet cross sections (b)
in $e^+p$ (solid) and $e^-p$ (dashed) scattering as a function of $Q^2$.
}
\label{f_wq1j}
\end{figure}
%
%
\begin{figure}[ht]
  \centering
 \mbox{\epsfig{file=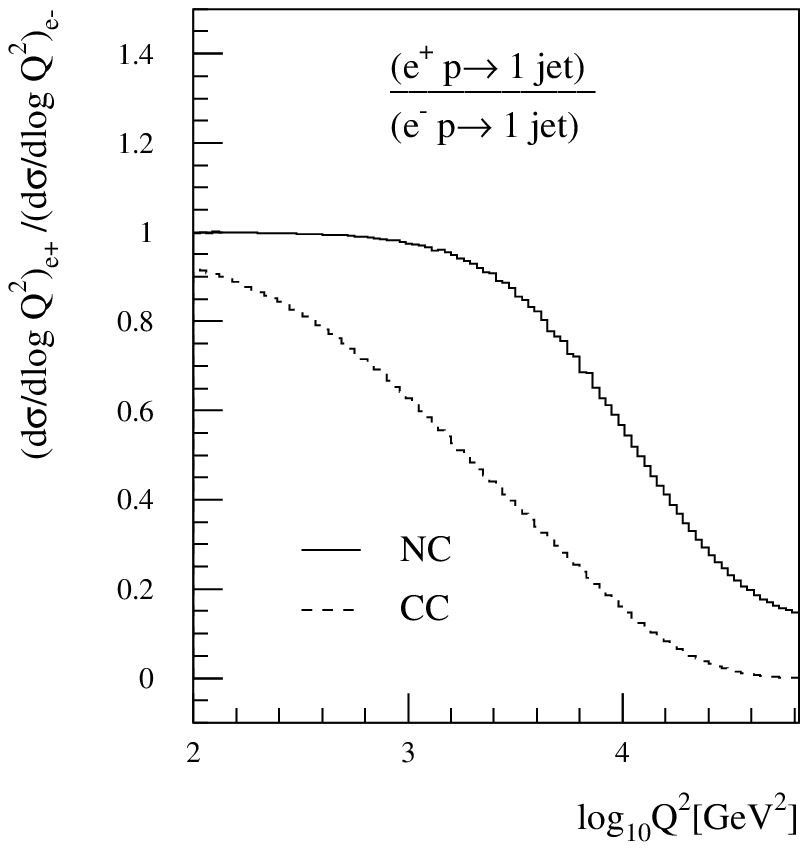,
               bbllx=15,bblly=250,bburx=290,bbury=520,
               width=0.5\linewidth}} 
\vspace*{-7mm}
\caption{
Ratio of  $e^+p$ to $e^-p$ 1-jet cross sections as a function of $Q^2$
for NC (solid) and CC (dashed) exchange.
}
\label{f_zwrel1j}
\end{figure}
%
%
\begin{figure}[ht]
  \centering
  \mbox{\epsfig{file=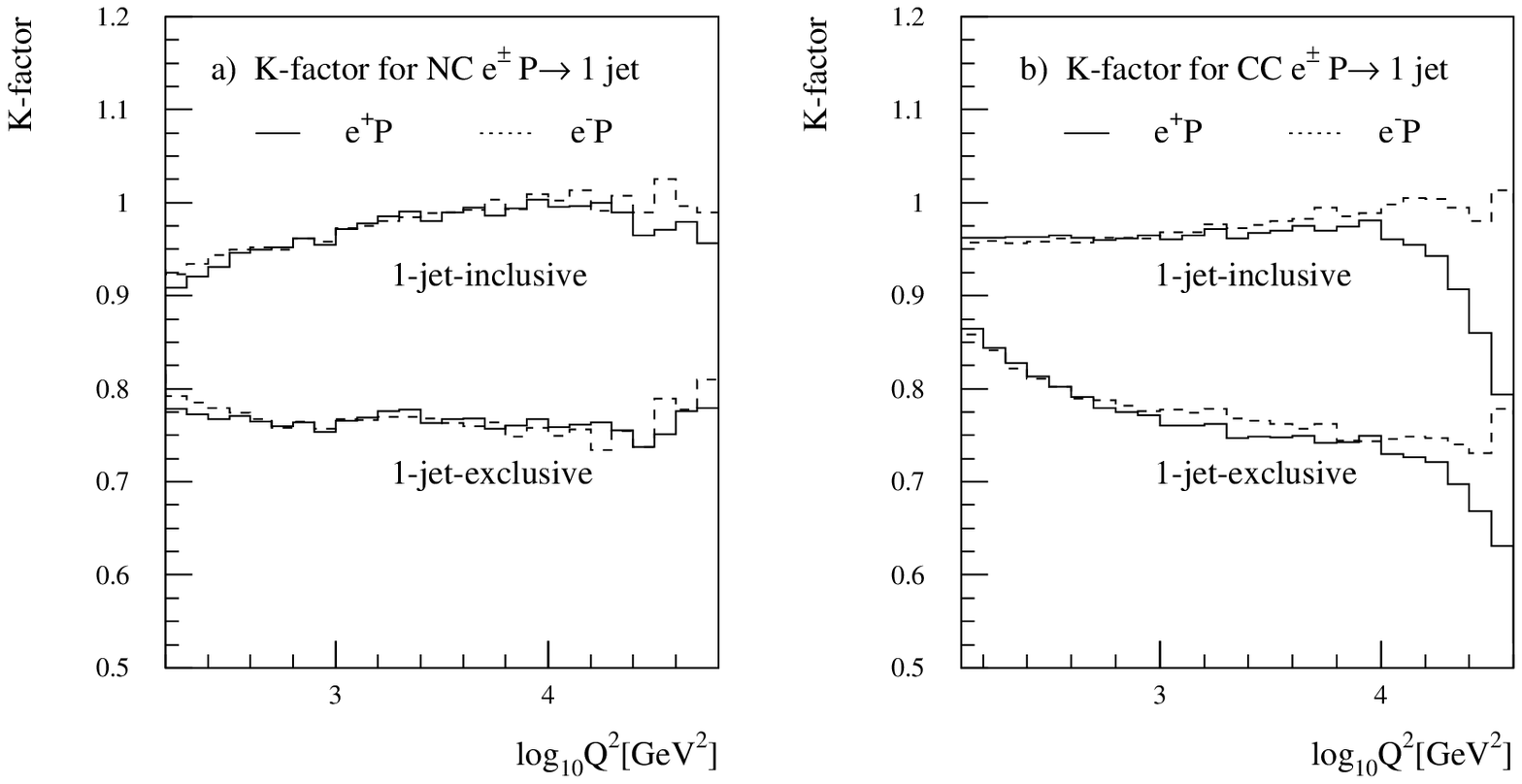,
         bbllx=0,bblly=240,bburx=540,bbury=540,
         width=0.95\linewidth}} 
\vspace*{-8mm}
\caption{
$Q^2$ dependence of the $K$-factor for NC (a)
and charged current (b)  1-jet inclusive and exclusive
cross sections in $e^+p$ (solid) and $e^-p$ (dashed) scattering.
The (LO) NLO  cross section, which enter the K-factor,
are calculated with LO (NLO) GRV parton distributions 
\protect\cite{grv} together
with the 1-loop (2-loop) formula for the strong coupling constant.
}
\label{f_zw1j_q_kfac}
\end{figure}
\enlargethispage{1cm}
\begin{figure}[ht]
  \centering
  \mbox{\epsfig{file=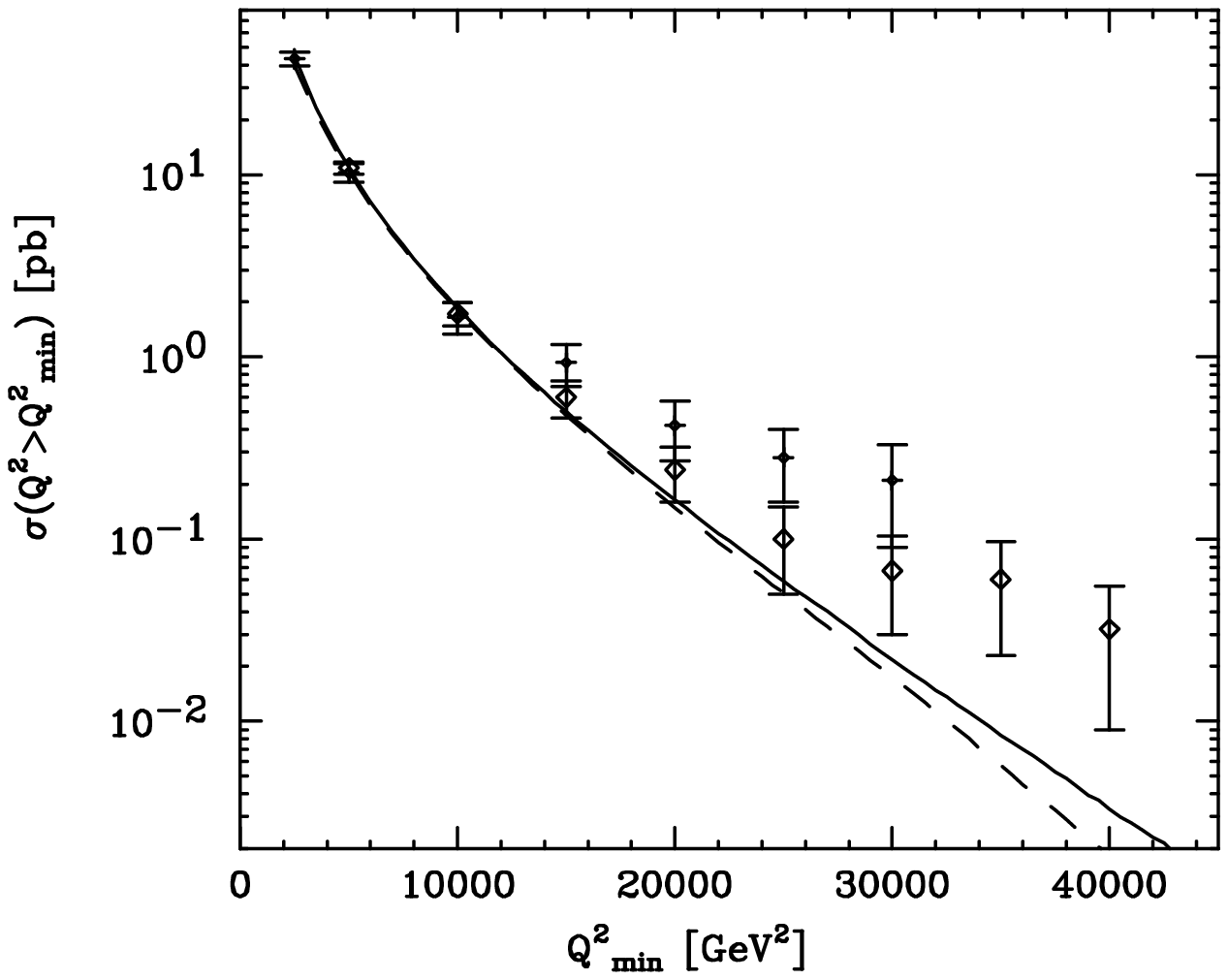,
         bbllx=0,bblly=210,bburx=460,bbury=550,
         width=0.75\linewidth}} 
\vspace*{-8mm}
\caption{
Integrated cross sections versus a minimum $Q^2$ for 
$e^+p\rightarrow e^+ X$ in NLO (solid line) and
integrated cross sections versus a minimum $Q^2$ for 
$e^+p\rightarrow e^+\, 1$-jet inclusive in NLO (dashed line).
In the dashed line, events are selected with
$E_T^{\mboxsc{lab}}(j)>15$ GeV, $E_T^{\mboxsc{lab}}(l^\prime)>25$ GeV, 
and $10^\circ<\theta(j)<145^\circ$.
The results are based on MRSA parton distribution functions
\protect\cite{mrsa}.
The data points correspond to H1 (solid) and ZEUS
(open) data as presented in \protect\cite{eps}.
}
\label{f_highq2}
\end{figure}
\clearpage
\section{Two-Jet Cross Sections\protect\vspace{1mm}}
\label{sec_twojet}
\subsection{Introductory Remarks: Gauge Boson Polarization Effects 
\protect\vspace{1mm}}
\label{twojetintro}
%
%
\enlargethispage{1cm}
The 2-jet final state in DIS provides various possibilities
for testing our understanding of perturbative QCD.
These include the measurement of the strong coupling constant 
(section~\ref{sec_alphas}),
the determination of the gluon density (section~\ref{sec_gluon})
and the study of event shape variables and power suppressed
corrections (section~\ref{eventshape}).
The 2-jet final state introduces also first sensitivity
to the non-diagonal ($m\neq m^\prime$) polarization density matrix elements
of the exchanged boson,\footnote{The 
lepton-hadron scattering process may be regarded as the scattering of a
polarized off-shell  gauge boson on the proton where the polarization of 
the gauge boson is tuned by the scattered lepton's momentum direction.}
\begin{equation}
h_{mm^{\prime}}= \epsilon_{\mu}^{\ast}(m)H^{\mu\nu}\epsilon_{\nu}(m^{\prime})
\hspace{1cm}
(m,m^{\prime}=+,0,-)
\label{hmndef}
\end{equation}
where
\begin{equation}
\epsilon_{\mu}(\pm) = \frac{1}{\sqrt{2}}(0;\pm1,-i,0)
\hspace{1cm}
\epsilon_{\mu}(0) = (0;0,0,1)
\end{equation}
denote the boson polarization vectors  
with the $z$ axis aligned along the boson-proton direction,
as defined for example in the HCM  frame.
$H^{\mu\nu}$ is the hadronic tensor for the  virtual boson-initial parton
QCD subprocess.
More inclusive observables like the usual DIS 
structure functions $F_2, F_3$ or $F_L$ are only sensitive
to the diagonal density matrix elements $h_{++},h_{--},h_{00}$.
One effect of the non-diagonal density
matrix elements is a nontrivial $\phi$ dependence of the jets around
the virtual boson-proton beam axis (for jets
defined in the HCM or the Breit frame) \cite{herai,georgie,ingelman}.
{\it In the absence of jet cuts in the laboratory frame},
the general structure of the 2-jet final
state is determined by the polarization effects of the exchanged
vector boson and the 2-jet cross section factorize
the following characteristic $y$ and $\phi$ dependence:
\begin{eqnarray}
d\sigma_{\mboxsc{had}}[2\mbox{-jet}]  \sim  \,\,\bigg\{  &
(1+(1-y)^2) \hspace{2mm} d\sigma^{F_2}[\mbox{2-jet}] \hspace{2mm}
-y^2        \hspace{2mm} d\sigma^{F_L}[\mbox{2-jet}] \hspace{2mm}
+y(2-y)     \hspace{2mm} d\sigma^{F_3}[\mbox{2-jet}] \nonumber \\[2mm]
&\hspace{-3.3cm}
-\,\cos\phi\,\sqrt{1-y}\, 
\bigg[\, (2-y) \hspace{2mm} d\sigma^{F_4}[\mbox{2-jet}] \hspace{2mm}
+           4y \hspace{2mm} d\sigma^{F_5}[\mbox{2-jet}] \, \bigg]
\hspace{2mm} + \, 2 \cos 2\phi\, (1-y)\, 
               \hspace{2mm} d\sigma^{F_6}[\mbox{2-jet}]\nonumber \\[2mm]
&\hspace{-3.7cm} +\,\sin\phi\, \sqrt{1-y}\,      \hspace{1mm}
\bigg[\, y\hspace{2mm}            d\sigma^{F_{7}}[\mbox{2-jet}] \hspace{2mm}
+(2-y)   \hspace{2mm}             d\sigma^{F_{8}}[\mbox{2-jet}] \,\bigg]
\hspace{2mm} +\,    \sin2\phi  \,(1-y)       \,   
                     \hspace{2mm} d\sigma^{F_{9}}[\mbox{2-jet}] 
\bigg\} 
\label{sigphi}
\end{eqnarray}
Here $\phi$ denotes the azimuthal angle of jet 1 around the
boson direction, where the lepton plane in the HCM or the Breit frame
defines $\phi=0^\circ$.
The helicity cross sections 
$d\sigma^{F_i}[\mbox{2-jet}]$ in Eq.~(\ref{sigphi})
are linearly related to the polarization density matrix elements
of the virtual boson in Eq.~(\ref{hmndef}) \cite{tom}:
\begin{eqnarray}
d\sigma^{F_2}[\mbox{2-jet}]&\sim&  h_{00}+h_{++}+h_{--}\label{hul}\nonumber \\
d\sigma^{F_L}[\mbox{2-jet}]&\sim&  h_{00} \hspace{4.45cm}         \nonumber \\
d\sigma^{F_3}[\mbox{2-jet}]&\sim&  h_{++} -h_{--}\label{heldens}  \nonumber \\
d\sigma^{F_4}[\mbox{2-jet}]&\sim&  h_{+0} +h_{0+} -h_{-0}-h_{0-}  \nonumber \\
d\sigma^{F_5}[\mbox{2-jet}]&\sim&  h_{+0} +h_{0+} +h_{-0}+h_{0-}  \nonumber \\
d\sigma^{F_6}[\mbox{2-jet}]&\sim&  h_{+-} +h_{-+} \hspace{3.05cm}  \nonumber \\
d\sigma^{F_7}[\mbox{2-jet}]&\sim&  h_{+-} -h_{-+}               \nonumber \\
d\sigma^{F_8}[\mbox{2-jet}]&\sim&  h_{+0} -h_{0+} +h_{-0}-h_{0-} \nonumber \\
d\sigma^{F_9}[\mbox{2-jet}]&\sim&  h_{+0} -h_{0+} -h_{-0}+h_{0-}  \nonumber
\end{eqnarray}
In the 1-photon exchange case one has only a contribution to the five
parity conserved helicity cross sections
$d\sigma^{F_{2,L,4,6,7}}[\mbox{2-jet}]$
whereas additional contributions
to $d\sigma^{F_{3,5,8,9}}[\mbox{2-jet}]$ originate from the 
axial vector couplings in the $Z$ and $W$ exchange.
To $\oas$ one populates only the so-called dispersive contributions
$d\sigma^{F_{2,L,3,4,5,6}}[\mbox{2-jet}]$. 
Analytical results for the partonic  helicity cross sections 
$d\hat{\sigma}_{q\rightarrow qg}^{F_{2,L,3,4,5,6}}[\mbox{2-jet}]$
and
$d\hat{\sigma}_{g\rightarrow q\bar{q}}^{F_{2,L,3,4,5,6}}[\mbox{2-jet}]$
are given in App.~\ref{helicross}.
Numerically small absorptive contributions 
$d\sigma^{F_{7,8,9}}[\mbox{2-jet}]$ come first in at $\oasz$ 
through the imaginary parts of the 1-loop contributions \cite{hagiwara}.

Without an (experimental) separation
of a quark, anti-quark or gluon jet, the $\cos\phi$ 
and $\sin\phi$ terms in Eq.~(\ref{sigphi}) are not observable
and are therefore averaged out. 
The  resulting $\cos 2\phi$ dependence, in LO,
is shown in Fig.~\ref{f_phi}a 
for dijet events in a cone scheme defined in the HCM with radius
$R=1$ and $p_T^{\mboxsc{HCM}}>5$ GeV.
The size  of the $\phi$ dependence in this normalized distribution
is rather insensitive on the $p_{T\,\mboxsc{min}}^{\mboxsc{HCM}}(j)$ 
requirements on the jets.
We find also very similar results for jets defined in the $k_T$ 
algorithm.

\begin{figure}[hbt]
\vspace*{1in}            
\begin{picture}(0,0)(0,0)
\includegraphics{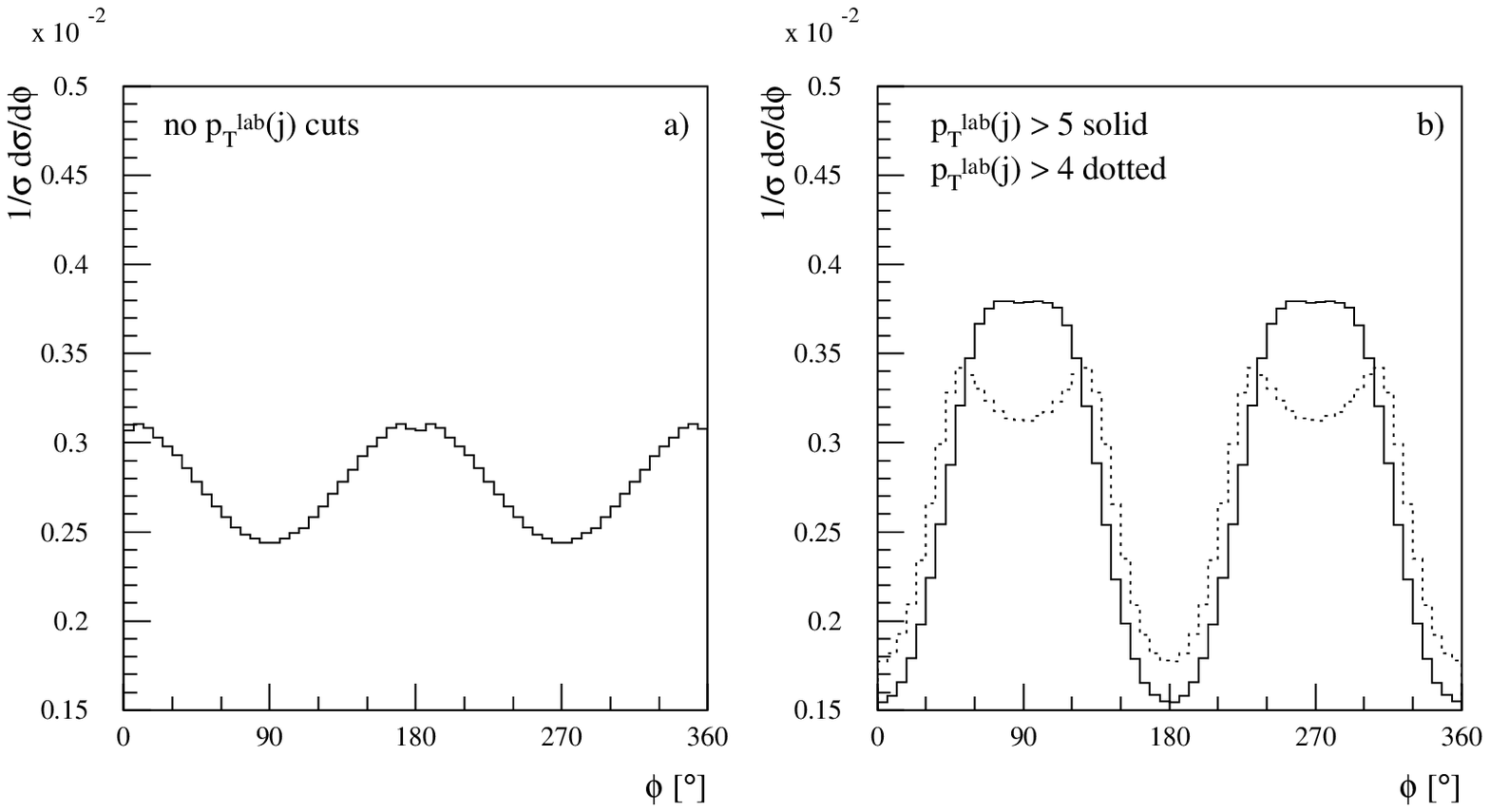}
\end{picture}
\vspace{6cm}
\caption{
(a) Normalized jet-azimuthal distribution in LO  around the virtual boson
direction. Jets are defined in a cone scheme  in
the HCM with $p_{T}^{\protect\mboxsc{HCM}}(j)>5$ GeV.
The $\phi$ integrated dijet cross section is 1465 pb;
(b) 
same as (a) but  with an additional cut of 
$p_T^{\protect\mboxsc{lab}}(j)>5$ 
(solid)
($p_T^{\protect\mboxsc{lab}}(j)>4$ (dotted)) GeV on the jets.
The $\phi$ integrated dijet cross section is 945 pb
(1146 pb).
All results are given in LO with MRSR1 \protect\cite{mrsr1}
parton distribution functions.
}
\label{f_phi}
\end{figure}

Averaging (integrating) over $\phi$ implies that only the
helicity cross sections $d\sigma^{F_{2,3,L}}[\mbox{2-jet}]$
in Eq.~(\ref{sigphi}) contributes to the dijet production cross section,
{\it i.e.}
$d\sigma^{F_{2,3,L}}[\mbox{2-jet}]$ are 
the 2-jet contributions to the inclusive structure 
functions $F_2,F_3$ and $F_L$, respectively.

In the presence of typical acceptance cuts on the jets in the
laboratory frame, the azimuthal distribution
is, however, dominated by  kinematic effects 
and the residual dynamical effects from the gauge boson
polarization are small \cite{phipaper}.
This is shown in Fig.~\ref{f_phi}b 
where an  additional cut of $p_T^{\mboxsc{lab}}(j)>5$ (solid)
($p_T^{\mboxsc{lab}}(j)>4$ (dotted)) GeV  on the jets
in the laboratory frame is imposed before boosting the
events to the HCM. Events with jets lying in
the leptonic plane (around $\phi=0^\circ$ and 
$\phi=180^\circ$) are preferredly rejected by the
$p_T^{\mboxsc{lab}}(j)$ cut.
The only remaining vestiges of the gauge boson polarization effects 
in the $\phi$ distribution are the dips at $\phi=90^\circ$ and
$270^\circ$ in the dashed curve in Fig.~\ref{f_phi}b.
A consequence of the (almost purely) kinematical $\phi$ dependence is
that the $\phi$ dependent part 
of the QCD matrix elements,
encoded in the helicity cross sections
$d\sigma^{F_{4,5,6,7,8,9}}$, contributes even to the dijet  
\underline{production} cross section.
Depending on the lab frame cuts, the production cross
section can be effected at the 5-8\% level \cite{phipaper}.
Therefore, the full helicity structure in the dijet matrix elements
has to be kept even for the calculation of dijet production
cross sections in the presence of typical lab frame acceptance cuts.
Since the LO and NLO analytical and numerical results, which will be presented
in the following, are based on helicity amplitudes
the full spin structure of the amplitude is kept and thus,
the full $\oasz$ corrections to all helicity cross sections 
$d\sigma^{F_i}[\mbox{2-jet}]$ in Eq.~(\ref{sigphi}) 
are effectively taken into account.

\subsection[NLO  Two-Jet Cross Sections 
(One-Photon Exchange)\protect\vspace{1mm}]
{NLO  Two-Jet Cross Sections (One-Photon Exchange)}
\label{sec_ana_twojet}
%
%
The lowest order \oas\ contributions to the 2-jet
cross section arises from the 2-parton final state processes in
Eqs.~(\ref{qtoqg}) and (\ref{gtoqqbar}) (see Fig.~\ref{f_atobc}).
\begin{figure}[hbt]
\vspace*{1cm}            
\begin{picture}(0,0)(0,0)
\includegraphics{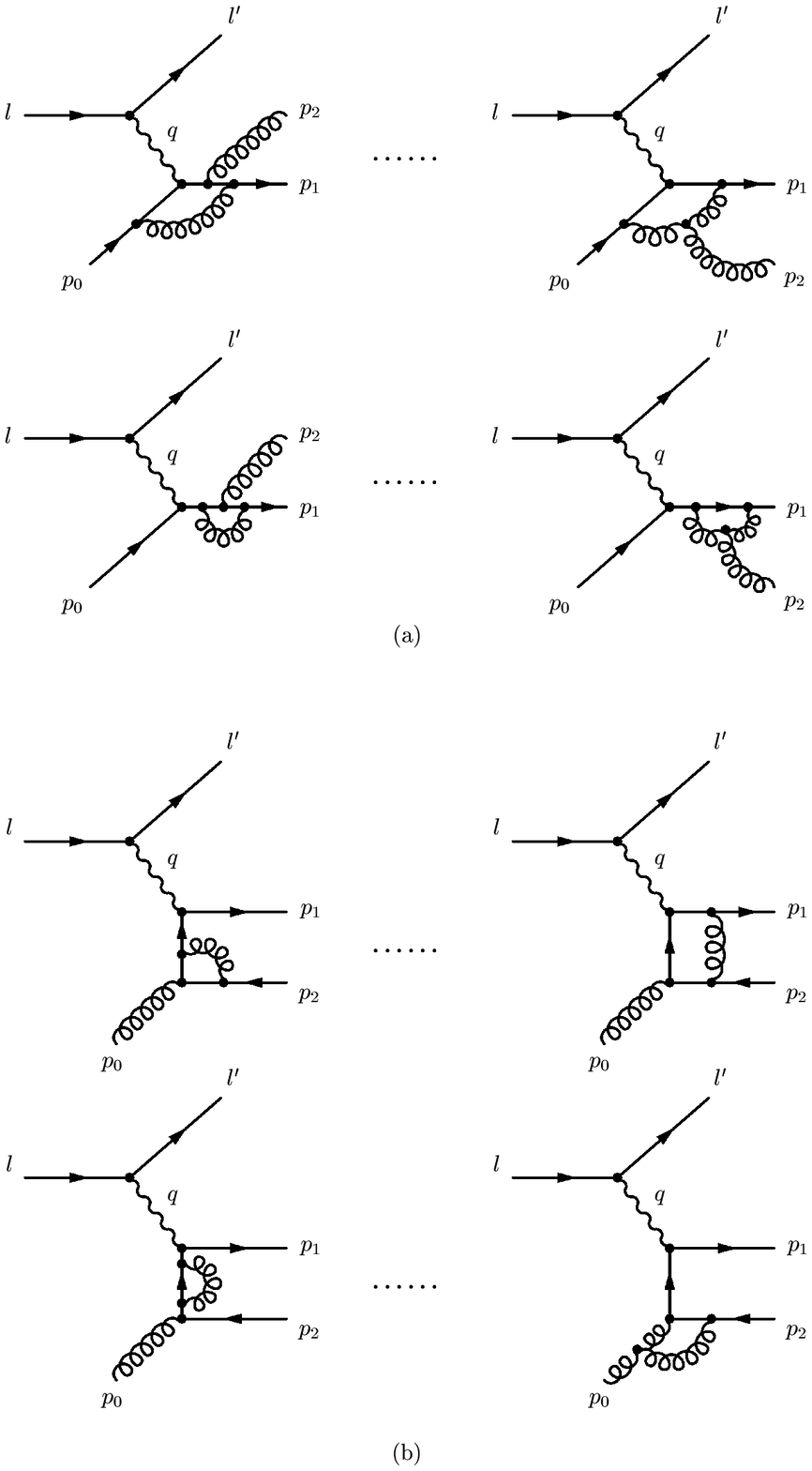}
\end{picture}
\vspace{18cm}
\caption{
Generic virtual gluon correction diagrams 
to the 2-parton final state processes 
in Fig.~\protect\ref{f_atobc}a (a) and 
Fig.~\protect\ref{f_atobc}b (b).
}
\label{f_virt}
\end{figure}
Analytical results for dijet cross sections up to NLO
will be presented in this section. 

The NLO \oasz\, 2-jet cross section receives contributions
from the 1-loop corrections to the Born subprocesses
in Eqs.~(\ref{qtoqg},\ref{gtoqqbar}) (see Fig.~\ref{f_virt})
and from the integration over the unresolved region
(defined by a given jet algorithm) of the 3-parton final
state tree level matrix  elements (Fig.~\ref{f_atobcd})
\begin{eqnarray}
e(l)+q(p_0)&\rightarrow& e(l^\prime) + q(p_1) + g(p_2) + g(p_3)
\label{qtoqgg} \\[1mm]
e(l)+q(p_0)&\rightarrow& e(l^\prime) + q(p_1) + q(p_2) + \bar{q}(p_3)
\label{qtoqqqbar} \\[1mm]
e(l)+g(p_0)&\rightarrow& e(l^\prime) + q(p_1) + \bar{q}(p_2) + g(p_3)
\label{gtoqqbarg}
\end{eqnarray}
and the corresponding antiquark processes with $q\leftrightarrow \bar{q}$.

\begin{figure}[hbt]
\vspace*{1cm}            
\begin{picture}(0,0)(0,0)
\includegraphics{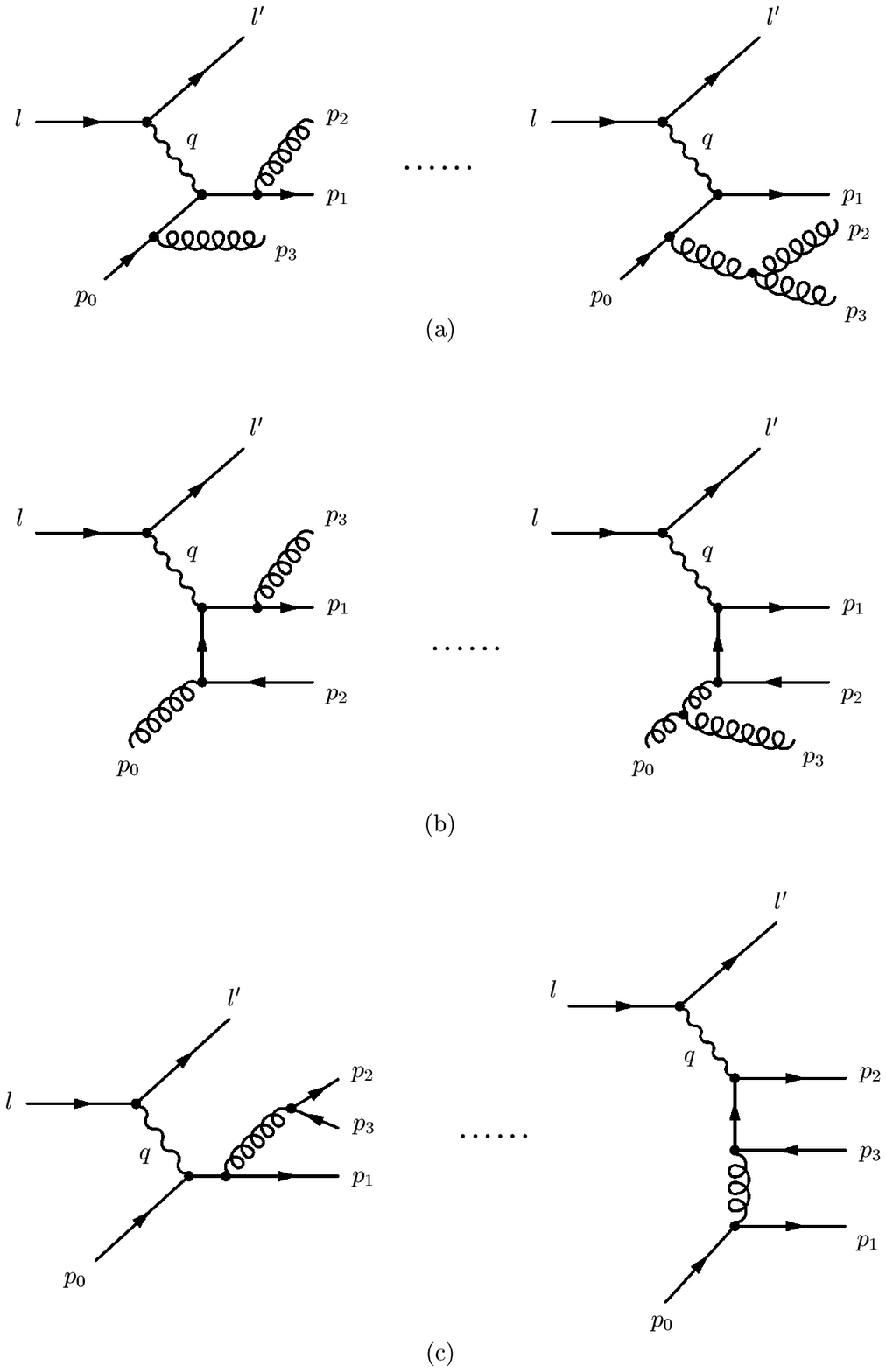}
\end{picture}
\vspace{16cm}
\caption{
Generic \protect\oasz\ tree diagrams:
$eq \rightarrow eqgg$ (a), $eg\rightarrow eq\bar{q}g$ (b),
and $eq\rightarrow eqq\bar{q}$ (c).
Only two out of eight diagrams are shown for each class.}
\label{f_atobcd}
\end{figure}

According  to Eqs.~(\ref{sig_sum},\ref{sig_virt}) 
the NLO two-jet cross section is given by
\begin{eqnarray}
\sigma_{\mboxsc{had}}^{\mboxsc{NLO}}[2\mbox{-jet}]
&=& 
\sigma_{\mboxsc{had}}^{\mboxsc{LO}}[2\mbox{-jet}]\nonumber \\[1mm]
&+&
\sigma_{\mboxsc{had}}^{\mboxsc{NLO, final}}[2\mbox{-jet}]
\label{sig_sum2j} \\[1mm]
&+&
\sigma_{\mboxsc{had}}^{\mboxsc{NLO, crossing}}[2\mbox{-jet}]\nonumber\\[1mm]
&+&
\sigma_{\mboxsc{had}}^{\mboxsc{NLO, hard}}[2\mbox{-jet}]\nonumber
\end{eqnarray}
where 
the general structure of the
hadronic cross sections on the r.h.s. of Eq.~(\ref{sig_sum2j})
is defined in Eqs.~(\ref{hadlo},\ref{hadfinal},\ref{hadcross},\ref{hadlo1})
with $n=2$, respectively.
We specify the relevant parity conserving (pc) partonic cross sections
for these hadronic cross section contributions
in the subsequent sections.
The final formula for the hadronic NLO 2-jet cross section in terms
of these partonic results is  given in Eq.~(\ref{twojet}).
\clearpage
%
%
\subsubsection[{Partonic cross sections for
$\sigma_{\protect\mboxsc{had}}^{\protect\mboxsc{LO}}$[2-jet]  and
$\sigma_{\protect\mboxsc{had}}^{\protect\mboxsc{NLO, crossing}}$[2-jet]
\protect\vspace{1mm}}]{
$\hat{\sigma}^{\protect\mboxsc{LO}}_{q\rightarrow qg}$ and
$\hat{\sigma}^{\protect\mboxsc{LO}}_{g\rightarrow q\bar{q}}$ in
$\sigma_{\protect\mboxsc{had}}^{\protect\mboxsc{LO}}$[2-jet]  and
$\sigma_{\protect\mboxsc{had}}^{\protect\mboxsc{NLO, crossing}}$[2-jet]}
%
%
According to Eqs.~(\ref{hadlo},\ref{hadcross}) the LO partonic
cross sections $\hat{\sigma}^{\protect\mboxsc{LO}}_{q\rightarrow qg}$ and
$\hat{\sigma}^{\protect\mboxsc{LO}}_{g\rightarrow q\bar{q}}$ 
for the subprocesses in Eqs.~(\ref{qtoqg},\ref{gtoqqbar}) and
Fig.~\ref{f_atobc}
enter the hadronic cross sections 
$\sigma_{\mboxsc{had}}^{\mboxsc{LO}}$[2-jet]  and
$\sigma_{\mboxsc{had}}^{\mboxsc{NLO, crossing}}$[2-jet].
These partonic cross sections are already given in terms of  compact
analytical expressions for the squared matrix elements 
$|M^{(\mboxsc{pc})}_{q\rightarrow qg}|^2$ and
$|M^{(\mboxsc{pc})}_{g\rightarrow q\bar{q}}|^2$ 
in Eqs.~({\ref{sig_qtoqg}-\ref{m_gtoqqbar}).
%
%
\subsubsection[{Partonic cross sections for
$\sigma_{\protect\mboxsc{had}}^{\protect\mboxsc{NLO, final}}$[2-jet]
\protect\vspace{1mm}}]{
$\hat{\sigma}^{\protect\mboxsc{NLO}}_{q\rightarrow qg}$ and
$\hat{\sigma}^{\protect\mboxsc{NLO}}_{g\rightarrow q\bar{q}}$ in
$\sigma_{\protect\mboxsc{had}}^{\protect\mboxsc{NLO, final}}$[2-jet]}
%
%
According to Eq.~(\ref{finalstruc}) the NLO (finite) partonic
cross section 
\begin{eqnarray}
\as^2\, \hat{\sigma}^{\mboxsc{NLO}}_{q\rightarrow qg}\,\,
&=&
(4\pi\alpha_s^2)\,e_q^2\,\sigma_0\,
\,
\left[\,
\,|M^{(\mboxsc{pc})}_{q\rightarrow qg}|^2
\,\,
{\cal{K}}_{q\rightarrow qg}(\smin,s,t,u,\mu_R) 
\,+\,{\cal {F}}^{(\mboxsc{pc})}_{q\rightarrow qg}
\right]
\label{m_nlo_qtoqg}
\end{eqnarray}
in the hadronic cross section 
$\sigma_{\protect\mboxsc{had}}^{\protect\mboxsc{NLO, final}}$[2-jet]}
combines the virtual 1-loop corrections in Fig.~\ref{f_virt}a 
to the Born process 
in Eq.~(\ref{qtoqg}) with the singular integrals over 
the {\em final state} unresolved phase space
region of the quark initiated three parton final state subprocesses 
in Eqs.~(\ref{qtoqgg},\ref{qtoqqqbar}) (see Fig.~\ref{f_atobcd}a,b).

Similarly, the NLO partonic cross section
\begin{eqnarray}
\as^2\, \hat{\sigma}^{\mboxsc{NLO}}_{g\rightarrow q\bar{q}}\,\,
&=&
(4\pi\alpha_s^2)\,e_q^2\,\sigma_0\,
\,
\left[\,
|M^{(\mboxsc{pc})}_{g\rightarrow q\bar{q}}|^2
\,\,
{\cal{K}}_{g\rightarrow q\bar{q}}
(\smin,\tilde{s},\tilde{t},\tilde{u},\mu_R) 
\,+\,{\cal {F}}^{(\mboxsc{pc})}_{g\rightarrow q\bar{q}}
\right]
\label{m_nlo_gtoqqbar}
\end{eqnarray}
in $\sigma_{\protect\mboxsc{had}}^{\protect\mboxsc{NLO, final}}$[2-jet]
combines the virtual 1-loop corrections in Fig.~\ref{f_virt}b
to the Born process 
in Eq.~(\ref{gtoqqbar}) with the singular integrals over 
the { final state} unresolved phase space
region of the gluon initiated  three parton final state subprocess
in Eq.~(\ref{gtoqqbarg}) (see Fig.~\ref{f_atobcd}c).

The dynamical ${\cal{K}}_{q\rightarrow qg}$ factor in 
Eq.~(\ref{m_nlo_qtoqg}), which multiplies the 
lowest order ${\cal{O}}(\alpha_s^1)$ squared matrix element 
$|M^{(\mboxsc{pc})}_{q\rightarrow qg}|^2$ in Eq.~(\ref{m_qtoqg}),
depends on $\smin$ and the invariant masses of the hard partons 
\begin{equation}
s= -2p_0.p_1 \hspace{1cm}
t= 2p_1.p_2 \hspace{1cm}
u= -2p_0.p_2
\label{stu_q}
\end{equation}
${\cal{K}}_{q\rightarrow qg}$
may be crossed from the analogous ${\cal{K}}$ factor
in $e^+e^-\rightarrow$ 2 partons as given in Eq.~(4.31) with $n=1$
in Ref.~\cite{giele1}\footnote{We have checked that the result
in Eq.~(4.31) in \protect\cite{giele1}
agrees  (after appropriate changes in the notations and
taking the same soft and collinear limits)
with the results in Ref.~\cite{kramer,ert}.}. 
One finds ($N=3$ is the number of colors):
\begin{eqnarray}
{\cal{K}}_{q\rightarrow qg}(\smin,s,t,u,\mu_R)&=&
\left(\frac{N}{2\pi}\right)
\left[
- \ln^2\left(\frac{|t|}{\smin}\right)
- \ln^2\left(\frac{|u|}{\smin}\right)
- \frac{\pi^2}{6} \nonumber \right. \\
&&
\left.
+\frac{3}{4} \ln\left(\frac{|t|}{\smin}\right) 
+\frac{3}{4} \ln\left(\frac{|u|}{\smin}\right) 
+ \frac{29}{9}  - \frac{5n_f}{9N}
\, + \, {\cal{O}}(\smin) 
\right]
\label{rqtoqg} \\
&&
\,-\,\,
\frac{1}{8}\,{\cal{K}}_{q\rightarrow q}(\smin,|s|)
\,\,+\,\,
\alpha_s(\mu_R)\,\,b_0\,\,\ln\left(\frac{\mu_R^2}{\smin}\right)
 \nonumber 
\end{eqnarray}
where dynamical factor 
${\cal{K}}_{q\rightarrow q}(\smin,|s|)$ is given in Eq.~(\ref{r_qtoq}).
Terms proportional to $\smin$ have been neglected in 
Eq.~(\ref{rqtoqg}).
Note that the $\overline{\mbox{MS}}$ 
renormalization of the ultraviolet divergencies
in the virtual corrections introduce also an explicit 
logarithmic dependence on the renormalization scale $\mu_R$ in
${\cal{K}}_{q\rightarrow qg}$.
The $\ln(\mu_R^2/\smin)$ term factorizes the usual
1-loop QCD beta function $b_0$
\begin{equation}
b_0 = \frac{3}{2\pi}\,\left(\frac{11}{6}-\frac{1}{9}n_f\right)
\end{equation}
where $n_f$ denotes the number of flavors.

The dynamical ${\cal{K}}_{g\rightarrow q\bar{q}}$ factor 
in Eq.~(\ref{m_nlo_gtoqqbar}), which multiplies 
$|M^{(\mboxsc{pc})}_{g\rightarrow q\bar{q}}|^2$ in Eq.~(\ref{m_gtoqqbar})
can be obtained from Eq.~(\ref{rqtoqg}) by crossing:
\begin{equation}
{\cal{K}}_{g\rightarrow q\bar{q}}(\smin,\tilde{s},\tilde{t},\tilde{u},\mu_R) 
= 
{\cal{K}}_{q\rightarrow qg}(\smin,\tilde{s},\tilde{t},\tilde{u},\mu_R) 
-\left(\frac{N}{2\pi}\right) \frac{5\pi^2}{9}
\label{rgtoqqbar}
\end{equation}
with
\begin{equation}
\tilde{s}= 2p_1.p_2 \hspace{1cm}
\tilde{t}= -2p_0.p_1 \hspace{1cm}
\tilde{u}= -2p_0.p_2
\label{tilde_stu}
\end{equation}
As already mentioned in section \ref{unresolved}, the 
${\cal{K}}_{q\rightarrow qg}$ and
${\cal{K}}_{g\rightarrow q\bar{q}}$ factors in
Eqs.~(\ref{rqtoqg},\ref{rgtoqqbar})
include also the crossing of a pair
of collinear partons with an invariant mass smaller than $\smin$
from the final state to the initial state. 
This ``wrong'' contribution is replaced by the
correct collinear initial state configuration by adding the appropriate
crossing function contribution to the hadronic cross section
as given in Eq.~(\ref{twojet}). The crossing function contributions,
which factorize the Born matrix elements,
take also into account the corresponding factorization of the
initial state singularities as described in section~\ref{crossing}.

The functions ${\cal{F}}_{q\rightarrow qg}^{(\mboxsc{pc})}$  and 
${\cal{F}}_{g\rightarrow q\bar{q}}^{(\mboxsc{pc})}$ 
in Eqs.~(\ref{m_nlo_qtoqg},\ref{m_nlo_gtoqqbar}) denote the
{\em finite} parts of the interference of the 1-loop 
amplitudes in Fig.~\ref{f_virt} with the Born amplitudes for the processes in 
Eqs.~({\ref{qtoqg},\ref{gtoqqbar}), which do not factorize the
corresponding squared  Born  matrix elements. 
The 1-loop amplitudes
can be obtained by crossing  the 1-loop helicity amplitude
in $e^+e^-\rightarrow 3$ partons,
which is given in  Appendix A of \cite{giele1}
in terms of the Weyl-van der Waerden spinors
\footnote{A summary of notations and rules for calculations in
the Weyl-van der Waerden spinor basis are described in Appendix 
\ref{helinotation} (for more details see Refs.~\cite{berendsgiele,weyl}).}.
From these results, one can derive helicity
dependent results for the DIS functions 
${\cal{F}}_{q\rightarrow qg}^{(\mboxsc{pc})}$  and 
${\cal{F}}_{g\rightarrow q\bar{q}}^{(\mboxsc{pc})}$
in Eqs.~(\ref{m_nlo_qtoqg}) and (\ref{m_nlo_gtoqqbar}), respectively.
Let us introduce the notation
\begin{equation}
{\cal{F}}^{(\mboxsc{pc})}_{a\rightarrow \mboxsc{2 partons}}
(\lambda_0;\lambda_1;\lambda_2;\lambda_l;\lambda_{l^\prime})
\equiv
{\cal{F}}^{(\mboxsc{pc})}_{a\rightarrow \mboxsc{2 partons}}
(p_0,\lambda_0;p_1,\lambda_1;p_2,\lambda_2;l,\lambda_l;
l^\prime,\lambda_{l^\prime})
\end{equation}
where $\lambda_i\,$ ($\lambda_l,\lambda_{l^\prime}$) denote
the helicities of the partons (leptons)
with momenta $p_i$ ($l,l^\prime$) as defined in 
Eqs.~(\ref{qtoqg}) and (\ref{gtoqqbar}).
In the following, we will drop the spin label 
on the helicity labels. Thus $(\pm)$ means $(\pm 1/2)$ in the case
of lepton and quark helicities, and $(\pm 1)$  in the case
of gluon helicities, respectively.
The superscript (pc) refers again to the
vector coupling at the leptonic and hadronic vertex in the 1-photon 
exchange.
One helicity combination for the function 
${\cal{F}}_{q\rightarrow qg}^{(\mboxsc{pc})}$ is 
\footnote{In order to keep our notation reasonably concise, we shall 
always drop the momentum labels that are not important for the
argument at hand.}:
\begin{eqnarray}
{\cal{F}}_{1,q\rightarrow qg}^{(\mboxsc{pc})}=
{\cal{F}}_{q\rightarrow qg}^{(\mboxsc{pc})}(+;+;+;-;-)
&=&
64
\left(\frac{N}{2\pi}\right) \,\,
\frac{\bra l^\prime p_0 \ket^2  \bra l^\prime l \ket
     }{ \bra p_1 p_2 \ket \bra p_2 p_0 \ket } \,\,
\bigg[  \nonumber \\
&&\hspace{1.35cm}
\left(\alpha_{1}-\frac{\bar{\alpha}_{1}}{N^2}\right)
\frac{\bra p_0 l^\prime\ket \bra p_1 p_0\ket \bra p_1 l \ketc 
           }{ \bra p_1 p_2\ket \bra p_2 p_0\ket }  
\label{f1qdef} \\
&&
\hspace{1cm}+
\left(\beta_{1}-\frac{\bar{\beta}_{1}}{N^2}\right)
\frac{\bra p_0 l^\prime\ket \bra p_2 p_0\ket \bra p_2 l \ketc
            }{ \bra p_1 p_2\ket \bra p_2 p_0\ket }
\nonumber \\
&&
\hspace{1cm}+ 
\left(\gamma_{1}-\frac{\bar{\gamma}_{1}}{N^2}\right)
\frac{- \bra p_2 l \ketc \bra p_1 l^\prime\ket \bra p_1 p_2 \ketc
            }{ \bra p_1 p_2\ket \bra p_2 p_0 \ketc } \nonumber
\bigg]^\dagger
\end{eqnarray}
Here, $\bra xy \ket$ ($\bra xy \ketc$)
denotes a spinor inner product as defined
in Eq.~(\ref{sip1def}) ((Eq.~\ref{sip2def})),
where $x$ and $y$ are the undotted (dotted) 
spinors associated with the corresponding four momentum vectors
as defined in Eqs.~(\ref{hs1}-\ref{hs4})\footnote{Note that we do 
not distinguish in our notation
between the four momenta and associated spinor letters.
Thus $(x.y)$ stands for a four momenta scalar product and
$\bra xy\ket$ stands for a spinor inner product.}.
Note that the overall factor in Eq.~(\ref{f1qdef})
is proportional to the Born helicity amplitude in
Eq.~(\ref{b1_def}).
The coefficients $\alpha_{i},\beta_{i},\gamma_{i},
\bar{\alpha}_{i},\bar{\beta}_{i},\bar{\gamma}_{i}$ 
are given in Eqs.~(\ref{alphadef}-\ref{gammadef})
of Appendix~\ref{oneloopheli}.
The complete function 
${\cal {F}}^{(\mboxsc{pc})}_{q\rightarrow qg}$ in 
Eq.~(\ref{m_nlo_qtoqg})  is  given by the sum of 
all eight nonvanishing helicity combinations, 
\begin{equation}
{\cal {F}}^{(\mboxsc{pc})}_{q\rightarrow qg}
= \sum_{i=1}^8\,
{\cal {F}}^{(\mboxsc{pc})}_{i,q\rightarrow qg}
\label{fcalqdef}
\end{equation}
where the seven remaining 
helicity contributions are obtained from
Eq.~(\ref{f1qdef}) by  parity and charge conjugation relations:
\begin{eqnarray}
{\cal{F}}_{2,q\rightarrow qg}^{(\mboxsc{pc})}&\equiv&
{\cal{F}}_{q\rightarrow qg}^{(\mboxsc{pc})}(+;+;+;+;+) 
={\cal{F}}_{1,q\rightarrow qg}^{(\mboxsc{pc})}
(l   \leftrightarrow -l^\prime)
\label{f2qdef}\\
{\cal{F}}_{3,q\rightarrow qg}^{(\mboxsc{pc})}&\equiv&
{\cal{F}}_{q\rightarrow qg}^{(\mboxsc{pc})}(-;-;+;-;-)
={\cal{F}}_{1,q\rightarrow qg}^{(\mboxsc{pc})}
(p_0 \leftrightarrow -p_1) 
\label{f3qdef} \\
{\cal{F}}_{4,q\rightarrow qg}^{(\mboxsc{pc})}&\equiv&
{\cal{F}}_{q\rightarrow qg}^{(\mboxsc{pc})}(-;-;+;+;+)
={\cal{F}}_{1,q\rightarrow qg}^{(\mboxsc{pc})}
(l   \leftrightarrow -l^\prime;
                 p_0 \leftrightarrow -p_1)  
\label{f4qdef}
\end{eqnarray}
and ${\cal{F}}^{(\mboxsc{pc})}_{i+4,q\rightarrow qg}
={\cal {F}}^{(\mboxsc{pc})}_{i,q\rightarrow qg}$
($i=1,2,3,4$) where ${\cal{F}}^{(\mboxsc{pc})}_{i+4,q\rightarrow qg}$ denote 
the contributions with all helicities reversed
in ${\cal{F}}^{(\mboxsc{pc})}_{i,q\rightarrow qg}$.
The interchange $(x\leftrightarrow y)$ on the r.h.s. of
Eqs.~(\ref{f2qdef}-\ref{f4qdef})
requires the exchange of the four momenta as well as the exchange
of the associated spinors in a spinor inner product.
Note that the coefficients 
$\alpha_i,\beta_i,\gamma_i,\bar{\alpha}_i,\bar{\beta}_i,\bar{\gamma}_i,$ 
for $i=3,4,7,8$ do also
change under the exchange of $(p_0 \leftrightarrow -p_1)$
(see Appendix~\ref{oneloopheli}).

The helicity dependent results for the gluon-initiated
process can be obtained from Eq.~({\ref{f1qdef}-\ref{f4qdef}) 
through crossing. One has:
\begin{equation}
{\cal{F}}^{(\mboxsc{pc})}_{i,g\rightarrow q\bar{q}}
(p_0,\lambda_0;\lambda_1;p_2,\lambda_2;\lambda_l;
\lambda_{l^\prime})=
-\frac{3}{8}
{\cal{F}}^{(\mboxsc{pc})}_{i,q\rightarrow qg}
(-p_2,-\lambda_2;\lambda_1;-p_0,-\lambda_0;\lambda_l;
\lambda_{l^\prime})
\label{fgcross}
\end{equation}
where the factor 3/8 takes into account the difference in the initial state
color average.
The spinor inner products with negative momentum spinor components
are defined in Eqs.~(\ref{n1}-\ref{n4}).
The complete function 
${\cal {F}}^{(\mboxsc{pc})}_{g\rightarrow qg}$ in 
Eq.~(\ref{m_nlo_gtoqqbar})  is  given by the sum of 
all eight nonvanishing helicity combinations, 
\begin{equation}
{\cal {F}}^{(\mboxsc{pc})}_{g\rightarrow q\bar{q}}
= \sum_{i=1}^8\,
{\cal {F}}^{(\mboxsc{pc})}_{i,g\rightarrow q\bar{q}}
\label{fcalgdef}
\end{equation}

\newpage
\subsubsection[{Partonic cross sections for
$\sigma_{\protect\mboxsc{had}}^{\protect\mboxsc{NLO, hard}}$[2-jet]
\protect\vspace{1mm}}]{
$\hat{\sigma}^{\protect\mboxsc{LO}}_{q\rightarrow qgg}$,
$\hat{\sigma}^{\protect\mboxsc{LO}}_{q\rightarrow q\bar{q}q}$ and
$\hat{\sigma}^{\protect\mboxsc{LO}}_{g\rightarrow q\bar{q}g}$ 
in $\sigma_{\protect\mboxsc{had}}^{\protect\mboxsc{NLO, hard}}$[2-jet]}
\label{threeparton}
%
%
Let us finally discuss the \oasz\ tree level matrix elements for the 
subprocesses in Eqs.~(\ref{qtoqgg}-\ref{gtoqqbarg}).
Some generic Feynman diagrams for each subprocess are shown in 
Fig.~\ref{f_atobcd}.
In order to be consistent with the notation used for the
representation of the helicity dependent functions
${\cal{F}}_{q\rightarrow qg}^{(\mboxsc{pc})}$  and 
${\cal{F}}_{g\rightarrow q\bar{q}}^{(\mboxsc{pc})}$
in the previous section, we will present
analytical results for the helicity amplitudes 
of the three level contributions in Eqs.~(\ref{qtoqgg}-\ref{gtoqqbarg})
also in   the Weyl-van der Waerden spinor basis \cite{heraii}.
The implementation of the matrix elements in \docuname, however, is
based on the results in Ref.~\cite{HZ}, which we have numerically checked
against the \oasz\ three level results in Ref.~\cite{tom}.

According to Eq.~(\ref{hadlo1}), the NLO contribution from these real
emission processes is given by
\begin{eqnarray}
\sigma_{\mboxsc{had}}^{\mboxsc{NLO, hard}}[2\mbox{-jet}] 
&=& \sum_a\int_0^1d\eta\,
\int \,d{\mbox{PS}}^{(l^\prime+3)}\,\,
f_a(\eta,\mu_F)\,\,
\label{hadlo_n2}\\
&& \hspace{-2cm}\alpha_s^{2}(\mu_R)\,\,
\hat{\sigma}^{\mboxsc{LO}}_{a\rightarrow 3\,\,\mboxsc{partons}}
(l+p_0\rightarrow l^\prime+p_1+p_2+p_3)\,\,
\prod_{i<j;\,0}^{3}\Theta(|s_{ij}| - \smin)\,\,
J_{2\leftarrow 3}(\{p_i\})
 \nonumber
\end{eqnarray}
Using the helicity formalism in the Weyl-van der Waerden spinor basis
as reviewed in Appendix~\ref{helinotation}
analytical results for the matrix elements
can be given in fairly concise forms.

We start with the 
partonic cross section for the $q\rightarrow qgg$ process 
in Eq.~(\ref{qtoqgg})
\bq
\as^2\,\, \hat{\sigma}^{\mboxsc{LO}}_{q\rightarrow qgg}\,\,
=
\sigma_0\,\,e_i^2\, 
(4\pi\alpha_s(\mu_R))^2\,
\,|M^{(\mboxsc{pc})}_{q\rightarrow qgg}|^2
\label{sig_qtoqgg}
\eq
with $\sigma_0$ defined in Eq.~(\ref{sigma0def}).
Denoting the helicity amplitude for this  process by
\begin{equation}
e_q\,b^{vV}_{q\rightarrow qgg}(\lambda_0;\lambda_1;\lambda_2;\lambda_3;
                    \lambda_l;\lambda_{l^\prime}) 
\equiv 
e_q\,b^{vV}_{q\rightarrow qgg}(p_0,\lambda_0;
                    p_1,\lambda_1;
                    p_2,\lambda_2;
                    p_3,\lambda_3;
                    l,\lambda_l;
                    l^{\prime},\lambda_{l^\prime}) 
\label{hqtoqgg}
\end{equation}
where $\lambda_i\,$ ($\lambda_l,\lambda_{l^\prime}$) denote
the helicities of the partons (leptons)
with momenta $p_i$ ($l,l^\prime$) as defined in Eq.~(\ref{qtoqgg}),
the matrix element squared $|M^{(\mboxsc{pc})}_{q\rightarrow qgg}|^2$
can be obtained by summing up all (numerically) squared helicity amplitudes
\begin{equation}
(4\pi\alpha)^2(4\pi\alpha_s)^2\frac{e_q^2}{Q^4}
|M^{(\mboxsc{pc})}_{q\rightarrow qgg}|^2=
e_q^2\frac{1}{3}\frac{1}{2}
\sum_{\mboxsc{16 heli}}|b^{vV}_{q\rightarrow qgg}|^2
\label{bvvdef}
\end{equation}
where the initial state color average factor 1/3 
and the factor 1/2 for two identical final state particles
on the r.h.s. of Eq.~(\ref{bvvdef})  are included in 
$|M^{(\mboxsc{pc})}_{q\rightarrow qgg}|^2$.
The lower case $v$ (upper case $V$) in the superscript of the
helicity amplitudes in Eq.~(\ref{hqtoqgg})
stands for the vector current coupling at the leptonic (hadronic) vertex
and hence the squared helicity amplitudes are labeled by the superscript (pc).
There are $2^4$ nonvanishing helicity amplitudes
$b^{vV}_{q\rightarrow qgg}
(\lambda_0;\lambda_1;\lambda_2;\lambda_3;\lambda_l;\lambda_{l^\prime})$
involved in this process
\footnote{We drop again the spin label 
on the helicity labels in the following. 
Thus $(\pm)$ means $(\pm 1/2)$ in the case
of lepton and quark helicities, and $(\pm 1)$  in the case
of gluon helicities, respectively.}.
However, it is sufficient 
to write down  only two of the 16 helicity amplitudes. 
We will again drop the momentum labels that are not important
for the argument at hand.
The case of equal gluon helicities lead to a very compact expression
\cite{heraii}:
\begin{eqnarray}
b^{vV}_{q\rightarrow qgg}(+;+;+;+;+;+)&=&\!\!\!
2ie^2g_s^2\frac{\bra l p_0 \ket^2}{\bra l l^\prime \ket}
\left\{ \frac{T^aT^b}{\bra p_1p_2 \ket \bra p_2p_3   \ket \bra p_3p_0   \ket}
+      \frac{T^bT^a}{\bra p_1p_3 \ket \bra p_3p_2   \ket \bra p_2p_0   \ket}
\right\}
\label{hqtoqgg1}
\end{eqnarray}
where $T_a$ denotes a SU(3) color matrix and
$e$ and $g_s$ are the electromagnetic and strong QCD
coupling constants, respectively ($e^2/4\pi=\alpha$ and
$g_s^2/4\pi=\alpha_s$).

The amplitude for  opposite gluon helicities is:
\begin{eqnarray}
b^{vV}_{q\rightarrow qgg}(+;+;-;+;+;+)&=&
\frac{ie^2g_s^2}{(l.l^\prime)(p_2.p_3)\,
                \bra p_0p_3 \ket
                \bra p_1p_2 \ketc
                \bra p_1p_3 \ket
                \bra p_0p_2 \ketc }\nonumber \\[1mm]
&&   \hspace{-8mm}       \cdot \bigg\{\hspace{3mm}
                   - T^bT^a(p_2.p_3)\bigg( \bra lp_0 \ket
                     \bra ll^\prime \ketc
                   - \bra p_2p_0  \ket
                     \bra p_2l^\prime \ketc \bigg) \nonumber \\[1mm]
&&           \cdot \,\,\,\,\,
                   \bigg( \bra l^\prime l \ket
                     \bra l^\prime p_1 \ketc
                   + \bra p_3 l \ket
                     \bra p_3 p_1 \ketc \bigg)            \nonumber \\[2mm]
&&-\,\,
           \frac{T^aT^b(p_1.p_3)-T^bT^a 
           (p_1.p_3+p_2.p_3)}{(p_1+p_2+p_3)^2}
         \label{hqtoqgg2} \\[1mm]
&&            \cdot  \,\,\,\,\,
                     \bra p_1p_3\ketc
                     \bra p_0 l\ket
                     \bra p_0p_2\ketc
                     \bra p_0p_3\ket
                     \bigg(\bra ll^\prime\ketc
                     \bra lp_2\ket
                   +  \bra p_0l^\prime \ketc
                     \bra  p_0p_2\ket   \bigg)             \nonumber \\[2mm]
&& +\,\,
            \frac{T^aT^b(p_0.p_2)-T^bT^a (p_0.p_2-p_2.p_3)}{(-p_0+p_2+p_3)^2}
                     \nonumber \\[1mm]
&&            \cdot \,\,\,\,\,
                     \bra p_1l^\prime\ketc
                     \bra p_0 p_2\ket
                     \bra p_1p_2\ket
                     \bra p_1p_3\ket
                     \bigg(\bra l^\prime l\ket
                     \bra l^\prime p_3\ketc
                   + \bra p_1l \ket
                     \bra  p_1p_3\ketc  \bigg)             \nonumber \\[2mm]
&& +\,\,              
                   2(T^aT^b-T^bT^a)(p_1.p_3)(p_0.p_2)
                     \bra p_1l^\prime\ketc
                     \bra p_0 l\ket   
                   \,\,\,\,\,  \bigg\}
\nonumber
\end{eqnarray}
In these equations, $\bra xy \ket$  ($\bra xy \ketc$) denotes again
a spinor inner product as defined
in Eq.~(\ref{sip1def}) ((Eq.~\ref{sip2def})),
where $x$ and $y$ are the undotted (dotted) 
spinors associated with the corresponding four momentum vectors
as defined in Eqs.~(\ref{hs1}-\ref{hs4}).

The remaining 14 helicity amplitudes can be obtained from 
Eqs.~(\ref{hqtoqgg1},\ref{hqtoqgg2}) by parity, 
charge conjugation and crossing.
One has:
\begin{eqnarray}
b^{vV}_{q\rightarrow qgg}
                   (-\lambda_0;
                    -\lambda_1;
                    -\lambda_2;
                    -\lambda_3;
                    -\lambda_l;
                    -\lambda_{l^\prime}) 
&=&
(b^{vV}_{q\rightarrow qgg}(
                    \lambda_0;
                    \lambda_1;
                    \lambda_2;
                    \lambda_3;
                    \lambda_l;
                    \lambda_{l^\prime}) )^\ast
\\
b^{vV}_{q\rightarrow qgg}
                   (\lambda_0;
                    \lambda_1;
                    \lambda_2;
                    \lambda_3;
                    l,-;
                    l^{\prime},-) 
&=&
-
b^{vV}_{q\rightarrow qgg}(
                    \lambda_0;
                    \lambda_1;
                    \lambda_2;
                    \lambda_3;
                    -l^\prime,+;
                    -l,+) 
\\
b^{vV}_{q\rightarrow qgg}
                   (p_0,-;
                    p_1,-;
                    \lambda_2;
                    \lambda_3;
                    \lambda_l;
                    \lambda_{l^\prime}) 
&=&
-
b^{vV}_{q\rightarrow qgg}(
                    -p_1,+;
                    -p_0,+;
                    \lambda_2;
                    \lambda_3;
                    \lambda_l;
                    \lambda_{l^\prime}) 
\end{eqnarray}
The spinor inner products with negative momentum spinor components
are defined in Eqs.~(\ref{n1}-\ref{n4}).
Finally, the helicity amplitudes for the anti-quark initiated process
can be obtained through crossing
\begin{equation}
b^{vV}_{\bar{q}\rightarrow\bar{q}gg}
(p_0,\lambda_0;
 p_1,\lambda_1;
 \lambda_2;
 \lambda_3;
 \lambda_l;
 \lambda_{l^\prime})
=
-b^{vV}_{q\rightarrow qgg}
(-p_1,-\lambda_1;
 -p_0,-\lambda_0;
 \lambda_2;
 \lambda_3;
 \lambda_l;
 \lambda_{l^\prime})
\label{antiquark}
\end{equation}

The partonic cross section for the
gluon-initiated three parton final state
tree level contributions in Eq.~(\ref{gtoqqbarg})
(see Fig.~\ref{f_atobcd}c for a generic set of diagrams)
\bq
\as^2\,\, \hat{\sigma}^{\mboxsc{LO}}_{g\rightarrow q\bar{q}g}\,\,
=
\sigma_0\,\,e_i^2\, 
(4\pi\alpha_s(\mu_R))^2\,
|M^{(\mboxsc{pc})}_{g\rightarrow q\bar{q}g}|^2
\label{sig_gtoqqbarg}
\eq
is closely related to the partonic cross section in Eq.~{bvvdef}
with $\sigma_0$ defined in Eq.~(\ref{sigma0def}).
$|M^{(\mboxsc{pc})}_{g\rightarrow q\bar{q}g}|^2$
can be obtained by summing up all (numerically) squared helicity amplitudes
for the diagrams in Fig.~\ref{f_atobcd}c:
\begin{equation}
(4\pi\alpha)^2(4\pi\alpha_s)^2\,\,\frac{e_q^2}{Q^4}\,\,
|M^{(\mboxsc{pc})}_{g\rightarrow q\bar{q}g}|^2=
\,e_q^2\,\frac{1}{8}\,
\sum_{\mboxsc{16 heli}}|b^{vV}_{g\rightarrow q\bar{q}g}|^2
\label{bvvgdef}
\end{equation}
where the initial state color average factor 1/8 on the r.h.s.
in Eq.~(\ref{bvvgdef}) is included in
$|M^{(\mboxsc{pc})}_{g\rightarrow q\bar{q}g}|^2$.
All $2^4$ nonvanishing helicity amplitudes
$b^{vV}_{g\rightarrow q\bar{q}g}
(\lambda_0;\lambda_1;\lambda_2;\lambda_3;\lambda_l;\lambda_{l^\prime})$
involved in this process
can be obtained by crossing from the helicity amplitudes 
$b^{vV}_{q\rightarrow qgg}$ via
\begin{equation}
b^{vV}_{g\rightarrow q\bar{q}g}
(p_0,\lambda_0;\lambda_1;p_2,\lambda_2;\lambda_3;\lambda_l;
\lambda_{l^\prime})=
-
{b}^{vV}_{q\rightarrow qgg}
(-p_2,-\lambda_2;\lambda_1;-p_0,-\lambda_0;\lambda_3;\lambda_l;
\lambda_{l^\prime})
\label{gluoncross}
\end{equation}
The matrix elements for the four quark subprocess
in Eq.~(\ref{qtoqqqbar}) have a more complicated structure.
One has to take into account the two possibilities that all quark
flavors are equal $(f=f^\prime$) 
and that one has two different pairs of quark flavors
($f\neq f^\prime$).
Depending on the charge coupling factors, color factors 
and weight factors with the quark and anti-quark parton densities,
it is useful to split the squared matrix element  into different
subclasses denoted by $D,D^\prime,E,F$ \cite{heraii}.
The details of this classification including a listing
of the allocation of various Feynman diagram contributions
and expressions for the corresponding helicity amplitudes are given 
in Ref.~\cite{heraii} and we will not repeat them here.
The four classes have the following folding structure with the
parton densities:
\begin{eqnarray}
\as^2\,\,
\sum_{q,\bar{q}}  \hat{\sigma}^{\mboxsc{LO}}_{q\rightarrow qq\bar{q}}\,\,
\otimes\,\,f_q 
&=&
\sigma_0\,\,
(4\pi\alpha_s(\mu_R))^2\,\nonumber \\
&&
\bigg\{ \sum_{f=1}^{n_f} e_f^2 \left(  f_q+ f_{\bar{q}}
\right) \otimes \left(n_{f'} \cdot |{ M^{(\mboxsc{pc})}}|^2_D
+|{ M^{(\mboxsc{pc})}}|^2_E \right)\nonumber\\
  &  & +\sum_{f=1}^{n_f}\big( \sum_{f'=1}^{n_{f'}} e_{f'}^2\big)
\left( f_q+ f_{\bar{q}} \right)
\otimes |{ M^{(\mboxsc{pc})}}|^2_{D'}
\label{fourquark}\\
  &  & +\sum_{f=1}^{n_f} \big(\sum_{f'=1}^{n_{f'}} e_{f'}\big)e_f
\left( f_q-  f_{\bar{q}} \right) \otimes
|{ M^{(\mboxsc{pc})}}|^2_F \bigg\} \nonumber\\[1mm]
&\equiv& 
\sigma_0\,(4\pi\alpha_s(\mu_R))^2\,\,
\{\mbox{4-quark term}\}^{(\mboxsc{pc})} \nonumber
\end{eqnarray}
The symbol $\otimes$ stands for the folding in of the parton
densities.
\newpage
\subsubsection{The Hadronic Two-Jet Cross Section \protect\vspace{1mm}}
%
%
Based on the results presented in the previous sections we are now in
the position to specify the
hadronic {\it exclusive} 2-jet cross section in 1-photon exchange
up to NLO in full detail. 
According to Eq.~(\ref{sig_sum2j}) one has
\\[5mm]
\fbox{\rule[-5cm]{0cm}{13.2cm}\mbox{\hspace{16.2cm}}}
\\[-13.5cm]
\begin{eqnarray}
\displaystyle
\sigma_{\mboxsc{had}}[\mbox{2-jet}] &=&
\nonumber\\
&&\hspace{-25mm}
\int_0^1 d\eta\,
\int\,d{\mbox{PS}}^{(l^\prime+2)}\,\,\sigma_0\,\,(4\pi\alpha_s(\mu_R))\,\,
\bigg[  \nonumber  \\ 
&&\,\,[\sum_{i=q,\bar{q}}e_i^2 f_i(\eta,\mu_F)]\,\,
|M^{(\mboxsc{pc})}_{q\rightarrow qg}|^2\,
\left(
1+\alpha_s(\mu_R)\,{\cal{K}}_{q\rightarrow qg}(\smin,s,t,u,\mu_R)\,
\right)\nonumber \\
&+&\,\,
[\sum_{i=q,\bar{q}}e_i^2 f_i(\eta,\mu_F)]\,\,
\alpha_s(\mu_R)\,\,{\cal{F}}^{(\mboxsc{pc})}_{q\rightarrow qg}
\nonumber\\ 
&+&\,\,[\sum_{i=q,\bar{q}}e_i^2\, 
\,C_i^{\overline{\mboxsc{MS}}}(\eta,\mu_F,\smin)]\,
\alpha_s(\mu_R)\,
\,\,|M^{(\mboxsc{pc})}_{q\rightarrow qg}|^2\,   \nonumber\\
&+& \,\,\,\,(\sum_{i=q}e_i^2 ) f_g(\eta,\mu_F)\,\,
|M^{(\mboxsc{pc})}_{g\rightarrow q\bar{q}}|^2\,\,
\left(
1+\alpha_s(\mu_R)\,{\cal{K}}_{g\rightarrow q\bar{q}}
(\smin,\tilde{s},\tilde{t},\tilde{u},\mu_R)\right)\nonumber \\
&+&\,\,
\,\,(\sum_{i=q}e_i^2 ) f_g(\eta,\mu_F)\,\,
\alpha_s(\mu_R)\,{\cal{F}}^{(\mboxsc{pc})}_{g\rightarrow q\bar{q}}
\nonumber\\ 
&+&\,\,      (\sum_{i=q}e_i^2 ) 
\,C_g^{\overline{\mboxsc{MS}}}(\eta,\mu_F,\smin)\,
\alpha_s(\mu_R)\,
\,\,|M^{(\mboxsc{pc})}_{g\rightarrow q\bar{q}}|^2\,
\bigg]\,\,
J_{2\leftarrow 2}(\{p_i\})
\label{twojet} \\
&&\hspace{-28mm}
+\,\,
\int_0^1 d\eta\,
\int
\,d\mbox{PS}^{(l^\prime+3)}\,\,\sigma_0\,\,(4\pi\alpha_s(\mu_R))^2
\bigg[ \nonumber \\
&&[\sum_{i=q,\bar{q}}e_i^2 f_i(\eta,\mu_F)]\,\,
\,\,|M^{(\mboxsc{pc})}_{q\rightarrow qgg}|^2
\nonumber\\
&+&\,\,(\sum_{i=q}e_i^2 )\,\, f_g(\eta,\mu_F)\,\,
\,\,|M^{(\mboxsc{pc})}_{g\rightarrow q\bar{q}g}|^2 \nonumber\\
&+&\hspace{2cm}
\{\mbox{4-quark term}\}^{(\mboxsc{pc})}
\hspace{0.3cm} \nonumber
\bigg]
\hspace{0.2cm} 
\prod_{i<j;\,0}^{3}\Theta(|s_{ij}| - \smin)\,\,
J_{2\leftarrow 3}(\{p_i\})
\nonumber\\ \nonumber
\end{eqnarray}
Table~{\ref{tab_eqs}} lists  the equations with definitions and
analytical expressions for the quantities in Eq.~(\ref{twojet}).
\begin{table}[htb]
\caption{
References to the analytical expressions and definitions
for the  quantities in Eq.~(\protect\ref{twojet}).}
\label{tab_eqs}
\vspace{2mm}
\begin{tabbing}{cc}
xxxxxxxxxxxxxxxxxxxxxxxxxxxxxxxxxxxxxxxxxxx   \=   \kill 
\hspace*{4cm} 
$d{\mbox{PS}}^{(l^\prime+2)}$
\> Eq.~(\ref{phasespace})        \\[2mm]
\hspace{4cm}
$\sigma_0 =\frac{1}{4p_0.l}\, \frac{1}{4}\,\frac{(4\pi\alpha)^2}{Q^4}$  
\> Eq.~(\ref{sigma0def})        \\[2mm]
\hspace{4cm} 
$|M^{(\mboxsc{pc})}_{q\rightarrow qg}|^2$
\> Eq.~(\ref{m_qtoqg})        \\[2mm]
\hspace{4cm} 
${\cal{K}}_{q\rightarrow qg}(\smin,s,t,u,\mu_R)$
\> Eq.~(\ref{rqtoqg})        \\[2mm]
\hspace{4cm} 
${\cal{F}}^{(\mboxsc{pc})}_{q\rightarrow qg}$
\> Eqs.~(\ref{f1qdef}-\ref{f4qdef},\ref{alphadef}-\ref{gammadef})        \\[2mm]
\hspace{4cm} 
$C_q^{\overline{\mboxsc{MS}}}(\eta,\mu_F,\smin)$
\> Eqs.~(\ref{crossf_uv},\ref{crossf_s})        \\[2mm]
\hspace{4cm} 
$|M^{(\mboxsc{pc})}_{g\rightarrow q\bar{q}}|^2$
\> Eq.~(\ref{m_gtoqqbar})        \\[2mm]
\hspace{4cm} 
${\cal{K}}_{g\rightarrow q\bar{q}}
(\smin,\tilde{s},\tilde{t},\tilde{u},\mu_R)$
\> Eq.~(\ref{rgtoqqbar})        \\[2mm]
\hspace{4cm} 
${\cal{F}}^{(\mboxsc{pc})}_{g\rightarrow q\bar{q}}$
\> Eq.~(\ref{fcalgdef})        \\[2mm]
\hspace{4cm} 
$C_g^{\overline{\mboxsc{MS}}}(\eta,\mu_F,\smin)$
\> Eq.~(\ref{crossf_g})        \\[2mm]
\hspace{4cm} 
$J_{2\leftarrow 2}(\{p_i\})$
\> Eq.~(\ref{jnndef})        \\[2mm]
\hspace{4cm} 
$d\mbox{PS}^{(l^\prime+3)}$  
\> Eq.~(\ref{phasespace})        \\[2mm]
\hspace{4cm} 
$|M^{(\mboxsc{pc})}_{q\rightarrow qgg}|^2$
\> Eqs.~(\ref{bvvdef}-\ref{antiquark})        \\[2mm]
\hspace{4cm} 
$|M^{(\mboxsc{pc})}_{g\rightarrow q\bar{q}g}|^2$
\> Eqs.~(\ref{bvvgdef},\ref{gluoncross})        \\[2mm]
\hspace{4cm} 
$\{\mbox{4-quark term}\}^{(\mboxsc{pc})}$
\> Eq.~(\ref{fourquark})        \\[2mm]
\hspace{4cm} 
$J_{2\leftarrow 3}(\{p_i\})$
\> Eq.~(\ref{jnn1def})        
\end{tabbing}
\end{table}
%
%
The 2-jet {\it inclusive} cross section is defined via
Eq.~(\ref{twojet}) by replacing
$J_{2\leftarrow 3}(\{p_i\})$ in the last line by 
($J_{2\leftarrow 3}(\{p_i\})+J_{3\leftarrow 3}(\{p_i\}))$,
{\it i.e.} the 2-jet inclusive cross section is defined
as the sum of the NLO 2-jet exclusive cross section (as defined
in Eq.~(\ref{twojet})) plus the LO three jet cross section.
Note, that a separation in terms proportional to
$f_q$, $C_q^{\overline{\mboxsc{MS}}}$ and $f_g$, $C_g^{\overline{\mboxsc{MS}}}$ 
in Eq.~(\ref{twojet}) does {\it not} correspond
to a separation into quark and gluon initiated
processes in NLO, since the crossing functions
$C_q^{\overline{\mboxsc{MS}}}$ and $C_g^{\overline{\mboxsc{MS}}}$
mix the quark and gluon initiated processes
(see Eqs.~(\ref{crossf_s},\ref{crossf_g})).

Eq.~(\ref{twojet}) includes all relevant information  to construct
a Monte Carlo program for the numerical evaluation of the fully differential
NLO 2-jet cross section in DIS.
In particular  all ``plus prescriptions'' associated with the 
factorization of the initial state collinear
divergencies
are absorbed in the crossing functions $C_q^{\overline{\mboxsc{MS}}}$ which  is
very useful for a Monte Carlo approach.

Note that the second integral over the bremsstrahlung matrix elements
is restricted to regions
where all partons are resolved, {\it i.e.} any pair of partons has
$|s_{ij}|>\smin$.
As mentioned already, 
the resolution parameter $\smin$ is an arbitrary theoretical parameter 
and any measurable quantity should not depend on it.
The bremsstahlung contribution
grows with $\ln^2\smin$ and $\ln\smin$
with decreasing $\smin$.
This logarithmic growth is exactly cancelled by the
explicit   $-\ln^2\smin$ and $-\ln\smin$ terms
in 
${\cal{K}}_{q\rightarrow qg}(\smin,s,t,u,\mu_R)$
and
${\cal{K}}_{g\rightarrow q\bar{q}}(\smin,\tilde{s},\tilde{t},\tilde{u},\mu_R)$
(see Eqs.~(\ref{rqtoqg},\ref{rgtoqqbar})) and 
the crossing functions $C_a^{\overline{\mboxsc{MS}}}(\eta,\mu_F,\smin)$ 
(see sect.~\ref{crossing})
once $\smin$ is small enough for the soft and collinear approximations
in 
${\cal{K}}_{q\rightarrow qg},
{\cal{K}}_{g\rightarrow q\bar{q}}$ and
$C_a^{\overline{\mboxsc{MS}}}$ to be valid.

A powerful test of the numerical program is the $\smin$ independence of 
the final result. Fig.~\ref{fig_smin_two_jet}
shows the inclusive dijet cross section as a 
function of $\smin$ for four jet algorithms (see section~\ref{sec_jetdef}). 
One observes
that for values smaller than 0.1~GeV$^2$ the results are indeed  
independent of $\smin$. The strong $\smin$ dependence of the NLO 
cross sections for larger values shows that the soft 
and collinear approximations used in the phase space region $s_{ij}<\smin$
are no longer valid, {\it i.e.} terms of ${\cal{O}}(\smin)$ and
${\cal{O}}(\smin\ln \smin)$ become important.
In general, one wants to choose $\smin$ as large as possible to avoid large
cancellations between the virtual+collinear+soft part ($s_{ij}<\smin$)
and the hard part of the phase space  ($s_{ij}>\smin$).
Note that factor 10 cancellations occur between the effective 2-parton and 
3-parton final states at the lowest $\smin$ values in 
Fig.~\ref{fig_smin_two_jet}
and hence very high 
Monte Carlo statistics is required for these points. 
$\smin$ independence 
is achieved at and below $\smin=0.1$~GeV$^2$ and we choose this value for 
our further studies in this section.
%
%
\setlength{\unitlength}{0.7mm}
\begin{figure}[tb]               \vspace*{-7cm}
\hspace*{1cm}
\begin{picture}(150,165)(-30,1)
\mbox{\epsfxsize8.0cm\epsffile[70 250 480 550]{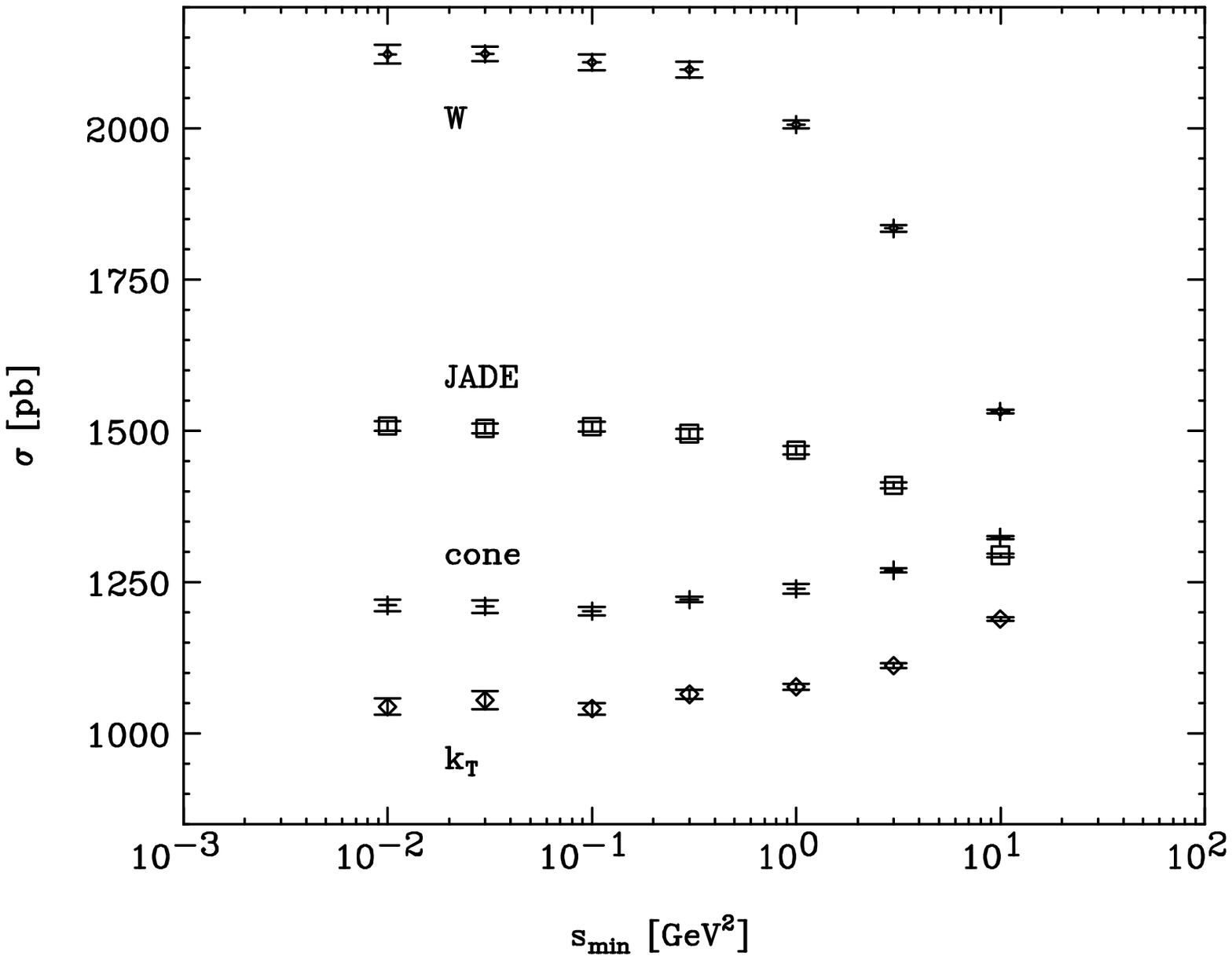}}
\end{picture}
\vspace*{2.6cm}
\caption{
Dependence of the inclusive two-jet cross section
in the $k_T$, cone, JADE, and the
$W$-scheme on $\protect\smin$, the two-parton resolution parameter. 
Partons are recombined in the $E$-scheme. 
Error bars represent statistical errors of the Monte Carlo program. 
$\protect\smin$ independence 
is achieved for $\protect\smin\lsimfig 0.1$~GeV$^2$. 
}\label{fig_smin_two_jet}
\end{figure}

\subsection{Numerical Results\protect\vspace{1mm}}
\label{num2jet}
\subsubsection{Charm and Bottom Mass Effects
\protect\vspace{1mm}}
\label{masseffects}
\enlargethispage{1cm}
The analytical results in section~\ref{sec_ana_twojet}
are presented in the limit of massless partons and leptons.
In particular the $b$- and $c$-quark masses $m_b,m_c$  
have been neglected.
How well justified are these approximations?
In Fig.~\ref{f_mass} we compare LO
massless 2-jet cross sections ($m_{u,d,s,c,b}=0$, solid line) with
results where $m_b$  and $m_c$ are set to 4.5~and~1.5~GeV, respectively
(dotted line), in the boson-gluon fusion subprocess of Eq.~(\ref{gtoqqbar}).
In Figs.~\ref{f_mass}a-d, jets are defined in a
cone scheme (in the lab frame) with
$p_T^{\protect\mboxsc{lab}},p_T^{\mboxsc{B}}>5$ GeV.
Neglecting the $b$ and $c$ quark mass 
introduces an error of about 5\% at low $Q=5$ GeV,
which  decreases with increasing $Q$ (see Figs.~\ref{f_mass}a,b).
Comparing the results for the subprocess $eg\rightarrow eb\bar{b}$
for $m_b=4.5$ GeV with $m_b=0$ alone (see Figs.~\ref{f_mass}c,d)
shows that the massless approximation is still very
bad for $b$ quarks. However, the absolute contribution
from this subprocess is small enough to 
justify the massless treatment.

The transverse momentum of the softest jet in the laboratory
frame is shown in Figs.~\ref{f_mass}e,f for the massless and
massive ($m_b=4.5$ GeV and $m_c=1.5$ GeV) case, 
for $Q^2>10$ GeV$^2$.
Sufficiently high transverse momenta ($\gsimfig 5$ GeV)
are required to reduce the error made by the
massless approximation below 5\%.

Similar studies of effects of  a finite $b$ and $c$ quark mass can 
be easily performed with \docuname\ for any infrared safe observable.

\begin{figure}[htb]
  \centering
 \mbox{\epsfig{file=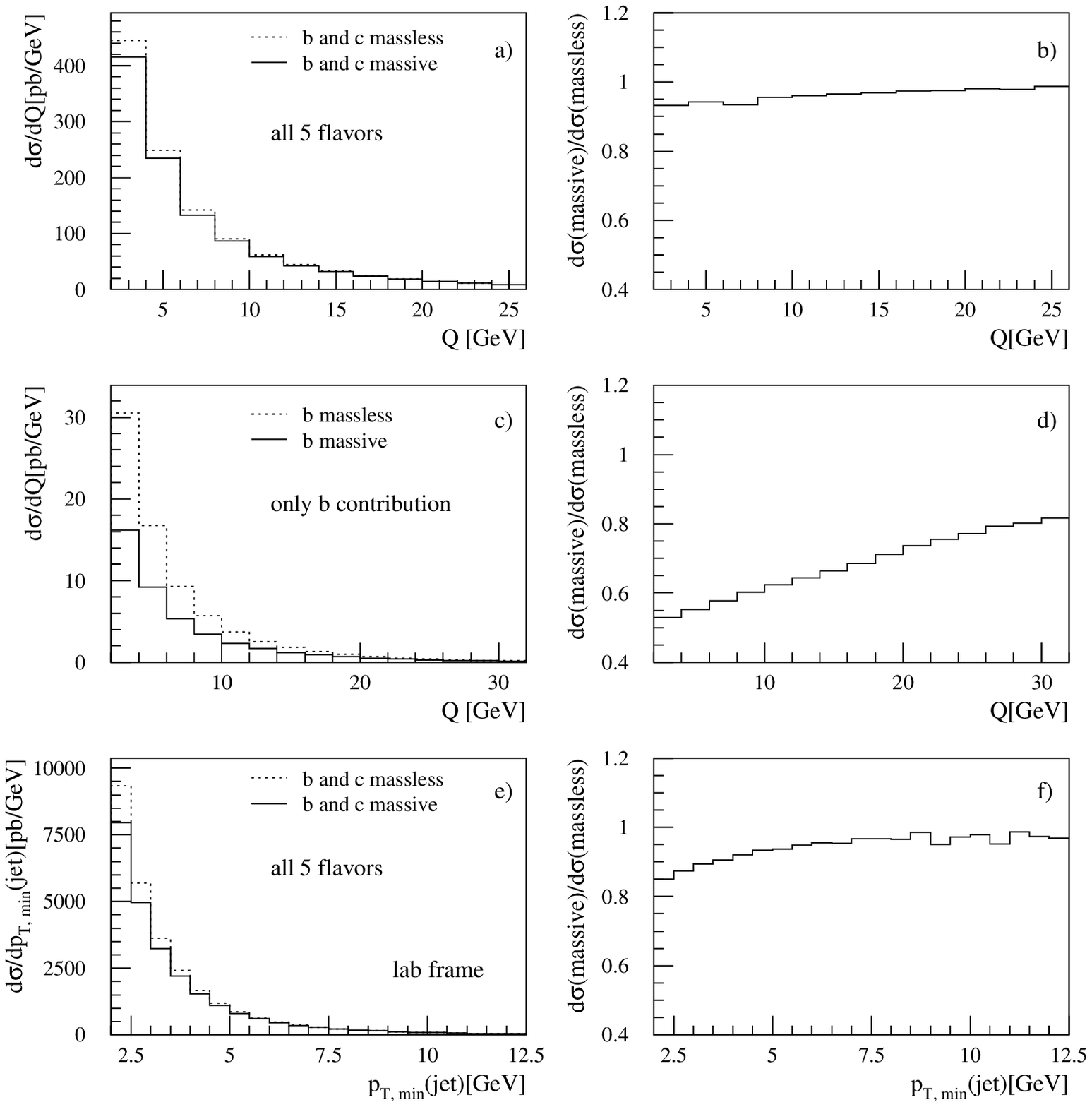,bbllx=20,bblly=145,
               bburx=550,bbury=680,width=0.95\linewidth}} 
\vspace*{-5mm}
\caption{
(a) $Q$ distribution of the dijet cross section with
$m_b=4.5$ GeV,  $m_c=1.5$ GeV  and all other flavors massless (solid) 
compared to the  dijet cross section , where the masses of all
five flavors  are neglected.
Events were selected with $y>0.04$, $E^\prime_e>10$ GeV.
Jets were reconstructed using a cone algorithm in the lab frame 
with $\Delta R=1$ and $|\eta^{\mboxsc{lab}}<3.5|$.
In addition, jets are required to have a minimum 
$p_T^{\protect\mboxsc{lab}},p_T^{\mboxsc{B}}>5$ GeV;
(b) the ratio of the curves in a);
(c,d) same as above  for the $b$-quark subprocesses
$eg\rightarrow eb\bar{b}$ alone;
(e) transverse momentum distribution 
for the jet with minimal transverse momentum.
Results with $m_b=4.5$ GeV,  $m_c=1.5$ GeV  
and all other flavors massless (solid)  are
compared to the results, where the masses of all
five flavors  are neglected. 
Events are selected for $Q^2>10$ GeV$^2$;
(f)  the ratio of the curves in e).
}
\label{f_mass}
\end{figure}
\newpage
\clearpage

\subsubsection{The Characteristic Scale in DIS Multi-Jet Production
\protect\vspace{1mm}}
\label{scalechoice}
Jet production in DIS is a multi-scale problem
and it is not {\it a priori} clear at which scale $\alpha_s$
and the parton densities are probed.
Clearly, the chosen scale  should be characteristic 
for the QCD portion of the process at hand. For dijet invariant 
masses, $m_{jj}$, below $Q$ we are in the DIS limit and $Q$ 
is expected to be the relevant scale. For large dijet invariant masses, 
however, $m_{jj}> Q$, the situation is more like in dijet production at 
hadron colliders and the jet transverse momenta $p_T^B$
(w.r.t. the boson-proton direction) set the physical scale of the 
process\footnote{The applicability of fixed order perturbation theory for
$n$-jet production ($n\ge 2$) 
requires large jet-jet invariant masses in addition to 
relatively large transverse momenta $p_T^B$ of the jets w.r.t. 
proton direction in the Breit or HCM frame.
The high 
$p_T^B$ requirement can also be replaced by a
cut on $k_T^B$ as defined below. Sufficiently high $k_T^B$ 
imply already  a  separation of the perturbative jets from the
proton remnant jet.}.
A variable which interpolates between these two limits is the sum of 
jet $k_T$s in the Breit frame, $\sum_j \,k_T^B(j)$,
where $k_T^B(j)$ and $p_T^B(j)$, 
the parton transverse momentum, are related by
\bq
(k_T^B(j))^2=2E_j^2(1-\cos\theta_{jp}) = 
      {2\over 1+\cos\theta_{jp}}(p_T^B(j))^2
\label{ktdef}
\eq
Here $\theta_{jp}$ is the 
angle between the parton and proton directions in the Breit frame. 
For LO 1-jet production, {\it i.e.}, in the naive parton model limit,
$k_T^B(j)=Q$, whereas
$\sum_j k_T^B(j)$  corresponds
to the sum of jet transverse momenta $p_T^B(j)$ in multi-jet production, 
when $Q$ becomes small compared to $p_T^B(j)$.
Thus the average $k_T^B(j)$ of the jets, $\langle k_T^B \rangle$,
appears to be the natural scale
for the short distance part of the  multi-jet production cross section
in DIS\footnote{
In addition to the momentum transfer of the hard scattering process, 
which represents a hard scale, jet cross sections also involve a 
a (softer) scale defining
the typical ``jet cone''. The perturbative expansion parameters
therefore are $\alpha_s\ln^2 y_{\mbox{\,\tiny{IR}}}$ and 
$\alpha_s\ln y_{\mbox{\,\tiny{IR}}}$
rather than the coupling  $\alpha_s(\mu_R)$ alone.
Here, $y_{\mbox{\,\tiny{IR}}}$ is a quantity 
such as a ratio of the two scales.
The applicability  of fixed order perturbation theory 
in jet physics requires that also
$\alpha_s\ln^2 y_{\mbox{\,\tiny{IR}}}$ and 
$\alpha_s\ln y_{\mbox{\,\tiny{IR}}}$ 
are sufficiently small.}.
There is in general
a qualitative difference between scale choices tied to 
$\langle k_T^B \rangle$ versus
scales related to $Q$, in particular on an event-by-event basis.
For example, even events with very large $\langle k_T^B \rangle$ 
are dominated by the small $Q$ region (see Fig.~1c in Ref.~\cite{hera_ws}
and Figs.~2,3,4 in Ref.~\cite{rom}).

A good measure of the improvement of a  NLO over a LO prediction
is provided by the residual scale dependence of the cross section. 
As an example we use the cone algorithm (implemented in the lab frame)
with $\Delta R=1$ and $p_T^{\mboxsc{lab}}(j)>5$ GeV. 
We find qualitatively  very similar results for the $k_T$
scheme implemented in the Breit frame.
The following set of kinematical cuts is imposed 
on the final state lepton and jets:
We require 40~GeV$^2<Q^2<2500$ GeV$^2$,
$0.04 < y < 1$, an energy cut of $E(e^\prime)>10$~GeV on the scattered 
electron, and a cut on the pseudo-rapidity $\eta=-\ln\tan(\theta/2)$
of the scattered lepton and jets of $|\eta|<3.5$. 
We used the parton distribution functions set MRS D-$^\prime$
\protect\cite{mrsdmp}.

Fig.~\ref{f_scale1} 
shows the scale dependence of the dijet cross section in LO and NLO. 
%
%
\begin{figure}[htb]
\vspace{7.8cm}
\begin{picture}(7,7)
\includegraphics{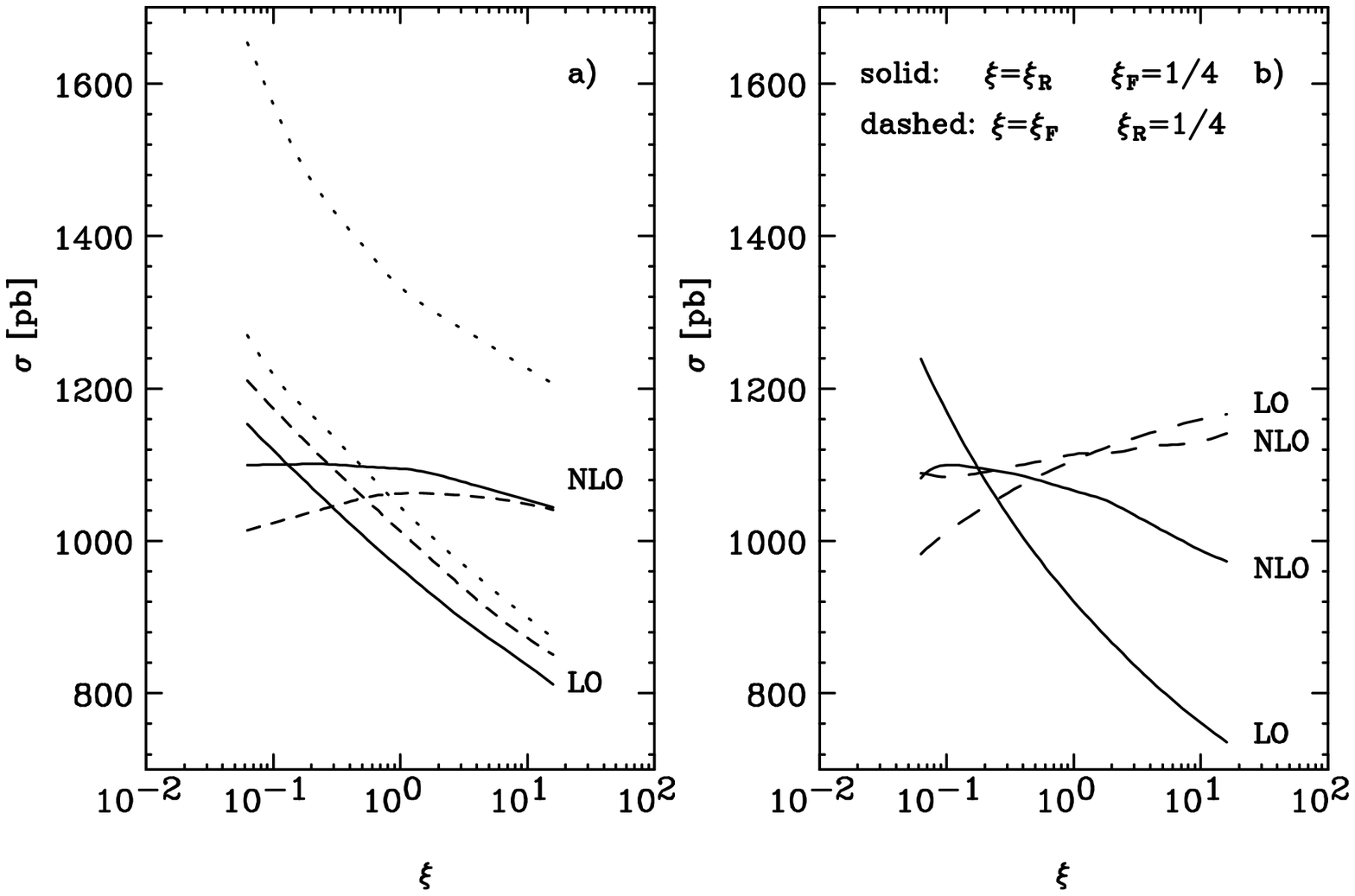}
\end{picture}
\vspace{1cm}
\caption{
(a) Dependence of the 2-jet exclusive cross 
section in the cone scheme on the  scale factor $\xi$.
The solid curves  are for $\mu_R^2=\mu_F^2=\xi\;(\sum_i\;k_T^B(i))^2$,
the dashed curves for $\mu_R^2=\mu_F^2=\xi\;(\sum_i\;p_T^B(i))^2$
and the dotted curves for $\mu_R^2=\mu_F^2=\xi\;Q^2$.
For the  solid (dashed) curves in (b), only
$\xi_R=\xi (\xi_F=\xi)$ is varied but $\xi_F=1/4 (\xi_R=1/4)$
is fixed, with $(\sum_i\;k_T^B(i))^2$ chosen as the basic scale.
Results are shown for the LO and NLO calculations.
}
\label{f_scale1}
\end{figure}
For scales related to $k_T^B$, $p_T^B$ and $Q$ the 
scale factor $\xi$  in Fig.~\ref{f_scale1}a is defined via
\begin{equation}
\mu_R^2 = \mu_F^2= \xi\;(\sum_j \,k_T^B(j))^2\, \hspace{1cm}
\mu_R^2 = \mu_F^2= \xi\;(\sum_j \,p_T^B(j))^2\, \hspace{1cm}
\mu_R^2 = \mu_F^2= \xi\;Q^2,
\label{xidef}
\end{equation}
respectively. 

The scale dependence of the dijet cross section
does not markedly improve in NLO for $\mu^2=\xi Q^2$.
This is shown by the dotted curves in Fig.~\ref{f_scale1}a 
where  the dependence of the 2-jet 
cross section  on the scale factor $\xi$ is shown.

For scales related to  $\sum_j \,p_T^B(j)$ the uncertainty from the variation
of the renormalization and factorization scale is markedly reduced
compared to the LO predictions (dashed curves in Fig.~\ref{f_scale1}).
The LO variation by a factor 1.54 is reduced to a 1.05 \% variation at NLO when 
both scales are varied simultaneously over the plotted range.

The resulting $\xi$ dependence for the natural scale choice
$ \mu_R^2 = \mu_F^2 = \xi\;(\sum_i \,k_T^B(i))^2$
is shown  as the solid lines in Fig.~\ref{f_scale1}a.
In this case, the NLO 2-jet cross section is essentially independent on
$\xi$ for $\xi<2$.
Hence, the theoretical uncertainties due to the scale variation
are very small suggesting a precise determination of 
$\alpha_s(\langle k_T^B\rangle)$ for different $\langle k_T^B\rangle$ bins
in dijet production, where 
\begin{equation}
<k_T^B>=\frac{1}{2}\,\, (\sum_{j=1,2} \,k_T^B(j))
\label{akt}
\end{equation}

Fig.~\ref{f_scale1}b shows the $\xi=\xi_R$ dependence 
of LO and NLO cross sections at fixed $\xi_F=1/4$ (solid curve)
and the  $\xi=\xi_F$ dependence at fixed $\xi_R=1/4$ (dashed curve).
Here $\xi_R$ and $\xi_F$ are defined via
\begin{equation}
\mu_R^2 =  \xi_R\;(\sum_j \,k_T^B(j))^2\, \hspace{1cm}
\mu_F^2 =  \xi_F\;(\sum_j \,k_T^B(j))^2
\label{xi1def}
\end{equation}
The $\xi_R$ and $\xi_F$ dependencies are fairly different.
The improvement in the scale dependence by the NLO corrections
is dominated by the improvement 
in the  renormalization scale $\mu_R$ (solid lines in Fig.~\ref{f_scale1}b).

\subsubsection{$K$-Factors and Recombination Scheme Dependence
\protect\vspace{1mm}}
\label{kfactors}
The importance of higher order corrections
and recombination scheme dependencies  
of the 2-jet cross sections is shown  in
Table~\ref{table_dijet} 
for the  four jet algorithms described in section~\ref{sec_jetdef}:
i) a cone algorithm (defined in the lab) 
with $\Delta R=1$ and $p_{T}^{\mboxsc{lab}}(j)>5$~GeV,
ii) the $k_T$ algorithm
with  $E_T^2=40$~GeV$^2$
and $y_{cut}=1$,
iii) the $W$-scheme 
with $y_{\mboxsc{cut}}=0.02$ and
iv) the ``JADE'' algorithm  which 
with $y_{\mboxsc{cut}}=0.02$.
In addition, jets 
must have transverse momenta of at least 2~GeV in the lab and the 
Breit frame for each jet algorithms.
Selection cuts and parameters are used
as described before Eq.~(\ref{xidef}).
Furthermore the renormalization scale  and
the factorization scale  are set to
$\mu_R=\mu_F= 1/2\,\sum_j \,p_T^B(j)$
(which gives very similar results as the suggested scale choice
$\mu_R=\mu_F= 1/2\,\sum_j \,k_T^B(j)$ in 
section~\ref{scalechoice}).

\begin{table}[b]
\caption{Two-jet cross sections in DIS at HERA. Results are given at LO and
NLO for the four jet definition schemes and acceptance cuts described in 
the text. The 2-jet inclusive cross section at NLO is given for three 
different recombination schemes.
}\label{table_dijet}
\vspace{2mm}
\begin{tabular}{lccccc}
        \hspace{2.3cm}
     &  \mbox{2-jet }
     &  \mbox{2-jet exclusive}
     &  \mbox{2-jet inclusive}
     &  \mbox{2-jet inclusive} 
     &  \mbox{2-jet inclusive} \\
     &  \mbox{LO}
     &  \mbox{NLO} ($E$)
     &  \mbox{NLO} ($E$)
     &  \mbox{NLO} ($E0$)
     &  \mbox{NLO} ($P$)\\
\hline\\[-3mm]
\mbox{cone} & 1107~pb & 1047~pb & 1203~pb & 1232 pb  & 1208 pb \\
$k_T$       & 1067 pb & 946 pb  & 1038 pb & 1014 pb  & 944 pb \\
$W$         & 1020 pb & 2061 pb & 2082 pb & 1438 pb  & 1315 pb \\
\mbox{JADE} & 1020 pb & 1473 pb & 1507 pb & 1387 pb  & 1265 pb \\
\end{tabular}
\end{table}
While the higher order corrections and recombination scheme
dependencies shown in table~\ref{table_dijet} in the cone
and $k_T$ schemes are small, very large corrections appear in the $W$-scheme. 
In addition, the large effective $K$-factor 
(defined as $K=\sigma_{NLO}/\sigma_{LO}$) 
of 2.04 (2.02) for the 2-jet inclusive
(exclusive) cross section in the $W$-scheme depends strongly 
on the recombination scheme  which is used in the
clustering algorithm. 
Such large dependencies are subject to potentially large higher order
uncertainties, since the recombination dependence is only simulated at tree
level in the NLO calculation.

The large corrections and recombination scheme dependencies 
in particular in the $W$ scheme 
can partly be traced to large single jet masses 
(compared to their energy
in the parton center of mass frame).
\setlength{\unitlength}{0.7mm}
\begin{figure}[hbt]               \vspace*{-2cm}
\begin{picture}(150,165)(-30,1)
\mbox{\epsfxsize10.0cm\epsffile[78 222 480 650]{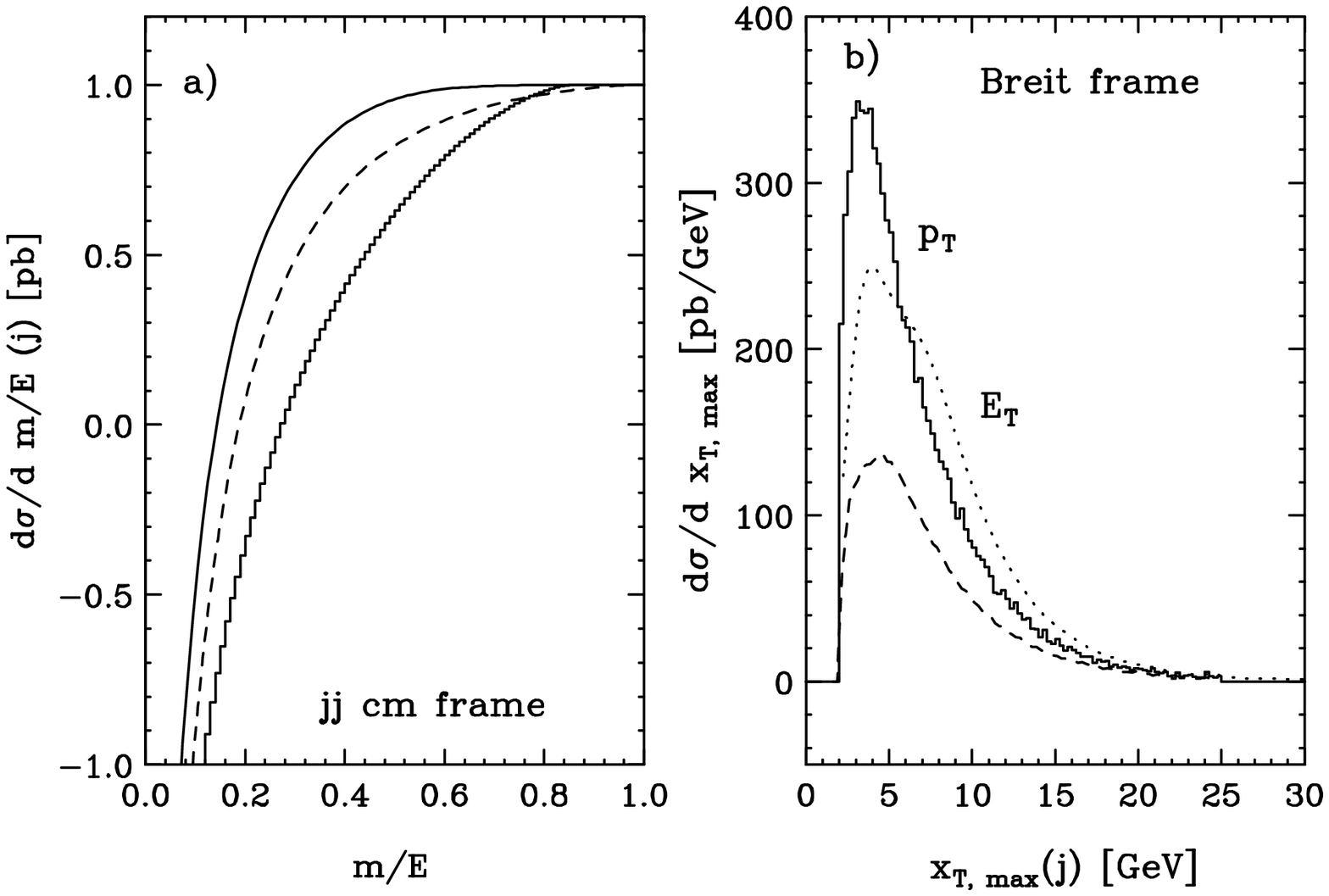}}
\end{picture}
\caption{
Single jet mass effects at NLO.
(a) Fraction of events in the cone scheme (solid curve), 
$k_T$ scheme (dashed curve), and $W$-scheme (histogram) with all jet 
mass to energy ratios below $m/E$, where $E$ is the corresponding 
jet's energy in the parton center of mass frame. Negative values at small 
$m/E$ are due to virtual contributions at $m/E=0$.
(b) NLO transverse momentum ($x_T=p_T$, solid histogram) 
   and transverse-energy  distribution ($x_T=E_T$, dotted curve)
   for the jet with largest $p_T$ and $E_T$ in the Breit frame, for
   the $W$-scheme. The dashed curve shows the LO result
   where both distributions are identical. From Ref.~\protect\cite{plb1}.
}
\label{f_fig5}
\end{figure}
This effect is investigated in Fig.~\ref{f_fig5},
where the fraction of events is shown with at least one jet being more 
massive than  $m/E$. Here $m$ is the invariant mass and $E$ the energy of 
the most massive of the jets in the parton center of mass frame.
Fig.~\ref{f_fig5}a shows that
50 \% of the events in the NLO cross section for the $W$ scheme
(with the $E$ recombination scheme)
have a massive jet with $m/E > 0.44$, while substantially smaller 
values are found in the other jet schemes.
The very large median value of $m/E$ in the $W$-scheme implies that at NLO 
we are dealing with very different types of jets than at LO, and this 
difference accounts for the large $K$-factor.
At NLO at least one of the $W$-scheme jets 
extends over a large solid angle, it is a massive,
slow moving object in the center of mass frame and, hence, very different from
the pencil-like, massless objects called jets at LO. The typically small 
relativistic $\gamma$-factor of these jets has large kinematic effects. 
For example the difference between transverse energy and transverse momentum 
distributions of the jets,  which are shown in Fig.~\ref{f_fig5}b, 
becomes quite 
pronounced in the $W$-scheme, an effect which is much smaller in the $k_T$ and 
cone schemes. 

In the JADE-algorithm the $K$-factor is reduced from 
1.48 in the $E$-scheme to 1.36 and 1.24 in the $E0$ and 
$P$-schemes.
For the  cone ($k_T$) scheme this recombination scheme
dependence is reduced to the 3\% (10\%) level.

\subsubsection{Characteristic NLO Effects in Differential Distributions
\protect\vspace{1mm}}
\label{nloeffects}
The effective $K$-factors close to unity which are found in the previous
sections (see table~\ref{table_dijet}) for jets defined in the cone and 
$k_T$ schemes could, in principle, be a coincidence arising from compensating 
effects in different phase space regions. It is important, therefore, to also 
compare LO and NLO distributions, in particular for those variables which
define the acceptance region. 

In this section we discuss such effects for a selection of distributions:
the transverse momenta of the jets and the scattered lepton,
jet pseudo-rapidities and the dijet-azimuthal angular
decorrelations.
For the additional variables,  Bjorken-$x$ and jet-jet invariant masses,
  see also Fig.~\ref{fig_gluon}.

\setlength{\unitlength}{0.7mm}
\begin{figure}[htb]               \vspace*{-2cm}
\begin{picture}(150,165)(-30,1)
\mbox{\epsfxsize10.0cm\epsffile[78 222 480 650]{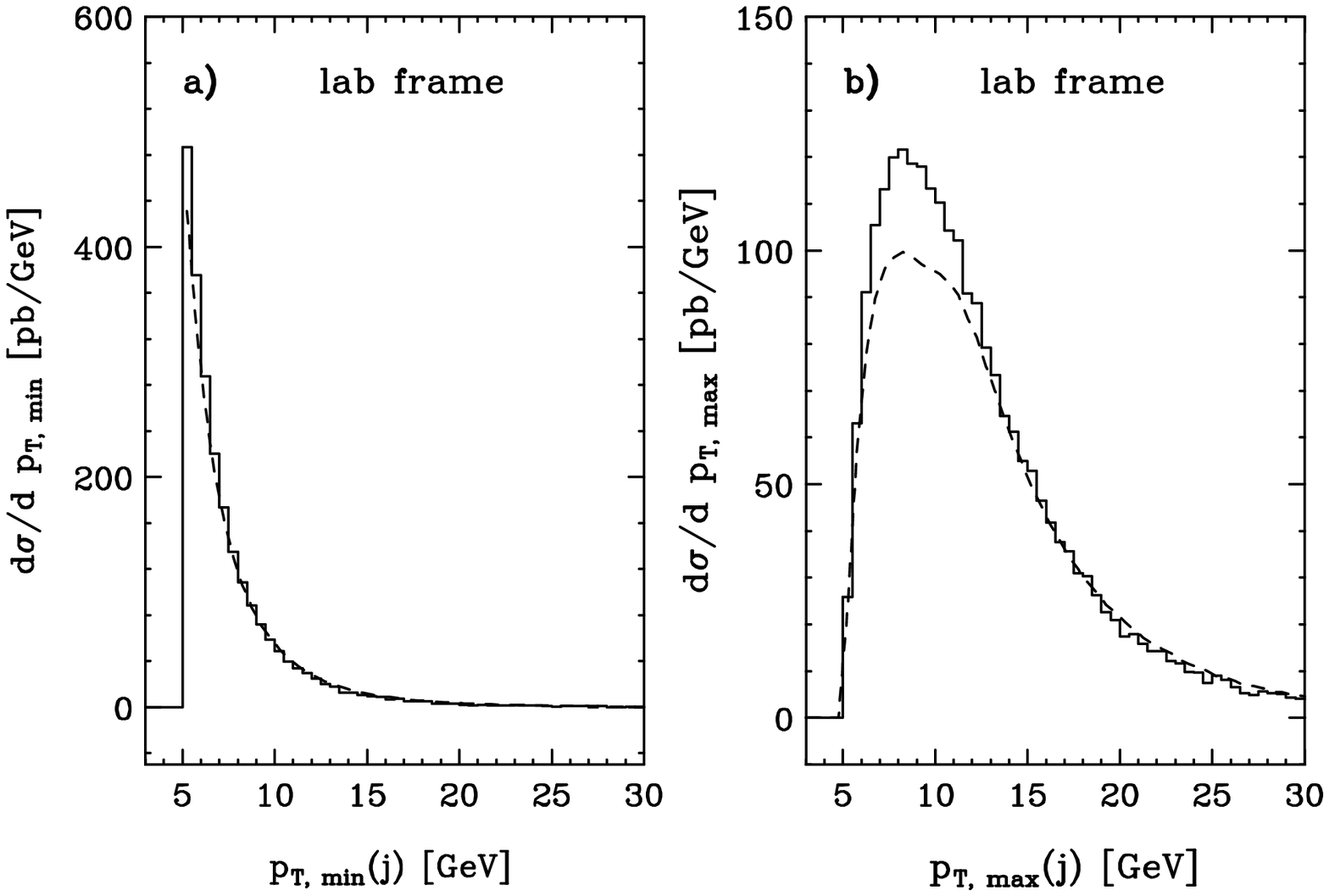}}
\end{picture}
\caption{
Transverse momentum distribution in the lab frame
for the jet with (a) minimal and (b) maximal transverse momentum.
Results are shown for the 2-jet inclusive cross
section in the cone scheme in LO (dashed curves)
and NLO  order (histograms).
From Ref.~\protect\cite{plb1}.
}
\label{f_fig3}
\end{figure}

The transverse momentum distributions of the 
softest and the hardest jet in the laboratory frame are shown in 
Fig.~\ref{f_fig3}, for the cone scheme. 
The fairly small NLO corrections allow for reliable theoretical predictions.
In general the largest radiative corrections are observed at small jet $p_T$,
as evidenced by the shape change in the 
$p_{T,\mboxsc{max}}$ distribution of Fig.~\ref{f_fig3}b. The 
predictions are therefore expected to become more reliable 
for higher jet transverse momenta.
A potential problem is the very steep
$p_{T,\,\mboxsc{min}}^{\mboxsc{lab}}$ distribution,
which, via the cut at 5~GeV, introduces a strong sensitivity to the 
correct matching of the parton $p_T$ and the measured jet $p_T$.
However, this is a general problem for all jet algorithms, {\it i.e.}
the jet rate falls very rapidly as the required energy scale of the jets 
is increased.

\setlength{\unitlength}{0.7mm}
\begin{figure}[htb]               \vspace*{-1.3cm}
\begin{picture}(150,165)(-30,1)
\mbox{\epsfxsize10.0cm\epsffile[78 222 480 650]{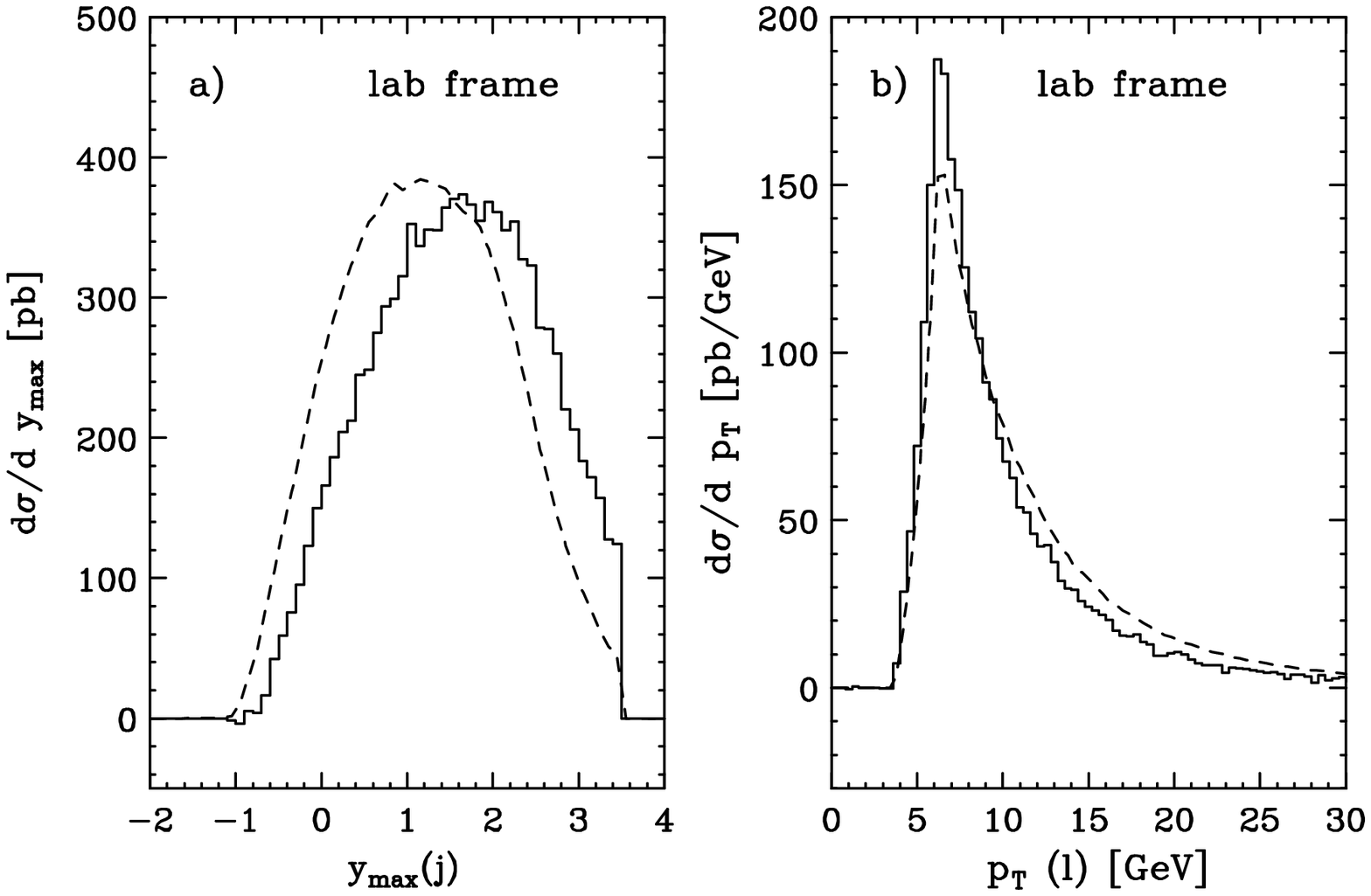}}
\end{picture}
\caption{
Rapidity distribution of the most forward jet (a) and 
transverse momentum distribution of the scattered electron (b)
in the lab frame. Results are shown for the  $k_T$ scheme in LO
(dashed curves) and NLO (histograms) for the 
2-jet inclusive cross section.
From Ref.~\protect\cite{plb1}.
}
\label{f_fig4}
\end{figure}

A more critical case is shown in Fig.~\ref{f_fig4} 
where, for jets defined in the 
$k_T$-scheme, the jet rapidity and the electron transverse momentum in the 
lab frame are shown. At NLO jets are produced somewhat more 
forward (in the proton direction) than at LO, see Fig.~\ref{f_fig4}a.
Hence, the rapidity cut at $|\eta_j|=3.5$ has a stronger effect in NLO,
which partially explains the relatively low 
$K$-factor of 0.97 in the $k_T$-scheme.

Another observable which exhibits rather large NLO corrections is the 
electron transverse momentum distribution in Fig.~\ref{f_fig4}b. 
The electron $p_T$ becomes considerably 
softer at NLO, with an effective $K$-factor above 
unity at small $p_T(\ell)$ and $K<1$ in the high transverse momentum 
region. In view of these shape changes the overall small change at NLO
has to be considered a coincidence, tied to the choice of $p_T(\ell)$ 
range. Since the electron transverse 
momentum and the $Q^2$ of the event are very closely related, a similar 
change in the size of radiative corrections is obtained by choosing 
different $Q^2$ bins. Very similar effects on the $y_{\mboxsc{max}}(j)$ and 
$p_T(l)$ distributions are also observed in the other jet definition schemes.
As a result, a judicious choice of phase space region could generate 
very large or small $K$-factors which would indicate that, in these phase 
space regions, even the NLO calculation is fraught with large uncertainties.
To avoid such potential problems,
one should investigate the effect of the higher order corrections
on those variables which are used to define kinematical cuts.

NLO effects in the
azimuthal angular distribution of dijet events around the virtual boson
proton direction will be discussed in the remaining part of this section.
The general structure of the $\phi$ distribution is discussed in 
section~\ref{twojetintro} together with
LO predictions for the $\phi$ distributions  in Fig.~\ref{f_phi}.
At LO, both jets are exactly back-to-back in $\phi$.
The NLO tree level contributions in
Eqs.~(\ref{qtoqgg}-\ref{gtoqqbarg}) imply however, that the two jets
are no longer necessarily back-to-back in $\phi$ like in LO, {\it i.e.}
one expects $\Delta\phi=|\phi(\mbox{jet 1})-\phi(\mbox{jet 2})|
\neq 180^\circ$.
A deviation from $\Delta\phi=180^\circ$ can arise 
for example in the cone scheme 
if one of the three final state partons 
is well separated in $\Delta R$ from the other two partons
but does not pass the acceptance cuts (like the
$p_T(j)>5$ GeV cut),
whereas the remaining two partons pass  all
jet requirements. Thus the event will be accepted
as a 2-jet event where the jets are, however, no longer balanced in $\phi$.

\begin{figure}[htb]
\vspace*{1in}            
\begin{picture}(0,0)(0,0)
\includegraphics{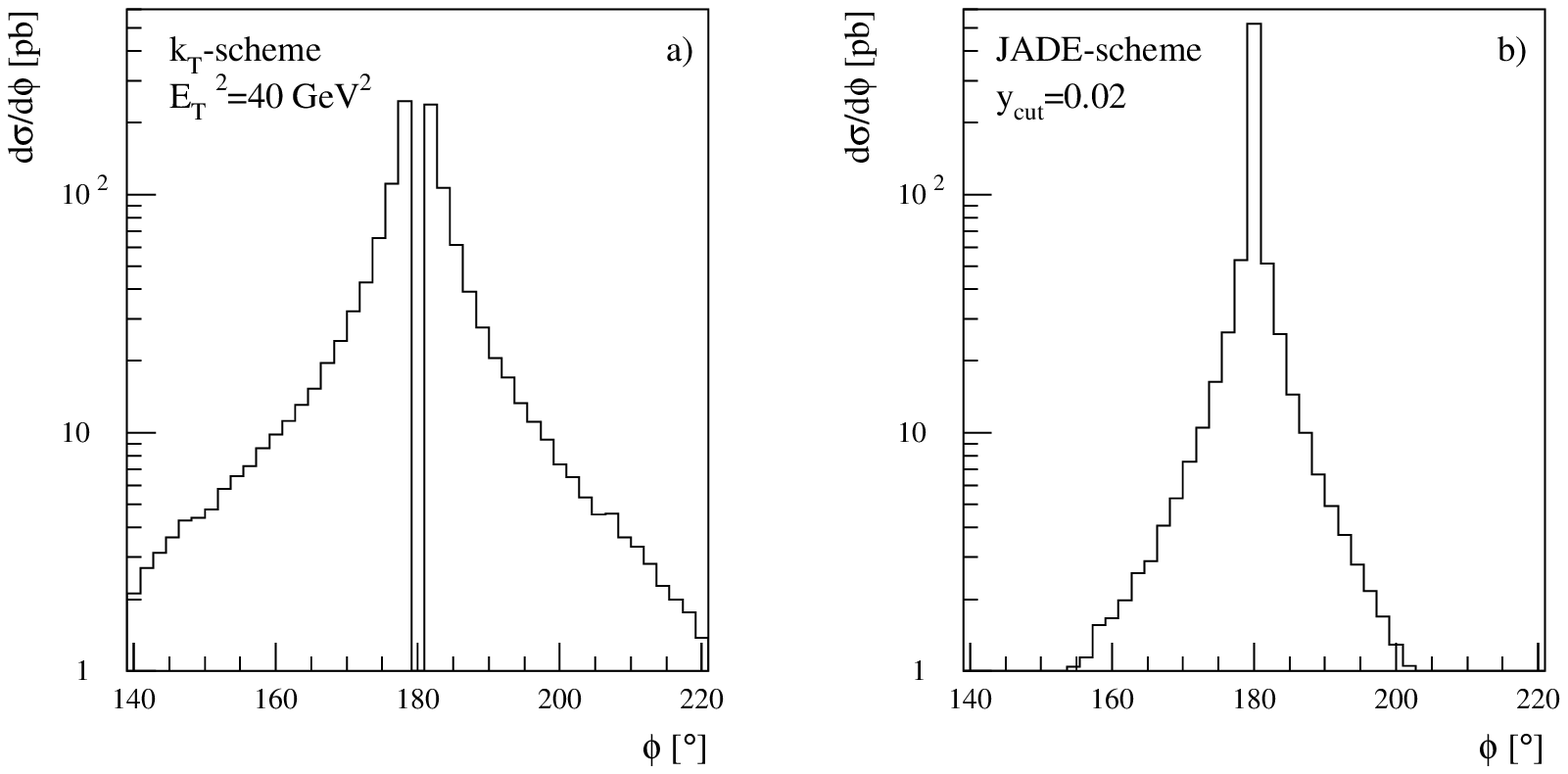}
\end{picture}
\vspace{6cm}
\caption{
Dijet azimuthal decorrelation $\Delta \phi$ in NLO for jets defined in (a)
the $k_T$ scheme with $E_T^2=40$ GeV$^2$ and
(b) the JADE scheme with $y_{\protect\mboxsc{cut}}=0.02$.
Decorrelation effects for jets defined in the cone scheme in the HCM
are similar to the results in (a).
From Ref.~\protect\cite{phipaper}.
}
\label{f_decorel}
\end{figure}
The effect is shown in Fig.~\ref{f_decorel} for 
the $k_T$ jet algorithm
with $E_T^2=40$ GeV$^2$ (Fig.~\ref{f_decorel}a) and
the JADE scheme 
with $y_{\mboxsc{cut}}=0.02$
(Fig.~\ref{f_decorel}b).
The corresponding NLO (LO) cross sections are
1350 pb (1240 pb) in the $k_T$ scheme and 1570 pb (970 pb) in the JADE scheme.
Fig.~\ref{f_decorel} illustrates that the decorrelation effect through the
NLO corrections depends strongly on the chosen jet algorithm \cite{phipaper}.
The decorrelation is larger in the $k_T$ scheme (or a cone scheme)
than in the JADE scheme.
Note that the central bin around 180$^\circ$ 
in Fig.~\ref{f_decorel}a has a negative weight showing
that the fixed NLO predictions 
are not infrared safe for $\Delta\phi$ close to 180$^\circ$.
This effect is caused by the negative contributions from the
virtual corrections, which contribute only to this bin
due to the Born kinematics. 
For the JADE scheme the negative contributions in the central bin
are already overcompensated by the positive tree-level contributions in 
Eqs.~(\ref{qtoqgg}-\ref{gtoqqbarg}).
Since the decorrelation effect is larger in Fig.~\ref{f_decorel}a,
one has either to choose wider bins in $\Delta\phi$ for the $k_T$ scheme
to arrive at a positive result in the central bin 
or alternatively one would have to use resummation techniques to
obtain a reliable perturbation expansion close to $\Delta\phi$=180$^\circ$.
The small asymmetry in the $\Delta\phi$ decorrelation
in Fig.~\ref{f_decorel} is caused
by our fixed ordering of jet~1 and jet~2,
{\it i.e.} it is assumed, that one can separate
a quark, anti-quark and a gluon jet.
The distributions  would be perfectly symmetric
without the latter assumption.

\newpage
\subsubsection{Comparisons with Experimental Results
\protect\vspace{1mm}}
\label{compare}
\enlargethispage{1cm}
In order to compare experimental jet cross sections and distributions
directly with NLO parton predictions, the measured jet cross sections
are  corrected to the parton level, {\it i.e.} corrections for both detector
effects and hadronization have to be  applied.
Such corrections can be made with models 
incorporating parton showers/dipol chains and a hadronization phase.
Various results presented at the DIS96 \cite{dis96} and DIS97 \cite{dis97}
({\it  e.g.} in Refs.~\cite{rom_exp,jose1,tancredi_rom,daiva,wobisch,weber})
show that the present data are well described by the
ARIADNE \cite{ariadne} program and are reasonably well described by LEPTO
\cite{lepto} or HERWIG \cite{herwig}
(see also Ref.~\cite{grindhammer}).
However, the relationship between the NLO partons and 
ARIADNE/LEPTO/HERWIG partons is not completely clear at the moment
and this introduces one of the main uncertainties in various
attempts to determine the strong coupling constant or
the gluon density in the proton (see the following two sections).

The most striking feature in the  comparison
of  corrected data and NLO parton predictions for dijet cross sections
is that the integrated theoretical dijet cross sections/rates
and distributions
in cone schemes and the $k_T$ scheme
tend to be lower than the measured corrected experimental dijet cross
sections{\footnote{NLO dijet cross sections
based on the JADE scheme are however in good agreement 
with the corrected data. They will be  discussed in the next section.}. 
With a coupling constant and parton densities
which describe the total DIS cross sections,  
the calculations are up to 30-40\% below the
measurements.
%
%
\begin{figure}[hb]
  \centering
  \mbox{\epsfig{file=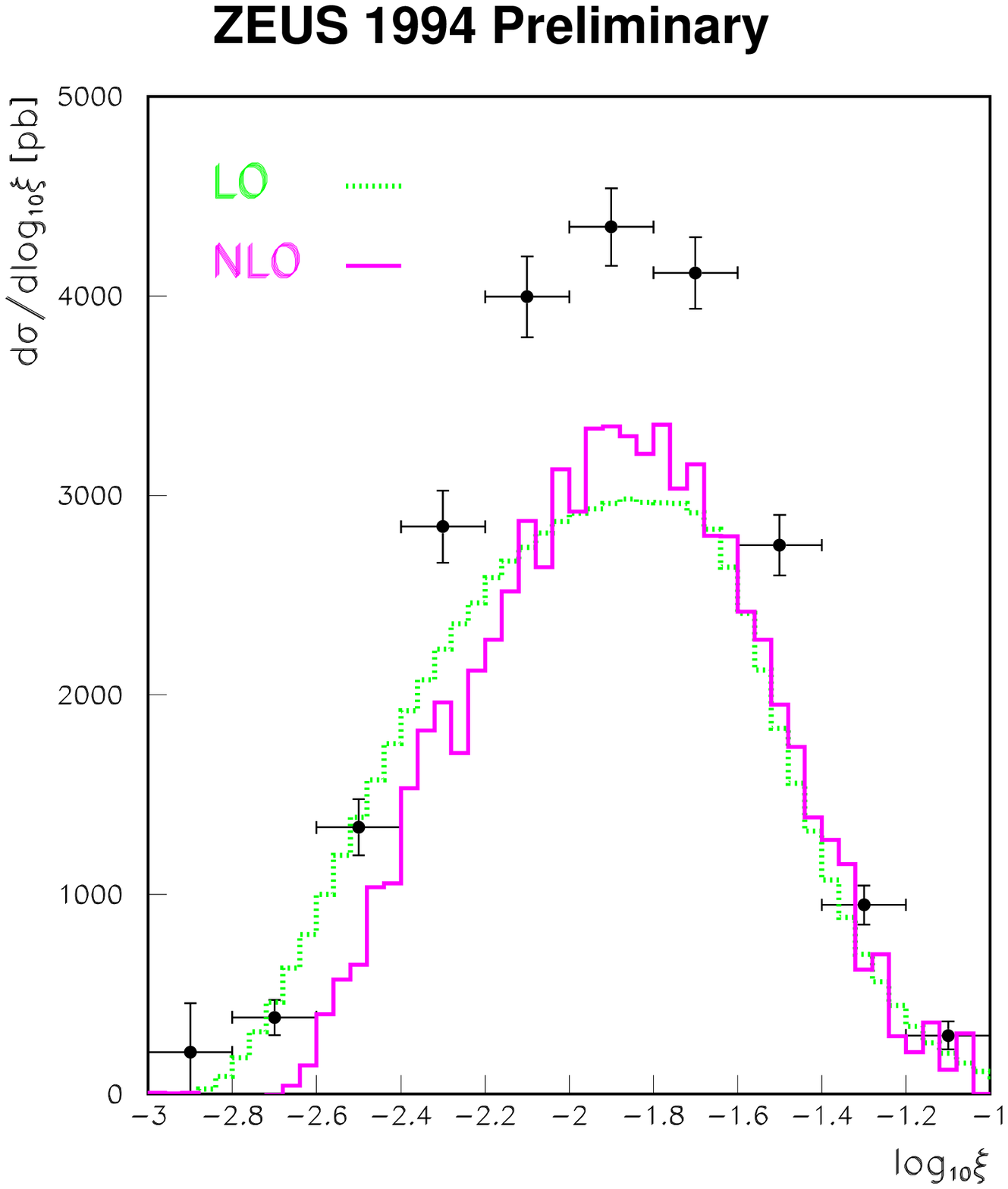,bbllx=10,bblly=0,bburx=590,bbury=590,
         width=0.50\linewidth}} 
\vspace*{-5mm}
\caption{
ZEUS preliminary dijet cross section as a function of $\xi$,
the fractional momentum of the  incoming parton,
compared to LO and NLO predictions. 
Events were selected with
7 GeV$^2<Q^2<100$ GeV$^2, y>0.04$, and $E_e^\prime>10$ GeV.
Jets were reconstructed using a cone algorithm with radius $R=1$ in the
laboratory frame.
In addition, jets are required to have a minimum 
$p_T^{\protect\mboxsc{lab}},p_T^{\protect\mboxsc{B}}>4$ GeV
and $|\eta^{\protect\mboxsc{lab}}(j)|<2$.
From Ref.~\protect\cite{daiva}.
}
\label{fig_daiva1}
\end{figure}
%
%

As an example, Fig.~\ref{fig_daiva1} compares
a preliminary dijet cross section from ZEUS, corrected to the parton level,
to NLO and LO predictions based on \docuname\ \cite{daiva}.
The cross section is measured as a function of $\xi=x(1+s_{jj}/Q^2)$,
the momentum fraction of the parton entering the hard scattering process
($s_{jj}$ denotes the 
invariant mass squared of the two jets).
The data are about 30\% higher than the NLO calculation and these differences
persist after taking into account 
various sources of systematic uncertainties like
variations in calorimeter energy scale, 
jet energy resolution, the Monte Carlo used to correct to the parton level,
the input parton densities or the factorization/factorization scale 
\cite{daiva}.
The disagreement holds  when the jet $p_T$ is raised to 6 GeV
\cite{daiva}.
A similar excess of measured dijet cross sections
within a kinematical range of 5 GeV$^2<Q^2<100$ GeV$^2$
above NLO predictions 
has been reported by the H1 collaboration \cite{wobisch}.
We will comment on these problems  in section~\ref{conclusions}.

\enlargethispage{1cm}
On the other hand, it is interesting to observe that
 the shape of the cross sections is well described 
by the NLO calculations. This is shown in 
Fig.~\ref{fig_daiva2} 
where normalized, measured 
preliminary differential dijet cross sections based on 1994 ZEUS data
are compared with LO and NLO predictions.
Fig.~\ref{fig_daiva2} shows
that NLO corrections are clearly needed on the theoretical side 
to describe the observed shape of the distributions.
Similar results are found when normalized
jet transverse momenta distributions in the laboratory frame 
and the HCM  are compared to NLO predictions \cite{daiva}.

%
%
\setlength{\unitlength}{0.7mm}
\begin{figure}[htb]               \vspace*{-2cm}
\begin{picture}(150,165)(-30,1)
\mbox{\epsfxsize10cm\epsffile[78 222 480 650]{f_daiva2.ps}}
\end{picture}
\vspace*{3.5cm}
\caption{
ZEUS preliminary normalized dijet cross section as a function of 
$Q^2$ (a), Bjorken $x$ (b), and the rapidity distribution
of the most forward (c) and most backward (d) jet in the
laboratory frame compared to LO and NLO predictions. 
Events were selected within the same range and jet definition as
in Fig.~\protect\ref{fig_daiva1}.
From Ref.~\protect\cite{daiva}.
}
\label{fig_daiva2}
\end{figure}
%
%

\clearpage

\subsubsection{The Determination of the Strong Coupling Constant
\protect\vspace{1mm}}
\label{sec_alphas}
\enlargethispage{0.5cm}
The above mentioned discrepancy between the NLO predictions and corrected
data for dijet events in the cone and $k_T$ schemes prevented so far 
a NLO $\alpha_s$ determination in these schemes.
The situation looks much better when jets are defined in a
JADE type clustering algorithm.
Due to the very large dependence of the NLO dijet cross sections on the
recombination prescription and on the exact definition of the
jet resolution mass 
(see section~\ref{kfactors} for a detailed discussion of these effects)
it is  essential that the theoretical calculations are
exactly  matched to the experimental definitions when using these cluster
algorithms. 
Rosenbauer and Trefzger~\cite{rom_exp}
find indeed similarly large differences in the experimental
jet cross sections, which are in good agreement with 
\docuname\ predictions,
when the data are processed with  
exactly the same jet resolution mass
and recombination prescription as used in the theoretical calculation.
The  2-jet rate, for example, in the $W$ scheme
(with $E$ recombination) and in the JADE scheme 
(with $P$ recombination) for corrected
ZEUS data are $18.6\pm 0.7 \%$ and $8.6\pm 0.5 \%$, respectively.
The corresponding 
NLO predictions from \protect\docuname\ for the same kinematics and the same
jet definitions  are $17.9 \% $ and 8.6 \%
(see T.~Trefzger \cite{rom_exp}).

A recent measurement of the differential dijet rate 
in the JADE algorithm (with $E$ recombination)  
has been performed by the H1 collaboration \cite{weber}.
Fig.~\ref{fig_j2} shows the distribution of the
resolution parameter $y_2$ for unfolded H1 data and NLO predictions
for different values of $\Lms{4}$.

\begin{figure}[hbt]
  \centering
  \mbox{\epsfig{file=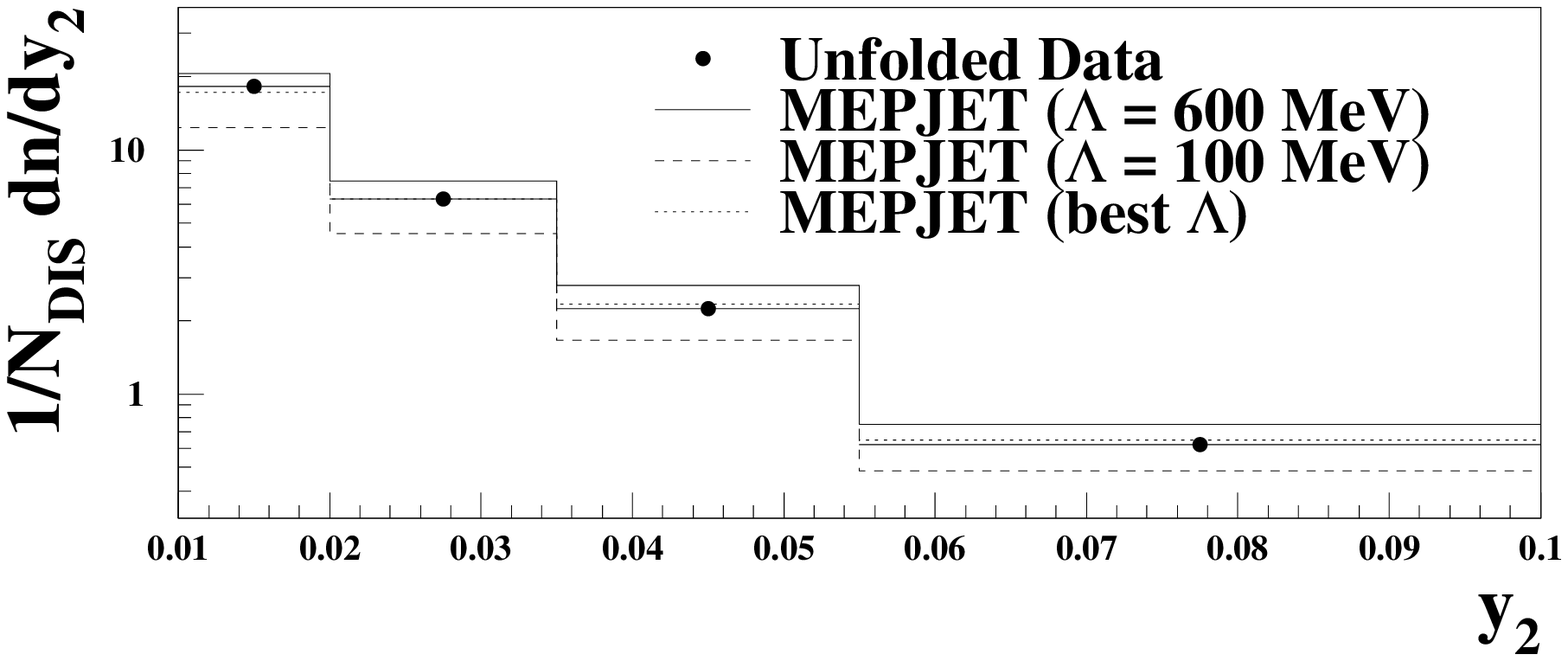,bbllx=0,bblly=0,bburx=595,bbury=265,
         width=0.90\linewidth}} 
\caption{
Distribution of $y_2$ as explained in the text
for unfolded data compared to NLO predictions for
different values of $\Lms{4}$.
Events were selected with
$Q^2>200$ GeV$^2$ and $W^2>5000$ GeV$^2$.
Clusters
 were required to have a polar angle $>7^\circ$ in the laboratory frame.
From Ref.~\protect\cite{weber}.
}
\label{fig_j2}
\end{figure}

For each event $y_2$ is defined as the value of $y_{\mboxsc{cut}}$
in the JADE algorithm where a 1-jet event
switches to a 2-jet event and hence
all events are considered as 2-jet events.
Neglecting the small fraction of 3-jet
events in the JADE scheme, the distribution of this quantity
is equivalent to 
$1/N_{\mboxsc{DIS}} d\sigma[2{\mbox{-jet}}]/dy_{\mboxsc{cut}}$.
$N_{\mboxsc{DIS}}$ is the total number of DIS events.
All $y_2$ bins in Fig.~\ref{fig_j2} show a strong sensitivity to
$\Lms{4}$ and thus to $\alpha_s(m_Z)$. The data and  NLO calculations
are in excellent agreement for one given value of $\Lms{4}$.
There is so far no published result on the actual value for $\alpha_s(m_Z)$
based on this analysis. The precision which can be expected from the 
measurement is of the order of 5-10\% \cite{tancredi}.

Previous programs \cite{disjet,projet} were limited to a $W$ type 
algorithm\footnote{DISJET \cite{disjet} and PROJET \cite{projet}
are largely based on the
fact that the calculation of the jet resolution mass squared,
$M_{ij}^2$, can be done in a
lorentz invariant way, {\it i.e.} as in the $W$ scheme.
Only in LO does this agree with the JADE
definition, defined in the lab frame,
which has been used in the experimental analysis in
\cite{exp_as}.}
and are not flexible enough to take into account
the effects of single jet masses (see
Fig.~\ref{f_fig5} in  section~\ref{nloeffects})
or differences between (massless) recombination schemes. 
In addition, approximations were made 
to the matrix elements in these programs which are not 
valid\footnote{The
problems in Refs.~\protect\cite{disjet,projet}  
are in fact much more severe than
being insensitive to massless recombination scheme prescriptions
\cite{disaster}.
The DISJET result for  the 2-jet inclusive cross sections
discussed in table~\ref{table_dijet}
is for example close to the Born result of 1020 pb,
which is far below the range of the  results based on
massless $E0$ and $P$ recombination schemes
in the $W$ or JADE Algorithmus  with MEPJET.} in large regions
of phase space~\cite{plb1}.
These problems are reflected in inconsistent values for $\alpha_s(M_Z^2)$ 
[ranging from 0.114 to 0.127 in the H1 analysis \cite{exp_as},
(see K. Rosenbauer \cite{rom_exp})], when these programs are used to analyze
the data with different (massless)
recombination schemes. Because of these problems,
the older programs cannot be used for precision studies at NLO in their 
present form. In order to reduce theoretical errors,
previous analyses \cite{exp_as} should be repeated with 
\docuname\ or  similar flexible Monte Carlo programs.
A first reanalysis, with \docuname, of H1 data by K. Rosenbauer
yields a markedly lower central value, $\alpha_s(M_Z^2)=0.112$ \cite{rom_exp},
which is {\it independent} of the recombination scheme (used in both 
data and theory), and the $\alpha_s(\mu_R^2)$ extracted from 
different kinematical bins follows nicely the expectation from the
renormalization group equation (in contrast to results based on 
{\large\sc projet}).
A similar reanalysis of the ZEUS data has also been performed 
by T. Trefzger \cite{rom_exp}, also with \docuname.

\subsubsection{The Determination of the Gluon Density
\protect\vspace{1mm}}
\label{sec_gluon}
\enlargethispage{1cm}
HERA opens  also a  new window to measure the proton structure functions,
in particular the gluon distribution, in a completely new kinematic
region. The accessible range in the Bjorken-scaling variable $x$ 
can be extended by more than two orders of magnitude towards low $x$ 
compared to previous fixed target experiments.
Dijet production in DIS at HERA in principle allows for a direct
measurement of the gluon density in the proton 
(via  $e g \rightarrow eq\bar{q}$).
The gluon initiated subprocess is the dominant contribution
to dijet production at small Bjorken $x$ (see below).

Let us first investigate theoretically the potential feasibility 
of the parton density determination
$f_i(x_i,\mu_F)$ in DIS dijet production \cite{krakau},
where $x_i$ denotes the fractional momentum 
of the  incoming parton $i$ ($i=q,g$).
In dijet production, the
fractional momentum $x_i$ and Bjorken $x$ differ substantially. 
Denoting as ${s_{jj}}$
the invariant mass squared of the produced dijet system, and considering 
2-jet exclusive events only, the two are related by
\begin{equation}
x_i = x \,\left(1+\frac{{s_{jj}}}{Q^2}\right)
\label{xi_def}
\end{equation}
For the following studies we use 
a cone scheme algorithm defined in the lab frame with $\Delta R=1$ in a 
$Q^2$ range of  $5<Q^2< 2500$ GeV$^2$. 
Jets are required to have transverse momenta of at least 5 GeV in the
laboratory frame and in the Breit frame.
In addition, we require 
$0.04 < y < 1$, an energy cut of $E(e^\prime)>10$~GeV on the scattered 
electron, and a cut on the pseudo-rapidity $\eta=-\ln\tan(\theta/2)$
of the scattered lepton and jets of $|\eta|<3.5$. 
The LO (NLO) results are based on 
the LO (NLO) parton distributions from GRV \cite{grv} together with
the 1-loop (2-loop) formula 
for the strong coupling constant.
With these parameters, one obtains 
2890 pb (2846 pb) for the LO (NLO) 2-jet exclusive cross section.

In order to investigate 
the feasibility of the parton density determination
at low $x$, Fig.~\ref{fig_gluon}a  shows the Bjorken $x$ distribution of the
2-jet exclusive cross section in the cone scheme.
The gluon initiated subprocess clearly dominates the Compton process
for small $x$ in the LO predictions. The effective $K$-factor close to unity
for the total exclusive dijet cross section is a consequence of
compensating effects in the low $x$ ($K>$ 1) and high $x$ ($K<$1) regime.
\begin{figure}[htb]
\vspace*{-6cm}
\begin{center}
\begin{picture}(150,165)(-30,1)
\mbox{\epsfxsize8.0cm\epsffile[78 222 480 650]{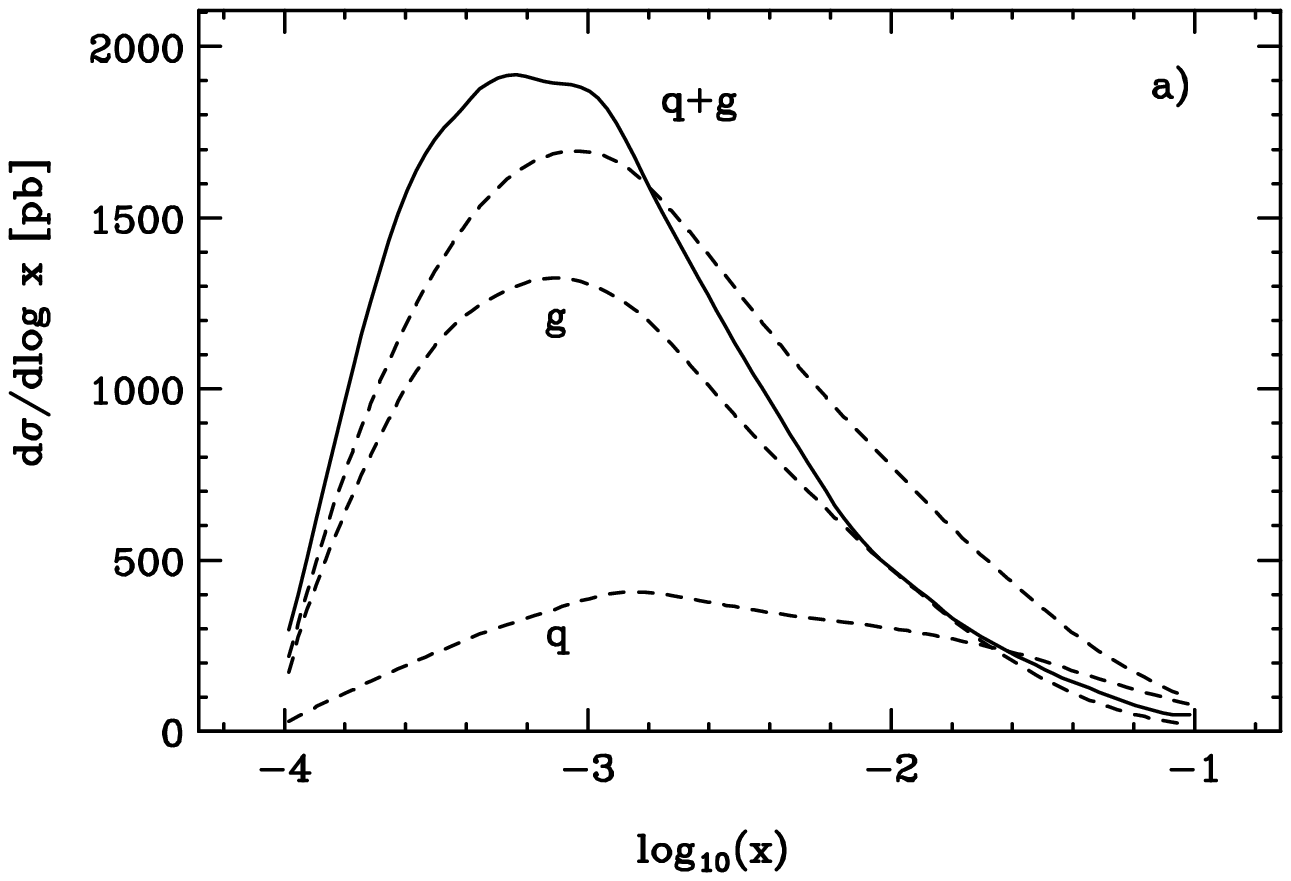}}
\end{picture}
\end{center}
\mbox{}\\[-8cm]
\begin{center}
\begin{picture}(150,165)(-30,1)
\mbox{\epsfxsize8.0cm\epsffile[78 222 480 650]{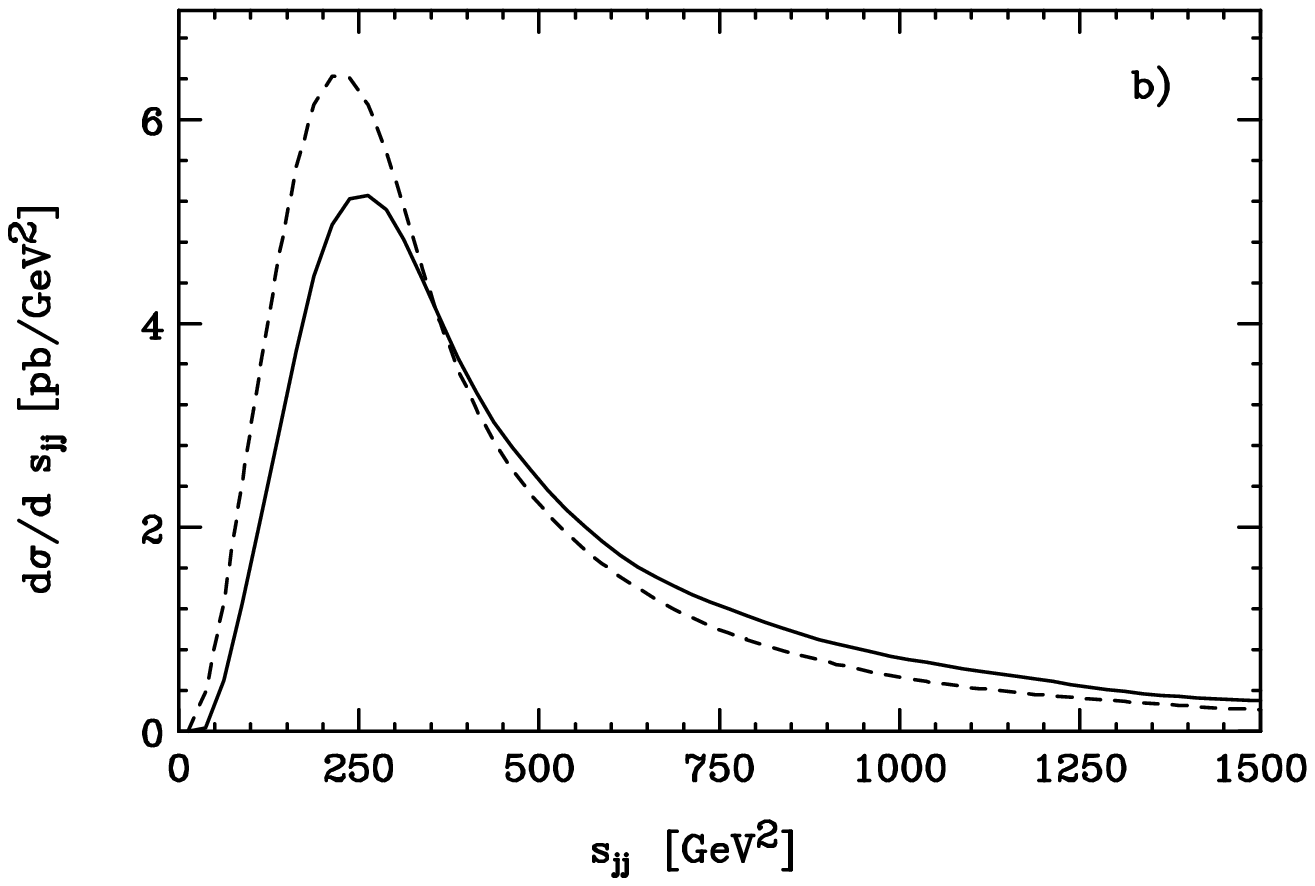}}
\end{picture}
\end{center}
\mbox{}\\[-8cm]
\begin{center}
\begin{picture}(150,165)(-30,1)
\mbox{\epsfxsize8.0cm\epsffile[78 222 480 650]{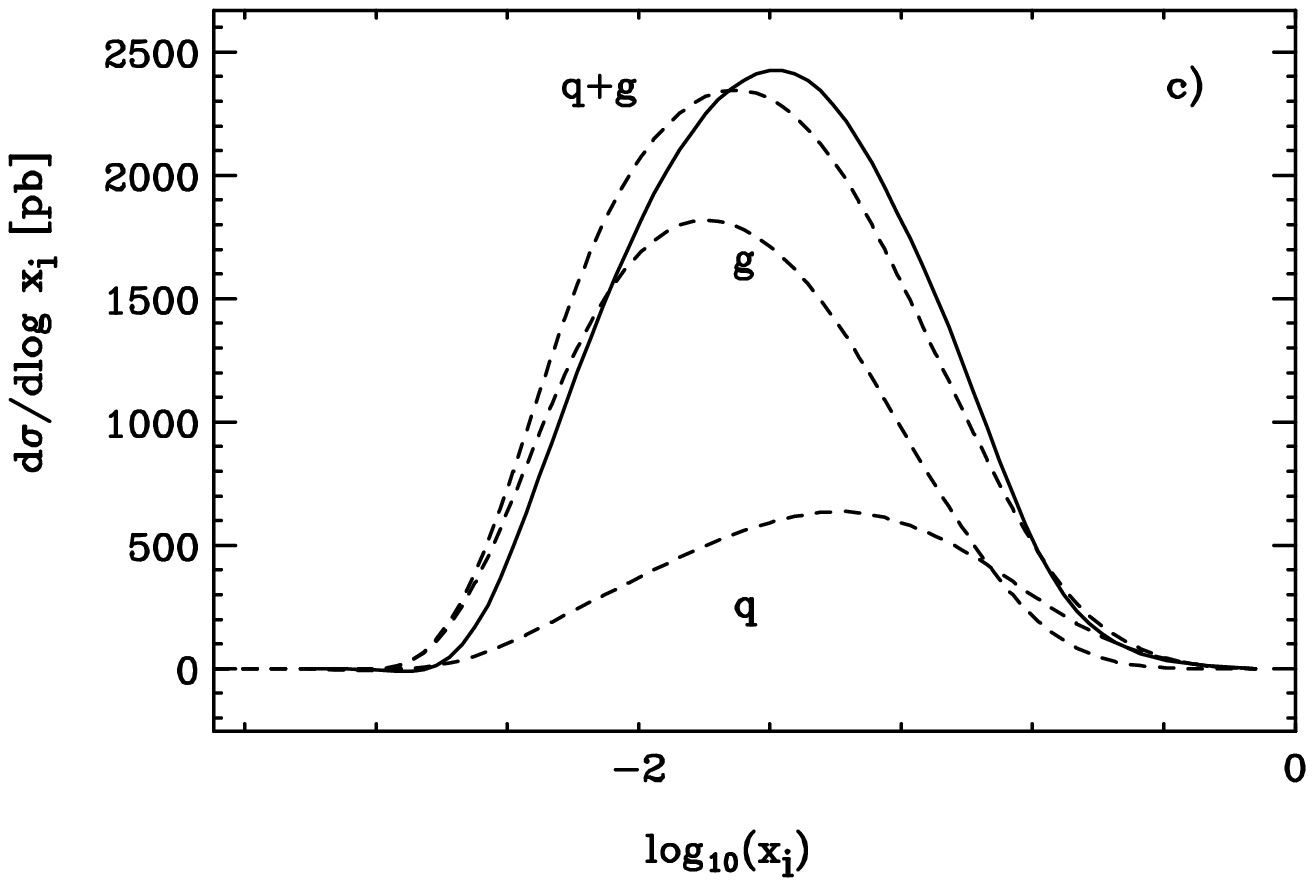}}
\end{picture}
\end{center}
\vspace*{-1.5cm}
\caption{
(a) Dependence of the exclusive 2-jet cross section
in the cone scheme on Bjorken $x$ for the
quark and gluon initiated subprocesses and for the 
sum. Both LO (dashed) and NLO (solid) results are shown;
(b) Dijet invariant mass distribution in LO (dashed) and in NLO (solid);
(c) Same as a) for the $x_i$ distribution, $x_i$ representing the 
momentum fraction of the incident parton at LO.
From Ref.~\protect\cite{krakau}.
}
\label{fig_gluon}
\end{figure}
For the isolation of parton structure functions we are interested in the 
fractional momentum $x_i$ which is related to Bjorken $x$ and the
the invariant mass squared of the produced dijet system $s_{jj}$
by Eq.~(\ref{xi_def}).
The $s_{jj}$ distribution of Fig.~\ref{fig_gluon}b exhibits rather large NLO
corrections as well. The invariant mass squared of the two jets is considerable
larger at NLO than at LO (the mean value of $s_{jj}$ rising to 570~GeV$^2$ 
at NLO from 470~GeV$^2$ at LO).

The large NLO corrections to the $x$ and $s_{jj}$ distributions
have a compensating effect on the $x_i$ distribution 
in Fig.~\ref{fig_gluon}c, which shows
very similar shapes at LO and NLO. 
These fairly small NLO corrections to the
$x_i$ distribution point towards a reliable extraction of $g(x_i,\mu_F)$
via DIS dijet production.
Note that the $x_i$ distribution is markedly shifted towards
larger values compared to the $x$ distribution, which demonstrates that
a determination of the gluon density $g(x_g,\mu_F)$ for very small $x_g$
values ($\lsimfig 10^{-3}$) is not feasible.

At LO a direct determination  of the gluon
density is possible from the LO $x_g$ distributions, 
after subtraction of the 
calculated Compton subprocess. 
Such a LO gluon analysis has been performed by the H1 collaboration 
\cite{h1_gluon}.
This simple picture is modified in NLO,
however, and the effects of Altarelli-Parisi splitting and low $p_T$ partons
need to be taken into account to determine the structure
functions at a well defined factorization scale $\mu_F$. 

\enlargethispage{1cm}
In principle, a direct extraction of the parton distributions $f_i(x_i,\mu_F)$
could be done using a NLO program. However,
this would imply re-generating numerical tables for parton distribution
functions for each new trial value of NLO  input distributions,
and then re-running the jet program for these values. This procedure 
is  extremely time-consuming.
In fact, much of the calculation-- such as the jet algorithm and
experimental cuts, the numerical integration over real emission, 
the balancing of real and virtual contributions--is  independent of the
precise form of the parton distribution functions.
Thus it would be useful to re-organize the calculation so as to minimize 
the amount of computational work needed for each trial value of the initial 
parton distributions in a fit to experimental data.
One method to determine the gluon density
in NLO based on the use of Mellin transforms is presented in Ref.~\cite{gluon}.
A more general approach ({\it e.g.} fully flexible in the choice 
of the factorization scale) which is also applicable for the NLO calculation 
based on the crossing function technique is presented in Ref.~\cite{kosower}.
These formalism reorganize the NLO calculations in such a way
that  most of the time consuming computations are done only
once. We refer the reader to \cite{gluon,kosower} for more details.

A first study by the ZEUS collaboration
of the experimental sensitivity  of dijet events
to the gluon density in the $x_g$ range between $10^{-3}$ and
$10^{-1}$ using a cone algorithm in the laboratory frame
has been reported by J.~Repond \cite{jose1}.
The slope  of the gluon density $x_g g(x_g,\mu_F)\propto x^\lambda$
is primarily defined by the exponent $\lambda$ and a fit of
ZEUS 1994 data to the slope of the  NLO predictions 
in Fig.~\ref{fig_daiva1} yielded 
$\lambda_{\mboxsc{meas}}=-0.38\pm 0.04\pm 0.18$ at $Q^2=4$ GeV$^2$
\cite{jose1,daiva}.

\clearpage
\subsubsection{The Internal Structure of Jets in DIS
\protect\vspace{1mm}}
\label{intstruc}
The shape of the transverse energy distribution of particles
within a jet gives interesting information about the process by which hard
partons are confined into jets of hadrons. An experimental
investigation of this internal jet structure has recently been presented 
by the ZEUS collaboration \cite{martinez}
for jets defined in a cone scheme in the lab frame with radius $R=1$.
Clearly, most of these jets are quark jets,
since this analysis does not require two or more jets in the final state.
The analyzed jet shape, $\Psi(r)$, is defined as the average fraction of 
the jet's transverse energy that lies inside an inner coaxial radius $r$:
\begin{equation}
\Psi(r)=\frac{1}{N_{jets}}\sum_{jets}\frac{E_T(r)}{E_T(R)},
\hspace{1cm}
E_T(r)=\int_0^r dr^\prime dE_T(r^\prime)/dr^\prime
\end{equation}
Clearly one has $\Psi(R)=1$, with $\Psi(r)$ rising monotonically
in $r$, indicating that a lower fraction of the jet energy lies in 
the sub-cone of radius $r$. Fig.~\ref{fig_jetshape}
compares the jet shape $\Psi(r)$ for jets  in DIS 
with those from $e^+e^-$ annihilation and $p\bar{p}$ experiments
for jets defined in a cone scheme with $R=1$.
DIS jets are selected
with transverse energies in the range of $37$ GeV
$<E_T^{\mboxsc{lab}}(j)<$ 45 GeV. Similarly the OPAL, CDF and D0
data are shown for minimum jet transverse energy ranges around 40 GeV.
One observes that the DIS and 
$e^+e^-$ jet profiles fall less steeply
compared to the $p\bar{p}$ jets for decreasing $r$,
{\it i.e.} the $p\bar{p}$ jets are broader, with only about  50\% of their
transverse energy within a sub-cone with radius of 0.2,
whereas the DIS and $e^+e^-$ jets contain about
70\% of their transverse energy within this sub-cone  radius.
This is qualitatively as expected, since the $p\bar{p}$ jets at these
$E_T$ values are  preferredly gluon jets, which should be broader
than the dominantly  produced quark jets in DIS and $e^+e^-$
(see {\it e.g.} Ref.~\cite{esw} for a detailed discussion).
\begin{figure}[hb]
  \centering
  \mbox{\epsfig{file=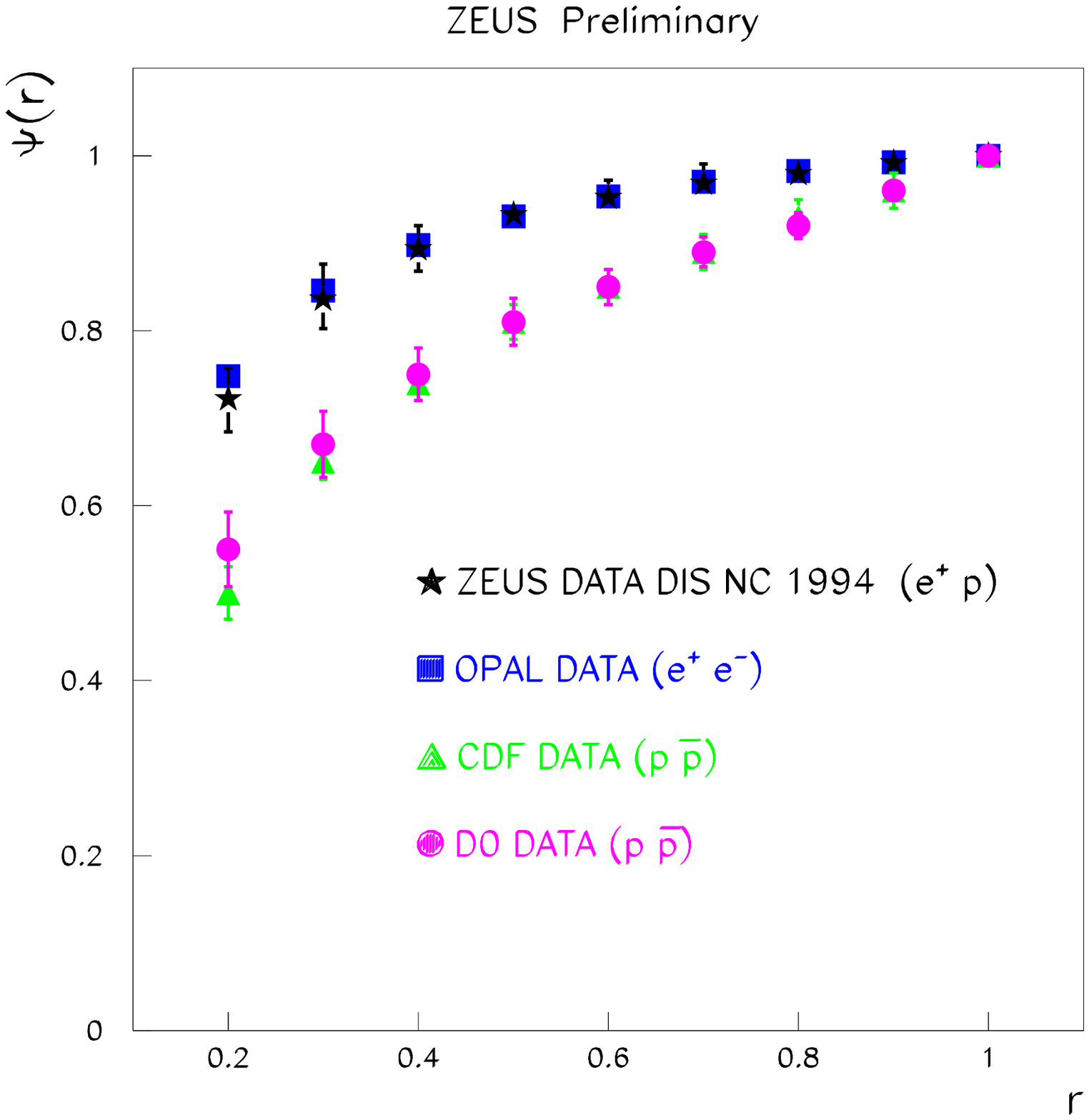,
        bbllx=15,bblly=128,bburx=581,bbury=685,
        width=0.45\linewidth}} 
\caption{
Comparison of jet shape measurements from ZEUS (DIS), OPAL $(e^+e^-)$,
CDF and D0 $(p\bar{p})$.
The jet energy ranges are 37$<E_T^{\protect\mboxsc{lab}}(j)<$ 45 GeV,
35 GeV $<E(j)$, $40<E_T(j)<$ 60 GeV and
$45<E_T(j)<$ 70 GeV, respectively.
From Ref.~\protect\cite{martinez}.
}
\label{fig_jetshape}
\end{figure}

First theoretical sensitivity to the partition of particles
in a jet and therefore the internal jet structure 
$\Psi(r)$ is obtained by
NLO corrections, which imply that a jet 
may consist of two partons.
The theoretical sensitivity to the internal jet structure is,  
however, only simulated at tree level 
in a NLO calculation and is therefore subject to potentially 
large higher order corrections.
A theoretical comparison with the DIS results in
Fig.~\ref{fig_jetshape} 
has not been done yet.

\subsubsection{Event Shape Variables and Power Corrections
\protect\vspace{1mm}}
\label{eventshape}
Another approach to studying the characteristics of the hadronic
final state in DIS is to use event shape variables,
which are already well-established and  successfully applied
in $e^+e^-$ experiments \cite{eventshape}
and which are expected to be
relatively insensitive to soft gluon
emission and collinear parton branching.
Restricting the DIS final state to the current hemisphere
in the Breit frame provides a similar kinematical configuration
(for fixed $Q$) as in $e^+e^-$ annihilation.

The H1 collaboration has recently presented a measurement of four
event shape variables in DIS for momentum transfers $7<Q^2<100$ GeV$^2$
in the current region of the Breit frame \cite{h1event}.
The measured quantities are
two version of thrust, $T_C$ and $T_Z$, the
jet broadening $B_C$ and the jet mass $\rho_C$:
\begin{eqnarray}
T_C&=&\mbox{max}\frac{\sum_h|\vec{p}_h\cdot\vec{n}_T|}{\sum_h|\vec{p}_h|}
\hspace{1cm}(\vec{n_T}\equiv\mbox{thrust axis})\label{tcdef}\\
T_Z&=&\frac{\sum_h|\vec{p}_h\cdot\vec{n}|}{\sum_h|\vec{p}_h|}
=\frac{\sum_h|p_{zh}|}{\sum_h|\vec{p}_h|}
\hspace{1cm}(\vec{n}\equiv\mbox{boson axis})\\
B_C&=&\frac{\sum_h|\vec{p}_h\times\vec{n}|}{2\sum_h|\vec{p}_h|}
=\frac{\sum_h|p_{\perp h}|}{2\sum_h|\vec{p}_h|}
\hspace{1cm}(\vec{n}\equiv\mbox{boson axis})\\
\rho_C&=&\frac{(\sum_h\vec{p}_h)^2}{Q^2} \label{shapes}
\end{eqnarray}
The $Q$ dependence of the  mean values of these event shapes
is shown in Fig.~\ref{fig_h1event} for H1 data.

Recent theoretical developments in the understanding of infrared renormalon
contributions 
together with complete \oasz\ QCD corrections
allow the first steps towards a direct  comparison of theory and these  data 
without invoking hadronization models. 
The renormalon contributions
for the event shape variables in
Eqs.~(\ref{tcdef}-\ref{shapes})
can be modelled by a  characteristic $1/Q$ dependence.
These power corrections have recently 
been calculated in Ref.~\cite{dasgupta}.
The size of the $1/Q$ corrections is characterized in their model by
a single nonperturbative parameter $\bar{\alpha}_0$.

The data in Fig.~\ref{fig_h1event} have  been fitted
to NLO  theory, based on the {\large \sc disent} program,
 with $\alpha_s(m_Z)$ as a free parameter
plus the calculated power corrections with ${\bar{\alpha}}_0$ as a
free parameter.
All data  are nicely fitted by the
$1/Q$ dependence in the power suppressed corrections.
The result for the fit to the nonperturbative parameter
$\bar{\alpha}_0$ for  $T_C,T_Z$ and $\rho_C$ is consistent 
with a single value of ${\bar{\alpha}}_0
=0.491\pm 0.003$(exp)$^{\mboxsc{+0.079}}_{\mboxsc{--0.042}}$(theory)
\cite{h1event}. The simultaneous fit to the strong coupling constant
yields $\alpha_s(m_Z)=0.123$ for $T_C$,
$\alpha_s(m_Z)=0.115$ for $T_Z$ and
$\alpha_s(m_Z)=0.130$ for $\rho_C$, 
with an experimental and theoretical error of about
$\pm 0.003^{+0.007}_{-0.005}$ in each case.
The common fit for the strong coupling constant
yields $\alpha_s(m_Z)=0.118
\pm 0.001$(exp)$^{\mboxsc{+0.007}}_{\mboxsc{--0.006}}$(theory).
In the case of the jet broadening $B_C$, the calculation
of the power corrections is subject to larger theoretical
uncertainties and was not included in the global fit.

A comment is in order regarding the definition of the  mean values,
shown in Fig.~\ref{fig_h1event}, which includes 
for the thrust variables (for example)
the regions $T_C, T_Z \rightarrow 1$.
The limits $T\rightarrow 1$ correspond to the 1-jet configuration
and this phase space region is
not reliably predicted by the fixed order \oasz\ calculations,
but is subject to potentially large resummation effects.
Thus, we strongly recomment to exclude the region ($\approx 0.95<T<1$)
in the definition of the mean values.

\begin{figure}[hb]
  \centering
  \mbox{\epsfig{file=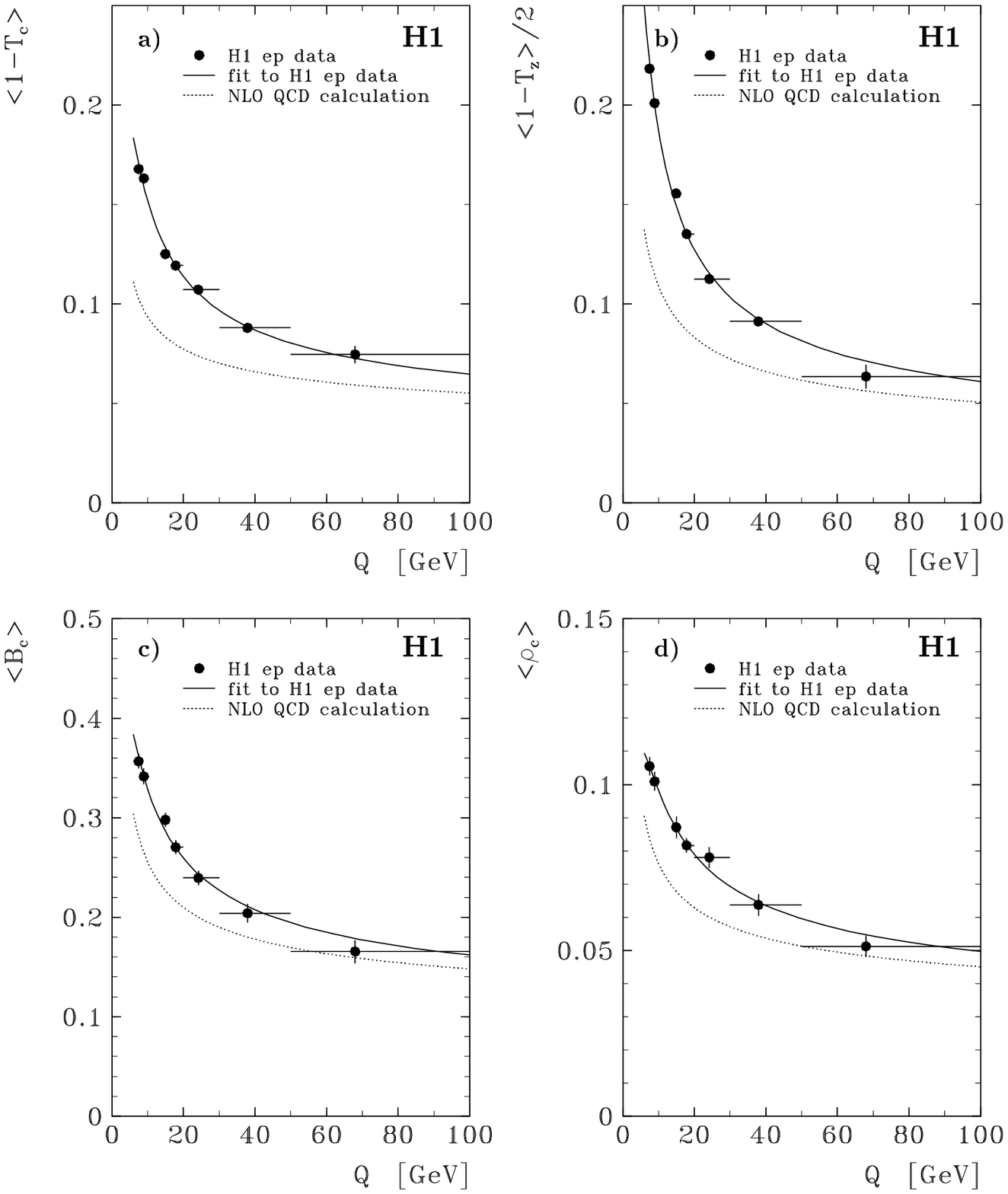,
         bbllx=0,bblly=120,bburx=580,bbury=785,
         width=0.80\linewidth}} 
\caption{
Mean values of event shape variables as a function of Q.
The values are  (a) $\langle 1-T_C \rangle$,
(b) $\langle 1-T_Z \rangle/2$, (c)
the jet broadening  $\langle B_C \rangle$ and (d) the jet
mass  $\langle \rho_C \rangle$.
The dotted line indicates the 
{\large \sc disent} NLO calculation.
The full line indicates the fit incorporating power corrections.
From Ref.~\protect\cite{h1event}.}
\label{fig_h1event}
\end{figure}
\clearpage
\subsection[{NLO  Two-Jet Cross Sections Including
$Z$ and $W$ Exchange\protect\vspace{1mm}}]{NLO  
Two-Jet Cross Sections Including
{\boldmath{$Z$}} and {\boldmath{$W$}} Exchange}
\label{twojetz}
Electroweak effects in DIS 2-jet production  become
important for momentum transfers $Q^2\gsimfig 2500$ GeV$^2$.
Analytical and numerical results will be presented
in the the subsequent sections. 

\subsubsection{Matrix Elements and Coupling Factors\protect\vspace{1mm}}}
Analytical results for the hadronic 2-jet cross section up to \oasz\
including $Z$ and $W$ exchange can be obtained 
by suitable replacements in Eq.~(\ref{twojet}) as specified below.
The $Q^2$ dependent electroweak coupling factors $A_f(Q^2)$ and
$B_f(Q^2)$ for $Z$ ($W$) exchange 
are defined in Eqs.~(\ref{afdef},\ref{bfdef}) (Eq.~(\ref{abccdef})).
Furthermore,  $\tilde\Delta f_q(\eta)$
($\tilde\Delta  C_q^{\overline{\mboxsc{MS}}}(\eta)$)
denote again the differences between the quark and antiquark parton
densities (crossing functions) as defined in
Eq.~(\ref{deltaqdef}) (Eq.~(\ref{deltacdef})).

The NC NLO hadronic  2-jet cross section including
all $\gamma-\gamma, \gamma-Z$ and $Z-Z$ contributions
can be obtained from Eq.~(\ref{twojet})
by the following replacements:
\begin{eqnarray}
\mbox{third line:}&& 
(|M^{(\mboxsc{pc,pv})}_{q\rightarrow qg}|^2
\mbox{\,\,are defined in Eqs.~(\ref{m_qtoqg},\ref{m_qtoqgpv})})
\nonumber  \\[2mm] \mbox{}
[\sum_{i=q,\bar{q}}e_i^2 f_i(\eta)]
\,\,|M^{(\mboxsc{pc})}_{q\rightarrow qg}|^2 
&\longrightarrow&
[\sum_{i=q,\bar{q}} A_i(Q^2) f_i(\eta)]
\,\,|M^{(\mboxsc{pc})}_{q\rightarrow qg}|^2 
+
[\sum_{i=q} B_i(Q^2) 
\tilde\Delta f_i(\eta)]
\,\,|M^{(\mboxsc{pv})}_{q\rightarrow qg}|^2 \nonumber \\[2mm]
\mbox{fourth line}:&&
({\cal{F}}^{(\mboxsc{pc,pv})}_{q\rightarrow qg}
 \mbox{\,\,are defined in Eqs.~(\ref{fcalqdef},\ref{fcalqpvdef})})
 \nonumber  \\[2mm] \mbox{}
[\sum_{i=q,\bar{q}}e_i^2 f_i(\eta)]\mbox{}
\,\,{\cal{F}}^{(\mboxsc{pc})}_{q\rightarrow qg}
&\longrightarrow&
[\sum_{i=q,\bar{q}} A_i(Q^2) f_i(\eta)]\mbox{}
\,\,{\cal{F}}^{(\mboxsc{pc})}_{q\rightarrow qg} 
+
[\sum_{i=q} B_i(Q^2) 
\tilde\Delta f_i(\eta)]\mbox{}
\,\,{\cal{F}}^{(\mboxsc{pv})}_{q\rightarrow qg} \nonumber \\[2mm]
\mbox{fifth line}:&& 
(|M^{(\mboxsc{pc,pv})}_{q\rightarrow qg}|^2 
\mbox{\,\,are defined in Eqs.~(\ref{m_gtoqqbar},\ref{m_gtoqqbarpv})})
\nonumber  \\[2mm] \mbox{}
[\sum_{i=q,\bar{q}}e_i^2 C_i^{\overline{\mboxsc{MS}}}(\eta)]\mbox{}
|M^{(\mboxsc{pc})}_{q\rightarrow qg}|^2 
&\rightarrow&\!
[\sum_{i=q,\bar{q}} A_i(Q^2) C_i^{\overline{\mboxsc{MS}}}(\eta)]\mbox{}
|M^{(\mboxsc{pc})}_{q\rightarrow qg}|^2 
+
[\sum_{i=q} B_i(Q^2) 
\tilde\Delta C_i^{\overline{\mboxsc{MS}}}(\eta)]
|M^{(\mboxsc{pv})}_{q\rightarrow qg}|^2 \nonumber \\[2mm]
\mbox{sixth line}:&& \nonumber  \\[2mm] \mbox{}
(\sum_{i=q}e_i^2 ) f_g(\eta)
\,\,|M^{(\mboxsc{pc})}_{g\rightarrow q\bar{q}}|^2
&\longrightarrow&
(\sum_{i=q}A_i(Q^2) ) f_g(\eta)
\,\,|M^{(\mboxsc{pc})}_{g\rightarrow q\bar{q}}|^2 
+
(\sum_{i=q}B_i(Q^2)) f_g(\eta)
\,\,|M^{(\mboxsc{pv})}_{g\rightarrow q\bar{q}}|^2 \nonumber \\[2mm]
\mbox{seventh line}:&& 
({\cal{F}}^{(\mboxsc{pc,pv})}_{g\rightarrow q\bar{q}}
\mbox{\,\,are defined in Eqs.(\ref{fcalgdef},\ref{fcalgpvdef})})
\nonumber  \\[2mm] \mbox{}
(\sum_{i=q}e_i^2 ) f_g(\eta)
\,\,{\cal{F}}^{(\mboxsc{pc})}_{g\rightarrow q\bar{q}}
&\longrightarrow&
(\sum_{i=q}A_i(Q^2) ) f_g(\eta)
\,\,{\cal{F}}^{(\mboxsc{pc})}_{g\rightarrow q\bar{q}} 
+
(\sum_{i=q}B_i(Q^2)) f_g(\eta)
\,\,{\cal{F}}^{(\mboxsc{pv})}_{g\rightarrow q\bar{q}} \nonumber\\[2mm]
\mbox{eigth line}:&& \nonumber  \\[2mm] \mbox{}
(\sum_{i=q}e_i^2 )
C_g^{\overline{\mboxsc{MS}}}(\eta)
|M^{(\mboxsc{pc})}_{g\rightarrow q\bar{q}}|^2
&\longrightarrow&
(\sum_{i=q}A_i(Q^2) ) 
C_g^{\overline{\mboxsc{MS}}}(\eta)
|M^{(\mboxsc{pc})}_{g\rightarrow q\bar{q}}|^2\,
+
(\sum_{i=q}B_i(Q^2)) 
C_g^{\overline{\mboxsc{MS}}}(\eta)
|M^{(\mboxsc{pv})}_{g\rightarrow q\bar{q}}|^2\, \nonumber \\[2mm]
\mbox{tenth line}:&& 
(|M^{(\mboxsc{pc,pv})}_{q\rightarrow qgg}|^2 \mbox{\,\,are defined 
in Eqs.(\ref{bvvdef},\ref{baadef})})
\nonumber  \\[2mm] \mbox{}
[\sum_{i=q,\bar{q}}e_i^2 f_i(\eta)]
\,\,|M^{(\mboxsc{pc})}_{q\rightarrow qgg}|^2
&\longrightarrow&
[\sum_{i=q,\bar{q}} A_i(Q^2) f_i(\eta)]
\,\,|M^{(\mboxsc{pc})}_{q\rightarrow qgg}|^2 
+
[\sum_{i=q} B_i(Q^2) 
\tilde\Delta(f_i(\eta)]
\,\,|M^{(\mboxsc{pv})}_{q\rightarrow qgg}|^2 \nonumber \\[2mm]
\mbox{eleventh line}:&& 
(|M^{(\mboxsc{pc,pv})}_{g\rightarrow q\bar{q}g}|^2 \mbox{\,\,are defined 
in Eqs.(\ref{bvvgdef},\ref{baagdef})})\nonumber  \\[2mm] \mbox{}
(\sum_{i=q}e_i^2 ) f_g(\eta)
\,\,|M^{(\mboxsc{pc})}_{g\rightarrow q\bar{q}g}|^2 
&\longrightarrow&
(\sum_{i=q}A_i(Q^2) ) f_g(\eta)
\,\,|M^{(\mboxsc{pc})}_{g\rightarrow q\bar{q}g}|^2 
+
(\sum_{i=q}B_i(Q^2)) f_g(\eta)
\,\,|M^{(\mboxsc{pv})}_{g\rightarrow q\bar{q}g}|^2 \nonumber\\[2mm]
\mbox{twelfth line}:&& 
\nonumber  \\[2mm] \mbox{}
\{\mbox{4-quark term}\}^{(\mboxsc{pc})}
&\longrightarrow&
\{\mbox{4-quark term}\}^{(\mboxsc{Z})} \nonumber
\end{eqnarray}
We will now specify the new parity violating functions
${\cal{F}}^{(\mboxsc{pv})}_{q\rightarrow qg},
{\cal{F}}^{(\mboxsc{pv})}_{g\rightarrow q\bar{q}},
|M^{(\mboxsc{pv})}_{q\rightarrow qgg}|^2,
|M^{(\mboxsc{pv})}_{g\rightarrow q\bar{q}g}|^2$
and  
$\{\mbox{4-quark term}\}^{(Z)}$  in the
previous formulae. The corresponding parity conserved quantities 
(with vector current couplings
of the exchanged boson at the leptonic and hadronic vertex)
have been given before (see table~\ref{tab_eqs})  
in terms of helicity amplitudes 
in the
Weyl-van der Waerden spinor formalism.
The calculation of the 
parity violating terms involves the replacement of the vector current
by the axial vector current at either or both the
leptonic and hadronic vertices:
\begin{equation}
\gamma^{\mu}\rightarrow \gamma^\mu\gamma^5
\label{replace}
\end{equation}
In terms of the Weyl-van der Waerden representation 
Eq.~(\ref{replace}) corresponds
to the replacement (see Eqs.~(\ref{repgam},\ref{repgam5}))
\begin{equation}
i\sigma^{\mu \dot{A}B}-i\sigma^{\mu}_{\dot{B}A}
\rightarrow
i\sigma^{\mu \dot{A}B}+i\sigma^{\mu}_{\dot{B}A}
\end{equation}
Thus the helicity amplitudes change sign every time a term
$\sigma^{\mu}$ contributes with lower spinor indices.
This will depend on the fermionic lepton ($\lambda_l$)
and quark ($\lambda_1$) helicities.
For the leptonic vertex 
one  obtains the relation for the helicity amplitudes 
%
%
\begin{equation}
b^{aV}_{a\rightarrow n  \mboxsc{\,\,\, partons}}
=(-)^{1/2-\lambda_l}b^{vV}_{a\rightarrow  n \mboxsc{\,\,\,partons}};
\hspace{1cm}
b^{aA}_{a\rightarrow n \mboxsc{\,\,\, partons}}
=(-)^{1/2-\lambda_l}b^{vA}_{a\rightarrow n \mboxsc{\,\,\, partons}}
\label{helilep1}
\end{equation}
The lower case superscript $v(a)$ refers to the vector (axial vector)
coupling at the letonic vertex whereas upper case letters stand for the
corresponding quantities at the hadronic vertex.
Eq.~(\ref{helilep1}) is true irrespective of the
associated parton process $a\rightarrow n$ partons.

An analogous relation holds true when one replaces
the hadronic vector current by the axial hadronic vector
current ($V\rightarrow A$ in the superscript)
in the two quark-one gluon 
and two quark-two gluon processes
$q\rightarrow qg, q\rightarrow qgg $ and
$g\rightarrow q\bar{q}, g\rightarrow q\bar{q}g$:
\begin{equation}
b^{vA}_{q\rightarrow qg(g)}=(-)^{1/2-\lambda_1}b^{vV}_{q\rightarrow qg(g)}
;\hspace{1cm}
b^{vA}_{g\rightarrow q\bar{q}(g)}
=(-)^{1/2-\lambda_1}b^{vV}_{g\rightarrow q\bar{q}(g)}
\label{helihad1} 
\end{equation}
Based on the relations in  
Eqs.~(\ref{helilep1},\ref{helihad1}) the parity violating three level 
matrix elements, which multiply the coupling factor $B_f(Q^2)$
in the NC scattering, are defined via:
\begin{equation}
(4\pi\alpha)^2(4\pi\alpha_s)^2\frac{e_q^2}{Q^4}
|M^{(\mboxsc{pv})}_{q\rightarrow qgg}|^2=
e_q^2\frac{1}{3}\frac{1}{2}
\sum_{\mboxsc{16 heli}}
{\cal{R}} (b^{vV}_{q\rightarrow qgg}{b^{aA}_{q\rightarrow qgg}}^\ast)
\label{baadef}
\end{equation}
and
\begin{equation}
(4\pi\alpha)^2(4\pi\alpha_s)^2\,\,\frac{e_q^2}{Q^4}\,\,
|M^{(\mboxsc{pv})}_{g\rightarrow q\bar{q}g}|^2=
e_q^2\,\,\frac{1}{8}\,
\sum_{\mboxsc{16 heli}}
{\cal{R}}
 (b^{vV}_{g\rightarrow q\bar{q}g}{b^{aA}_{g\rightarrow q\bar{q}g}}^\ast)
\label{baagdef}
\end{equation}
where $b^{vV}_{q\rightarrow qgg}$ and 
$b^{vV}_{g\rightarrow q\bar{q}g}$ are given in
Eqs.~(\ref{hqtoqgg1}--\ref{antiquark}) and
Eq.~(\ref{gluoncross}), respectively.
The superscript (pv) refers to the interference of the vector
with the axial vector current in these contributions.

The parity violating functions
${\cal{F}}^{(\mboxsc{pv})}_{q\rightarrow qg}$ and
${\cal{F}}^{(\mboxsc{pv})}_{g\rightarrow q\bar{q}}$
are also determined by the relations in Eqs.~(\ref{helilep1},\ref{helihad1}).
With the definition of the parity conserved (pc)
helicity dependent quantities in 
Eqs.~(\ref{f1qdef}-\ref{f4qdef}), one finds:
\begin{eqnarray}
{\cal{F}}_{i,q\rightarrow qg}^{(\mboxsc{pv})}&=&
-
{\cal{F}}_{i,q\rightarrow qg}^{(\mboxsc{pc})}
\hspace{2cm}i=1,4,5,8 \nonumber \\
{\cal{F}}_{i,q\rightarrow qg}^{(\mboxsc{pv})}&=&
\,\,\,\,
{\cal{F}}_{i,q\rightarrow qg}^{(\mboxsc{pc})}
\hspace{2cm}i=2,3,6,7
\end{eqnarray}
and
\begin{equation}
{\cal {F}}^{(\mboxsc{pv})}_{q\rightarrow qg}
= \sum_{i=1}^8\,
{\cal {F}}^{(\mboxsc{pv})}_{i,q\rightarrow qg}
\label{fcalqpvdef}
\end{equation}
\begin{equation}
{\cal {F}}^{(\mboxsc{pv})}_{g\rightarrow q\bar{q}}
= \sum_{i=1}^8\,
{\cal {F}}^{(\mboxsc{pv})}_{i,g\rightarrow q\bar{q}}
\label{fcalgpvdef}
\end{equation}
where
${\cal{F}}^{(\mboxsc{pv})}_{i,g\rightarrow q\bar{q}}$
are obtained by the relation in Eq.~(\ref{fgcross}) from
${\cal {F}}^{(\mboxsc{pv})}_{i,q\rightarrow qg}$.

The helicity amplitudes with axial vector couplings for
the four quark process cannot be obtained by the simple
rule given in Eq.~(\ref{helihad1}) because
the hadronic axial vector coupling can occur at either of the
two quark lines.
The  matrix elements  including the
electroweak coupling structure and folding structure
with the parton densities,$\{\mbox{4-quark term}\}^{(Z)}$,
are given in Eqs.~(59-61) in ref.~\cite{heraii}
and we will not repeat them here.

Analytical results for the CC hadronic 2-jet cross section
including the electroweak coupling structure
are presented next.
Based on the electroweak factors for 
CC $e^\pm p$ scattering  in Eqs.~(\ref{abccdef},\ref{chiwdef})
the hadronic 2-jet cross section
for $e^-p$ CC scattering is obtained from Eq.~(\ref{twojet})
by the following replacements:
\begin{eqnarray}
\mbox{third line:}&& 
(|M^{(\mboxsc{pc$\pm$ pv})}_{q\rightarrow qg}|^2 
\mbox{are defined in Eqs.~(\ref{qtoqgsum},\ref{qtoqgdiff})})
\nonumber  \\[2mm] \mbox{}
[\sum_{i=q,\bar{q}}e_i^2 f_i(\eta)]
\,\,|M^{(\mboxsc{pc})}_{q\rightarrow qg}|^2 
&\longrightarrow&
\chi_W^2
\bigg[
(\sum_{i=u,c}  f_i(\eta))
\,\,|M^{(\mboxsc{pc+pv})}_{q\rightarrow qg}|^2 
+
(\sum_{i=\bar{d},\bar{s}} f_i(\eta))
\,\,\,|M^{(\mboxsc{pc--pv})}_{q\rightarrow qg}|^2 \bigg]
\nonumber\\[2mm]
\mbox{fourth line}:&&
({\cal{F}}^{(\mboxsc{pc$\pm$ pv})}_{q\rightarrow qg} 
\mbox{\,\,are defined below})
 \nonumber  \\[2mm] \mbox{}
[\sum_{i=q,\bar{q}}e_i^2 f_i(\eta)]\mbox{}
\,\,{\cal{F}}^{(\mboxsc{pc})}_{q\rightarrow qg}
&\longrightarrow&
\chi_W^2
\bigg[
(\sum_{i=u,c}  f_i(\eta))
\,\,{\cal{F}}^{(\mboxsc{pc+pv})}_{q\rightarrow qg}
+
(\sum_{i=\bar{d},\bar{s}} f_i(\eta))
\,\,\,{\cal{F}}^{(\mboxsc{pc--pv})}_{q\rightarrow qg}\bigg]
\nonumber\\[2mm]
\mbox{fifth line}:&& \nonumber  \\[2mm] \mbox{}
[\sum_{i=q,\bar{q}}e_i^2 C_i^{\overline{\mboxsc{MS}}}(\eta)]\mbox{}
|M^{(\mboxsc{pc})}_{q\rightarrow qg}|^2 
&\rightarrow&
\chi_W^2
\bigg[
(\sum_{i=u,c}  C_i^{\overline{\mboxsc{MS}}}(\eta))\mbox{}
\,\,|M^{(\mboxsc{pc+pv})}_{q\rightarrow qg}|^2 
+
(\sum_{i=\bar{d},\bar{s}}  C_i^{\overline{\mboxsc{MS}}}(\eta))\mbox{}
\,\,\,|M^{(\mboxsc{pc--pv})}_{q\rightarrow qg}|^2 \bigg]
\nonumber\\[2mm]
\mbox{sixth line}:&& 
|M^{(\mboxsc{pc$\pm$ pv})}_{g\rightarrow q\bar{q}}|^2 
\mbox{ are defined in Eqs.~(\ref{qtoqqbardiff},\ref{qtoqqbarsum})}
\nonumber  \\[2mm] \mbox{}
(\sum_{i=q}e_i^2 ) f_g(\eta)
\,\,|M^{(\mboxsc{pc})}_{g\rightarrow q\bar{q}}|^2
&\longrightarrow&
2\chi_W^2 f_g(\eta)\,\,
|M^{(\mboxsc{pc--pv})}_{g\rightarrow q\bar{q}}|^2 
\nonumber \\[2mm]
\mbox{seventh line}:&& 
({\cal{F}}^{(\mboxsc{pc--pv})}_{g\rightarrow q\bar{q}}
\mbox{\,\,is defined below})
\nonumber  \\[2mm] \mbox{}
(\sum_{i=q}e_i^2 ) f_g(\eta)
\,\,{\cal{F}}^{(\mboxsc{pc})}_{g\rightarrow q\bar{q}}
&\longrightarrow&
2\chi_W^2 f_g(\eta)\,\,
\,\,{\cal{F}}^{(\mboxsc{pc--pv})}_{g\rightarrow q\bar{q}}
\nonumber \\[2mm]
\mbox{eigth line}:&& \nonumber  \\[2mm] \mbox{}
(\sum_{i=q}e_i^2 )
C_g^{\overline{\mboxsc{MS}}}(\eta)
|M^{(\mboxsc{pc})}_{g\rightarrow q\bar{q}}|^2
&\longrightarrow&
2\chi_W^2 C_g^{\overline{\mboxsc{MS}}}(\eta)\,\,
|M^{(\mboxsc{pc--pv})}_{g\rightarrow q\bar{q}}|^2 
\nonumber \\[2mm]
\mbox{tenth line}:&& 
(|M^{(\mboxsc{pc,pv})}_{q\rightarrow qgg}|^2 \mbox{\,\,are defined below})
\nonumber  \\[2mm] \mbox{}
[\sum_{i=q,\bar{q}}e_i^2 f_i(\eta)]
\,\,|M^{(\mboxsc{pc})}_{q\rightarrow qgg}|^2
&\longrightarrow&
\chi_W^2
\bigg[
(\sum_{i=u,c}  f_i(\eta))
\,\,|M^{(\mboxsc{pc+pv})}_{q\rightarrow qgg}|^2 
+
(\sum_{i=\bar{d},\bar{s}} f_i(\eta))
\,\,\,|M^{(\mboxsc{pc--pv})}_{q\rightarrow qgg}|^2 \bigg]
\nonumber\\[2mm]
\mbox{eleventh line}:&& 
(|M^{(\mboxsc{pc--pv})}_{g\rightarrow q\bar{q}g}|^2 \mbox{\,\,is defined below})
\nonumber  \\[2mm] \mbox{}
(\sum_{i=q}e_i^2 ) f_g(\eta)
\,\,|M^{(\mboxsc{pc})}_{g\rightarrow q\bar{q}g}|^2 
&\longrightarrow&
2\chi_W^2 f_g(\eta)\,\,
|M^{(\mboxsc{pc--pv})}_{g\rightarrow q\bar{q}g}|^2 
\nonumber \\[2mm]
\mbox{twelfth line}:&& 
(\{\mbox{4-quark term}\}^{(W^-)} \mbox{\,\,is defined below})
\nonumber  \\[2mm] \mbox{}
\{\mbox{4-quark term}\}^{(\mboxsc{pc})}
&\longrightarrow&
\{\mbox{4-quark term}\}^{(W^-)} \nonumber
\end{eqnarray}
For CC $e^+p$ scattering one has to replace
$\sum_{i=u,c}$ ($\sum_{i=\bar{d},\bar{s}}$) by
$\sum_{i=\bar{u},\bar{c}}$ ($\sum_{i=d,s}$), respectively, 
in the previous equations and 
$|M^{(\mboxsc{pc--pv})}_{g\rightarrow q\bar{q}}|^2,
{\cal{F}}^{(\mboxsc{pc--pv})}_{g\rightarrow q\bar{q}},
|M^{(\mboxsc{pc--pv})}_{g\rightarrow q\bar{q}g}|^2$
by
$|M^{(\mboxsc{pc+pv})}_{g\rightarrow q\bar{q}}|^2,
{\cal{F}}^{(\mboxsc{pc+pv})}_{g\rightarrow q\bar{q}}$,\\$
|M^{(\mboxsc{pc+pv})}_{g\rightarrow q\bar{q}g}|^2$, respectively,
in the $6^{\mboxsc{th}},7^{\mboxsc{th}},8^{\mboxsc{th}}$ and
$11^{\mboxsc{th}}$ line.
The sum and differences of the parity conserved and parity violating terms
can be  obtained from previous results via:
${\cal{F}}^{(\mboxsc{pc$\pm$ pv})}_{q\rightarrow qg}$ 
from Eq.~(\ref{fcalqdef}) and Eq.~(\ref{fcalqpvdef}),
${\cal{F}}^{(\mboxsc{pc$\pm$ pv})}_{g\rightarrow q\bar{q}}$
from  Eq.~(\ref{fcalgdef}) and Eq.~(\ref{fcalgpvdef}),
$|M^{(\mboxsc{pc$\pm$ pv})}_{q\rightarrow qgg}|^2$
from Eq.~(\ref{bvvdef}) and Eq.~(\ref{baadef}) and 
$|M^{(\mboxsc{pc$\pm$ pv})}_{g\rightarrow q\bar{q}g}|^2$
from Eq.~(\ref{bvvgdef}) and Eq.~(\ref{baagdef}).

The  matrix element  
for $\{\mbox{4-quark term}\}^{(W)}$ can be derived from the results
given in Eqs.~(59-61) in ref.~\cite{heraii} and we will not 
repeat this here.

\newpage

\subsubsection[{Numerical Results Including $Z$ 
and $W$ Exchange \protect\vspace{1mm}}]{Numerical 
Results Including {\boldmath{$Z$}} 
and {\boldmath{$W$}} Exchange}
\enlargethispage{1cm}
The following numerical studies
show effects of the $Z$ and $W$ exchange in
2-jet  cross sections up to \oasz.
The kinematics and parameters are chosen as discussed at the beginning
of section~\ref{numzw1}. Only 
the renormalization scale  and
the factorization scale are changed to
$\mu_R=\mu_F= 1/2\,\sum_j \,k_T^B(j)$ (see section~\ref{scalechoice}).

Fig.~\ref{f_zq2j} shows the $Q^2$ distribution of the 2-jet cross sections 
for NC $e^+ p$ (a) and $e^- p$  scattering.
Results are given for complete NC $\gamma^\ast$ and $Z$ exchange
(solid), for pure $\gamma^\ast$ (dot-dashed) and for
pure $Z$ (dotted) exchange. 
The $Q^2$ distribution of the CC 2-jet cross section
is compared to the complete NC 2-jet cross section in Fig.~\ref{f_wq2j}a.
The 2-jet cross sections are suppressed by a factor $\approx$ 5-10
compared to the 1-jet results shown in 
Fig.~\ref{f_zq1j}a,c and Fig.~\ref{f_wq1j}a.
The $Q^2$-dependence of the 
2-jet rate, defined as the ratio of the 2-jet to the 1-jet
cross section, is shown in Fig.~\ref{f_wq2j}b
for $e^+p$ scattering in NC (solid) and CC (dashed) exchange.
The relative importance of the electroweak
effects in 2-jet events is very similar as already found
in the 1-jet case. Thus, the ratios in Eq.~(\ref{ewratio})
for dijet events ($n=2$) are very similar to the results shown 
Fig.~\ref{f_zq1j}b,d and Fig.~\ref{f_wq1j}b and Fig.~\ref{f_zwrel1j}.

\begin{figure}[htb]
  \centering
 \mbox{\epsfig{file=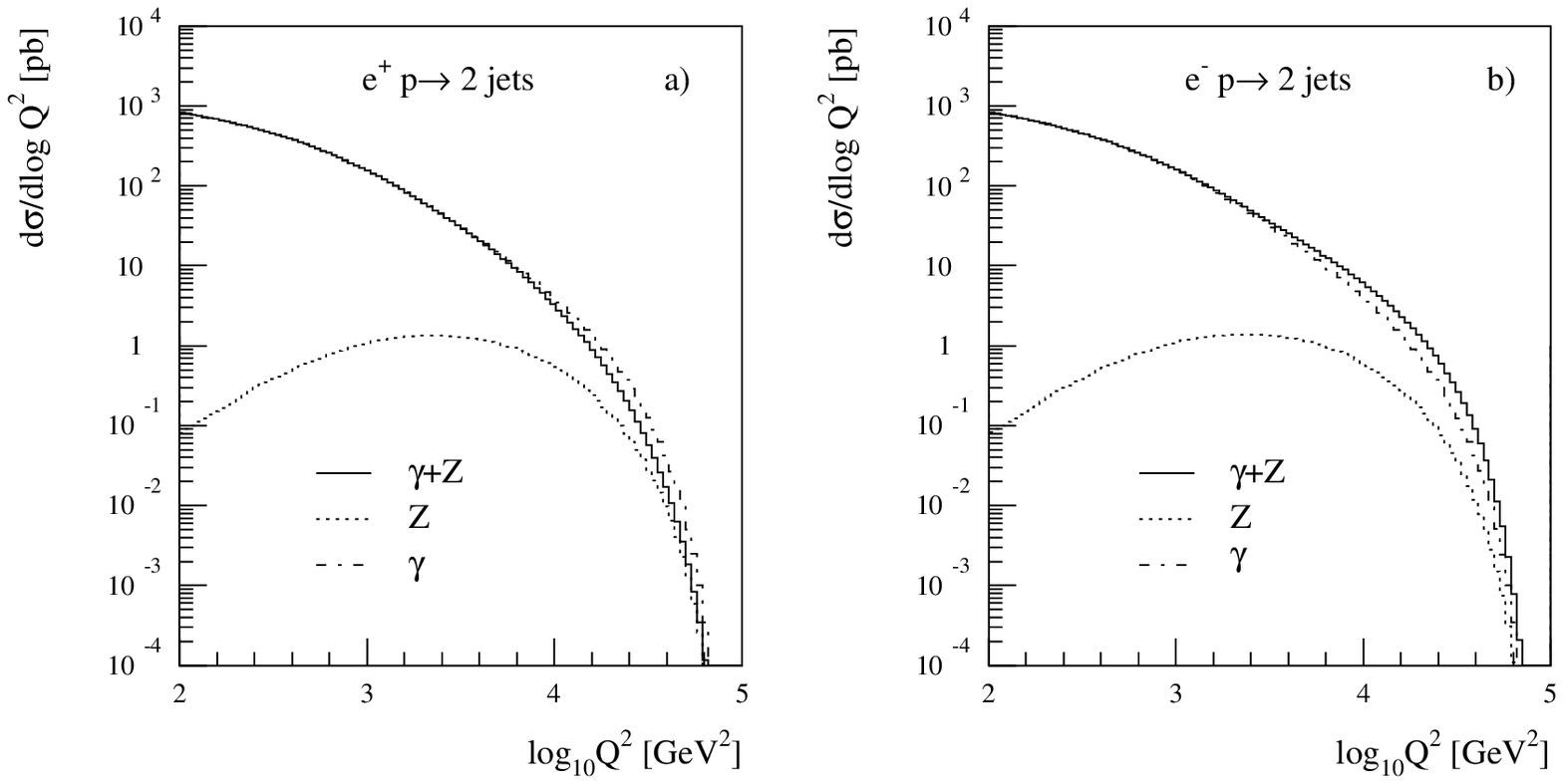,bbllx=0,bblly=250,
               bburx=550,bbury=550,width=0.95\linewidth}} 
\vspace*{-0.8cm}
\caption{
$Q^2$ dependence of the 2-jet cross section 
for  $e^+p$ (a) and $e^-p$ (b)  scattering in
NC $\gamma^\ast/Z$ (solid), $\gamma^\ast$ (dot-dashed) 
and pure $Z$ (dotted) exchange.
Jet are defined in a cone scheme in the lab frame
with radius $R=1$. Results are shown in LO with
MRSR1 parton distribution functions \protect\cite{mrsr1}. 
Additional cuts are explained in the text.
}
\label{f_zq2j}
\end{figure}
%
%
\begin{figure}[htb]
  \centering
 \mbox{\epsfig{file=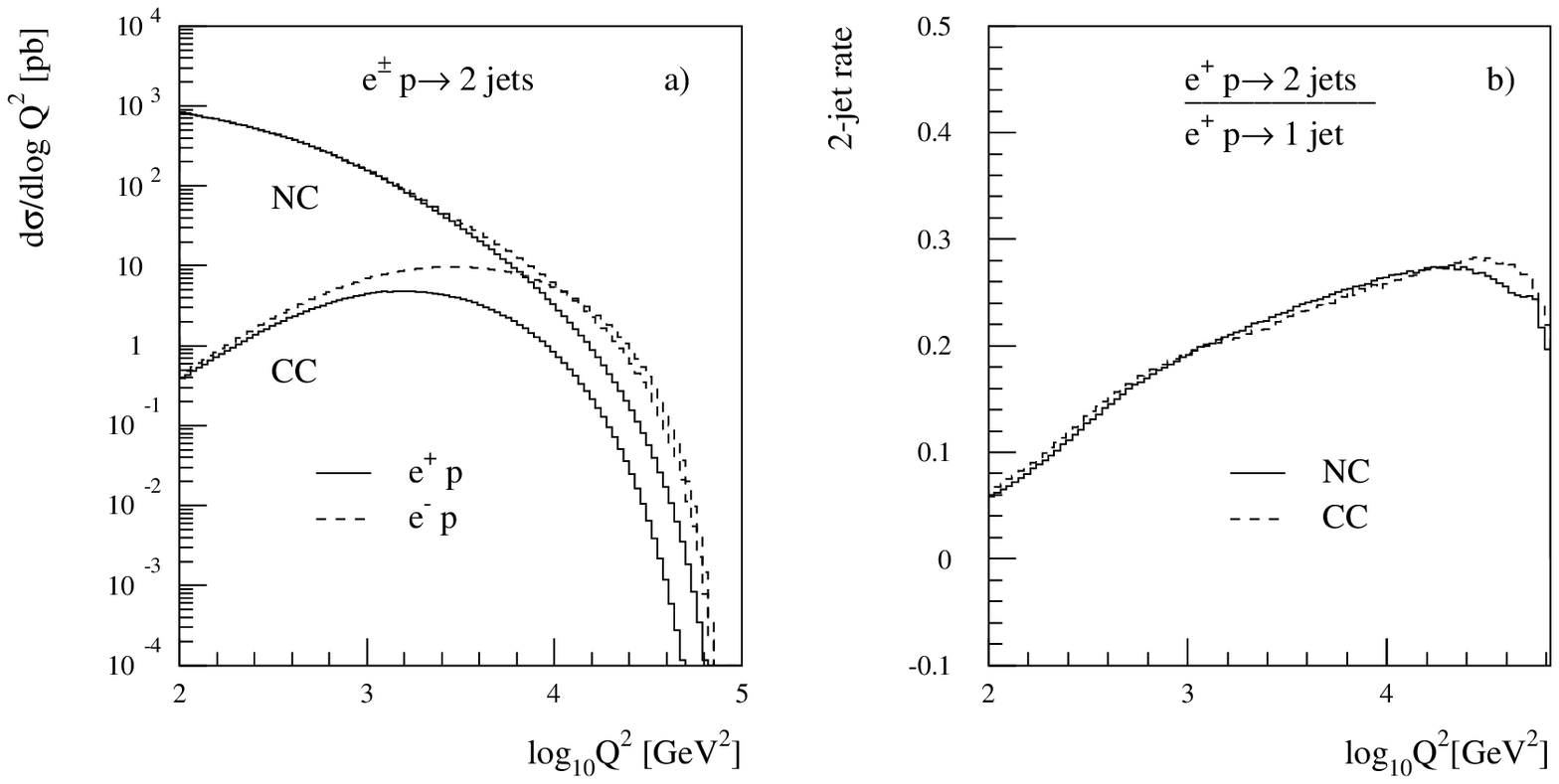,bbllx=0,bblly=250,
               bburx=550,bbury=550,width=0.95\linewidth}} 
\vspace*{-0.8cm}
\caption{
a) $Q^2$ dependence of the  CC 
and NC 2-jet cross section for
$e^+p$ (solid) and $e^-p$ (dashed) scattering.
b) 2-jet rate $\sigma(\protect\mbox{2-jet})[Q^2]/
               \sigma(\protect\mbox{1-jet})[Q^2]$ for NC (solid) and CC 
(dashed) exchange. 
}
\label{f_wq2j}
\end{figure}
%
%

\begin{figure}[t]
  \centering
 \mbox{\epsfig{file=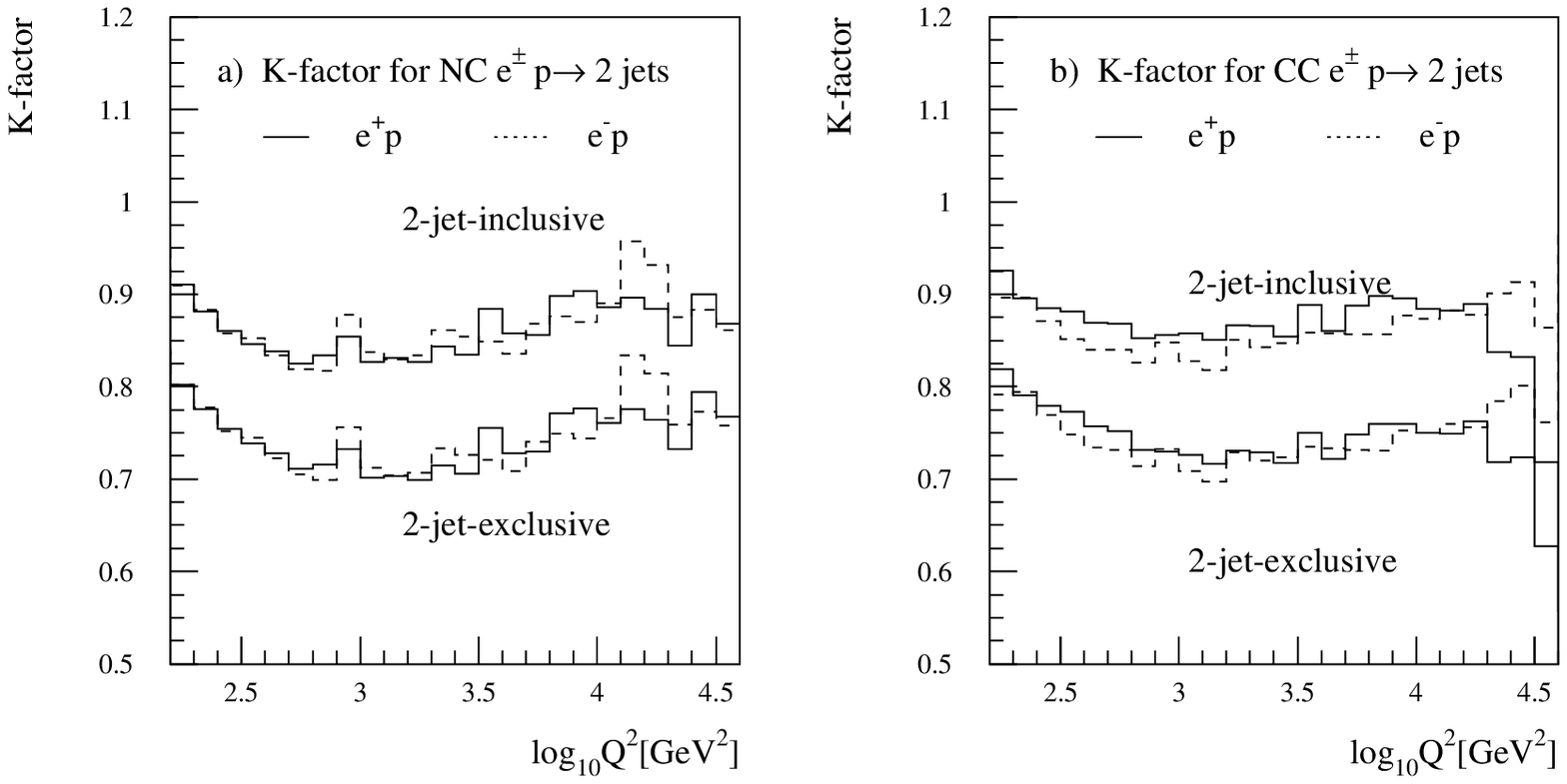,bbllx=0,bblly=250,
               bburx=550,bbury=550,width=0.95\linewidth}} 
\caption{
$Q^2$ dependence of the $K$-factor for NC a)
and charged current b)  2-jet inclusive and exclusive
cross sections in $e^+p$ (solid) and $e^-p$ (dashed) scattering.
}
\label{f_zw2j_q_kfac}
\end{figure}

The $K$-factors for the NC and CC 2-jet 
cross sections 
are shown in Fig.~\ref{f_zw2j_q_kfac}.
We find a $K$-factor of about 0.8-0.9 (0.7-0.8) in the
NC 2-jet-inclusive (2-jet-exclusive) case (Fig.~\ref{f_zw2j_q_kfac}a)
independent on $e^+p$ or $e^-p$ scattering.
The $K$-factors for the CC case shown 
in Fig.~\ref{f_zw2j_q_kfac}b
are similar in size, besides for 
$e^+p$ scattering at very high $Q^2$, where 
we find large (negative) corrections.
A similar effect was observed in the $K$-factor for the
1-jet cross section shown in Fig.~\ref{f_zw1j_q_kfac}.
As in the 1-jet case, the CC cross section for $e^+p\rightarrow 2$ jet
becomes very small at these high $Q^2$ values
(see Fig.~\ref{f_wq2j}) and resummation
effects due to the effectively strongly restricted phase space
are expected to become important.
We find very similar $K$-factors for jets defined in the $k_T$
scheme with $E_T^2=40$ GeV$^2$.
\clearpage
\newpage
\section{Three- and Four-Jet Cross Sections\protect\vspace{1mm}}
\label{sec_34jet}
Higher jet multiplicities, {\it i.e.} 3-jet and 4-jet final states in DIS,
provide further interesting laboratories of perturbative QCD
and will be discussed in the following.

\subsection{Matrix Elements and Leading Order Cross
 Sections\protect\vspace{1mm}}
\label{sec_34jetamt}
The calculation of NLO 2-jet exclusive cross sections, which was discussed 
in the previous section, involves integrating the tree level squared matrix 
elements for the subprocesses $eq\to eqgg,eq\to eqq\bar q$  and
$eg\to eq\bar{q}g$ (Eqs.~(\ref{qtoqgg}-\ref{gtoqqbarg}))
over those collinear and infrared regions of phase space where exactly two
jets are resolved. 

The cross section for 3-jet events, at LO, is obtained
by integrating the same squared matrix elements over the phase space region 
where all three final state partons are resolved as jets. Representative 
Feynman graphs for processes with three-parton final states are shown in 
Fig.~\ref{f_atobcd}.

Similarly, 4-jet cross sections in LO (${\cal O}(\alpha_s^3)$) are obtained
by integrating the squared tree-level matrix elements for the subprocesses
\begin{eqnarray}
e(l)+q(p_0)&\rightarrow& e(l^\prime) + q(p_1) + g(p_2) + g(p_3) + g(p_4)
\label{qtoqggg} \\[1mm]
e(l)+q(p_0)&\rightarrow& e(l^\prime) + q(p_1) + q(p_2) + \bar{q}(p_3) + g(p_4)
\label{qtoqqqbarg} \\[1mm]
e(l)+g(p_0)&\rightarrow& e(l^\prime) + q(p_1) + \bar{q}(p_2) + g(p_3) + g(p_4)
\label{gtoqqbargg} \\[1mm]
e(l)+g(p_0)&\rightarrow& e(l^\prime) + q(p_1) + \bar{q}(p_2) + 
q(p_3) + \bar{q}(p_4) 
\label{gtoqqbarqqbar} 
\end{eqnarray}
(and crossing related ones)
over the resolved 4-jet phase space region.
Representative Feynman graphs for these processes with four-parton final 
states are shown in Fig.~\ref{f_atobcde}.
\begin{figure}[htb]
  \centering
 \mbox{\epsfig{file=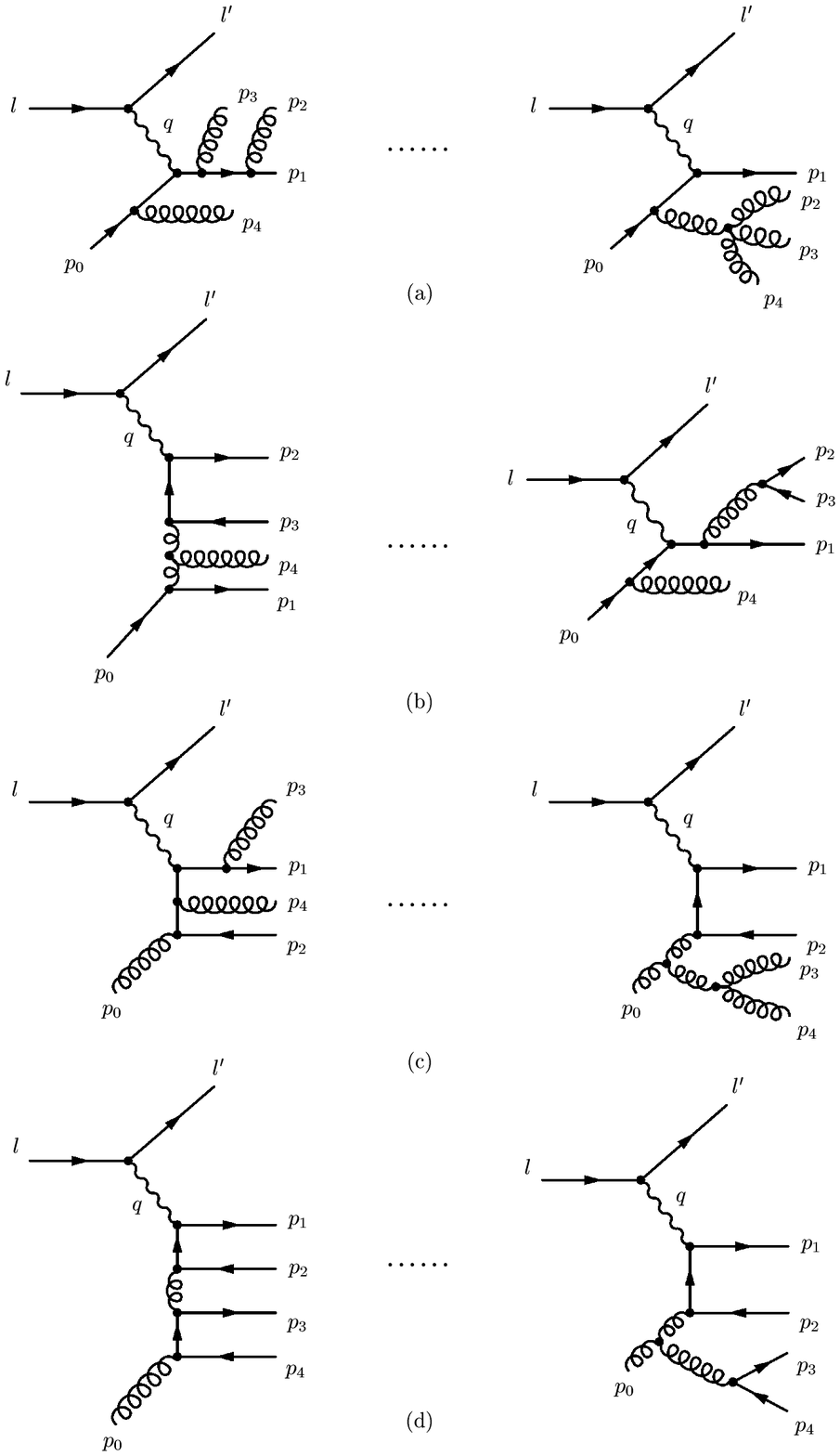,bbllx=100,bblly=80,
               bburx=580,bbury=750,width=0.9\linewidth}} 
\caption{
Generic ${\cal{O}}(\alpha_s^3)$ tree level Feynman diagrams for 
a) $eq \rightarrow eqggg$, b) $eq\rightarrow eqq\bar{q}g$, 
c) $eg\rightarrow q\bar{q}gg$, d) $eg\rightarrow q\bar{q}q\bar{q}$.
Only two out of 54 (48) diagrams are shown for classes a,c (b,d).}
\label{f_atobcde}
\end{figure}

The hadronic 3- and 4-jet cross sections are thus obtained by 
the parton model formula Eq.~(\ref{hadlo}) for $n=3,4$:
\begin{eqnarray}
\sigma_{\mboxsc{had}}^{\mboxsc{LO}}[3,4\mbox{-jet}] 
&=& \sum_a\int_0^1d\eta
\int d{\mbox{PS}}^{(l^\prime+3,4)} 
f_a(\eta,\mu_F)\,\,\label{hadlo34} \\
&& \alpha_s^{(2,3)}(\mu_R)\,\,
\hat{\sigma}^{\mboxsc{LO}}_{a\rightarrow 3,4\,\, \mboxsc{partons}}
(l+p_0\rightarrow l^\prime+p_1\ldots p_{3,4})\,\,
J_{3,4\leftarrow 3,4}(\{p_i\})\nonumber
\end{eqnarray}
%
%
where
$d\mbox{PS}^{(l^\prime+3,4)}$
stands for the  $4,5$ particle Lorentz-invariant  phase-space measure
in Eq.~(\ref{phasespace}).
The jet algorithm $J_{n \leftarrow n}$  ($n=3,4$)
defines the resolution for the 3- and 4-jet phase space region
where exactly $3,4$ jets are resolved according to the given
jet definition scheme.

$\hat{\sigma}^{\mboxsc{LO}}_{a\rightarrow 3,4\,\, \mboxsc{partons}}$
denotes the LO differential cross section for the  $3$-parton  
and $4$-parton  
final state processes above with $\as$ set to one.

These partonic cross sections  are, as mentioned above, only needed
in non-singular resolved regions 
and thus they can be calculated in four dimensions. 
The method of choice for their 
evaluation then is the direct numerical calculation
of amplitudes for fixed polarizations of the external particles
\cite{hzoriginal,calkul,zxu,berendsgiele,HZ,heraii}.
For the processes of Eqs.~(\ref{qtoqggg}-\ref{gtoqqbarqqbar})
the amplitudes were first given in Ref.~\cite{HZ},
using the helicity basis in~\cite{hzoriginal}.
The formalism 
is based on the decomposition 
of Dirac spinors into two-component Weyl spinors of chirality $\tau$, 
$\psi(\bar p,\bar \sigma)_\tau$.
The Weyl spinors are evaluated numerically for 
a given physical momentum $\bar p=Sp=(\bar p^0,\bar p_x,\bar p_y,\bar p_z)
=(\bar p^0,{\bf \overline{p}})$ and helicity $\bar\sigma/2=S\sigma/2$.
Here $S$ is a sign factor distinguishing external anti-fermions ($S=-1$) from
fermions ($S=+1$). For massless fermions chirality and helicity are related
by $\bar\sigma=S\tau$ and the Weyl spinors are given by
\bq
\psi(\bar p,\bar \sigma)_\tau = S\;\delta_{\sigma,\tau}\sqrt{2\bar p^0}
\chi_{\sigma_i}(\bar p)
\eq
with 
\ba
\chi_+(p) & = & {1\over \sqrt{ 2|{\bf p}|(|{\bf p}|+p_z)}}
\left(\begin{array}{c}
|{\bf p}| + p_z \\
p_x + ip_y
\end{array}\right) \,, \\
\chi_-(p) & = & {1\over \sqrt{ 2|{\bf p}|(|{\bf p}|+p_z)}}
\left(\begin{array}{r}
-p_x + ip_y \\
|{\bf p}| + p_z
\end{array}\right) \,.
\ea
The multiplication of strings of $\gamma$-matrices (from vertices and 
propagators) onto Dirac spinors is then reduced to successive two-by-two 
matrix multiplication onto Weyl spinors, which is performed numerically. 
Based on this modular approach, 
full expressions for the matrix elements for the partonic processes in
Eqs.~(\ref{qtoqggg}-\ref{gtoqqbarqqbar}),
including their decomposition into
orthogonal color tensors and the rules for calculating the final color factors,
are given in Ref.~\cite{HZ} and we will not repeat them here.

In sections~\ref{sec_onejet} and \ref{sec_twojet} we have separately discussed
parity-conserving single-photon-exchange contributions, proportional to
$|M^{\rm (pc)}|^2$, and the modifications needed when parity violating effects
from $Z$ and $W$-exchange need to be considered. 
Within the helicity amplitude formalism,
the parity conserving pieces are simply obtained by setting
the $Z$ couplings  to zero  and then summing 
the squared amplitudes over external polarizations. The 
parity-violating contributions to the squared amplitudes can be obtained 
by taking appropriate differences of left- and righthanded quarks and leptons.
Similarly, only contributions of lefthanded fermions  are needed
for CC processes. 
For the numerical implementation of the LO 3- and 4-jet matrix elements
in \docuname\,  including $\gamma^\ast$ and/or $Z$ exchange,
we have used this approach.

$\O(\alpha_s^4)$ matrix elements for 5-jet production in DIS
could be obtained by crossing from the matrix elements 
for the process  $p\bar{p}\rightarrow $Z$ + 4$ jets \cite{z4jets}.
We have used the {\large \sc madgraph} program \cite{madgraph}
in Ref.~\cite{z4jets} to generate the helicity
amplitudes for the very large number of contributing subprocesses
in this case\footnote{Each of these subprocesses involves a
large number of feyman diagrams, 
{\it e.g.} the  process $ eq\rightarrow  eqgggg$ would
involve for example 516 diagrams!}.
We have not implemented these matrix elements in the present
version of \docuname~2.1.

\newpage
\subsection{Numerical Results\protect\vspace{1mm}}
\label{num34}
Fig.~\ref{f_zq4j} shows the $Q^2$ distribution of the 
3-jet (a) and 4-jet (b) cross section
for NC $e^+ p$  scattering. 
Results are given for complete NC $\gamma^\ast$ and $Z$ exchange
(solid), for pure $\gamma^\ast$ (dot-dashed) and for
pure $Z$ (dotted) exchange. 

The kinematics and parameters are chosen as
discussed at the beginning of section~\ref{numzw1}.
The renormalization and factorization  scales in this section are set to
\begin{equation}
\mu_R=\mu_F=\xi \sum_j k_j^B(j)
\end{equation}
with $\xi=1$ for 1-jet, $\xi=1/2$ for 2- and 3-jets,
and $\xi=1/4$ for 4-jets results (see section~\ref{scalechoice}).

Comparing Figs.~\ref{f_zq4j}a,b with Fig.~\ref{f_zq2j}a and Fig.~\ref{f_zq1j}a
one observes the expected $\alpha_s$-hierarchy of the cross sections, namely 
$\sigma$[4-jet]$>\sigma$[3-jet]$>\sigma$[2-jet]$>\sigma$[1-jet].
However, the relative importance of the electroweak
effects is, as already mentioned in section~\ref{numzw1}, 
almost independent on the jet multiplicity.
Thus, the ratios in Eq.~(\ref{ewratio})
for 3-jets and 4-jets  events ($n=3,4$) are very similar to the results shown 
in Figs.~\ref{f_zq1j}b,d, Fig.~\ref{f_wq1j}b and Fig.~\ref{f_zwrel1j}.

\begin{figure}[htb]
  \centering
 \mbox{\epsfig{file=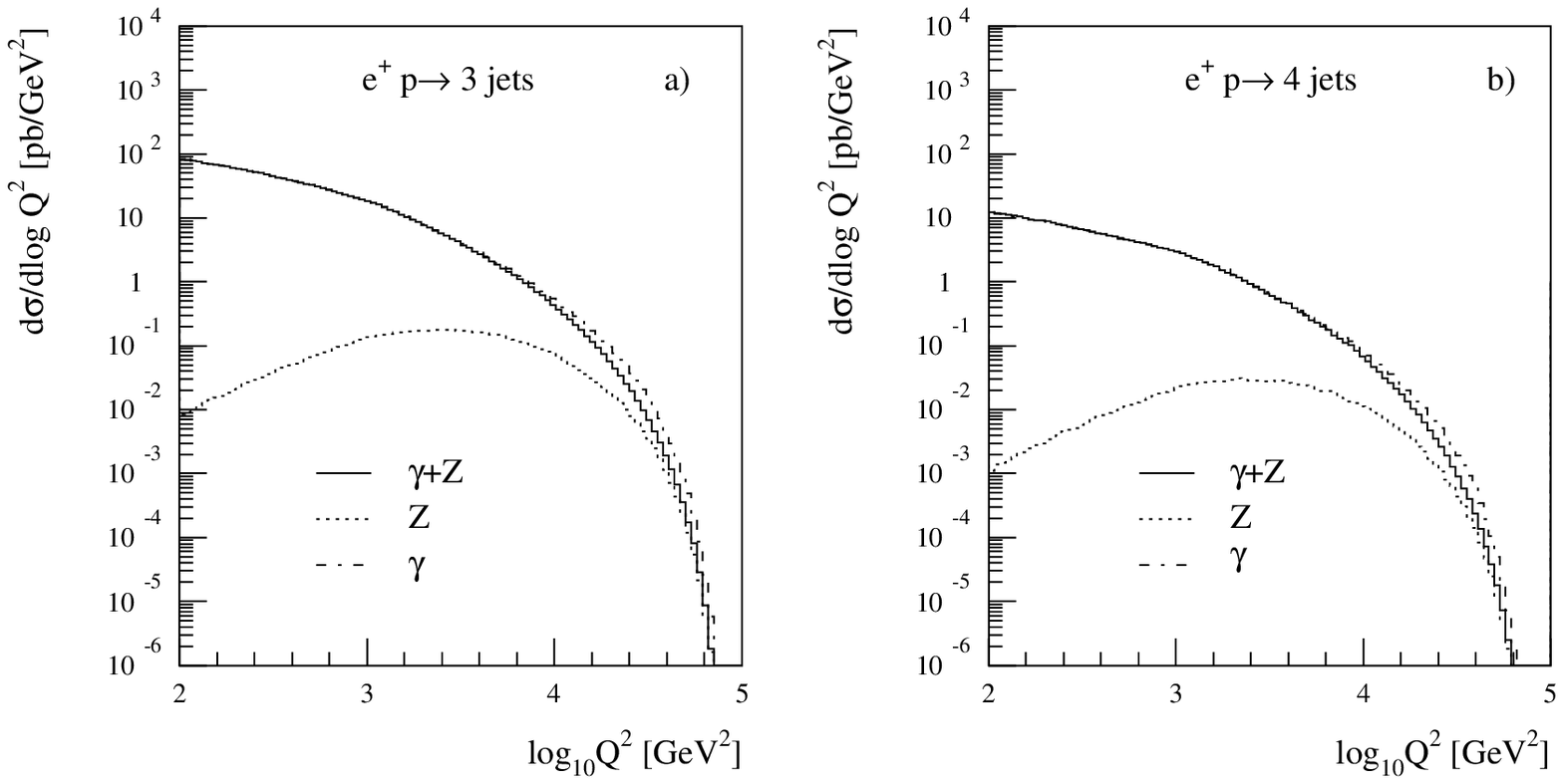,bbllx=0,bblly=250,
               bburx=550,bbury=550,width=0.95\linewidth}} 
\caption{
$Q^2$ dependence of the 3-jet (a)
and 4-jet (b)  cross section  for  $e^+p$ scattering in
NC $\gamma^\ast/Z$ (solid), $\gamma^\ast$ (dot-dashed) 
and pure $Z$ (dotted) exchange.
Jets are defined in a cone scheme in the lab frame
with radius $R=1$. Results are shown in LO with
MRSR1 parton distribution functions \protect\cite{mrsr1}. 
Additional cuts are explained in the text.
}
\label{f_zq4j}
\end{figure}
%
%

Numerical results for $n$-jet rates ($n=1,2,3,4$)
\begin{equation}
R_n(y_{\mboxsc{cut}})=\frac{\sigma[n\mbox{-jet}](y_{\mboxsc{cut}})}{
                        \sum_{i=1}^4\sigma[i\mbox{-jet}](y_{\mboxsc{cut}})}
\label{rndef}
\end{equation} 
are shown in Fig.~\ref{f_ycut_dep}
as a function of $y_{\mboxsc{cut}}$  using both the 
JADE (a) and the $k_T$ (b) algorithm with $E_T^2=40$ GeV$^2$
(see section~\ref{sec_jetdef}). 
Here, events were selected with 40~GeV$^2<Q^2<2500$ GeV$^2$,
$0.04 < y < 1$, an energy cut of $E(e^\prime)>10$~GeV on the scattered 
electron, and a cut on the pseudo-rapidity $\eta=-\ln\tan(\theta/2)$
of the scattered lepton and jets of $|\eta|<3.5$. 
With the currently available high luminosity at HERA, a experimental comparison
of this $y_{\mboxsc{cut}}$ dependence should be feasible in the
near future  allowing for  interesting QCD tests with
these multi-jet final states.

\begin{figure}[htb]
  \centering
 \mbox{\epsfig{file=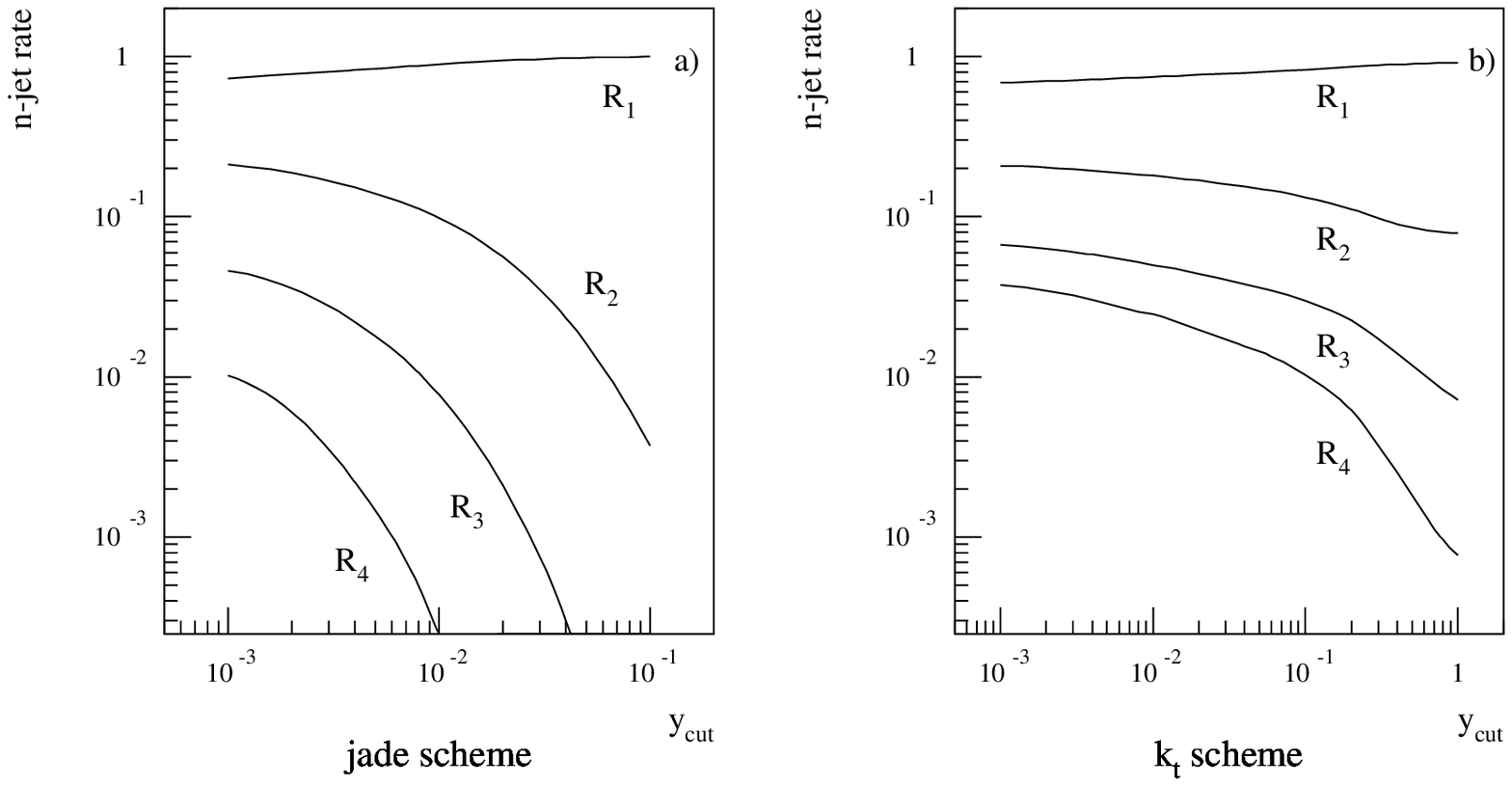,bbllx=0,bblly=250,
               bburx=550,bbury=540,width=0.95\linewidth}} 
\caption{
1-jet, 2-jet, 3-jet and 4-jet rates $R_n$ in Eq.~(\protect\ref{rndef})
as a function of $y_{\protect\mboxsc{cut}}$.
Jets are defined in the JADE scheme (a) and the $k_T$ scheme (b).
Results are shown in LO with
MRSR1 parton distribution functions \protect\cite{mrsr1}. 
Additional parameters are given in the text.
}
\label{f_ycut_dep}
\end{figure}
%
%
\clearpage
\newpage
\section{Forward Jet Production in the low {\boldmath{$x$}} 
regime\protect\vspace{1mm}}
\label{sec_forward}


Deep-inelastic scattering  at HERA provides for an ideal place to
probe strong interaction dynamics at small Bjorken $x$. 
One focus of interest 
has been the rise of the structure function 
$F_2(x,Q^2)$\cite{lowxF2} for large
$1/x$. One would like to identify 
the power law growth, $1/x^{\alpha_p-1}$, as 
predicted by the Balitsky-Fadin-Kuraev-Lipatov (BFKL)~\cite{bfkl} evolution 
equation. This evolution equation resums all leading $\alpha_s \ln{1/x}$
terms, as opposed to the more  standard 
Dokshitzer-Gribov-Lipatov-Altarelli-Parisi (DGLAP) equation~\cite{dglap}
which resums all leading $\alpha_s \ln{Q^2}$ terms.
Unfortunately, the  measurement of $F_2$ in the HERA range is probably 
too inclusive to discriminate between BFKL and the conventional DGLAP 
dynamics~\cite{viele}. 

A more sensitive test of BFKL dynamics at small $x$ is expected from deep 
inelastic scattering with a measured forward jet (in the proton direction) 
and $p_T^2(j)\approx Q^2$ \cite{mueller,allen1}. The idea is to study DIS events 
which contain an identified jet of longitudinal momentum 
fraction $x_{jet}=p_z(jet)/E_{proton}$ which is large compared to Bjorken $x$. 

A typical Feynman graph which contributes to the parton evolution
is shown in Fig.~\ref{fig:feyn}. 
The $x_i$ denote the momentum fractions (relative to the 
incoming proton) of the incident virtual partons and $p_{Ti}$ is the 
transverse momentum of emitted parton $i$. In the axial gauge, such 
``ladder-type'' diagrams with strong ordering in transverse 
momenta, $Q^2\approx p_{Tn}^2 \gg \ldots \gg p_T(j)^2$ but only soft 
ordering for the longitudinal fraction $x_1 > x_2>\ldots> x_n\approx x$ 
are the source of the leading log $Q^2$ contribution. 
In the BFKL approximation transverse momenta are no longer ordered along 
the ladder while there is a strong ordering in the fractional 
momentum $x_n \ll x_{n-1}\ll\ldots \ll x_1\approx x_{jet}$. When
tagging a forward jet with $p_T(j)\approx Q$ this leaves little room for
DGLAP evolution while the condition $x_{jet}\gg x$ leaves BFKL evolution 
active. This leads to an enhancement of the forward jet production cross 
section proportional to $(x_{jet}/x)^{\alpha_P -1}$ over the DGLAP 
expectation.

\begin{figure}[htb]
  \centering
  \mbox{\epsfig{file=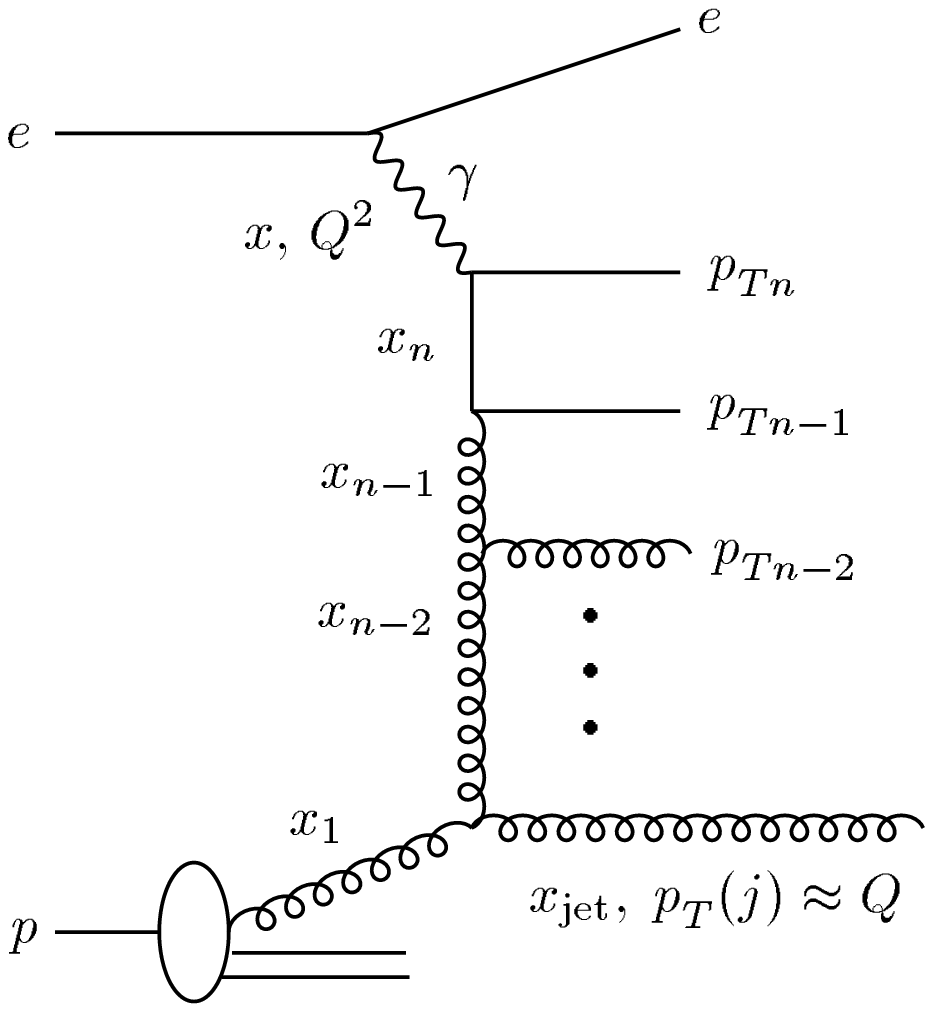,
         bbllx=120,bblly=260,bburx=400,bbury=560,width=0.50\linewidth}} 
\caption{
Ladder diagram contributing to jet production in DIS. The forward jet
could arise from the bottom gluon. 
}
\label{fig:feyn}
\end{figure}

Fig.~\ref{fig:feyn} shows that the multiple $t$-channel gluon propagators 
which lead to BFKL resummation first appear at ${\cal O}(\alpha_s^3)$. 
A conventional QCD calculation at ${\cal O}(\alpha_s^2)$ or below contains no
resummation terms and should be considered as a background to the detection 
of BFKL dynamics. A full 1-loop calculation of this ``fixed order'' 
background, at ${\cal O}(\alpha_s^2)$, is presented in this section
\cite{prl,hera_forward}
The kinematical region populated by these events and the scale 
dependence of the fixed order perturbative QCD cross sections are discussed.
We also present LO $\alpha_s^3$ predictions
for forward jet cross sections with three additional hard jets,
which gives some indications about the
possible size of NNLO $\O(\alpha_s^3)$ corrections.

Numerical results below will be presented both for tree level and 
NLO simulations. The LO parton 
distributions of Gl\"uck, Reya and Vogt~\cite{grv} together with the 
1-loop formula for the strong coupling constant are used for the
tree level 1-jet and 2-jet results. At ${\cal O}(\alpha_s^2)$
all cross sections are determined using the NLO GRV parton distribution 
functions $f(x_1,\mu_F^2)$ and the two loop formula for $\alpha_s(\mu_R^2)$
in Eq.~(\ref{asnlo}).
With this procedure the 2-jet inclusive rate at ${\cal O}(\alpha_s^2)$ 
is simply given as the 
sum of the NLO 2-jet and the LO 3-jet exclusive cross sections. 
Unless otherwise stated, both the 
renormalization and the factorization scales are tied to the sum of 
parton $k_T$'s in the Breit frame,
$\mu_R = \mu_F = {1\over 2} \sum_i k_T^B(i)$, where 
$k_T^B$ is defined in Eq.~(\ref{ktdef}).
Our NLO Monte Carlo runs have a relative statistical error of about $1\%$.

We are interested in events with a forward jet (denoted ``$j$'') 
with $p_T(j)\approx Q$ and $x_{jet}\gg x$. Such events are selected 
by requiring
\ba
0.5 & < & p_T^2(j)/Q^2 < 4\; , \nonumber  \\
x_{jet}=p_z(j)/E_p &>& 0.05\;,\qquad {\rm while}\qquad x<0.004\; , 
\label{eq:fj}
\ea
in the laboratory frame. In order to facilitate a comparison with 
forthcoming H1 data, we impose additional cuts which closely model the 
H1 selection  of such events. Jets are defined in the cone 
scheme (in the laboratory frame) with $\Delta R = 1$ and a
pseudo-rapidity of  $|\eta|<3.5$. 
Here $\eta=-\ln\tan(\theta/2)$ denotes the pseudo-rapidity of a jet.
Unless noted otherwise, all jets must have transverse momenta of at least 
4~GeV in both the laboratory and the Breit frames. The forward jet 
must be in the angular range $6.3^o < \theta(j) < 20^o$ and have
transverse momentum $p_T^{lab}(j)>5$~GeV. 
Additional selection cuts are $Q^2>8~$GeV$^2$, $0.1 < y < 1$, as well 
as $E(l^\prime)>11$~GeV and $160^o < \theta(l^\prime) < 173.5^o$ for the 
energy and lab angle of the scattered lepton.
The energies of the incoming electron and proton are set to 27.5~GeV
and 820~GeV, respectively.

Numerical results for the multi-jet cross sections are shown in 
Table~\ref{table_for1}. Without the requirement of a forward jet 
(right-hand column) the cross sections show the typical decrease 
with increasing jet multiplicity which is expected in a well-behaved 
QCD calculation. 
The requirement of a forward 
jet with large longitudinal momentum fraction
($x_{jet}>0.05$) and restricted transverse momentum ($0.5<p_T^2(j)/Q^2<4$)
severely restricts the available phase space. In particular one finds that 
the 1-jet cross section vanishes at ${\cal O}(\alpha_s^0)$, due to the 
contradicting $x<0.004$ and $x_{jet}>0.05$ 
requirements: this forward jet kinematics is impossible for one single
massless parton in the final state. 

\begin{table}[t]
\vspace{3mm}
\caption{Cross sections for $n$-jet events
in DIS at HERA at order $\alpha_s^0$, $\alpha_s$, and $\alpha_s^2$. 
The jet multiplicity includes the forward jet.
The $p_T>4$~GeV requirements for central jets 
are replaced by the condition $k_T^B>4$~GeV in the second column.
No $p_T^B$ cut is imposed in the 1-jet case at ${\cal O}(\alpha_s^0)$
and the factorization scale is fixed to $Q$.
See text for further details.
}\label{table_for1}
\vspace{2mm}
\begin{tabular}{l|rr|r}
        \hspace{0.8cm}
     &  \mbox{with forward jet} &
     &  \mbox{without forward jet } \\
     &  \mbox{$p_T^B,p_T^{lab}>4$~GeV}
     &  \mbox{$k_T^B>4$~GeV}
     &  \mbox{$p_T^B,p_T^{lab}>4$~GeV}\\
\hline\\[-3mm]
\mbox{${\cal O}(\alpha_s^0)$: 1 jet}
                    & 0    pb & 0    pb & 8630 pb    \\
\mbox{\mbox{${\cal O}(\alpha_s)$}: 2 jet }
                    & 18.9 pb & 22.4 pb & 2120 pb    \\
\mbox{${\cal O}(\alpha_s^2)$: 1 jet inclusive} 
                    & 100 pb & 100 pb &           \\
\mbox{\phantom{${\cal O}(\alpha_s^0)$:} 2 jet inclusive} 
                    & 83.8 pb & 98.3 pb & 2400 pb    \\
\mbox{\phantom{${\cal O}(\alpha_s^0)$:} 2 jet exclusive}  
                    & 69.0 pb & 66.8 pb & 2190 pb    \\
\mbox{\phantom{${\cal O}(\alpha_s^0)$:} 3 jet }   
                    & 14.8 pb & 31.5 pb & 210  pb    \\
\mbox{\mbox{${\cal O}(\alpha_s^3)$:} 4 jet }   
                    & 2.8 pb &  5.2 pb    & 23  pb    \\
\end{tabular}
\end{table}

Suppose now that we had performed a full ${\cal O}(\alpha_s^2)$ calculation 
of the DIS cross section, which would contain 3-parton final states at tree 
level, 1-loop corrections to 2-parton final states and 2-loop corrections to
1-parton final states. These 2-loop contributions would vanish identically,
once $x\ll x_{jet}$ is imposed. The remaining 2-parton and 
3-parton differential cross sections, however, and the cancellation of
divergences between them, would be the same as those entering 
a calculation of 2-jet inclusive rates. These elements are already 
implemented in the \docuname\ program which therefore can be used to determine 
the inclusive forward jet cross section within the 
cuts of Eq.~(\ref{eq:fj}). At ${\cal O}(\alpha_s^2)$ this 
cross section is obtained from the cross section for 2-jet inclusive events 
by integrating over the full phase space of the additional
jets, without any cuts on their transverse momenta or pseudo-rapidities. 
Numerical results are shown in the third row of Table~\ref{table_for1}. 

The table exhibits some other remarkable features of forward jet events:
the ${\cal O}(\alpha_s^2)$ 2-jet inclusive cross section exceeds the 
${\cal O}(\alpha_s)$ 2-jet cross section 
by more than a factor four and the 3-jet rate at ${\cal O}(\alpha_s^2)$ is 
about as large as the 2-jet rate at ${\cal O}(\alpha_s)$. These 
characteristics can be understood in terms of the kinematics of forward 
jet events. Kinematics puts severe constraints on the ``recoil system'', 
the part of the final state in the $\gamma$-parton collision of 
Fig.~\ref{fig:feyn} which is left after taking out the forward jet. 
For $x\ll x_{jet}$, a high invariant mass hadronic system must be 
produced by the photon-parton collision. For small scattering angle but large
energy, $E(j)$, of the forward jet this condition translates into 
\bq\label{eq:kinem}
M^2+2E(j)m_T\;e^{-y} \approx \hat{s}_{\gamma,parton}
                 \gsimfig Q^2\left({x_{jet}\over x}-1\right) \gg Q^2\; .
\eq
Here $m_T=\sqrt{M^2+p_T^2}$ and $y$ are the transverse mass and rapidity 
of the partonic recoil system. Eq.~(\ref{eq:kinem}) implies that $m_T$ 
must be large, the more 
so the larger the ratio $x_{jet}/x$. On the other hand, the
transverse momentum, $p_T$, of the recoil system is fixed by momentum 
conservation, $p_T = |{\bf q}_T +{\bf p}_T(j)|$, and the cross section is
largest when the transverse momenta of both the virtual photon, $q_T$, and 
of the forward jet are small. Thus, a large recoil transverse mass is 
most easily achieved by two or more partons which create a subsystem of
large invariant mass, $M$, and some of these partons will manifest 
themselves as fairly hard hadronic jets. 

\begin{figure}[tb]
\vspace*{0.5in}            
\hspace{-1cm}
\begin{picture}(0,0)(0,0)
\includegraphics{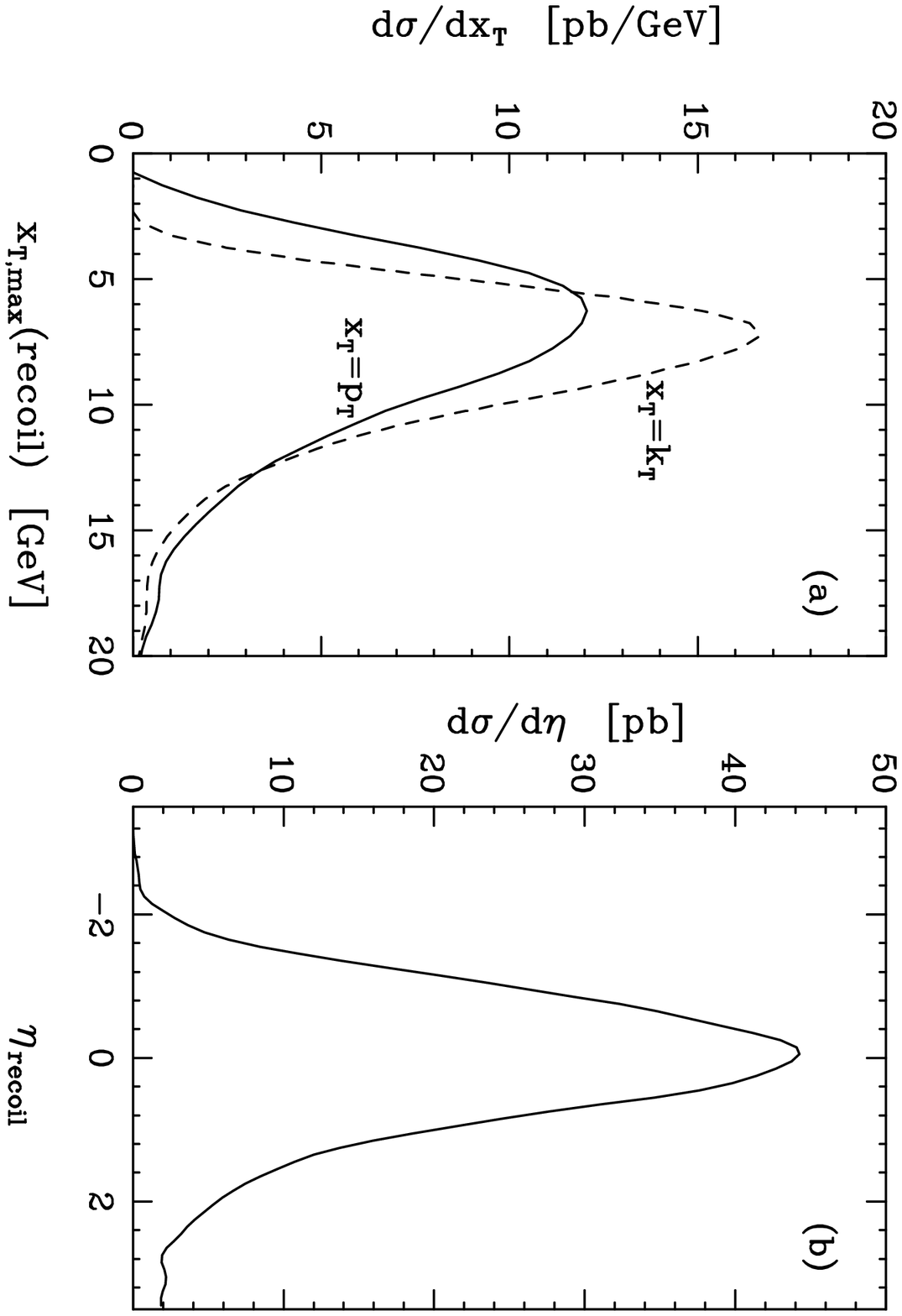}
\end{picture}
\vspace{8.5cm}
\caption{
Characteristics of the highest transverse momentum ``jet'' in the recoil 
system, i.e. excluding the forward jet. Distributions shown are  
(a) $d\sigma/dp_T$ in the lab frame (solid line) and $d\sigma/dk_T$ in the 
Breit frame (dashed line) and (b) the jet's pseudo-rapidity distribution in 
the laboratory frame. All distributions are calculated at order $\alpha_s^2$.
Jet transverse momentum cuts have been relaxed to $p_T^{lab},p_T^B>1$~GeV.
From Ref.~\protect\cite{prl}.
\label{fig:recoil}
}
\end{figure}

This fact is demonstrated in 
Fig.~\ref{fig:recoil} where the transverse momentum and the pseudo-rapidity 
distributions of the recoil jet with the 
highest $p_T^{lab}$ are shown, subject only to a nominal requirement 
of $p_T^{lab},p_T^B>1$~GeV. Over 80\% of all forward jet events contain at
least one second jet, with $p_T^{lab}\gsimfig 4$~GeV, which typically 
falls into the central part of the detector. This fraction rises to 98\% 
when the $k_T^B$ of a jet is used instead of its transverse
momentum (dashed line in Fig.~\ref{fig:recoil}(a) and second column of
Table~\ref{table_for1}). Beyond illuminating the properties of the recoil 
system~\cite{BFKLjets}, this observation intuitively explains why we are 
able to calculate the 1-jet inclusive forward jet cross section with a 
program which simulates dijet events at NLO: the forward jet kinematics 
almost always leads to at least two identifiable jets. 


Another important information in Table~\ref{table_for1} is
the relatively small fraction of ${\cal O}(\alpha_s^3)$
4-jet events. The 4-jet cross section is
considerably smaller than the LO ${\cal O}(\alpha_s^2)$ 
3-jet cross section, also for forward jet events in column 2.
(Recall that the LO \oasz\ 3-jet cross section with a forward jet
is about as large as the \oas\ 2-jet cross section.)
This is an indication that the fixed order NNLO  ${\cal O}(\alpha_s^3)$
corrections to the 1-jet inclusive cross section
might be moderate, at least considerably smaller compared
to the NLO ${\cal O}(\alpha_s^2)$ corrections discussed before.
As a consequence, the strong enhancement of the forward jet cross section
would require some new dynamical mechanism, as could {\it e.g.}
be provided by the BFKL evolution.

How reliable is the determination of the forward jet cross section 
at ${\cal O}(\alpha_s^2)$?
The importance of higher order corrections can be estimated by studying 
the dependence of the cross section on the choice of factorization and
renormalization scales, $\mu_F$ and $\mu_R$. Our standard choice is 
$\mu_R^2 = \mu_F^2 = {1\over 4}\left( \sum_i k_T^B(i) \right)^2$. In 
Fig.~\ref{fig:scale} we investigate variations of the 2-jet inclusive 
cross section when changing this scale by a factor,
$\xi$, (solid lines)
\bq
\mu_R^2 = \mu_F^2 = \xi\;{1\over 4} \left( \sum_i k_T^B(i) \right)^2 \; ,
\eq
and we also consider renormalization and factorization scales which are
proportional to the photon virtuality, $Q^2$, (dashed lines)
\bq
\mu_R^2 = \mu_F^2 = \xi\; Q^2  \; .
\eq
\begin{figure}[t]            
\begin{picture}(0,0)(0,0)
\hspace{-1cm}
\includegraphics{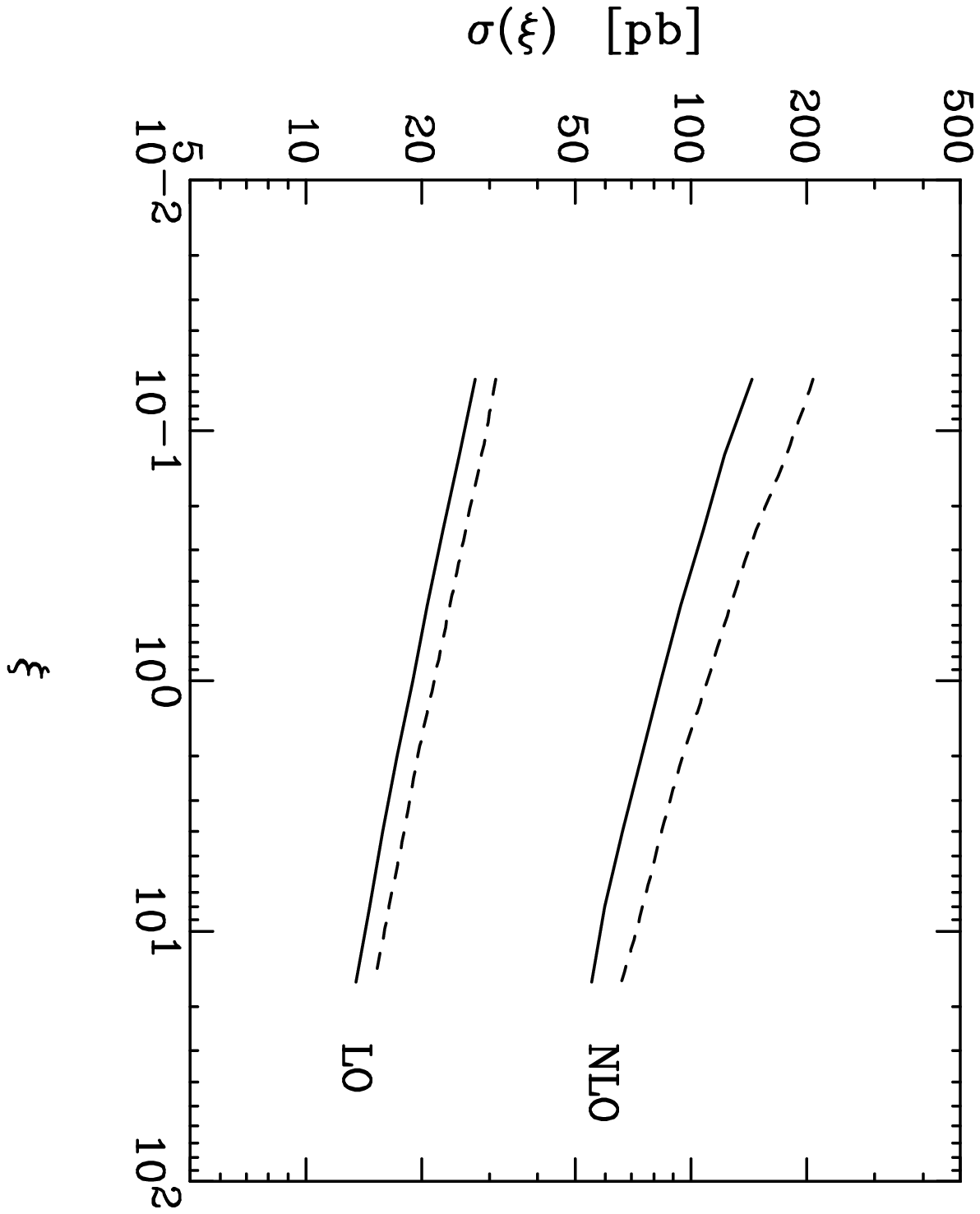}
\end{picture}
\vspace{7.6cm}
\caption{
Scale dependence of the 2-jet inclusive 
cross section with a forward jet satisfying $x_j>0.05$ 
and $p_{T}(j)>5$~GeV in the lab frame (see text for additional cuts). Results 
are shown for $\mu_R^2=\mu_F^2=\xi Q^2$ (dashed lines) 
and $\mu_R^2=\mu_F^2=\xi (0.5\sum k_T)^2$ (solid lines), 
at ${\cal O}(\alpha_s)$ (lower curves)
and at ${\cal O}(\alpha_s^2)$ (upper curves).
From Ref.~\protect\cite{prl}.
\label{fig:scale}
}
\end{figure}

Two striking features of the forward jet cross section become apparent
in this comparison. The large effective $K$-factor, $K\approx 5$, was
already noted in Table~\ref{table_for1}. In addition one finds that the scale 
dependence is at least as strong at ${\cal O}(\alpha_s^2)$ (NLO) as 
at ${\cal O}(\alpha_s)$ (LO). Both features are
closely related. The smallness of the
LO 2-jet compared to the NLO 2-jet inclusive cross section means that
at least three final state partons are required to access the relevant part
of the phase space. This three-parton cross section, however, has only been 
calculated at tree level and is subject to the typical scale uncertainties
of a tree level calculation. Thus, even though we have performed a full 
${\cal O}(\alpha_s^2)$ calculation of the forward jet cross section,
including all virtual effects, our calculation effectively only gives a LO
estimate and large corrections may be expected from
higher order effects, like the gluon ladders in Fig.~\ref{fig:feyn}.

The size of these corrections may be estimated by comparing to BFKL 
calculations or to existing experimental results. The H1 Collaboration has 
published such a measurement which was made during the 1993 HERA run with 
incident electron and proton energies of $E_e=26.7$~GeV 
and $E_p=820$~GeV~\cite{H1result}. The acceptance cuts used for this 
measurement differed somewhat from the ones described before.  Because 
of the lower luminosity in this early HERA run, the $x_{jet}$ cut on the 
forward jet was lowered and defined in terms of the jet energy,
\bq
x_{jet}=E(j)/E_p > 0.025\;,  \label{eq:H1cuta}
\eq
and its angular range was chosen slightly 
larger, $6^o < \theta(j) < 20^o$.
Scattered electrons must satisfy $E(l^\prime)>12$~GeV and 
$160^o < \theta(l^\prime) < 173^o$. Finally the 
Bjorken-$x$ and $Q^2$ ranges were chosen as $0.0002<x<0.002$ and 
5~GeV$^2<Q^2<100$~GeV$^2$. Within these cuts H1 has measured cross sections
of $709\pm 42\pm 166$~pb for $0.0002<x<0.001$ and $475\pm 39\pm 110$~pb 
for $0.001<x<0.002$. These two data points, normalized to bin sizes of 0.0002,
are shown as diamonds with error bars in Fig.~\ref{fig:h1comp}. Also included
(dashed histogram) is a recent calculation of the BFKL cross 
section~\cite{bartelsH1}.
%
\begin{figure}[tb]
  \centering
  \mbox{\epsfig{file=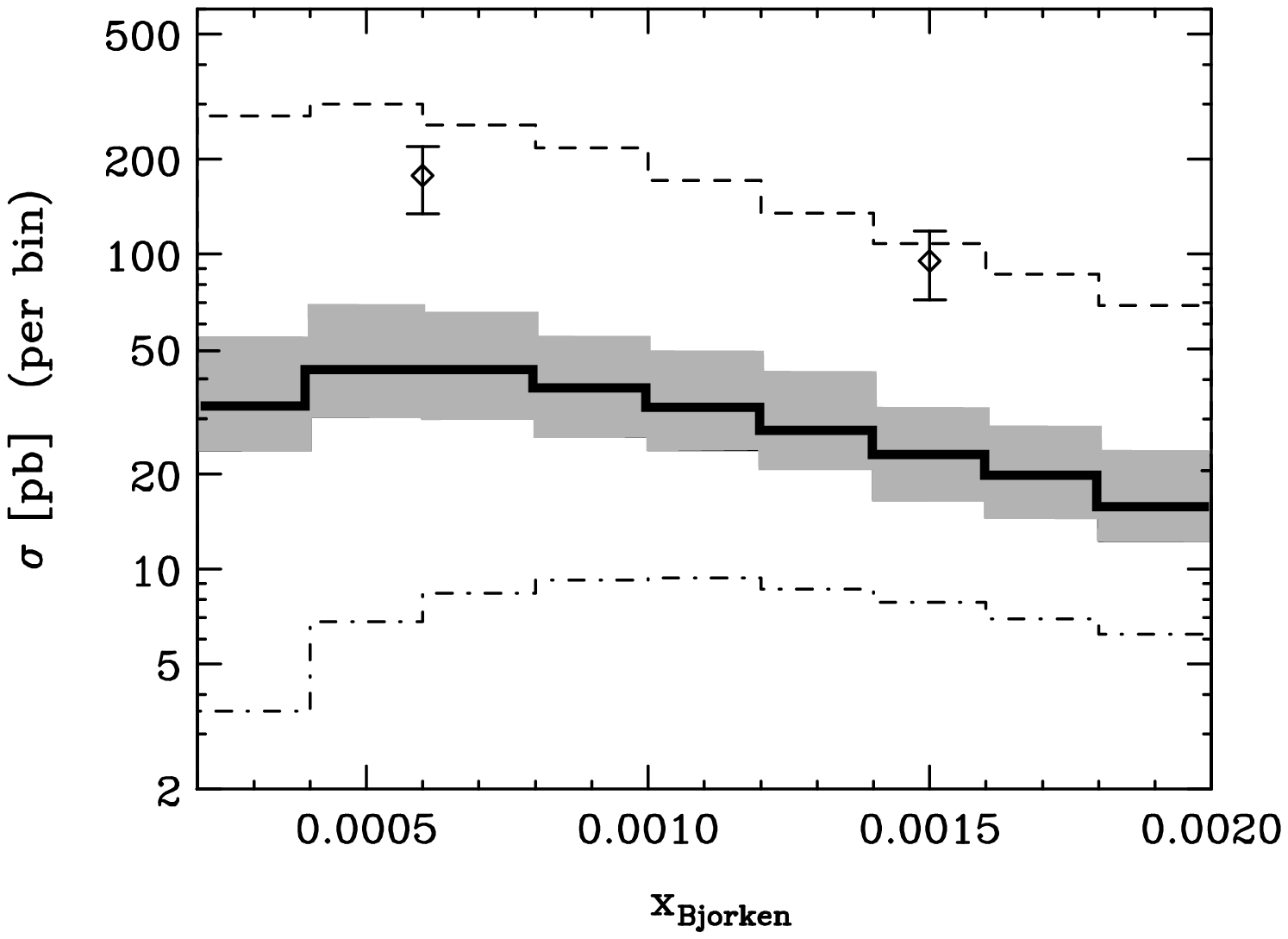,
         bbllx=60,bblly=240,bburx=550,bbury=600,width=0.80\linewidth}} 
\vspace*{-5mm}
\caption{
Forward jet cross section at HERA as a function of Bjorken $x$ within the H1 
acceptance cuts~\protect\cite{H1result} (see text). 
The solid (dash-dotted) histogram gives the NLO (LO) \docuname\ result for  
the scale choice $\mu_R^2=\mu_F^2=\xi(0.5\sum k_T)^2$ with $\xi=1$. 
The shaded area shows the uncertainty of the NLO prediction, corresponding
to a variation of $\xi$ between 0.1 and 10. The BFKL result of 
Bartels et al.~\protect\cite{bartelsH1} is shown as the dashed 
histogram. The two data points with error bars correspond to the H1 
measurement~\protect\cite{H1result}.
\label{fig:h1comp}
From Ref.~\protect\cite{prl}.
}
\end{figure}

As shown before, the \docuname\ program allows to calculate the full 1-jet 
inclusive forward jet cross section for $x\ll x_{jet}$. 
Results are shown in Fig.~\ref{fig:h1comp} at ${\cal O}(\alpha_s)$ 
(LO, dash-dotted histogram) and 
at ${\cal O}(\alpha_s^2)$ (NLO, solid histogram).
While the BFKL results~\cite{bartelsH1} agree well with the H1 data, the 
fixed order perturbative QCD calculations clearly fall well below the 
measured cross section, even when accounting for variations of the 
factorization and renormalization scales. The measured cross section is
\begin{figure}[tb]
  \centering
 \mbox{\epsfig{file=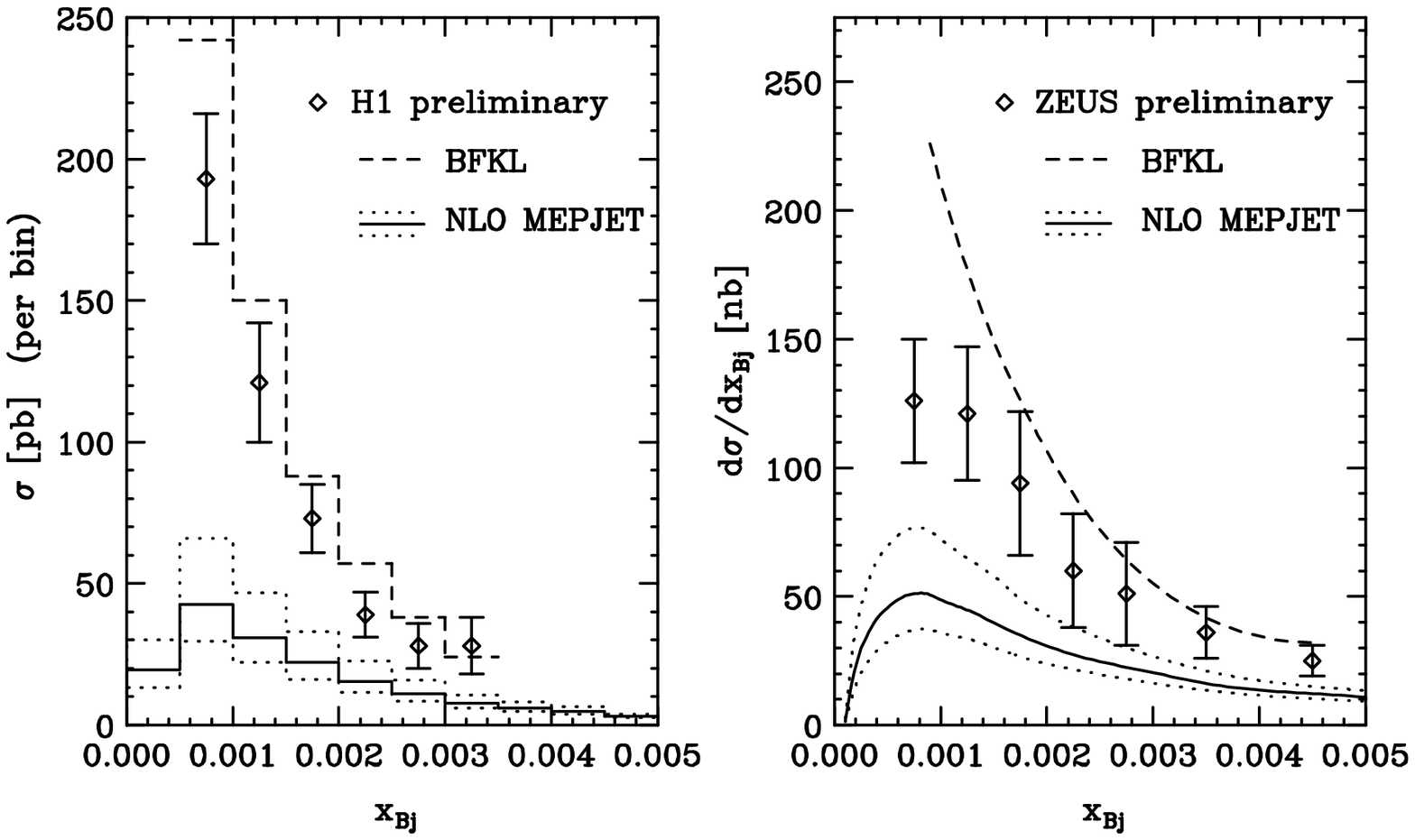,
         bbllx=0,bblly=200,bburx=592,bbury=550,
         width=1.0\linewidth}} 
\vspace*{-1.5cm}
\caption{
Forward jet cross section at HERA as a function of Bjorken $x$ within (a)
the H1 \protect\cite{wobisch} and (b) the ZEUS \protect\cite{zeus_fw_dis97}
acceptance cuts. 
The BFKL result of Bartels et al.~\protect\cite{bartelsH1} 
is shown as the dashed line.
The solid and dotted lines give the the NLO  \docuname\ result for  
the scale choice $\mu_R^2=\mu_F^2=\xi(0.5\sum k_T)^2$ with $\xi=0.1,1$ and 10,
which provides a measure for the uncertainty of the NLO prediction.
\label{f_fw_dis97}
}
\end{figure}
%
%
%
a factor 4 above the NLO expectation. The shape of the NLO prediction, 
on the other hand, is perfectly compatible with the H1 results, and not very 
different from the BFKL curve in Fig.~\ref{fig:h1comp}. At LO
a marked shape difference is still observed, which can be traced directly 
to the kinematical arguments given before: according to Eq.~(\ref{eq:kinem})
the transverse mass of the recoil system must increase proportional
to $x_{jet}/x$ and this requires increased transverse momentum of the 
forward jet at LO. Thus, at LO, the expected cross section falls rapidly at 
small $x$, an effect which is avoided when additional partons are available
in the final state to balance the overall transverse momentum.

We conclude from Fig.~\ref{fig:h1comp} 
that the  H1 data show evidence for BFKL dynamics in 
forward jet events via an enhancement in the observed forward jet cross 
section above ${\cal O}(\alpha_s^2)$ expectations. 
The variation of the cross section with $x$, 
on the other hand, is perfectly compatible with either BFKL dynamics or 
NLO QCD. 

Thus more decisive shape tests are very important. 
In Fig.~\ref{f_fw_dis97}
recent H1 \cite{wobisch}  and ZEUS \cite{zeus_fw_dis97}
data are compared with BFKL and 
fixed oder QCD predictions.
The conditions $p_T(j)\approx Q$ and $x_{jet}\gg x$
are satisfied in the two experiments by slightly different selection cuts.
H1 selects events with a forward jet of $p_T(j)>3.5$ GeV
in the angular range $6^o < \theta(j) < 20^o$ with
\begin{equation}
0.5<p_T(j)^2/Q^2<2,
\hspace{2cm}
x_{jet}=E_{jet}/E_{proton}>0.035
\end{equation}
while ZEUS triggers on somewhat harder jets of $p_T(j)>5$ GeV and
$\eta(j)<2.4$ with
\begin{equation}
0.5<p_T(j)^2/Q^2<4,
\hspace{2cm}
x_{jet}=p_{z}(j)/E_{proton}>0.035
\end{equation}
Clearly, both experiments observe substantially more forward jet events
than expected from NLO QCD. A very rough estimate of the uncertainty
of the NLO calculation is provided by the two dotted lines which correspond to
variations by a factor 10 of the renormalization and factorization scales
$\mu_R^2$ and $\mu_F^2$.
The result from the BFKL calculation (dashed lines) agrees again better 
with the data, but here the overall normalization is uncertain and the
agreement may be fortuitous.
Also, we recall that both experiments observe more centrally
produced dijet events than predicted by NLO QCD calculations
(see section~\ref{sec_gluon}). Whatever mechanism is responsible for the 
enhancement in central jet production may also play a role in the enhanced 
forward jet cross section. Clearly these issues must be resolved before the 
evidence for BFKL dynamics can be elevated to the status of discovery.
\clearpage
\newpage
\section{Jet Production in Polarized {\boldmath{$ep$}} Scattering
\protect\vspace{1mm}}
\label{poljets}
%
\subsection{Introduction\protect\vspace{1mm}}

After the confirmation of the surprising EMC result,
that quarks carry  only a little fraction of the 
nucleon spin
the spin structure of a longitudinally proton
is  actively  being studied theoretically and
experimentally   by several
fixed target experiments
at CERN, DESY and SLAC~\cite{ro}.
So far only the inclusive polarized structure functions
$g_1$ and $g_2$ have been measured.
These measurements, however, do not  allow to distinguish between
the role  of quarks and gluon distributions.
The measurement of the polarized gluon distribution 
$\Delta g(x_g,Q^2)$
has become the key experiment
in order to understand the 
QCD properties of the spin of the nucleon.
We study here the possible direct measurement of $\Delta g(x_g,\mu_F)$
from dijet events at a HERA collider,
in the scenario where both the electron and the proton
beam  are polarized.
As in the unpolarized case, the gluon distribution  enters 
the 2-jets production cross section at LO (see Fig.~\ref{f_atobc})
thus suggesting  such a direct measurement.

For {\it polarized} lepton hadron scattering, the hadronic ($n$-jet) 
cross section is obtained from Eq.~(\ref{sighaddef}) with the replacements
$(\sigma_{\mboxsc{had}}, f_a\,\, , \hat{\sigma}_a)\rightarrow
 (\Delta\sigma_{\mboxsc{had}}, \Delta f_a, \Delta\hat{\sigma}_a)$:
\begin{equation}
d\Delta\sigma_{\mboxsc{had}}[n\mbox{-jet}] =
\sum_a \int d\eta \,\,\Delta f_a(\eta,\mu_F)\,\,\,
\as^n(\mu_R) \,\, \Delta \hat{\sigma}_a(p_0=\eta P, \mu_R, \mu_F)
\label{sighaddefpol}
\end{equation}
Here, the polarized hadronic cross section is defined as
\begin{equation}
d\Delta\sigma_{\mboxsc{had}}[n\mbox{-jet}] \equiv
d\sigma_{\mboxsc{had}}^{\uparrow\downarrow}[n\mbox{-jet}] 
-
d\sigma_{\mboxsc{had}}^{\uparrow\uparrow}[n\mbox{-jet}] 
\end{equation}
where the left arrow in the superscript denotes the polarization of the 
incoming lepton with respect to the direction of its momentum.
The right arrow stands for the polarization of the proton parallel 
or anti-parallel to the polarization of the incoming lepton.
The polarized parton distributions are defined by
$\Delta f_a(x_a,\mu_F^2)
\equiv f_{a \uparrow}(x_a,\mu_F)-f_{a \downarrow}(x_a,\mu_F)$.
Here, $f_{a \uparrow} (f_{a \downarrow})$ denotes 
the probability to find a parton $a$ 
in the longitudinally polarized  
proton whose spin is aligned (anti-aligned) to the proton's spin.
$\Delta\hat{\sigma}^a$ denote the corresponding polarized
partonic cross sections for the corresponding polarized
subprocesses  listed in Eq.~(\ref{processes}).

First discussions about jet
production in polarized lepton-hadron scattering
can be found in Ref.~\cite{ziegler}, where jets were defined in
the JADE scheme (see section~\ref{sec_jetdef} 
for center of mass energies of 20~GeV,
(which is about the energy of the fixed-target  EMC  experiment
at CERN  with a polarized muon beam of enery around 220 GeV).
Although this energy is  too small to observe clear jet structures, 
the studies in Ref.~\cite{ziegler} demonstrated alredy the unique possibility
for a measurement of the polarized gluon density from dijet events in
polarized DIS.
Prospects for measuring the polarized gluon distribution at HERA energies 
have been discussed first in  \cite{feltesse,heraws_pol}.
The results have been confirmed with the PEPSI program \cite{pepsi}.
We will extend these studies in the following
including a discussion of the QCD corrections to the polarized
inclusive and 1-jet cross sections and \oasz\ polarized 3-jet cross sections.

\subsection{Polarized Jet Cross Sections\protect\vspace{1mm}}
In complete analogy to the unpolarized case in Eq.~(\ref{onejet})
the polarized hadronic 1-jet {\it exclusive}
cross section in the 1-photon exchange up to $\O(\alpha_s)$ reads
\\[5mm]
\fbox{\rule[-5cm]{0cm}{8cm}\mbox{\hspace{16.2cm}}}
\\[-8.3cm]
\begin{eqnarray}
\displaystyle
\Delta\sigma_{\mboxsc{had}}[\mbox{1-jet}] &=&
\nonumber\\
&&\hspace{-15mm}
\int_0^1d\eta \int
\,d{\mbox{PS}}^{(l^\prime+1)}\,\,\sigma_0\,\,
\bigg[  \nonumber \\
&&  [\sum_{i=q,\bar{q}}e_i^2 \Delta f_i(\eta,\mu_F)\big]\,
\,\,\Delta |M^{(\mboxsc{pc})}_{q\rightarrow q}|^2\,
\left(1+\alpha_s(\mu_R)\,
{\cal{K}}_{q\rightarrow q}(\smin,Q^2)\right)\nonumber \\
&+&
\,
[\sum_{i=q,\bar{q}}e_i^2\, 
\,\Delta C_i^{\overline{\mboxsc{MS}}}(\eta,\mu_F,\smin)]\,\,
\alpha_s(\mu_R)\,\,\,\Delta |M^{(\mboxsc{pc})}_{q\rightarrow q}|^2\,
\,\bigg] J_{1\leftarrow 1}(\{p_i\}) 
\label{onejetpol}\\
&&\hspace{-20mm}
+
\int_0^1 d\eta \int
\,d\mbox{PS}^{(l^\prime+2)}\,\,\sigma_0\,\,
(4\pi\alpha_s(\mu_R))\,\,  
\bigg[    \nonumber \\
&&[\sum_{i=q,\bar{q}}e_i^2 \Delta f_i(\eta,\mu_F)] \,
\,\,\Delta |M^{(\mboxsc{pc})}_{q\rightarrow qg}|^2
\nonumber    \\ 
&+&
(\sum_{i=q}e_i^2 ) \Delta f_g(\eta,\mu_F)\,\,
\,\Delta |M^{(\mboxsc{pc})}_{g\rightarrow q\bar{q}}|^2
\bigg]
\,\,
\prod_{i<j;\,0}^{2}\Theta(|s_{ij}| - \smin)\,\,
J_{1\leftarrow 2}(\{p_i\})
\nonumber\\[5mm]\nonumber
\end{eqnarray}
where the Lorentz-invariant phase space measure
$d\mbox{PS}^{(l^\prime+n)}$
is defined in Eq.~(\ref{phasespace})
and the jet algorithms $J_{1\leftarrow 1}(\{p_i\})$ and
$J_{1\leftarrow 2}(\{p_i\})$ are described in Eq.~(\ref{jnndef})
and (\ref{jnn1def}), respectively.
The 1-jet {\it inclusive} cross section is defined via
Eq.~(\ref{onejetpol}) by replacing
$J_{1\leftarrow 2}(\{p_i\})$ in the last line by 
($J_{1\leftarrow 2}(\{p_i\})+J_{2\leftarrow 2}(\{p_i\}))$,
{\it i.e.} the 1-jet inclusive cross section is defined
as the sum of the NLO 1-jet exclusive cross section (as defined
in Eq.~(\ref{onejetpol})) plus the LO two jet cross section.
The dynamical ${\cal{K}}_{q\rightarrow q}$  factor
in Eq.~(\ref{onejetpol}) is the same as is 
unpolarized case in Eqs.~(\ref{r_qtoq}).
The polarized squared matrix elements   in Eq.~(\ref{onejetpol}) are
(see table~\ref{helitab}):
%
%
\begin{eqnarray}
\Delta |M^{(\mboxsc{pc})}_{q\rightarrow q}|^2&=&
|M^{(\mboxsc{pv})}_{q\rightarrow q}|^2=
32\,
\left[
(p_0.l)^2  \, -\,
(p_0.l^\prime)^2 \,
\right]
\,= \,8 \hat{s}^2\,\,(1-(1-y)^2)
\label{dm_qtoq} \\[2mm]
\Delta |M^{(\mboxsc{pc})}_{q\rightarrow qg}|^2
&=&
|M^{(\mboxsc{pv})}_{q\rightarrow qg}|^2 
=
\frac{128}{3}\,\,(l.l^\prime)\,\,
\frac{(l.p_0)^2-(l^\prime.p_0)^2-(l.p_1)^2+(l^\prime.p_1)^2
     }{(p_1.p_2)(p_0.p_2)}
\label{dm_qtoqg} \\[2mm]
\Delta |M^{(\mboxsc{pc})}_{g\rightarrow q\bar{q}}|^2
&=&
16\,\,(l.l^\prime)\,
\frac{(l^\prime.p_2)^2-(l.p_2)^2+(l^\prime.p_1)^2-(l.p_1)^2
     }{(p_0.p_1)(p_0.p_2)}
\label{dm_gtoqqbar} 
\end{eqnarray}
Color facors (including the initial state color average) 
are included in these results.

The structure of the polarized crossing functions
$\Delta C_{q,\bar{q}}^{\overline{\mboxsc{MS}}}(\eta,\mu_F,\smin)$
in Eq.~(\ref{onejetpol})
is  identical to the structure of the unpolarized crossing functions
discussed in section~\ref{crossing}:
\begin{equation}
\Delta C_{a}^{\overline{\mboxsc{MS}}}(x,\mu_F,\smin)=
\left(\frac{N}{2\pi}\right)
\left[ \Delta A_{a}(x,\mu_F)\ln\left(\smin/\mu_F\right)
+      \Delta B_{a}^{\overline{\mboxsc{MS}}}(x,\mu_F)\right]
\label{dcrossf}
\end{equation}
with
\begin{equation}
\Delta A_a(x,\mu_F) = \sum_p \Delta A_{p\rightarrow a}(x,\mu_F)
%
%
\hspace{1.5cm}
\Delta B_a^{{\overline{\mboxsc{MS}}}}(x,\mu_F) = \sum_p
\Delta B_{p\rightarrow a}^{{\overline{\mboxsc{MS}}}}(x,\mu_F)
\end{equation}
The sum runs again over $p=q,\bar{q},g$.
More specifically, the polarized  crossing functions for valence quarks
and sea quarks, which are needed in Eq.~(\ref{onejetpol}),
can be obtained from Eqs.~(\ref{crossf_uv},\ref{crossf_s}) by replacing
$A,B,C$ by $\Delta A,\Delta B,\Delta C$ respectively.

The finite  functions 
$\Delta A_{q\rightarrow q}(x,\mu_F)$ 
and 
$\Delta B_{q\rightarrow q}^{{\overline{\mboxsc{MS}}}}(x,\mu_F)$ 
can be obtained from  the r.h.s. of
Eqs.~(\ref{aqq},\ref{bqq}) with $f_q$ replaced by $\Delta f_q$.
The polarized $g\rightarrow q$ induced functions
$\Delta A_{g\rightarrow q}(x,\mu_F)$ 
and
$\Delta B_{g\rightarrow q}^{{\overline{\mboxsc{MS}}}}(x,\mu_F)$ 
are given by
\begin{eqnarray}
\Delta A_{g\rightarrow q} &=& \int_x^1 \frac{dz}{z}\,\, \Delta f_g(x/z,\mu_F)
\,\,\,\frac{1}{4}\,\,\,\Delta\hat{P}^{(4)}_{g\rightarrow q}(z)
\label{dagq}
\\
\Delta B_{g\rightarrow q}^{{\overline{\mboxsc{MS}}}} &=& 
\int_x^1 \frac{dz}{z}\,\,
\Delta f_g(x/z,\mu_F)
\,\,\,\frac{1}{4}\,\,\,\left\{ \Delta\hat{P}^{(4)}_{g\rightarrow q}(z)
\ln(1-z)-\Delta \hat{P}^{(\epsilon)}_{g\rightarrow q}(z)\right\}
\label{dbgq}
\end{eqnarray}
%
%
%
%
with the polarized Altarelli-Parisi kernels 
%
%
\begin{eqnarray}
\Delta\hat{P}^{(4)}_{g\rightarrow q}(z) &=& 
\frac{1}{3} 
\Delta P^{(4)}_{q\bar{q}\rightarrow g}(z) = 
\frac{2}{3}\,[2z-1] \\
\Delta \hat{P}^{(\epsilon)}_{g\rightarrow q}(z) &=& 
\frac{1}{3} 
\Delta P^{(\epsilon)}_{g\rightarrow q}(z) = 
0
\end{eqnarray}
%
%
%
%
%
For a detailed  derivation of the 
$\Delta A_{p\rightarrow a}$ and
$\Delta B^{\overline{\mboxsc{MS}}}_{p\rightarrow a}$ terms
in Eq.~(\ref{dcrossf})
we refer the reader to \cite{polcross}.
Finally, leading order polarized $n$-jet cross sections are obtained 
from the parton model formula Eq.~(\ref{hadlo}) with
the replacements
$(\sigma_{\mboxsc{had}}^{\mboxsc{LO}}, f_a\,\, , \
\hat{\sigma}^{\mboxsc{LO}}_{a\rightarrow n\,\, \mboxsc{partons}}) \rightarrow
(\Delta\sigma_{\mboxsc{had}}^{\mboxsc{LO}}, \Delta f_a, 
\Delta\hat{\sigma}^{\mboxsc{LO}}_{a\rightarrow n\,\, \mboxsc{partons}}) $.

\subsection{Numerical Results\protect\vspace{1mm}}
Eq.~(\ref{onejetpol}) includes all relevant information
for the calculation of the fully differential 1-jet or total \oas\
polarized cross section.

For the following numerical studies
we use again a cone algorithm 
defined in the lab frame with $\Delta R=1$
and $p_T^{\mboxsc{lab}}>5$ GeV.
Events are selected
in the $Q^2$ range of  $40<Q^2< 2500$ GeV$^2$ at HERA energies.
In addition, we require 
$0.3 < y < 1$,
an energy cut of $E(e^\prime)>5$~GeV on the scattered 
electron, and a cut on the pseudo-rapidity $\eta=-\ln\tan(\theta/2)$
of the scattered lepton and jets of $|\eta|<3.5$. 
The results are based on 
parton distributions from Gehrman and Stirling (GS)
\cite{gs} ``gluon set A'' together with
the 2-loop formula for the strong coupling constant.

With these parameters, one obtains 
23~pb for the LO 1-jet cross section 
$(\equiv \sigma_{\mboxsc{tot}}^{\mboxsc{LO}})$
and -10.4~pb for the NLO
1-jet inclusive cross section\footnote{The difference between
the 1-jet (inclusive)
cross section and the total $({\cal{O}}(\alpha_s))$ cross section
due to effects discussed in section~\protect\ref{numonejet}
are small and will be neglected in the following.}
$(\equiv \sigma_{\mboxsc{tot}}^{\mboxsc{NLO}})$

The origin of these extremely  large corrections is investigated in 
Fig.~\ref{f_pol1}. The Bjorken-$x$ dependence of the corresponding
cross sections  in Fig.~\ref{f_pol1}a shows that the corrections are 
dominated by events at small $x$.
\begin{figure}[htb]
  \centering
  \mbox{\epsfig{file=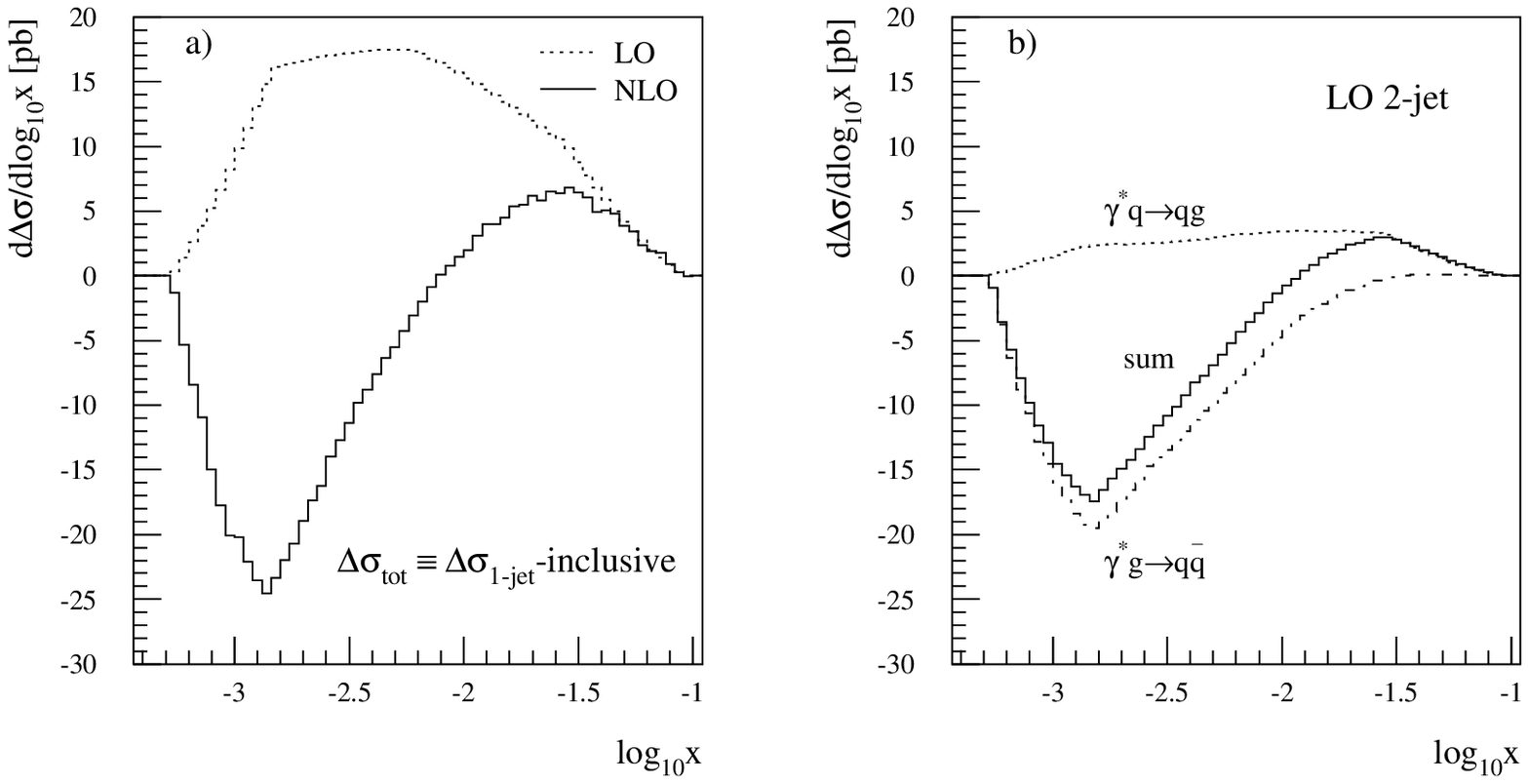,bbllx=0,bblly=270,bburx=550,bbury=550,
         width=0.95\linewidth}} 
\vspace*{-3mm}
\caption{
a) Dependence of the polarized LO and NLO 1-jet inclusive
cross section as a function of Bjorken $x$  with cuts as decribed
in the text. LO (NLO) results are based on
LO (NLO) ``gluon set A'' parton distributions  \protect\cite{gs};
b) LO 2-jet contribution
to the NLO 1-jet inclusive result in a). 
Jets are required to have  $p_T^{\protect\mboxsc{lab}}(j)>5$ GeV.
Results are shown for the quark and gluon initiated subprocesses 
alone and for the sum.
}
\label{f_pol1}
\end{figure}
As already mentioned the  \oas\ corrected 1-jet inclusive 
cross section (solid curve in Fig.~\ref{f_pol1}a) 
is defined as the sum of the NLO 1-jet exclusive
and the LO 2-jet cross section.
Fig.~1b shows the $x$ dependence of the hard LO 2-jet
contribution.
The negative corrections are entirely
due to the hard boson-gluon fusion subprocess (lower curve
in Fig.~\ref{f_pol1}b), which is negative for 
$x\lsimfig 0.025$, whereas the contribution from the
quark-initiated process is positive (but fairly small) 
over the whole kinematical range.
The important observation is that
the \oas\ corrections in Fig.~\ref{f_pol1}a
are dominated by these \underline{hard} 2-jet events,
and in particular by the large negative 
contribution from the boson-gluon fusion subprocess.
 
In order to compare the feasibility and the
sensitivity of the  measurement of the spin asymmetry at HERA
energies, Fig.~\ref{f_pol2}a compares the 
asymmetries 
\begin{equation}
\langle A_{\mboxsc{tot}} \rangle = 
\frac{\Delta\sigma^{\mboxsc{had}}_{\mboxsc{NLO}}[\mbox{tot}]}{
            \sigma^{\mboxsc{had}}_{\mboxsc{NLO}}[\mbox{tot}]}
\hspace{2cm}
\langle A_{\mboxsc{2-jet}} \rangle = 
\frac{\Delta\sigma^{\mboxsc{had}}[\mbox{2-jet}]}{
            \sigma^{\mboxsc{had}}[\mbox{2-jet}]}
\label{asymdef}
\end{equation}
as a function of $x$.
The unpolarized cross sections in the denumerators of Eq.~(\ref{asymdef})
are based on NLO GRV \cite{grv} parton distribution functions.
\begin{figure}[t]
  \centering
  \mbox{\epsfig{file=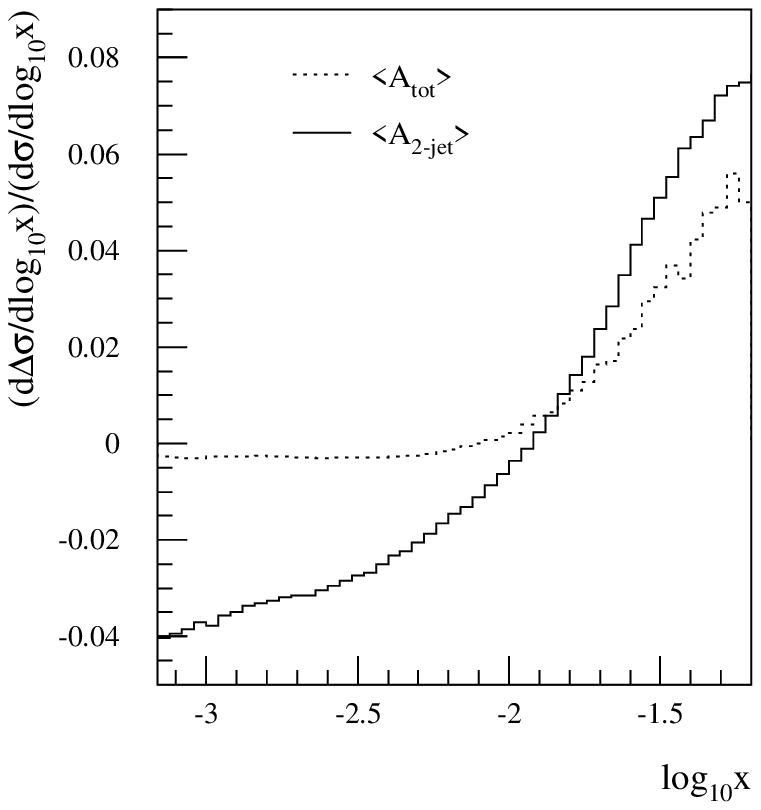,bbllx=0,bblly=400,bburx=300,bbury=650,
         width=0.5\linewidth}} 
\vspace*{-6mm}
\caption{
Asymmetries $\langle A_{\mboxsc{2-jet}} \rangle$
and $\langle A_{\mboxsc{tot}} \rangle$ in Eq.~(\protect\ref{asymdef})
as a function of $x$.
}
\label{f_pol2}
\end{figure}
One observes that the dijet asymmetry $\langle A_{\mboxsc{2-jet}} \rangle$
is much larger (up to 3-4\%) than the inclusive asymmetry
$\langle A_{\mboxsc{tot}} \rangle$ in the
low $x$ region, which is hardly (or even not at all)
constrained by currently available DIS data.
Thus, the dijet events from polarized electron and
polarized proton collisions at HERA 
are expected to provide the best measurement of the gluon polarization
distribution in the small $x$ regime.

For the isolation of polarized 
parton structure functions we are interested, however,  in the 
fractional momentum $x_a$ of incoming parton $a$ ($a=q,g$),
which is related to $x$ by 
$
x_a = x \,\left(1+{{s_{jj}}}/{Q^2}\right)
$
($s_{jj}$ denotes the 
invariant mass squared of the two jets, see section~\ref{sec_gluon}).
The corresponding $x_a$ distributions of the polarized 2-jet cross 
sections are shown in Fig.~\ref{f_pol3}a.

How reliable are these LO 2-jet results  in view of the
large QCD corrections to the 1-jet  cross section
in Fig.~\ref{f_pol1}a?
First preliminary results of a full \oasz\ calculation 
for the polarized dijet cross section show that
the QCD corrections are fairly small \cite{polcross}.
The reason for this is shown in
Fig.~\ref{f_pol3}b, where LO predictions
for the polarized \oasz\  3-jet cross sections are shown
as a function of $x$.
One observes a very similar shape for the gluon
and quark initiated subprocesses as already found 
for the 2-jet results in Fig.~\ref{f_pol1}b.
Moreover, the 3-jet cross sections are now suppressed
by about a factor five compared to the LO 2-jet cross sections
in Fig.~\ref{f_pol1}b.
\enlargethispage{1cm}
Note that the new gluon initiated subprocess $eg\rightarrow eq\bar{q}$ 
was responsible for the very large \oas\ corrections
in the 1-jet inclusive case. 
There is no such {\it new } contributing
subprocess starting at \oasz\ (see Fig.~\ref{f_pol3}b), which could
introduce  similarly large corrections to the 2-jet results.

It was also shown in Refs.~\cite{feltesse,heraws_pol} that
the 2-jet spin asymmetry is not washed out by hadronization
effects. 
Further  asymmetry distributions for several kinematic variables 
are considered in Ref.~\cite{maul}.

In  conclusion,  the dijets events  from polarized  electron
and polarized proton collisions at HERA can provide
a good measurement of the gluon polarization
distribution for $x_g < 0.2$, the region
where $x_g\Delta g(x_g)$ is expected
to show a  maximum.

\begin{figure}[hb]
\vspace*{1cm}
  \centering
  \mbox{\epsfig{file=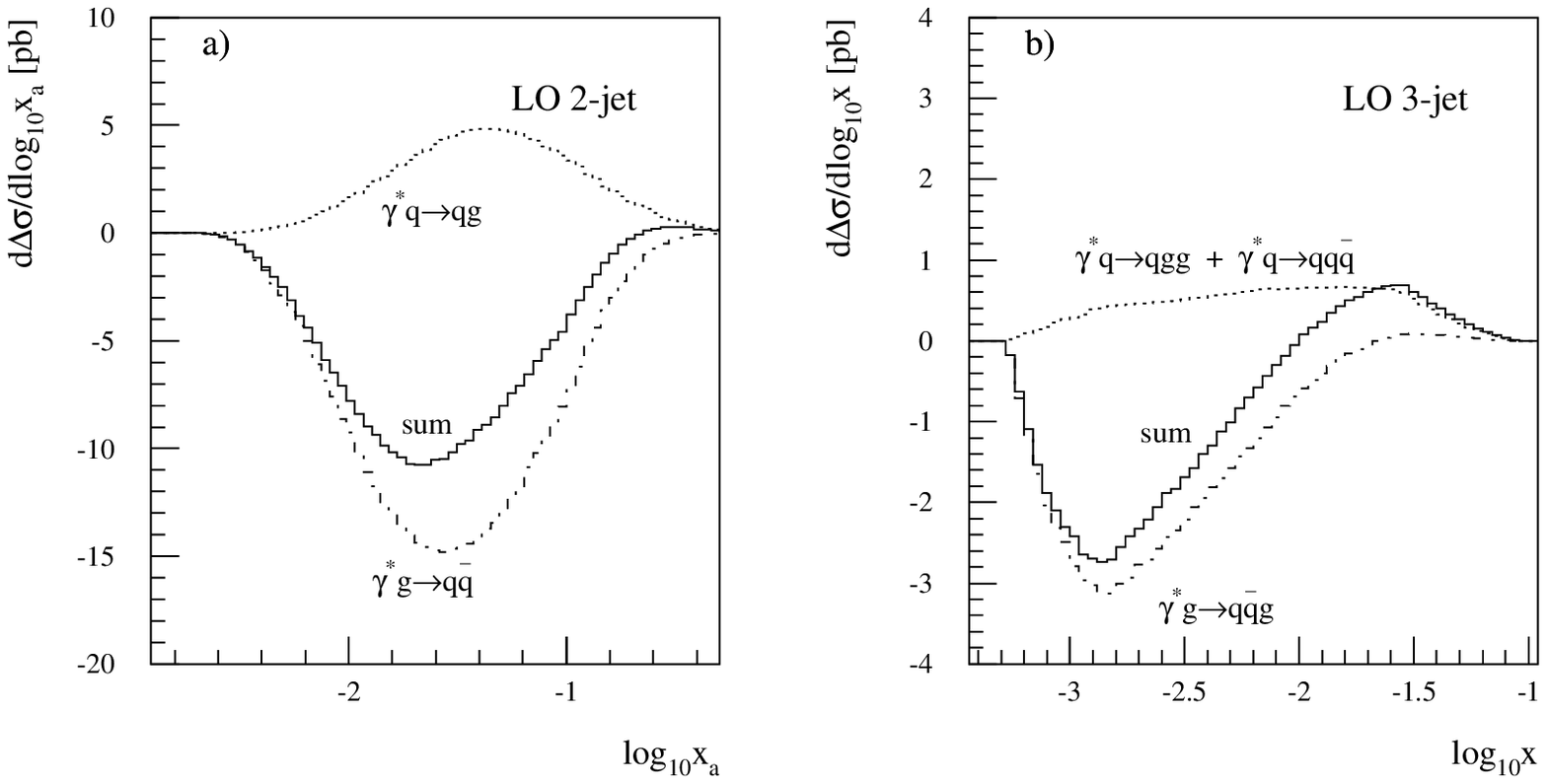,bbllx=40,bblly=400,bburx=540,bbury=630,
         width=0.8\linewidth}} 
\vspace*{-5mm}
\caption{
a)
Same as Fig.~\protect\ref{f_pol1}b for the $x_a$ ($a=q,g$) distribution, 
$x_a$ representing
the momentum fraction of the incident parton at LO;
b) \protect\oasz\ 3-jet 
contribution with $p_T^{\protect\mboxsc{lab}}(j)>5$ GeV.
Results are shown for the quark and gluon initiated subprocesses 
alone and for the sum.
}
\label{f_pol3}
\end{figure}
\clearpage
\newpage
\section{Conclusions and  Outlook}
\label{conclusions}

Significant progress in the development of fully flexible
NLO QCD calculations for a wide range of hadronic final
state variables in DIS allows for detailed comparisons
with increasingly precise data from HERA.
Such comparisons offer many interesting possibilities
for precision QCD tests.

A major area of study at HERA is dijet production, 
which can be  used for 
the measurement of the strong coupling constant 
and the determination of the gluon density.
The NLO Monte Carlo program \docuname\
allows to study these jet cross sections for arbitrary jet algorithms
(and event shape variables) including $Z$ and $W$ exchange effects.
Internal jet structure, parton recombination effects, 
and the effects of arbitrary acceptance cuts can  be simulated at the full
${\cal O}(\alpha_s^2)$ level.
We have presented a detailed description of the theoretical
framework underlying this calculation. All analytical formulae
necessary for the NLO calculation are provided.
In addition, a detailed overview of phenomenological studies
and comparisons of theoretical expectations
with a variety of recent experimental results at HERA
are presented.

A second fully flexible NLO calculation for 1- and 2-jet-like production
in the one-photon exchange approximation 
has become available with the {\large \sc disent} program \cite{disent}.
The general algorithm underlying this calculation is described 
in depth in Ref.~\cite{dipol}.
Very recently   a third calculation, {\large \sc disaster}++, has been
provided in Ref.~\cite{disaster}, but is again restricted to the
one-photon approximation.
Comparisons between these calculations, in the one-photon approximation,  
yielded satisfactory agreement so far~\cite{compare,disaster}.

\docuname\ also allows to calculate LO
cross sections up to ${\cal O}(\alpha_s^3)$, {\it i.e.} up to
4-jet final states for complete
neutral current exchange ($\gamma^\ast$ and/or $Z$).
In addition, bottom mass and charm mass effects
in LO 2-jet and 3-jet calculations can be taken into account
in the one-photon exchange case.
\docuname\ is also the only program that allos to calculate
jet cross sections in polarized $ep$ scattering up to \oasz.

The major results of the individual studies presented
in this review include:
\begin{itemize}
\item[i)]
The analysis of the ``high $Q^2$ (1-jet inclusive) events''  at HERA
is now supported by a fully versatile 1-loop calculation,
which includes all electroweak effects.
Taking  the  requirement of at least one jet into account
in the NLO calculation  increases   the
currently observed discrepancy between the SM
predictions and data. 
\item[ii)]
High jet transverse momenta in the laboratory and the Breit frame
with $p_{T,\mboxsc{min}}^{\mboxsc{lab}}(j)\approx 
p_{T,\mboxsc{min}}^B(j)\approx Q_{\mboxsc{min}}$
are a good criterion for the 
applicability of fixed order perturbation theory
in the calculation of  $n$-jet ($n\ge 2$) cross sections in DIS.
\item[iii)]
Large NLO effects in dijet cross sections and distributions are found
for some jet definition schemes (in particular the
$W$-scheme). Cone and $k_T$ schemes appear to be better suited for precision
QCD tests. 
\item[iv)]
The ``natural''  choice 
for both the renormalization $(\mu_R)$ and factorization
$(\mu_F)$  scales for $n$-jet production in DIS is the average  $k_T^B$
of the jets in the Breit frame, which suggests analyzing the data
in different $\langle k_T^B\rangle$ rather than $\langle Q \rangle$
intervals. There is in general
a qualitative difference between scale choices tied to 
$\langle k_T^B \rangle$ versus
scales related to $Q$.
\item[v)]
Effective  $K$-factors
close to unity could  be
a coincidence arising from compensating 
effects in different phase space regions. 
It is important, therefore, to also 
compare LO and NLO distributions, in particular for those variables which
define the acceptance region. 
As shown in section~\ref{nloeffects}, 
a judicious choice of phase space regions 
could generate very large or small $K$-factors.
To avoid such potential problems,
one should investigate the effect of the higher order corrections
on those variables which are used to define kinematical cuts.
\item[vi)]
Bottom and charm quark mass effects are small (below 5\%)
for sufficiently high $Q^2$ and jet transverse momenta.
\item[vii)]
Electroweak effects in NC $n$-jet production
are known to become important for $Q^2\gsimfig 2500$ GeV$^2$
and can affect the 1-photon exchange result by more than a factor
of two for very high $Q^2$.
While NC and CC effects  differ markedly for $e^+p$ and
$e^-p$ scattering,  the QCD $K$-factors are largely
independent of the initial state lepton.
\item[viii)]
In the presence of typical acceptance cuts on the jets in the
laboratory frame, the azimuthal distribution of dijet events
around the virtual boson proton direction
is dominated by  kinematic effects 
and the residual dynamical effects from gauge boson
polarization are small.
A consequence of this  kinematical $\phi$ dependence is
that the $\phi$ dependent parts 
of the QCD matrix elements contribute even to the dijet 
{\it production} cross section.
Therefore, it is essential that the full helicity structure 
in the  matrix elements is kept even for the calculation of dijet production
cross sections.
\item[ix)]
NLO dijet cross sections in the JADE scheme are in good agreement
with corrected data. A recent preliminary measurement of the differential
dijet rate for $Q^2>200$ GeV$^2$
show a strong sensitivity to $\alpha_s$ \cite{weber}.
The precision which can be expected from the 
measurement is of the order of 5-10\%.
\item[x)]
Studies for the
extraction of the gluon distribution function
from dijet events show that large NLO 
corrections are present in the Bjorken $x$ distribution,
while these effects are mitigated in the reconstructed Feynman $x$ ($x_i$)
distribution, thus aiding a reliable extraction of $g(x_i,\mu_F)$.
\item[xi)]
H1 data for various event shape variables can be nicely
fitted by NLO theory ({\large \sc disent}) plus 
calculated power corrections with ${\bar{\alpha}}_0$ as one
free parameter. The results are \cite{h1event}
\begin{eqnarray}
{\bar{\alpha}}_0
&=& 0.491\pm 0.003\,\mbox{(exp)}^{+0.079}_{-0.042}\, \mbox{(theory)}
\nonumber \\
\alpha_s(m_Z)
&=& 0.118 \pm 0.001\, \mbox{(exp)}^{+0.007}_{-0.006} \,\mbox{(theory)} \nonumber
\end{eqnarray}
\item[xii)]
For the study of BFKL evolution, 
in events with a forward ``Mueller''-jet, very large  
QCD corrections are found at ${\cal O}(\alpha_s^2)$.
These fixed order effects form an important background to the observation
of BFKL evolution at HERA. 
Both experiments, H1 and ZEUS, observe substantially more forward jet events
than expected from NLO QCD. 
Results from BFKL calculations (which suffer from fairly 
large uncertainties in the absolute normalization, however) agree better 
with the data.
More decisive shape tests of the $x$ distribution of the
forward jet cross section are mandatory before the evidence
for BFKL dynamics can be elevated to the status of discovery.
\item[xiii)]
Studies of the prospects for a direct measurement of the
polarized gluon density distribution $\Delta g$ from dijet
events at a HERA collider,
in the scenario where both the electron and the proton
beam  are polarized, show that the 2-jet spin asymmetry can reach 
a few percent and can thus provide a reliable extraction of 
$\Delta g(x_i,\mu_F)$. Whereas the QCD corrections to the total
polarized cross section are found to be very large,
only moderate corrections are expected for the dijet cross sections.
\end{itemize}
A striking observation in  current  comparisons 
of  corrected dijet data and NLO parton predictions 
is that the integrated theoretical dijet cross sections/rates and
distributions in cone schemes and the $k_T$ scheme
tend to be significantly lower  than the 
experimental dijet cross
sections (see section~\ref{compare}).
What is the reason for this discrepancy? 
One problem  is that jet transverse momenta 
are typically quite small, of order 4-8~GeV, and thus much lower than 
{\it e.g.} for TEVATRON jets.
Even for $Q^2>1000$ GeV$^2$, most of the jets are 
at $p_T^B<10$ GeV as long as no hard jet definition cuts are imposed.
This implies that effects of modelling the hadronization process
or ``corrections'' of data to the parton level are significant.
Unfortunately, these corrections
for nonperturbative effects are not on solid theoretical grounds.
One way to study and reduce the uncertainties  would be to implement
parton-shower activity and a hadronization phase
directly into a NLO event generator.
Such an extended NLO generator is presently not available.
Ideally one would perform a direct comparison of NLO parton level 
calculations with jet-data at the hadron level, 
only corrected for detector resolution effects.
Such a comparison is, however, only meaningful in a kinematic
regime where the hadronization corrections are expected to
be small, {\it i.e.} for large transverse momenta of the jets.

In the case of the event shape analyses of the H1 collaboration
\cite{h1event} (see section~\ref{eventshape})
the importance of the hadronization effects
is reflected by the large power corrections
needed to fit the data to NLO theory on the parton level.
A calculation of non-perturbative power corrections
for arbitrary dijet cross sections in DIS 
would  improve the theoretical understanding of the
above mentioned problems considerably but
is currently not available.

Clearly, the many facets of the hadronic final state production in DIS 
at HERA offer many possibilities to study perturbative and
non-perturbative QCD dynamics
and will continue to be a very interesting topic in the future.\\[3mm]
\enlargethispage{1cm} 
\noindent
{\bf Acknowledgments}\\[2mm]
It is a great pleasure to thank D.~Zeppenfeld for his
collaboration on work described in this review.
I thank D.~Kosower, A.~Martin, E.~Reya and B.~Webber
for various interesting discussions and communications and
S. Willfahrt for help with the figures.
I  have also  benefitted from many instructive conversations and
communications with my experimental colleagues at DESY
and in particular with 
C.~Berger,
A.~de~Roeck,
T.~Carli,
K.~Flamm,
T.~Haas,
J.~Hartmann,
M.~Kuhlen,
H.-U.~Martyn,
B.~Musgrave,
C.~Niedzbal,
K.~Rabbertz,
J.~Repond,
K.~Rosenbauer,
T.~Trefzger,
M.~Weber,
M.~Wobisch.
I also enjoyed the collaboration with
J.~Feltesse,
F.~Kunne,
M.~Maul and
A.~Sch\"afer on problems in polarized jet production.
I am also grateful to Mrs.~Frasure, M.~Weber and D. Zeppenfeld for their
careful reading of the manuscript and fruitful comments.
Above all, I am deeply indebted to my family for the patience
during the preparation of this work.
Last but not least, I thank J.H.~K\"uhn for encouraging me
to complete this work and for 
his continuous support.

\appendix
\newpage
\section{
Helicity Amplitudes in the
Weyl-van der Waerden Formalism\protect\vspace{1mm}}
\label{helinotation}
In this appendix we provide tree level (section~\ref{app_born})
and 1-loop (section~\ref{oneloopheli}) helicity amplitudes
for the subprocesses $eq\rightarrow eqg$ and $eg\rightarrow eq\bar{q}$
in the Weyl-van der Waerden spinor basis \cite{weyl,berendsgiele}.
Fairly compact analytical expressions for  the squared matrix elements
can be derived from these results.
Matrix elements for the three parton final state subprocesses 
in Eqs.~(\ref{qtoqgg}-\ref{gtoqqbarg}) are presented in
section~\ref{threeparton}.
In order to define the basic notation of the formalism, we first recapitulate
some of the basic notions of the  Weyl-van der Waerden helicity
formalism in section~\ref{a1}. We follow the conventions as introduced
in Ref.~\cite{berendsgiele}.

\subsection{Basic Notions\protect\vspace{1mm}}
\label{a1}
The first step in evaluating helicity amplitudes in the
Weyl-van der Waerden formalism is to replace spinors, 
$\gamma$ matrices, etc. in the ususal covariant 
Feynman amplitude representation to their
Weyl-van der Waerden two-spinor counterparts. 
We remind the reader that we take all partons and leptons massless.
In this section, we use upper case letters $P,Q,K$ for arbitrary
massless four momentum vectors and lower case letters 
$p,q,k$ for the spinors associated with them.\\[3mm]
{\em 1. helicity spinors}
\begin{eqnarray}
u_+(P)       &=& v_-(P)       \rightarrow  
\left( \begin{array}{l} p_B \\ 0  \end{array} \right)        \label{hs1}  \\
u_-(P)       &=& v_+(P)       \rightarrow                     
\left( \begin{array}{l} 0   \\ p^{\dot{B}}   \end{array} \right)       \\
\bar{u}_+(Q) &=& \bar{v}_-(Q) \rightarrow (0,-iq_{\dot{A}})\\
\bar{u}_-(Q) &=& \bar{v}_+(Q) \rightarrow (iq^{{A}},0)   \label{hs4}
\end{eqnarray}
where the subscripts $\pm$ on the spinors refer to the 
helicities of the fermions.\\[3mm]
{\em 2. $\gamma$-matrices, momenta and polarization vectors slashed}
\begin{eqnarray}
\gamma^\mu   &\rightarrow& i\sigma^{\mu\,\dot{A}B}
                          -i\sigma^{\mu}_{\dot{B}A}\label{repgam}       \\
\gamma^\mu\gamma_5   &\rightarrow& i\sigma^{\mu\,\dot{A}B}
                          +i\sigma^{\mu}_{\dot{B}A}\label{repgam5} \\
\not{\!\!K}  &\rightarrow & ik^{\dot{A}}k^B-ik_{\dot{B}}k_A
\hspace{2cm} \mbox{where} \hspace{2cm} K^2=0\\
\not{\!\epsilon_+^*}(K) &\rightarrow& 
\frac{\sqrt{2}}{\bra k g\ket}(ik_{\dot{B}}g_A-ik^{\dot{A}}g^B) 
\label{eps1def}\\
\not{\!\epsilon_-^*}(K) &\rightarrow&
\frac{\sqrt{2}}{\bra k g\ketc}(ig_{\dot{B}}k_A-ig^{\dot{A}}k^B) 
\label{eps2def}
\end{eqnarray}
where $\sigma^0$ is the unit matrix and $\sigma^i$ are the Pauli
matrices.  In Eqs.~(\ref{eps1def},\ref{eps2def}), $g$ denotes 
a gauge spinor related to any four-monentum which can be 
conveniently specified.  An optimal choice for the gauge spinor leads
to particularly compact expressions for the helicity amplitudes.
The polarization vectors in Eqs.~(\ref{eps1def},\ref{eps2def})
are normalize to $(-1)$. This normalization differs from 
the one used is \cite{berendsgiele}.
The dotted and undotted upper and lower case indices take the values
1 and 2. Tensor contractions are to be done only for upper-lower
pairs of dotted (undotted) indices, where a summation over 
repeated indices is understood. Terms that cannot be summed 
in this sense have to be dropped.
A contraction of 2 two-spinors associated with massless particles 
can be viewed as a complex-valued  scalar-product and will be referred
to as a  {\em spinor inner product}:
\begin{eqnarray}
\bra kp \ket      &:=& k_Ap^A              \label{sip1def}   \\
\bra kp \ketc &:=& k_{\dot{A}}p^{\dot{A}}  \label{sip2def}
\end{eqnarray}
Spinors with upper and lower indices are related by means of 
the two dimensional antisymmetric  ``metric tensor''
\begin{equation}
\varepsilon^{AB} = \varepsilon_{\dot{A}\dot{B}} = 
-\varepsilon_{AB} =-\varepsilon^{\dot{A}\dot{B}} =
-\varepsilon^{BA}=
\left(
\begin{array}{rr}
0  & 1 \\
-1 & 0 \\
\end{array}
\right) 
\end{equation}
via
\begin{eqnarray}
p^A\varepsilon_{AB}                    &=&  p_B      \\
p_{\dot{A}}\varepsilon^{\dot{A}\dot{B}} &=&  p^{\dot{B}} 
\end{eqnarray}
This leads to the relation
\begin{eqnarray}
\bra kp \ket      &:=& - \bra pk\ket      \\
\bra  kp \ketc &:=& - \bra pk\ketc 
\end{eqnarray}
One has to be very careful with the position and type of
indices. Thus one  has {\em e.g.}
$\varepsilon_{BA}p^A  = - p_B$.
Furthermore, 
if one has to evaluate spinor-inner-products involving momenta with
negative energy component (crossing), one can use:
\begin{eqnarray}
\bra -kp \ket      &=&  i \bra kp\ket   \label{n1}   \\
\bra -kp \ketc &=&  i \bra kp\ketc \\
\bra -k-p \ket     &=&  - \bra kp\ket      \\
\bra -k-p \ketc&=&  - \bra kp\ketc \label{n4}
\end{eqnarray}
When using Eqs.~(\ref{n1}-\ref{n4}) for a crossed process,
one also has to multiply the  amplitude  with a factor $(-i)^n$, 
where $n$ is the number of crossed fermions.

The following important identities
are useful in the evaluation of helicity amplitudes:
{\em scalar four product:}
\begin{eqnarray}
\bra pq\ket\bra pq\ketc &=& 2P.Q \label{scalar}
\label{sip_def}
\end{eqnarray}
{\em Schouten identities:}
\begin{eqnarray}
\varepsilon^{AB}\varepsilon^{CD} +
\varepsilon^{AC}\varepsilon^{DB} +
\varepsilon^{AD}\varepsilon^{BC} &=& 0 \\
\varepsilon^{\dot{A}\dot{B}}\varepsilon^{\dot{C}\dot{D}} +
\varepsilon^{\dot{A}\dot{C}}\varepsilon^{\dot{D}\dot{B}} +
\varepsilon^{\dot{A}\dot{D}}\varepsilon^{\dot{B}\dot{C}} &=& 0 
\end{eqnarray}
{\em Anti-commutator of $\gamma$ matrices}
\begin{eqnarray}
\sigma^\mu_{\dot{A}B}\sigma^{\nu \,\dot{C}B} +
\sigma^\nu_{\dot{A}B}\sigma^{\mu \,\dot{C}B} 
&=& 2 g^{\mu\nu}\delta_{\dot{A}}^{\dot{C}} \\
\sigma^{\mu\,\dot{B}A}\sigma^{\nu}_{\dot{B}C} +
\sigma^{\nu\,\dot{B}A}\sigma^{\mu}_{\dot{B}C} 
&=& 2 g^{\mu\nu}\delta_{C}^{A} 
\end{eqnarray}
{\em Fierz transformations}
\begin{eqnarray}
\sigma^{\mu\,\dot{A}B}\sigma_{\mu}^{\dot{C}D} 
&=& 2\varepsilon^{\dot{A}\dot{C}}\varepsilon^{BD}\\
\sigma^{\mu\,\dot{A}B}\sigma_{\mu\,\dot{C}D} 
&=& 2\delta^{\dot{A}}_{\dot{C}}\delta^{B}_{D}\\
\sigma^{\mu}_{\dot{A}B}\sigma_{\mu\,\dot{C}D} 
&=& 2\varepsilon_{\dot{A}\dot{C}}\varepsilon_{BD}
\end{eqnarray}
For the numerical evaluation of the helicity amplitudes one needs
a definite representation for the two-spinors and 
spinor-inner-products.
We use the following representation for the
complex valued spinor-inner-product $\bra kp \ket $:
\begin{equation}
\bra kp \ket = 
\sqrt{(K_0+K_3)(P_0-P_3)} e^{i\varphi(K)} \, 
-
\sqrt{(P_0+P_3)(K_0-K_3)} e^{i\varphi(P)}
\label{choice}
\end{equation}
where
\begin{equation}
\varphi(Q)=\left\{
\begin{array}{lll}
\mbox{arctan}\left(-\frac{Q_2}{Q_1}\right)       & , & Q_1 > 0 \\[3mm]
\mbox{arctan}\left(-\frac{Q_2}{Q_1}\right) + \pi & , & Q_1 < 0 \\
      -\frac{\pi}{2}      & , &  Q_1 =0\,\,\mbox{and}\,\,Q_2>0 \\[3mm]
\,\,\,\,\,    \frac{\pi}{2}      & , &  Q_1 =0\,\,\mbox{and}\,\,Q_2<0 \\[1mm]
\,\,\,\,\,\,    0                  & , &  Q_1 =0\,\,\mbox{and}\,\,Q_2=0 \\
\end{array}
\right.
\end{equation}
where $Q=(Q_0,Q_1,Q_2,Q_3)$ and $Q_0>0$.
%
%
\subsection{Tree Level Amplitudes for Two-Parton Final
 States\protect\vspace{1mm}}
\label{app_born}
%
%
In this part we provide a representation of the matrix elements
for the subprocesses $eq\rightarrow eqg$ and $eg \rightarrow eq\bar{q}$
which leads to the compact expressions for the corresponding
squared matrix elements as given in 
Eqs.~(\ref{m_qtoqg},\ref{m_gtoqqbar},\ref{m_qtoqgpv},\ref{m_gtoqqbarpv}).

Let us denote the helicity amplitude for the process
\begin{eqnarray}
e^-(l,\lambda_l)                  + q(p_0,\lambda_0)
&\rightarrow&
e^-({l^\prime},\lambda_{l^\prime}) + q(p_1,\lambda_1)
                                  + g(p_2,\lambda_2)
\end{eqnarray}
with 1-photon exchange by
\begin{equation}
e_q\,b^{vV}_{q\rightarrow qg}(\lambda_0;\lambda_1;\lambda_2;
                    \lambda_l;\lambda_{l^\prime}) 
\equiv 
e_q\,b^{vV}_{q\rightarrow qg}(p_0,\lambda_0;
                    p_1,\lambda_1;
                    p_2,\lambda_2;
                    l,\lambda_l;
                    l^{\prime},\lambda_{l^\prime}) 
\end{equation}
where each of the particles is labelled by its helicity $\lambda_i$ and
momentum $p_i$ indices.  
In contrast to the notation in section~\ref{a1}, we use here lower 
case letters for both the four momenta and the associated spinors in 
a spinor inner product.
We will also drop the spin label 
on the helicity labels. Thus $(\pm)$ means $(\pm 1/2)$ in the case
of lepton and quark helicities, and $(\pm 1)$  in the case
of gluon helicities, respectively.
The lower case $v$ (upper case $V$) in the superscript stands
for the vector current at the leptonic (hadronic) vertex.

It is sufficient to calculate just one helicity amplitude
explicitly. Based on the formalism outlined in section~\ref{a1},
one nonvanishing helicity combination can be written as
\begin{equation}
b^{vV}_{1,q} \equiv b^{vV}_{q\to qg}(+;+;+;-;-)=
-2\sqrt{2}\,\frac{e^2}{Q^2}\,g_s\, T^a_{ij}\,
\frac{\bra l^\prime p_0 \ket^2  \bra l^\prime l \ket
     }{ \bra p_1 p_2 \ket \bra p_2 p_0 \ket } \label{b1_def}
\end{equation}
where $T_a$ denotes a SU(3) color matrix and
$e$ and $g_s$ are the electromagnetic and strong QCD
coupling constants, respectively ($e^2/4\pi=\alpha$ and
$g_s^2/4\pi=\alpha_s$).
The seven other nonvanishing amplitudes 
are obtained by parity and charge conjugation relations
\begin{eqnarray}
b^{vV}_{q\rightarrow qg}
                   (-\lambda_0;
                    -\lambda_1;
                    -\lambda_2;
                    -\lambda_l;
                    -\lambda_{l^\prime}) 
&=&
(b^{vV}_{q\rightarrow qg}(
                    \lambda_0;
                    \lambda_1;
                    \lambda_2;
                    \lambda_l;
                    \lambda_{l^\prime}) )^\ast
\\
b^{vV}_{q\rightarrow qg}
                   (\lambda_0;
                    \lambda_1;
                    \lambda_2;
                    l,+;
                    l^{\prime},+) 
&=&
-
b^{vV}_{q\rightarrow qg}(
                    \lambda_0;
                    \lambda_1;
                    \lambda_2;
                    -l^\prime,-;
                    -l,-) 
\\
b^{vV}_{q\rightarrow qg}
                   (p_0,-;
                    p_1,-;
                    \lambda_2;
                    \lambda_l;
                    \lambda_{l^\prime}) 
&=&
-
b^{vV}_{q\rightarrow qg}(
                    -p_1,+;
                     -p_0,+;
                    \lambda_2;
                    \lambda_l;
                    \lambda_{l^\prime}) 
\end{eqnarray}
or more explicitly (see also table~\ref{helitab})
\begin{eqnarray}
b^{vV}_{2,q}&\equiv& b^{vV}_{q\to qg}(+,+,+,+,+)
=-b^{vV}_{1,q}(l\leftrightarrow -l^\prime)\\
b^{vV}_{3,q}&\equiv& b^{vV}_{q\to qg}(-,-,+,-,-)
=-b^{vV}_{1,q}(p_0\leftrightarrow -p_1)\\
b^{vV}_{4,q}&\equiv& b^{vV}_{q\to qg}(-,-,+,+,+)
=b^{vV}_{1,q}(l\leftrightarrow -l^\prime,
                \,p_0\leftrightarrow -p_1) \label{b4_def}
\end{eqnarray}
The remaining four amplitudes
$b^{vV}_{i+4,q}$ with $i=1,2,3,4$  with all helicities
reversed from CP invariance
\begin{equation}
b^{vV}_{i+4,q}(-\lambda_0;-\lambda_1;-\lambda_2;
              -\lambda_l;-\lambda_{l^\prime}) =
              (b^{vV}_{i,q}(\lambda_0;\lambda_1;\lambda_2;
               \lambda_l;\lambda_{l^\prime}) )^\ast
\end{equation} 
Using Eq.~(\ref{sip_def}), one can easily square each helicity amplitude
analytically, {\it e.g.}
\begin{equation}
|b^{vV}_{1,q}|^2  = 
4\cdot 2\,\frac{e^4}{Q^4}\,g_s^2\, |T^a_{ij}|^2\,\,
\frac{(l^\prime.p_0)^2(l.l^\prime)}{(p_1.p_2)(p_0.p_2)}
\end{equation}
Summing over the colors yield a color factor 
\begin{equation}
\sum_{ij}^{3}\sum_a^{8} |T^a_{ij}|^2 \,
= N \,C_F =4
\end{equation}
Summing over all eight nonvanishing helicity amplitudes yields
the following expression for the color averaged squared matrix element
\begin{equation}
e_q^2\,\sum_{i=1}^{8}\sum_{colors}\,|b^{vV}_{i,q}|^2 
= (4\pi\alpha)^2(4\pi\alpha_s)\frac{e_q^2}{Q^4}\,
|M^{(\mboxsc{pc})}_{q\to qg}|^2
\label{sumqtoqg}
\end{equation}
\bq
|M^{(\mboxsc{pc})}_{q\rightarrow qg}|^2=
\frac{128}{3}\,(l.l^\prime)\,
\frac{(l.p_0)^2+(l^\prime.p_0)^2+(l.p_1)^2+(l^\prime.p_1)^2
     }{(p_1.p_2)(p_0.p_2)}
\label{heli_mqtoqg}
\eq
The superscript pc ($\equiv$ parity conserving) reminds on the
vector current coupling at the leptonic and hadronic vertex.

The helicity amplitudes for the antiquark
and gluon initiated processes can be obtained from the quark
initiated amplitues through crossing (see table~\ref{helitab}
for explicit results)
\begin{eqnarray}
b^{vV}_{\bar{q}\rightarrow\bar{q}g}
(p_0,\lambda_0;
 p_1,\lambda_1;
 p_2;
 \lambda_l;
 \lambda_{l^\prime})
&=&
-b^{vV}_{q\rightarrow qg}
(-p_1,-\lambda_1;
 -p_0,-\lambda_0;
 \lambda_2;
 \lambda_l;
 \lambda_{l^\prime})\\
b^{vV}_{g\rightarrow q\bar{q}}
(p_0,\lambda_0;\lambda_1;p_2,\lambda_2;\lambda_l;
\lambda_{l^\prime})
&=& 
-
{b}^{vV}_{q\rightarrow qg}
(-p_2,-\lambda_2;\lambda_1;-p_0,-\lambda_0;\lambda_l;\lambda_{l^\prime})
\label{crossg}
\end{eqnarray}
Summing over all analytically squared helicity amplitudes
yields the parity conserving color averaged squared matrix element
for the $eg\rightarrow e q\bar{q}$ subprocess as given 
in Eq.~(\ref{m_gtoqqbar}).

In order to investigate also parity-violating
(pv) effects, which are introduced through the
additional $Z$ and $W$ exchange contributions,
one also needs the axial vector current
helicity amplitudes. This involves
the replacement of the vector current by the axial vector current
at either or both the leptonic and hadronic vertices.
This replacement if effected by
\begin{equation}
\gamma^{\mu}\rightarrow \gamma^\mu\gamma^5
\end{equation}
at the leptonic and/or hadronic current vertex.
In terms of the Weyl-van der Waerden representation this implies 
the replacement (see Eqs.~(\ref{repgam},\ref{repgam5}))
\begin{equation}
i\sigma^{\mu \dot{A}B}-i\sigma^{\mu}_{\dot{B}A}
\rightarrow
i\sigma^{\mu \dot{A}B}+i\sigma^{\mu}_{\dot{B}A}
\end{equation}
Thus the helicity amplitudes change sign every time a term
$\sigma^{\mu}$ contributes with lower spinor indices.
This will depend on the fermionic lepton and quark helicities.
For the leptonic vertex 
(the lower case superscript $v(a)$ refers to the vector (axial vector)
coupling at the letonic vertex)
one then obtains the relation
\begin{equation}
b^{aV}_{a\rightarrow n  \mboxsc{\,\,\, partons}}
=(-)^{1/2-\lambda_l}b^{vV}_{a\rightarrow  n \mboxsc{\,\,\,partons}};
\hspace{1cm}
b^{aA}_{a\rightarrow n \mboxsc{\,\,\, partons}}
=(-)^{1/2-\lambda_l}b^{vA}_{a\rightarrow n \mboxsc{\,\,\, partons}}
\label{helilep1a}
\end{equation}
The relation in Eq.~(\ref{helilep1a}) is true irrespective of the
associated parton process $a\rightarrow n$ partons.

An analogous relation holds true when one replaces
the hadronic vector current by the axial hadronic vector
current ($V\rightarrow A$ in the superscript)
in the two quark-one gluon processes
$q\rightarrow qg$ and $g\rightarrow q\bar{q}$:
\begin{equation}
b^{vA}_{q\rightarrow qg}=(-)^{1/2-\lambda_1}b^{vV}_{q\rightarrow qg}
;\hspace{1cm}
b^{vA}_{g\rightarrow q\bar{q}}
=(-)^{1/2-\lambda_1}b^{vV}_{g\rightarrow q\bar{q}}
\label{helihad1a} 
\end{equation}
The appropriate coupling factors of quarks ($v_f,a_f$)
and leptons ($v_e,a_e$) to the $Z$ and $W$ and
the appropriate propagator factor modifications 
($\chi_Z$ and $\chi_W$) are described in 
section~\ref{onejetz}.
When squaring the helicity amplitudes for
%
%
$
e_f b^{vV}\,+\,\chi_Z\, [v_ev_f b^{vV} + a_ev_f b^{aV}
                   +v_ea_f b^{vA} + a_ea_f b^{aA}]
$
%
%
one will have
$\gamma-\gamma,\gamma-Z$ and $Z-Z$ 
contributions\footnote{Note 
that $|b^{vV}|^2=|b^{vA}|^2=|b^{aV}|^2=|b^{aA}|^2$,
$(b^{vV}b^{aA\,\ast}+b^{vV}b^{aA\,\ast})
=(b^{vV}b^{aA\,\ast}+b^{vV}b^{aA\,\ast})$\, ,
$b^{vV}b^{vA}=0$ etc. (see Eqs.~(\ref{helilep1a},\ref{helihad1a})).}.
When one totals these contributions 
one obtains the $Q^2$-dependent coupling combinations
$A_f(Q^2)$ and $B_f(Q^2)$ as defined in
Eqs.~(\ref{afdef},\ref{bfdef}).
The coupling combination $A_f(Q^2)$ multiplies the parity-conserving (pc)
hadronic contribution whereas $B_f(Q^2)$ multiplies the parity violating
hadronic contributions:
\begin{equation}
\sum_{i=1}^8\sum_{colors}\left\{
A_f(Q^2)\,|b^{vV}_{i,q}|^2\otimes(q_f+\bar{q}_f)
+
B_f(Q^2)\,2{\cal{R}}(b^{vV}_{i,q}b^{aA\,\ast}_{i,q})\,
\otimes(q_f-\bar{q}_f)
\right\}
\end{equation}
where the symbol $\otimes$ stands for the folding in of the appropriate 
parton densities.
Eqs.~(\ref{helilep1a},\ref{helihad1a}) imply that:
\begin{eqnarray}
b_{i,q}^{aA}&=&-b_{i,q}^{vV}
\hspace{2cm}i=1,4,5,8 \nonumber \\
b_{i,q}^{aA}&=&\,\,\,\,b_{i,q}^{vV}
\hspace{2cm}i=2,3,6,7 \label{biaa}
\end{eqnarray}
In analogy to Eq.~(\ref{sumqtoqg}), we define 
$|M^{(\mboxsc{pv})}_{q\to qg}|^2$ via
\begin{equation}
 e_q^2\,\sum_{i=1}^{8}\sum_{colors}\,
2{\cal{R}}(b^{vV}_{i,q}b^{aA\,\ast}_{i,q})\,
\,=\, (4\pi\alpha)^2(4\pi\alpha_s)\frac{e_q^2}{Q^4}\,\,
|M^{(\mboxsc{pv})}_{q\to qg}|^2
\label{sum_mqtoqgpv}
\end{equation}
which yields the result (see also Eq.~\ref{m_qtoqgpv})
\bq
|M^{(\mboxsc{pv})}_{q\rightarrow qg}|^2=
\frac{128}{3}\,(l.l^\prime)\,
\frac{(l.p_0)^2-(l^\prime.p_0)^2-(l.p_1)^2+(l^\prime.p_1)^2
     }{(p_1.p_2)(p_0.p_2)}
\label{heli_mqtoqgpv}
\eq
where the minus signs in Eq.~(\ref{heli_mqtoqgpv}) originate
from the relations in Eqs.~(\ref{biaa}).

The squared parity violating matrix element for the
gluon intitialed process $eg\rightarrow e q\bar{q}$
is listed in Eq.~(\ref{m_gtoqqbarpv}).
The result can be obtained through crossing
from the previous results (see Eq.~(\ref{crossg})).
%
%
\subsection{One-loop  Amplitudes for Two-Parton Final 
States\protect\vspace{1mm}}
\label{oneloopheli}
In this appendix we provide analytical expressions
for the coefficients
$\alpha_{i},\beta_{i},\gamma_{i},
\bar{\alpha}_{i},\bar{\beta}_{i},\bar{\gamma}_{i}$ which are used in
the helicity dependent functions
${\cal{F}}_{q\rightarrow qg}^{(\mboxsc{pc})}$  and 
${\cal{F}}_{g\rightarrow q\bar{q}}^{(\mboxsc{pc})}$ 
in Eqs.~(\ref{f1qdef}-\ref{fcalgdef}).
The results can be obtained by crossing the 1-loop
results for the process $e^+e^-\rightarrow$ 3 partons
which are listed in  Appendix A of \cite{giele1}.
Some care must be taken in crossing these results 
since the argument of the dilog function 
$\mbox{Li}_2$, logarithms and spinor inner products
can be negative or greater than one.

The coefficients 
for the subprocess
$e(l)+q(p_0)\rightarrow e(l^\prime) + q(p_1) + g(p_2)$
are defined in terms of the scaled variables
\begin{equation}
y_s = \frac{s}{Q^2}<0\hspace{1cm}
y_t = \frac{t}{Q^2}>0\hspace{1cm}
y_u = \frac{u}{Q^2}<0
\label{ystu}
\end{equation}
with $s,t,u$ in Eq.~(\ref{stu_q}).
Results for $\alpha_{i},\beta_{i},\gamma_{i},
\bar{\alpha}_{i},\bar{\beta}_{i},\bar{\gamma}_{i}$ 
with $i=1,2,5,6$ are given in Eqs.~(\ref{alphadef}-\ref{gammadef})
below. The results for $i=3,4,7,8$ are obtained from these results
by the exchange $p_0\leftrightarrow -p_1$ 
(see Eqs.~(\ref{f3qdef},\ref{f4qdef}))
or in terms of the scaled variables by  
$y_t\leftrightarrow y_u$. 
\begin{eqnarray}
\alpha_{i} &=&- R(y_t,y_u) - \frac{y_u(4-3y_u)}{2(1-y_u)^2}\ln|y_u|
            - \frac{y_u}{2(1-y_u)}
            - \frac{3}{4}\ln|y_t| - \frac{3}{4}\ln|y_u|  
           \label{alphadef}\\[2mm]
\beta_{i} &=& - R(y_t,y_u) +  \frac{4-3y_u}{2(1-y_u)}\ln|y_u| + 1
           - \frac{3}{4}\ln|y_t| - \frac{3}{4}\ln|y_u| \\[2mm]
\gamma_{i} &=& \frac{y_u}{2(1-y_u)^2}\ln|y_u| + \frac{y_u}{2(1-y_u)} \\[2mm]
\bar{\alpha}_{i} &=& - R(y_s,y_t) - \frac{(1-y_s)^2}{y_t^2} R(y_s,y_u) -
             \frac{y_u}{y_t}\ln|y_s|\nonumber\\
         &&  - \frac{y_u(4-3y_u)}{2(1-y_u)^2} 
             + \frac{y_u^2}{y_t(1-y_u)}\ln|y_u|
             - \frac{y_u}{2(1-y_u)}
             - \frac{3}{2}\ln|y_s|\\[2mm]
\bar{\beta}_{i} &=& - R(y_s,y_t) + \left( \frac{y_s(1-y_s)}{y_t^2}
             + \frac{1}{y_t} \right)
              R(y_s,y_u)
              + \left( \frac{y_t}{(1-y_s)^2}
              + \frac{(1-y_u)}{y_t} \right)\ln|y_s|
                      \nonumber\\
                      &&   + \left( \frac{4-3y_u}{2(1-y_u)}
                      + \frac{y_u}{y_t}  \right)\ln|y_u|
             + \frac{y_t}{1-y_s}
             - \frac{3}{2}\ln|y_s|\\[2mm]
\bar{\gamma}_{i} &=& \frac{y_u}{y_t^2} R(y_s,y_u) - \left(\frac{y_t}{(1-y_s)^2} 
            - \frac{1}{y_t} \right)\ln|y_s|
            + \left( \frac{y_u}{2(1-y_u)^2}
            + \frac{y_u}{y_t(1-y_u)} \right)\ln|y_u|
                                         \nonumber\\
        && + \frac{y_u}{1-y_s} + \frac{y_u}{2(1-y_u)}
         \label{gammadef}
\end{eqnarray}
The function $R(x,y)$ is defined for 
arguments $x>0,y>0$ as
\begin{eqnarray}
R(x,y) &=& \ln(x) \ln(y) - 
      \ln(x) \ln(1-x) - 
      \ln(y) \ln(1-y) 
       -
      \mbox{Li}_2(x) - \mbox{Li}_2(y) + \frac{\pi^2}{6}
\end{eqnarray}
and for arguments  $x<0,y>0$ as 
\begin{eqnarray}
R(x,y) &=& - \mbox{Li}_2\left(\frac{1}{1-x}\right) - \frac{1}{2}\ln^2(1-x)
           + \mbox{Li}_2\left(\frac{1}{y}\right) + \frac{1}{2}\ln^2(y)
           + \ln(y)\ln\frac{|x|}{|1-y|}
\end{eqnarray}
The coefficients 
$\alpha_{i},\beta_{i},\gamma_{i},
\bar{\alpha}_{i},\bar{\beta}_{i},\bar{\gamma}_{i}$ 
with $i=1,2,5,6$
for the subprocess
$e(l)+g(p_0)\rightarrow e(l^\prime) + q(p_1) + \bar{q}(p_2)$
are  given by Eqs.~(\ref{alphadef}-\ref{gammadef}),
were the scaled variables are now defined via
\begin{equation}
y_s = \frac{\tilde{s}}{Q^2}>0\hspace{1cm}
y_t = \frac{\tilde{t}}{Q^2}<0\hspace{1cm}
y_u = \frac{\tilde{u}}{Q^2}<0
\end{equation}
with $\tilde{s},\tilde{t},\tilde{u}$
in Eq.~(\ref{tilde_stu}).
Results for $i=3,4,7,8$ are obtained from these by
$y_t\leftrightarrow y_u$.

\newpage
\section{The NLO One-Jet Cross Section without the
Crossing Function Formalism\protect\vspace{1mm}}
%
%
In this section we derive a formula for the hadronic
NLO 1-jet cross section without making use of the crossing 
function formalism.
The ``unfolding'' of the crossing functions 
demonstrates the advantage of this formalism in
the most simple case.

\subsection{The NLO Hadronic Cross Section\protect\vspace{1mm}}
Let us first rewrite Eq.~(\ref{onejet}) using more  DIS like standard variables
\cite{herai}. The $\O(\as^0)$ 1-jet cross section can be described
by two independent kinematical variables, {\it e.g.} $x$ and $y$
as defined in Eq.~(\ref{variables}),
whereas three more variables are needed to describe the
reactions in Eqs.~(\ref{qtoqg},\ref{gtoqqbar}).
Besides  the partonic center of mass energy squared 
\bq
\hat{s}=(p_0+l)^2=\eta s
\label{sp}
\eq
we choose the partonic scaling variables $x_p$ and $z$ 
\begin{eqnarray} 
x_p & = & \frac{Q^2}{2p_0.q} \hspace{1cm} (x<x_p \le 1)  \\
z   & = & \frac{p_0.p_1}{p_0.q} \hspace{1cm} (0\le z \le 1) 
\label{partonvar}
\end{eqnarray}
and $\phi$, the azimuthal angle between the parton plane 
$(\vec{p}_0,\vec{p}_1)$ and the lepton plane 
$(\vec{l},\vec{l^\prime})$ in the $\vec{p_0}+\vec{q}=0$ 
center of mass frame.
Using these variables, the  general phase space for a
$n$-parton final state in Eq.~({\ref{phasespace}) can be factorized as
\bq
d\mbox{PS}^{(l^\prime+n)}=
d\mbox{PS}^{(l^\prime)}d\mbox{PS}^{(n)}
\eq
where
\begin{equation}
d\mbox{PS}^{(l^\prime)}=
\frac{d^3l^\prime}{2E^\prime (2\pi)^3}
= 
\frac{y\hat{s}}{16 \pi^2}\, dy dx_p
\label{psl}
\end{equation}
\begin{equation}
d\mbox{PS}^{(n)}= (2\pi)^4 \delta^4(p_0+q-\sum_{i=1}^{n}p_i)
\prod_{i=1}^{n}\frac{d^3 p_i}{(2\pi)^3\,2E_i}
\label{psn}
\end{equation}
With $\eta=x/x_p$ (and taking into account the kinematical
constraints) the $\eta$ and phase space integrals
over an arbitrary function $F(\eta)$ can be written as
\bq
\int_0^1d\eta\,
\int
\,d{\mbox{PS}}^{(l^\prime+n)}\,\,\frac{1}{4p_0.l}\,\,F(\eta)
= 
\int dx\,\int dy\,\int_x^1\frac{d x_p}{x_p}\,
\int\,d{\mbox{PS}}^{(n)}\,\,\frac{y}{32 \pi^2}\,\,
F(\frac{x}{x_p})
\eq
Using in particular 
\begin{equation}
d\mbox{PS}^{(1)}= 
\frac{2\pi x_p}{Q^2}\,\delta(1-x_p)
\hspace{2cm}
%
%
d\mbox{PS}^{(2)}= 
\frac{1}{8\pi}\, dz \, d\phi
\label{ps1u2}
\end{equation}
the NLO 1-jet cross section
Eq.~(\ref{onejet}) can be rewritten as
\begin{eqnarray}
\frac{d\sigma_{\mboxsc{had}}[\mbox{1-jet}]}{dx dy} &=&
\frac{\pi\alpha^2}{4Q^4xs}\,\,
\bigg[\nonumber
\\
&&
[\sum_{i=q,\bar{q}}e_i^2 f_i(x,\mu_F)]\,
\,\,|M^{(\mboxsc{pc})}_{q\rightarrow q}|^2\,
\left(1+ \alpha_s(\mu_R)\,
{\cal{K}}_{q\rightarrow q}(\smin,Q^2)\right)\nonumber \\
&+&
\,[\sum_{i=q,\bar{q}}e_i^2\, 
\,C_i^{\overline{\mboxsc{MS}}}(x,\mu_F,\smin)]\,\alpha_s(\mu_R)
\,\,|M^{(\mboxsc{pc})}_{q\rightarrow q}|^2\,
\bigg]
\,\,
J_{1\leftarrow 1}(\{p_i\})
\label{onejetnew}\\
&&\hspace{-20mm}
+
\int_x^1 \frac{d x_p}{x_p}
\int
dz\,\frac{d\phi}{2\pi}\,\,\,
\frac{\alpha^2}{32Q^2xs}\,\,
(4\pi\alpha_s(\mu_R))\,\,\bigg[ \nonumber \\
&&
[\sum_{i=q,\bar{q}}e_i^2 f_i(\frac{x}{x_p},\mu_F)]\,\,
\,\,|M^{(\mboxsc{pc})}_{q\rightarrow qg}|^2
\nonumber\\
&+&
(\sum_{i=q}e_i^2 ) f_g(\frac{x}{x_p},\mu_F)\,\,
\,|M^{(\mboxsc{pc})}_{g\rightarrow qg}|^2
\bigg]\,\,
\prod_{i<j;0}^{2}\Theta(|s_{ij}| - \smin)\,\,
J_{1\leftarrow 2}(\{p_i\})
\nonumber
\end{eqnarray}
The matrix element squared 
$|M^{(\mboxsc{pc})}_{q\rightarrow qg}|^2$ and  
$|M^{(\mboxsc{pc})}_{g\rightarrow q\bar{q}}|^2$,
which are given in terms of compact expressions
in Eqs.~(\ref{m_qtoqg},\ref{m_gtoqqbar}), can also be 
expressed  in terms of $\hat{s},x_p,z$ and $\phi$ 
as shown in the following subsection.

As explained in Eq.~(\ref{sig_virt}), the complete
NLO virtual+soft+collinear contribution 
is given (after the factorization of the
initial state collinear divergencies into the  bare
parton densities) by the sum of
$d\sigma_{\mboxsc{had}}^{\mboxsc{NLO, final}}
+
d\sigma_{\mboxsc{had}}^{\mboxsc{NLO, crossing}}$,
{\it i.e.} by the sum of the 
${\cal K}_{q\rightarrow q}(\smin,Q^2)$ and the 
$C_q(x,\mu_F,\smin)$ contribution in Eq.~(\ref{onejetnew}).
To exhibit the full soft and collinear structure it is useful
to combine these terms again.
Replacing the $A$ and $B$ functions in
Eqs.~(\ref{crossf_uv},\ref{crossf_s}) by the r.h.s. of 
Eqs.~(\ref{aqq},\ref{agq},\ref{bqq},\ref{bgq}) 
yields
\begin{eqnarray}
\displaystyle
\frac{d\sigma_{\mboxsc{had}}[\mbox{1-jet}]}{dx dy} &=&
\frac{\pi\alpha^2}{4Q^4xs}\,\,
\bigg[       \nonumber \\
&&
[\sum_{i=q,\bar{q}}e_i^2 f_i(x,\mu_F)]\,
\,\,|M^{(\mboxsc{pc})}_{q\rightarrow q}|^2\,
\left(1+ \alpha_s(\mu_R)\,
\tilde{\cal{K}}_{q\rightarrow q}(\smin,Q^2)\right)\nonumber \\
&+&
\frac{\alpha_s(\mu_R)}{\pi}\,
\int_x^1\frac{d x_p}{x_p}
[\sum_{i=q,\bar{q}}e_i^2 f_i(\frac{x}{x_p},\mu_F)]\,\,
\,\,|M^{(\mboxsc{pc})}_{q\rightarrow q}|^2\,\,
\tilde{F}_{q\rightarrow q}(x_p) \nonumber \\
&+&
\frac{\alpha_s(\mu_R)}{\pi}\,
\int_x^1\frac{d x_p}{x_p}
(\sum_{i=q}e_i^2 ) f_g(\frac{x}{x_p},\mu_F)\,\,
\,\,|M^{(\mboxsc{pc})}_{q\rightarrow q}|^2\,\,
\tilde{F}_{g\rightarrow q}(x_p)
\bigg]\,\,
J_{1\leftarrow 1}(\{p_i\}) \label{nocrossing} \\
&&\hspace{-20mm}
+
\int_x^1 \frac{d x_p}{x_p}\int
dz\,\frac{d\phi}{2\pi}\,\,\,
\frac{\alpha^2}{32Q^2xs}\,\,
(4\pi\alpha_s(\mu_R))\,\,\bigg[  \nonumber \\
&&
[\sum_{i=q,\bar{q}}e_i^2 f_i(\frac{x}{x_p},\mu_F)]\,\,
\,\,|M^{(\mboxsc{pc})}_{q\rightarrow qg}|^2 \nonumber \\
&+&
(\sum_{i=q}e_i^2 ) f_g(\frac{x}{x_p},\mu_F)\,\,
|M^{(\mboxsc{pc})}_{g\rightarrow qg}|^2
\bigg]
\,\,
\prod_{i<j;\,0}^{2}\Theta(|s_{ij}| - \smin)\,\,
J_{1\leftarrow 2}(\{p_i\})
\nonumber
\end{eqnarray}
with
\begin{equation}
\tilde{{\cal{K}}}_{q\rightarrow q} = \frac{8}{9}
\left(\frac{N}{2\pi}\right)
\left[
-\ln^2\left( \frac{\smin}{Q^2}\right) 
- \frac{3}{2}\ln\left(\frac{\smin}{Q^2}\right)
-\frac{\pi^2 }{6}-\frac{9}{4}
+\frac{3}{4}\ln\left(\frac{\smin}{\mu_F^2}\right)
\, + \, {\cal{O}}(\smin)
 \right]
\end{equation}
\begin{equation}
\tilde{F}_{q\rightarrow q}(x_p) =
\frac{4}{3}
\left[
\frac{1}{2}(1+x_p^2)\left[
 \frac{1}{(1-x_p)_+}\ln\frac{\smin}{\mu_F^2}
+
\left(\frac{\ln(1-x_p)}{1-x_p}\right)_+
\right] + \frac{1}{2}(1-x_p)
\right]
\end{equation}
\begin{equation}
\tilde{F}_{g\rightarrow q}(x_p) =
\left[
\frac{1}{2}(x_p^2+(1-x_p)^2)\left(
\ln\left(\frac{\smin}{\mu_F^2}\right)
+ \ln(1-x_p) \right)
-\frac{1}{4}P^{(\epsilon)}_{g\to q}(x_p)
\right]
\label{ftildegq}
\end{equation}
The $(\,\,\,)_+$ prescriptions and 
$P^{(\epsilon)}_{g\to q})$ are defined in 
Eqs.~(\ref{plusdef},\ref{pgqepsi}), respectively.
Note that the $z$ integral over the 2 parton final state
in Eq.~(\ref{nocrossing})
matrix elements is restricted to regions where all partons are resolved by the
requirement that $|s_{ij}|>\smin$, whereas the first term in
Eq.~(\ref{nocrossing}) originates from the (analytical)
$z$-integration over the complementary region with 
$|s_{ij}|<\smin$.

The results in Eqs.~(\ref{nocrossing}-\ref{ftildegq})
can also be obtained within  the standard (but
much more tedious) approach by directly integrating the corresponding
$n$-dimensional matrix elements as given in \cite{herai}
over the phase space region $|s_{ij}|<\smin$,
adding the virtual contributions and factorizing the
remaining initial state collinear singularities
(using the $n$-dimensional Altarelli-Parisi splitting functions
in Eqs.~(\ref{pgqdef},\ref{pqqdef}) )
into the bare parton densities.
We have checked that such a calculation gives the same result
as above.

\subsection{LO Results for the Partonic Helicity Cross
 Sections\protect\vspace{1mm}}
\label{helicross}
In this section we provide analytical expressions for the
\oas\ squared matrix elements
$|M^{(\mboxsc{pc,pv})}_{q\rightarrow qg}|^2$ 
in Eqs.~(\ref{m_qtoqg},\ref{m_qtoqgpv})
and  
$|M^{(\mboxsc{pc,pv})}_{g\rightarrow q\bar{q}}|^2$
in Eqs.~(\ref{m_gtoqqbar},\ref{m_gtoqqbarpv})
in terms of the partonic DIS variables $x_p,z$ and $\phi$
(and the scaling variable $y$) as introduced in the previous section.
The characteristic
$y$ and $\phi$ dependent coefficients of the helicity cross sections
in section~\ref{twojetintro} factorize naturally with this set of variables.

Replacing the scalar products for the (massless) 2-parton final
by
\bq
\begin{array}{lll}
  2p_0.p_1 = \hat{s} y z \hspace{1cm} 
& 2p_0.p_2 = \hat{s} y (1-z) \hspace{1cm}
& 2p_0.q = \hat{s} y \\
  Q^2 = 2l.l^\prime=\hat{s}yx_p 
& 2p_1.p_2 = \hat{s} y (1-x_p) 
& 2p_1.q = \hat{s} y (1-x_p-z) \\
&
  2p_2.q = \hat{s}y(z-x_p)
& \\
\end{array}
\eq
yields 
\begin{eqnarray}
|M^{(\mboxsc{pc})}_{q\to qg}|^2 &=& 
    \,\,\,       [\mbox{Eq.~(\protect\ref{m_qtoqg})}] \nonumber \\
       &=&
            \frac{64Q^2}{y^2}\,\,\bigg\{
            \,\,\, (1+(1-y)^2)\,\,d\hat\sigma^{F_2}_{q\to qg}
            \,\,\,-y^2\,\,d\hat\sigma^{F_L}_{q\to qg}\\
       &&   \left.\hspace{5mm}
            \,\,\,-\sqrt{1-y}\,(2-y)\,\cos\phi\,\,
            d\hat\sigma^{F_4}_{q\to qg}              
           \,\,\,+\,\, 2(1-y)\,\cos 2\phi \,\,d\hat\sigma^{F_6}_{q\to qg}
            \right. \nonumber \\
      && \hspace{5mm}
         \,\,\,+\,\,y\sqrt{1-y}\sin\phi\,\,d\hat\sigma^{F_7}_{q\to qg} 
        \bigg\}\nonumber  \\[3mm]
|M^{(\mboxsc{pv})}_{q\to qg}|^2 &=& 
    \,\,\,       [\mbox{Eq.~(\protect\ref{m_qtoqgpv})}] \nonumber \\
       &=&
            \frac{64Q^2}{y^2}\,\,\bigg\{
            \,\,\, y(2-y)\,\,d\hat\sigma^{F_3}_{q\to qg}
            \,\,\,-4y\sqrt{1-y}\,\cos\phi \,\,d\hat\sigma^{F_5}_{q\to qg}\\
       &&   \hspace{5mm}
            \,\,\,+\,\, (2-y)\sqrt{1-y}\sin\phi\,\,d\hat\sigma^{F_8}_{q\to qg}
            \,\,\,+\,\, (1-y)\sin2\phi \,\,\hat\sigma^{F_9}_{q\to qg}  
        \bigg\}\nonumber
\end{eqnarray}
and analogously for the
gluon intitiated squared matrix elements in
Eqs.~(\ref{m_gtoqqbar},\ref{m_gtoqqbarpv}).

The LO partonic  helicity cross sections for the
quark initiated subprocess read:
\begin{eqnarray}
d\hat\sigma_{q\to qg}^{F_2}& =& 
\frac{1+x_p^2z^2}{(1-x_p)(1-z)}+(1-x_p)(1-z)+4x_p z \nonumber\\[2mm]
d\hat\sigma_{q\to qg}^{F_L}  & =& 4x_p z \nonumber\\[2mm]
d\hat\sigma_{q\to qg}^{F_3}  & =& d\hat\sigma_{q\to qg}^{U+L} 
                            -2(1-x_p)(1-z)-4x_pz  \nonumber       \\[2mm]
d\hat\sigma_{q\to qg}^{F_4}  & =& 
\frac{-4\sqrt{x_p z}\,((1-x_p)(1-z)+x_pz)}{\sqrt{(1-x_p)(1-z)}}
\nonumber\\[2mm]
d\hat\sigma_{q\to qg}^{F_5}  & =& \frac{\sqrt{x_pz}(1-z-x_p)}{
                            \sqrt{(1-x_p)(1-z)}}\\[2mm]
d\hat\sigma_{q\to qg}^{F_6}  & =& 2x_p z \nonumber\\[2mm]
d\hat\sigma_{q\to qg}^{F_7}  & =& 0\nonumber \\[2mm]
d\hat\sigma_{q\to qg}^{F_8}  & =& 0 \nonumber\\[2mm]
d\hat\sigma_{q\to qg}^{F_9}  & =& 0 \nonumber
\end{eqnarray}
and similar for the gluon initiated subprocess
\begin{eqnarray}
\hat\sigma_{g\to q\bar{q}}^{F_2}& =& 
\left(\frac{1}{z}+\frac{1}{1-z}-2\right)
(1-2x_p(1-x_p)) + 8x_p(1-x_p)\nonumber\\[2mm]
d\hat\sigma_{g\to q\bar{q}}^{F_L}  & =& 8x_p (1-x_p)\nonumber \\[2mm]
d\hat\sigma_{g\to q\bar{q}}^{F_3}  & =& (1-2x_p(1-x_p))
    \frac{(z-(1-z))}{z(1-z)}\nonumber\\[2mm]
d\hat\sigma_{g\to q\bar{q}}^{F_4}  & =& 
-4(2x_p-1)(2z-1)\sqrt\frac{x_p(1-x_p)}{z(1-z)} \nonumber\\[2mm]
d\hat\sigma_{g\to q\bar{q}}^{F_5}  & =& 
(1-2x_p)\sqrt{\frac{x_p(1-x_p)}{z(1-z)}}\\[2mm]
d\hat\sigma_{g\to q\bar{q}}^{F_6}  & =& 4x_p (1-x_p) \nonumber\\[2mm]
d\hat\sigma_{g\to q\bar{q}}^{F_7}  & =& 0 \nonumber\\
d\hat\sigma_{g\to q\bar{q}}^{8}  & =& 0 \nonumber\\[2mm]
d\hat\sigma_{g\to q\bar{q}}^{9}  & =& 0 \nonumber
\label{sigheli}
\end{eqnarray}
Note that $d\hat\sigma^{F_{7,8,9}}$ 
vanish in LO. Contributions to these helicity cross sections  first come
in at \oasz\ through the imaginary parts of the 
1-loop contributions \cite{hagiwara}.
\newpage
\section{Documentation of MEPJET 2.1\protect\vspace{1mm}}
\label{sec_docu}
\begin{center}
\begin{large}
\begin{bf}
\docuname \hspace{2mm}\version\hspace{2mm}: \\
A next-to-leading order event generator for\\ $ep\rightarrow n$ jets.
\\
\end{bf}
\end{large}
\end{center}
{\bf
  \center{Erwin Mirkes$^a$, Dieter Zeppenfeld$^b$
    and Stefan Willfahrt$^a$}\\[1cm]
}
{\it 
$^a$Inst. f\"ur Theor. Teilchenphysik, Univ. Karls\-ruhe,
D-76128 Karlsruhe, Germany\\
\vspace*{3mm}
$^b$Physics Department, University of Wisconsin, Madison WI 53706, USA\\[4mm]
}
Please send comments or suggestions to \\
mirkes@ttpux5.physik.uni-karlsruhe.de \,
or \\
dieter@pheno.physics.wisc.edu\\[4mm]
\noindent
{\large Brief description of the program}\\[2mm]
The NLO Monte Carlo program 
\docuname , \version\,\, allows to study jet cross sections
for arbitrary jet algorithms and arbitrary event shapes
in deep inelastic $e^-p$ and $e^+p$ scattering. 
The program is written in fortran. The required CERN
libraries can be found in a file ``makefile'' which is
provided together with the program.
A list of input parameters is written on a file
mepjet.dat, which will be specified below in detail.
\docuname\ returns the cross section for the LO or NLO cross section
plus a list of standard histograms.\\[2mm]
\docuname\ 2.1 allows for the calculation of 
\begin{itemize}
\item[i)]  NLO 1-jet and 2-jet cross sections for complete
     $\gamma^\ast$ and/or $Z$ and $W^\pm$  exchange
\item[ii)]
 LO 1,2,3,4 jet cross sections for $\gamma^\ast$ and/or $Z$ exchange
\item[iii)] LO 1,2,3 jet cross section for $W^\pm$ exchange
\item[iv)]  charm and bottom mass effects in LO 1,2,3 jet cross sections
            with $\gamma^\ast$ exchange
\item[v)]  NLO 1-jet cross sections with  $\gamma^\ast$ exchange 
     in polarized $ep$ scattering
\item[vi)] LO 1,2,3 jet cross sections with $\gamma^\ast$ exchange
     in polarized $ep$ scattering 
\end{itemize}


Note that a set of ``crossing functions'' (corresponding to the
chosen set of parton distribution function) is needed for the NLO runs.
A selection of  ``crossing functions'' sorted according to ``mode\_pdf''
are given in the subdirectory ``pdfcross'' together with
a program [make\_str\_pdf1.f] to generate your own set of crossing functions.
The  new (standard) $x-Q$  grid for these crossing functions
differs from the grid used for   versions prior to  \docuname ~2.0.
If you still want to use the old $x-Q$ grid
some trivial modifications have to be done
in the routine STR\_PDF1\_NF5. We recommend to use the
new grid since the size of the resulting data file and the
time to generate them is markedly reduced.
All NLO results are given in the $\overline{MS}$ renormalization 
and factorization scheme.
The number of flavors is fixed to $n_f=5$, {\it i.e.} gluons are 
allowed to split into five flavors of quarks\footnote{It would
be easy to modify the program to a variable number of flavors.
However, the applicability of fixed order perturbation theory
requires   hard scales
for the partonic processes  ($> 5$ GeV)
and thus,  $n_f$ should be fixed to 5 for  all practical applications.
}.

Kinematical information on the final state lepton, 
initial and final state partons and jets
in the lab frame, Breit frame and
hadronic center of mass frame (HCM)
are available at the end of the routine accept.f. 
Various acceptance cuts on the final state lepton and jets
are analyzed in this routine,
which might be modified by the user.
In addition, two parton momenta may be recombined to a jet
momentum in a NLO calculation according to a given
jet algorithm and recombination scheme.
The resulting NLO jet momenta are stored in an
three dimensional array  as described below.
After an event passes the required acceptance criteria,
the routine accept.f provides the following
kinematical information on leptons ($x=v$), initial and final
state partons ($x=p$) and jets ($x=p\_j$), which 
are encoded in three dimensional real*8 arrays
$$x(\mu=0:7,\mbox{ipart},\mbox{iframe}).$$
The third component iframe specifies the 
\begin{itemize}
\item lab frame (iframe=1)
\item Breit frame (iframe=2)
\item  HCM frame (iframe=3)
\end{itemize}
The  proton direction defines the $+z$-axis in each frame. 

For each particle/jet, $\mu=0:7$ encodes the following 
kinematical information in a given frame iframe:\\
\begin{tabular}{ll}
$\mu=0$     & energy\\
$\mu$=1,2,3 & components of three momentum $p_x$,$p_y$,$p_z$\\
$\mu$=4     & mass \\
$\mu$=5     & if \mbox{iframe=1}:\\
            & \hspace*{2mm} transverse momentum $p_T$ wrt to the
                            lepton-proton beam axis  \\
            & \mbox{iframe=2}:\\
            & \hspace*{2mm} $k_T= \sqrt{2\,E_j^2(1-\cos\theta_{jP})}$,
                            where the subscripts $j$ and $P$ denote the\\
            & \hspace*{2mm} parton and proton (all quantities defined in the
                            Breit frame)\\
            & \mbox{iframe=3}:\\
            & \hspace*{2mm} transverse momentum $p_T$ with respect to the
                            boson-proton  axis \\
            & \hspace*{2mm} as defined in the HCM frame 
                            (or in the Breit frame)\\  
$\mu$=6     & pseudo-rapidity 
              $\eta= -\ln(\tan(\theta/2))$, (also for mass $> 0$!)\\
$\mu$=7     & azimuthal angle $\phi$ \\
\end{tabular}\\[2mm]
%
\newpage
The different momentum arrays are\\[2mm]
\underline{\bf i)
parton array $p(\mu=0:7,\mbox{ipart},\mbox{iframe}):$}\\[2mm]
The second component $\mbox{ipart}$  is a particle flag 
which  specifies 
\begin{itemize}
\item the initial state parton ($\mbox{ipart}=1$)
\item the $\mbox{npart}$ final state partons 
      ($\mbox{ipart}=2,\ldots \mbox{npart+1}$)
\item  the proton remnant ($\mbox{ipart}=\mbox{np}+1$;\,
       $\mbox{np=5}$ is a constant)
\end{itemize}
{\underline{\bf{ii)}
lepton array $v(\mu=0:7,\mbox{ipart(=1,2)},\mbox{iframe}):$}\\[2mm]
Here, $\mbox{ipart}$ specifies
\begin{itemize}
\item \mbox{ipart=1}: incoming lepton
\item \mbox{ipart=2}: outgoing lepton
\end{itemize}
{\underline{\bf{iii)}
jet array $p\_j(\mu=0:7,\mbox{ipart},\mbox{iframe}):$}\\[2mm]
Here, $\mbox{ipart}$ specifies
\begin{itemize}
\item the initial state parton ($\mbox{ipart}=1$) (same as in $x=p$)
\item the $\mbox{njet}\leq \mbox{npart}$ observable jets
      ($\mbox{ipart}=2,\ldots \mbox{njet+1}$):\\
      - here the jet momenta can originate from recombined
        parton momenta  according to different jet schemes;
        at most one such recombination is allowed at NLO\\
      - partons (not jets!) outside the detector range which is defined by
        ypartmax\_def (see below) are eliminated and added to the
        proton remnant;\\
      - a pointer $\mbox{idef\_jet}(1:\mbox{njet})$ is defined
        to point to all njet final state jet vectors,
        which are ordered according to their transverse momenta in the 
        lab frame, {\it i.e.} \\
        $p\_j(6,\mbox{idef\_jet}(i+1),1) \,\,>\,\,
        p\_j(6,\mbox{idef\_jet}(i),1)$.
\item the proton remnant ($\mbox{ipart}=\mbox{np}+1$):\\
      - the remnant jet as in $p(\mu,\mbox{np+1},\mbox{iframe})$ 
        plus partons outside ypartmax\_def
\end{itemize}
{\underline{\bf{iv)} boson array  and proton array} 
\begin{itemize}
\item $q(\mu=0:3,\mbox{iframe})$:
             virtual boson  
\item $p_{\mboxsc{proton}}(\mu=0:3,\mbox{iframe})$ 
             proton momentum 
\end{itemize}
Note that $\mu$ runs only over $0(=E),1(=p_x),2(=p_y),3(=p_z)$.\\
The parameter \mbox{iframe} is defined as before.

\newpage
\noindent
{\Large \bf Input parameters} \\[2mm]
The input parameters for \docuname\ are written on a file
mepjet.dat.
Parameters to choose are: \\[2mm]
{\bf GLOBAL DEFINIIONS}
%
%
\begin{list}
{\arabic{bean})}{\usecounter{bean}\setlength{\rightmargin}{\leftmargin}}
\item    {\bf el }: real*8,\,\, [GeV] \\ lepton beam energy 
\item    {\bf ep }: real*8,\,\, [GeV] \\ hadron beam energy [GeV]
\item    {\bf ilepton }: integer $\in$ \{1,2\}\\
            initial lepton: $1=e^-,\,2=e^+$)
\item    {\bf iboson}: integer $\in$ \{1,2,3,4\}\\
            exchanged boson: $1=\gamma^*,\, 2=Z, 3=\gamma^*-Z, 4=W^\pm$.\\
            For polarized $ep$ scattering, iboson=1 only.
\item     {\bf jscheme}: integer $\in$ \{0,1,2,3,11,12,13,14,22,23\} \\
          jet definition scheme: \\
          0=total cross section (independent of the jet-algorithm)\\
          npartmin and njetmin must be 1, njetmax 2\\
          depending on iord=0,1, the total cross section
          is calculated in $O(\alpha_s^0)$ (the parton model result)
          or in $O(\alpha_s^1)$.
          The parameters [\ref{irjj}-\ref{ifw}] are not effective besides of
          [\ref{iypartmax}=ypartmax\_def], which defines the rapidity range
          for the hadrons. The parameter [\ref{iptmindef}=ptmin\_def]
          must be small, i.e. of the order of $10^{-2}$-1
          (see also section~\ref{numonejet} for a careful
          discussion about the relation of the total
          and 1-jet inclusive cross section).\\[2mm]
          1=cone algorithm defined in the lab frame. 
           (see also items 
           \ref{irjj},\ref{irjl}).\\[2mm]
          2=cone algorithm defined in the Breit frame
           (see also items  \ref{irjj},\ref{irjl}) .\\[2mm]
          3=cone algorithm defined in the HCM frame
           (see also items  \ref{irjj},\ref{irjl}).\\[2mm]
          11=W-scheme, i.e. the JADE algorithm with 
            $W^2$ as the jet resolution mass 
            (see also items \ref{iycutw},\ref{im2min} ).
            The resolution mass squared
            $M^2_{ij}=(p_i+p_j)^2$ is calculated for each pair
            of final state particles (including the remnant).
            If the pair with the smallest invariant mass squared is below 
            $y_{cut}W^2$, the pair is clustered according 
            to a recombination prescription defined in the next item.
            This process 
            is repeated until all 
            invariant masses are above $y_{cut} W^2$.\\[2mm]
          12=JADE algorithm, same as 11, but with 
            $M^2_{ij}= 2E_iE_j(1-\cos\theta_{ij})$,
            i.e. all explicit mass terms are neglected 
            in the resolution criterion.
            All quantities are defined in the lab frame.\\[2mm]
          13= same as 12, but with
            $M^2_{ij}= 2E_iE_j(1-\cos\theta_{ij})$
            evaluated in the Breit frame.\\[2mm]
          14= same as 12, but with
            $M^2_{ij}= 2E_iE_j(1-\cos\theta_{ij})$
            evaluated in the HCM frame.\\[2mm]
          21 
              No longer supported in Version 2.0 and up. \\[2mm]
          22=$k_T$ algorithm as defined in PLB285 (1992) 291;
             (see also items 
             \ref{iycutkt},\ref{iet2},\ref{im2minkt}).  \\
             The KTCLUS-routine written by M. Seymour is used
             for the clustering. Only the momenta returned by
             KTRECO with JET(I).NE.0 are defined as jets.
             Only for {\bf jrec}=1.\\[2mm]
          23=$k_T$ algorithm as defined  in the
            KTCLUS-routine.
            The KTCLUS-routine is used  for the clustering.
            The NJET momenta returned by the routine KTRECO 
            are defined as jet. Only for {\bf jrec}=1.

\item    {\bf jrec}: integer $\in$ \{1,2,3\}\\
          recombination scheme:\\
          1=$E$\\
          2=$E_0$\\
          3=$P$
\item    {\bf njetmin}:  integer $\in$ \{1,2,3\}\\
          minimum number of defined jets.
\item    {\bf njetmax}:  integer $\in$ \{\mbox{njetmin},\mbox{njetmin}+1\}\\
          maximum number of defined jets. For njetmax=njetmin+1, the 
          cross section for ``njetmin-inclusive'' events is calculated.
\item    {\bf npartmin}: integer $\in$ \{1,2,3\}\\
          minimal number of final state partons that 
          can be recombined to jets.          \\
          (normally equal to njetmin).          
\item    {\bf iord}:  integer $\in$ \{0,1\}\\
          the njetmin cross section is calculated in leading  (0) or 
          in next-to-leading (1) order.\\
          iord=1 allowed for njetmin=1,2.
\item    {\bf n2max1}: integer\\
         Log2 of number of points for VEGAS for the 
          LO calculation or for  the calculation of the 
          soft and collinear part.
          A reasonable choice is\\ n2max1 = 19 (if iord=0)
          or  n2max1 = 22-24 (if iord=1).
\item    {\bf iterations1}: integer\\
          number of iterations for VEGAS used for the LO calculation
          (if iord=0) or for  the calculation of the soft and collinear part 
          (if iord=1). A reasonable choice is iterations1= 4 or 5,
          if no precalculated grid is used.
\item    {\bf n2max2}: integer\\
          Only effective for iord=1.
          Log2 of number of points for VEGAS 
          for the calculation of the hard part of the NLO contribution.
          A reasonable choice is n2max1 = 25-27.
\item    {\bf iterations2}: integer\\
          Only effective for iord=1.
          Number of iterations for VEGAS used
          for the calculation of the hard part of the NLO contribution.
          A reasonable choice is iterations2= 4 or 5,
          if no precalculated grid is used.
\item    {\bf grid2}: character\\
           name of input grid for npart=2. 
\item    {\bf grid3}: character\\
            name of input grid for npart=3
\item    {\bf ihist}: integer $\in$ \{0,1,2 \}\\
            histogram switch:\\
            0: no histograms\\
            1: only *.hbook file\\
            2: all: *.hbook, *.top and in standard output
\item    {\bf epnj}:  nametrunc: file name for output files 
          (*.hbook and *.top)
\item    {\bf smin}: real*8, [GeV$^2$]\\
          cutoff to define the theoretical $s_{min}$ cone which is 
          introduced to separate the soft and collinear part from the
          hard part of the NLO cross section.
          smin must be small enough that the approximations in the soft
          and collinear part are valid. 
          A reasonable (save) choice is smin=0.1 GeV$^2$.
          Depending on the kinematics, it may be necessary 
          to choose a  smaller value for smin
          (see Figs.~\ref{fig_smin_one_jet} and
           \ref{fig_smin_two_jet} and the related discussions for
          a suitable choice of smin 1-jet and 2-jet production).
\item    {\bf ias}: integer $\in$ \{1,2\}\\
          Choose 1-loop (1) or 2-loop (2) formula for the 
          strong coupling constant $\alpha_s$ in LO. 
          For iord=1 (NLO), 2-loop $\alpha_s$ is always  used.
          If NLO parton distributions are used,
          ias should be set equal to two for the LO calculation (iord=0).
          The value of $\alpha_s$ is matched at the thresholds $q=m_q$
          and the number of flavours $n_f$ in $\alpha_s$ is
          currently fixed to $n_f=5$.
\item    {\bf ilam, dlam4}: ilam: integer $\in$ \{0,1\}, dlam4: real*8, [GeV]\\
          ilam=0: $\Lms{4}$ in $\alpha_s$ 
          is chosen according to its value in the parton
          densities\\
          ilam=1: choose $\Lms{4}$ by hand: $\Lms{4}=$dlam4
          (dlam4 is only effective for ilam=1!)
\item    {\bf iaem}: integer $\in$ \{1,2\}\\
          electromagnetic coupling constant $\alpha$:\\
          1: $\alpha=$ fixed=1/137\\
          2: $Q^2$ dependent  $\alpha$.
\item    {\bf mode-pdf}: integer $\in$ \{2\}\\
          choose parton distribution functions:\\
          0: x*(1-x) test function\\
          2: MRS D${}_-^\prime$ (Ntype=1, Ngroup=3, Nset=31)
             \cite{pdflib}\\
          3: MRSA (Ntype=1, Ngroup=3, Nset=38)\\
          4: MRSH (Ntype=1, Ngroup=3, Nset=35)\\
          5: MRSR1 (Ntype=1, Ngroup=3, Nset=53)\\
          6: MRSR2 (Ntype=1, Ngroup=3, Nset=54)\\
          11: GRV 94 (HO,NLO) MSbar (Ntype=1, Ngroup=5, Nset=6)\\
          12: GRV 94 (LO) (Ntype=1, Ngroup=5, Nset=5)\\
          201-204: pol. PDF's by Glueck, Reya, Stratmann,
                   {\it Phys. Rev.} {\bf D51}(1995) 3220.\\
          201: NLO set 1 ('standard' scenario)\\
          202: NLO set 2 ('valence' scenario)\\
          203: LO  set 3 ('standard' scenario)\\
          204: LO  set 4 ('valence' scenario)\\
          -200: PDF=constant in polmat.f\\
          210-215: pol. PDF's by Gehrmann, Stirling, \cite{gs}.
          210: LO  set gluon A\\
          211: LO  set gluon B\\
          212: LO  set gluon C\\
          213: NLO  set gluon A\\
          214: NLO  set gluon B\\
          215: NLO  set gluon C\\
          Further parametrizations can be easily  added.\\
          Note, that 
          one needs a set of ``crossing'' functions 
          for the NLO cross sections.
\item    {\bf iproc}:  integer $\in$ \{100,101,102,201,202,200\}\\
          101: only quark-initiated subprocesses\\
          102: only gluon-initiated subprocesses\\
          100: complete, {\it i.e.} sum of 101 and 102.\\
          200,201,202: analog for polarized $ep$ scattering.
\item    {\bf iuds}: integer  $\in$ \{0,1\}\\
           idus $\neq 1$ only allowed for LO 1,2,3 jet production
           with $\gamma^\ast$ exchange at the moment\\
           flavour switch for up, down and strange quarks:\\
           iuds=0: no u,d,s in the initial \underline{and} final state;\\
           iuds=1: allow for massless u,d,s in the initial \underline{and}
            final state.
\item    {\bf icharm}: integer  $\in$ \{0,1,2\}\\
           icharm $\neq 1$ only allowed for LO 1,2,3 jet production
           with $\gamma^\ast$ exchange at the moment\\
           icharm=0: no charm in the initial \underline{and} final state;\\
           icharm=1: allow for massless charm in the initial \underline{and}
           final state.\\
           icharm=2: allow for masssive charm-pair production.
\item    {\bf ibottom}: integer  $\in$ \{0,1,2\}\\
           ibottom $\neq 1$ only allowed for LO 1,2,3 jet production
           with $\gamma^\ast$ exchange at the moment\\
           flavour switch for bottom quarks:\\
           ibottom=0: no bottom in the initial \underline{and} final state;\\
           ibottom=1: allow for massless bottom in the initial \underline{and}
           final state\\
           ibottom=2: allow for masssive bottom-pair production.
\item    {\bf ar,br,cr}: real*8, [cr=GeV$^2$]\\
          choose renormalization scale $\mu_R$ according to\\
          $\mu_R^2$ = ar $\left(\sum_j p_T^B(j)\right)^2$ + br $Q^2$ 
          + cr $\left(\sum_j k_T^B(j)\right)^2$ \\
          $p_T^{B}(j)$ denotes the transverse momentum of the jets
          in the Breit (or hadronic center of mass) system;\\
          $(k_T^{B}(j))^2$ is defined by 
          $2\,E_j^2(1-\cos\theta_{jP})$, where all quantities are defined
          in the Breit frame.
\item    {\bf af,bf,cf}: real*8, [cf=GeV$^2$]\\
          choose factorization  scale $\mu_F$ according to\\
          $\mu_F^2$ = af $\left(\sum_j p_T^B(j)\right)^2$ 
          + bf $Q^2$ 
          + cf $\left(\sum_j k_T^B(j)\right)^2$ 
          \label{list1}
\end{list}
{\bf LAB FRAME CUTS ON THE FINAL STATE LEPTON}
\begin{list}
{\arabic{bean})}{\usecounter{bean}
\setcounter{bean}{29}
\setlength{\rightmargin}{\leftmargin}}
\item    {\bf elmin}: real*8 [GeV]\\
          minimum energy of  final lepton
\item    {\bf ptlmin}: real*8 [GeV]\\
          minimum transverse momentum  $p_T$  of final lepton
\item    {\bf ptlmax}: real*8 [GeV]\\
          maximum $p_T$  of final lepton
\item    {\bf ylmin}: real*8\\
           minimum pseudo-rapidity 
           $\eta=-\ln\tan(\theta_l/2)$  of final lepton.\\
           For ylmin~$< -100$, the standard H1 cuts (depending
           on $Q^2$) are imposed, i.e.\\
           $ -2.794 < \eta(l^\prime)$\, for $Q^2 < 100$ GeV$^2$
           (corresponding to $ \theta(l^\prime) < 173.^o$);\\
           $ -1.317 < \eta(l^\prime)$ \,for $Q^2 > 100$ GeV$^2$
           (corresponding to $ \theta(l^\prime) < 150.^o$);
\item    {\bf ylmax}: real*8\\
           maximum pseudo-rapidity  of final lepton.\\
           For ylmax~$> 100$, the standard cuts H1 (depending
           on $Q^2$) are imposed, i.e.\\
           $ -1.735 > \eta(l^\prime)$ \,for $Q^2 < 100$ GeV$^2$
           (corresponding to $ \theta(l^\prime) > 160.^o$);\\
           $ \,\,\,\,2.436  > \eta(l^\prime)$ \,for $Q^2 > 100$ GeV$^2$
           (corresponding to $ \theta(l^\prime) > 10.^o$);
\end{list}
{\bf INVARIANT GLOBAL CUTS BEFORE CLUSTERING}
\begin{list}
{\arabic{bean})}
{\usecounter{bean}
\setcounter{bean}{34}
\setlength{\rightmargin}{\leftmargin}}
\item    {\bf q2min}: real*8 [GeV$^2$] \\
           minimum $Q^2$ of the event
\item    {\bf q2max}: real*8 [GeV$^2$]\\
           maximum $Q^2$ of the event
\item    {\bf w2min}: real*8 [GeV$^2$]\\
           minimum (true) $W^2$ of the event
\item    {\bf w2max}: real*8 [GeV$^2$]\\
           maximum (true) $W^2$ of the event
\item    {\bf xbjmin}: real*8, $\in [0,1]$\\
           minimum value of x-Bjorken
\item    {\bf xbjmax}: real*8  $\in [0,1]$\\
           maximum value of x-Bjorken
\item    {\bf ybjmin}: real*8  $\in [0,1]$\\
           minimum value of y scaling variable
\item    {\bf ybjmax}: real*8  $\in [0,1]$\\
           maximum value of y scaling variable
\end{list}
{\bf CUTS IN THE CONE-SCHEME; \\ 
ONLY EFFECTIVE IF JSCHEME=1, 2, 3.}}
\begin{list}
{\arabic{bean})}
{\usecounter{bean}
\setcounter{bean}{42}
\setlength{\rightmargin}{\leftmargin}}
\item    {\bf rjjmin} real*8\\
           minimum jet-jet  separation 
           $\Delta R_{jj}=\sqrt{(\Delta\eta)^2+(\Delta\phi)^2}$,
           where the cone variables are the pseudo-rapidity $\eta$
           and the azimuthal angle $\phi$.
           \label{irjj}
\item    {\bf rjlmin} real*8\\
           minimum jet-lepton  separation $\Delta R_{jl}$.
           \label{irjl}
\end{list}
{\bf CUTS IN THE W-SCHEME (JADE);\\ ONLY EFFECTIVE IF JSCHEME=11,12,13,14}
\begin{list}
{\arabic{bean})}
{\usecounter{bean}
\setcounter{bean}{44}
\setlength{\rightmargin}{\leftmargin}}
\item    {\bf ycut}: real*8\\
          resolution parameter 
          $y_{cut}$ for minimal dijet mass$^2$ (including the remnant)
          \label{iycutw}
\item    {\bf m2min}: real*8 [GeV$^2$]\\
          fixed minimal dijet mass$^2$ (including the  remnant)     
         \label{im2min}
\end{list}
{\bf CUTS IN THE KT-SCHEME;\\ ONLY EFFECTIVE IF JSCHEME=22,23}
\begin{list}
{\arabic{bean})}
{\usecounter{bean}
\setcounter{bean}{46}
\setlength{\rightmargin}{\leftmargin}}
\item    {\bf ycut\_kt}:  real*8\\
           resolution parameter $y_{cut}$ as defined in the $k_T$-scheme 
          \label{iycutkt}
\item    {\bf et2}: real*8 [GeV$^2$]\\
           hard scattering scale squared as defined in the $k_T$-scheme.
           The hard scattering scale squared is set equal to 
           $-\mbox{et2}\,{Q^2}=|\mbox{et2}|\,{Q^2}$ 
           for negative et2.
           \label{iet2}
\item    {\bf m2minkt}:   real*8 [GeV$^2$]\\
           fixed minimal dijet mass squared (Breit frame) in $k_T$-scheme
          \label{im2minkt}
\end{list}
{\bf CUTS FOR THRUST DISTRIBUTIONS;\\ Not supported in version 2.0 and below}
\begin{list}
{\arabic{bean})}
{\usecounter{bean}
\setcounter{bean}{49}
\setlength{\rightmargin}{\leftmargin}}
\item    {\bf thrmin\_def}:  real*8\\
           minimum value for Thrust
\item    {\bf thrmax\_def}:  real*8\\
           maximum value for Thrust
\end{list}
{\bf INVARIANT CUTS ON THE JETS AFTER CLUSTERING}
\begin{list}
{\arabic{bean})}
{\usecounter{bean}
\setcounter{bean}{51}
\setlength{\rightmargin}{\leftmargin}}
\item    {\bf zpmin}: real*8  $\in [0,1]$\\
           minimum value of zp(i) scaling variable;\\
           Note, that for jrec=2, zp is no longer in [0,1].
\item    {\bf zpmax}: real*8  $\in [0,1]$\\
           maximum value of zp(i) scaling variable.
           Note, that for jrec=2, zp is no longer in [0,1].
\item    {\bf sjjmin\_def}: real*8  \\
           minimum jet-jet invariant mass squared.
\item    {\bf sjjmax\_def}: real*8  \\
           maximum jet-jet invariant mass squared.
\end{list}
{\bf LAB FRAME CUTS  ON THE JETS (AFTER CLUSTERING) AND PARTONS}
\begin{list}
{\arabic{bean})}
{\usecounter{bean}
\setcounter{bean}{55}
\setlength{\rightmargin}{\leftmargin}}
\item    {\bf ptmin\_def}: real*8 [GeV]\\
        minimum transverse momentum $p_T$ to define a jet.\\
        ptmin\_def must be larger than zero.
         \label{iptmindef}
\item    {\bf ymin\_def}:  real*8 \\
        minimum pseudo-rapidity  $\eta=-\ln\tan(\theta_j/2)$ on the jets.
\item    {\bf ymax\_def}:  real*8 \\
        maximum pseudo-rapidity  on the jets.
\item    {\bf ypartmax\_def}: real*8\\
         maximum pseudo-rapidity 
         on the partons before entering the clustering.
         The momentum of the parton is set to zero, if it is larger than
         ypartmax\_def.
         \label{iypartmax}
\end{list}
\enlargethispage{1cm}
{\bf BREIT FRAME CUTS  ON THE JETS (AFTER CLUSTERING)}
\begin{list}
{\arabic{bean})}
{\usecounter{bean}
\setcounter{bean}{59}
\setlength{\rightmargin}{\leftmargin}}
\item    {\bf ptminb\_def}:  real*8 [GeV]\\
         minimum transverse momentum  $p_T$  in the Breit frame 
\item    {\bf ptmaxb\_def}:  real*8 [GeV]\\
         maximum transverse momentum  $p_T$  in the Breit frame 
\item    {\bf yminb\_def}: real*8\\
          minimum jet  pseudo-rapidity in the Breit frame 
\item    {\bf ymaxb\_def}: real*8\\
          maximum jet pseudo-rapidity in the Breit frame 
\item    {\bf ktminb\_def}:  real*8 [GeV]\\
         minimum  $k_T(j)$  in the Breit frame 
\item    {\bf ktmaxb\_def}:  real*8 [GeV]\\
         maximum  $k_T(j)$  in the Breit frame 
\item    {\bf sum\_ktminb\_def}:  real*8 [GeV]\\
         minimum sum  $k_T(jet)$  in the Breit frame 
\item    {\bf sum\_ktmaxb\_def}:  real*8 [GeV]\\
         maximum sum  $k_T(jet)$  in the Breit frame 
\end{list}
{\bf FLAG FOR FORWARD JET PRODUCTION}
\begin{list}
{\arabic{bean})}
{\usecounter{bean}
\setcounter{bean}{67}
\setlength{\rightmargin}{\leftmargin}}
\item    {\bf  iforward}: integer\\
         0: no forward jet\\
         1: require one forward jet. The kinematical cuts
           imposed on this jet are read in from the file
           forward.dat (see below)
         \label{ifw}
\end{list}
\vspace{3mm}
{\bf CUTS FOR FORWARD JET PRODUCTION, IF IFORWARD=1}
\begin{itemize}
\item[a)]   {\bf xfj\_min}: real*8 [GeV]\\
          minimum long. momentum  fraction $p_z(j)/E_p$ of the forward jet
\item[b)]   {\bf efj\_min}: real*8 [GeV]\\
          minimum energy  fraction $E(j)/E_p$ of the forward jet
\item[c)] {\bf ymin\_fj\_def}: real*8 \\
          min. rapidity of the forward jet
\item[d)] {\bf ymax\_fj\_def}: real*8 \\
          max. rapidity of the forward jet
\item[e)] {\bf emin\_fj\_def}: real*8 [GeV] \\
           min. energy of the forwad jet
\item[f)] {\bf ptmin\_fj\_def }: real*8 [GeV] \\
           min. pt (lab) of the  forward jet
\item[g)] {\bf pt2\_ratio\_fj\_min  }: real*8 \\
           min. ratio for $p_T^2$(lab)/Q2 in the lab.
\item[h)] {\bf pt2\_ratio\_fj\_max  }: real*8 \\
           max. ratio for $p_T^2$(lab)/Q2 in the lab.

\end{itemize}
\enlargethispage{1cm}
The program is  available 
via anonymous ftp:
\begin{verbatim}
ftp ttpux2.physik.uni-karlsruhe.de
username:  anonymous
password:  (email address)
cd mirkes
get mepjet2_1.tar.gz
\end{verbatim}


\def\ap#1#2#3   {{\em Ann. Phys. (NY)} {\bf#1} (#2) #3}
\def\apj#1#2#3  {{\em Astrophys. J.} {\bf#1} (#2) #3}
\def\apjl#1#2#3 {{\em Astrophys. J. Lett.} {\bf#1} (#2) #3}
\def\app#1#2#3  {{\em Acta. Phys. Pol.} {\bf#1} (#2) #3}
\def\ar#1#2#3   {{\em Ann. Rev. Nucl. Part. Sci.} {\bf#1} (#2) #3}
\def\cpc#1#2#3  {{\em Computer Phys. Comm.} {\bf#1} (#2) #3}
\def\err#1#2#3  {{\it Erratum} {\bf#1} (#2) #3}
\def\ib#1#2#3   {{\it ibid.} {\bf#1} (#2) #3}
\def\jmp#1#2#3  {{\em J. Math. Phys.} {\bf#1} (#2) #3}
\def\ijmp#1#2#3 {{\em Int. J. Mod. Phys.} {\bf#1} (#2) #3}
\def\jetp#1#2#3 {{\em JETP Lett.} {\bf#1} (#2) #3}
\def\jpg#1#2#3  {{\em J. Phys. G.} {\bf#1} (#2) #3}
\def\mpl#1#2#3  {{\em Mod. Phys. Lett.} {\bf#1} (#2) #3}
\def\nat#1#2#3  {{\em Nature (London)} {\bf#1} (#2) #3}
\def\nc#1#2#3   {{\em Nuovo Cim.} {\bf#1} (#2) #3}
\def\nim#1#2#3  {{\em Nucl. Instr. Meth.} {\bf#1} (#2) #3}
\def\np#1#2#3   {{\em Nucl. Phys.} {\bf#1} (#2) #3}
\def\pcps#1#2#3 {{\em Proc. Cam. Phil. Soc.} {\bf#1} (#2) #3}
\def\pl#1#2#3   {{\em Phys. Lett.} {\bf#1} (#2) #3}
\def\prep#1#2#3 {{\em Phys. Rep.} {\bf#1} (#2) #3}
\def\prev#1#2#3 {{\em Phys. Rev.} {\bf#1} (#2) #3}
\def\prl#1#2#3  {{\em Phys. Rev. Lett.} {\bf#1} (#2) #3}
\def\prs#1#2#3  {{\em Proc. Roy. Soc.} {\bf#1} (#2) #3}
\def\ptp#1#2#3  {{\em Prog. Th. Phys.} {\bf#1} (#2) #3}
\def\ps#1#2#3   {{\em Physica Scripta} {\bf#1} (#2) #3}
\def\rmp#1#2#3  {{\em Rev. Mod. Phys.} {\bf#1} (#2) #3}
\def\rpp#1#2#3  {{\em Rep. Prog. Phys.} {\bf#1} (#2) #3}
\def\sjnp#1#2#3 {{\em Sov. J. Nucl. Phys.} {\bf#1} (#2) #3}
\def\spj#1#2#3  {{\em Sov. Phys. JEPT} {\bf#1} (#2) #3}
\def\spu#1#2#3  {{\em Sov. Phys.-Usp.} {\bf#1} (#2) #3}
\def\zp#1#2#3   {{\em Zeit. Phys.} {\bf#1} (#2) #3}
\def\thebibliography#1{\list
{[\arabic{enumi}]}{\settowidth\labelwidth{[#1]}\leftmargin\labelwidth
\advance\leftmargin\labelsep
\usecounter{enumi}}
\sloppy
\sfcode`\.=1000\relax}
\let\endthebibliography=\endlist

\end{document}